\documentstyle[12pt]{report}
\begin{document}
\def\om{\omega}
\def\omt{\tilde{\omega}}
\def\ti{\tilde}
\def\o{\Omega}
\def\Lm{\Lambda}
\def\bchi{\bar\chi^i}
\def\In{{\rm Int}}
\def\ba{\bar a}
\def\w{\wedge}
\def\ep{\epsilon}
\def\k{\kappa}
\def\Tr{{\rm Tr}}
\def\tr{{\rm tr}}
\def\ST{{\rm STr}}
\def\ss{\subset}
\def\ot{\otimes}
\def\bc{{\bf C}}
\def\br{{\bf R}}
\def\bt{{\bf T}}
\def\bz{{\bf Z}}
\def\bn{{\bf N}}
\def\de{\delta}
\def\mk{M_\k(\hat G)}
\def\mo{M_0(\hat G)}
\def\al{\alpha}
\def\la{\langle}
\def\ra{\rangle}
\def\G{{\cal G}}
\def\th{\theta}
\def\lm{\lambda}
\def\Th{\Theta}
\def\tt{\times_\Th}
\def\U{\Upsilon}
\def\e{\varepsilon}
\def\ve{\varepsilon}
\def\fe{{1\over \e}(1-{\rm e}^{-\e X^3})}
\def\ex{\js\e\bar XX}
\def\bq{\}_{qWZ}}
\def\thg{T^*\hat G}
\def\jp{{1\over 2}}
\def\tf{\hat t_{\infty}}
\def\tF{\hat T^{\infty}}
\def\ttf{\tilde t_{\infty}}
\def\ttF{\tilde T^{\infty}}
\def\zg{)_\G}
\def\hdg{\widehat{DG_0}}
\def\hlg{\widehat{LG_0}}
\def\lgc{LG_0^\bc}
\def\dgc{DG_0^\bc}
\def\ppp{{1\over 2\pi}}
\def\spp{{1\over 4\pi}}
\def\js{{1\over 4}}
\def\sp{{1\over 4\pi^2}}
\def\d{\partial}
\def\dz{\partial_-}
\def\dbz{\partial_+}
\def\LWD{Lu-Weinstein-Soibelman double}
\def\Dz{\partial_z}
\def\Dbz{\partial_{\bar z}}
\def\dto{\biggl({d\over ds}\biggr)_{s=0}}
\def\be{\begin{equation}}
\def\ee{\end{equation}}
\def\bea{\begin{eqnarray}}
\def\eea{\end{eqnarray}}
\def\D{{\cal D}}
\def\G{{\cal G}}
\def\H{{\cal H}}
\def\B{{\cal B}}
\def\E{{\cal E}}
\def\C{{\cal C}}
\def\A{{\cal A}}
\def\P{{\cal P}}
\def\R{{\cal R}}
\def\L{{\cal L}}
\def\wti{\widetilde}
\def\wh{\widehat}
\def\ttD{\tilde{\tilde D}}
\def\hhD{\hat{\hat D}}
\def\ttK{\tilde{\tilde K}}
\def\hhK{\hat{\hat K}}
\def\bbK{\bar{\bar K}}
\def\ttd{\tilde{\tilde \D}}
\def\hhd{\hat{\hat \D}}
\def\ttx{\tilde{\tilde x}}
\def\hhx{\hat{\hat x}}
\def\tty{\tilde{\tilde y}}
\def\hhy{\hat{\hat y}}
\def\tta{\tilde{\tilde \al}}
\def\hha{\hat{\hat \al}}
\def\ttb{\tilde{\tilde \b}}
\def\hhb{\hat{\hat \b}}
\def\ttxi{\tilde{\tilde \xi}}
\def\hhxi{\hat{\hat \xi}}
\def\ttyi{\tilde{\tilde \eta}}
\def\hhyi{\hat{\hat \eta}}
\def\hal{\hat\alpha}
\def\bl{\bar l}
\def\ptp{\stackrel{\otimes}{,}}
\def\hG{\hat {\cal G}}
\def\T{{\cal T}}
\def\F{{\cal F}}
\def\n{{1\over n}}
\def\dg{\dagger}
\def\si{\sigma}
\def\Si{\Sigma}
\def\RK{R^*_{P_K^{-1}}}
\def\l+{L_+G_0^\bc}
\def\+{D_+G_0^\bc}
\def\hdgc{\widehat{DG_0^\bc}}
\def\rn{\vert \alpha\vert^2}
\def\hlgc{\widehat{LG_0^\bc}}
\def\nl{{\nabla^L}}
\def\nr{{\nabla^R}}
\def\ot{\otimes}
\def\b{\beta}
\def\vr{\varrho}
\def\ga{\gamma}
\def\gc{G^{\bf C}}
\def\Gc{\G^{\bf C}}
\def\lw{D_{LWS}}
\def\LW{\D_{lws}}

\def\rdgc{\ ^\br\widehat {DG_0^\bc}}
\def\cdgc{\ ^\bc\widehat {DG_0^\bc}}
\def\rlgc{\ ^\br\widehat {LG_0^\bc}}
\def\clgc{\ ^\bc\widehat {LG_0^\bc}}

\def\st{\stackrel}
\def\stw{\stackrel{\w}{,}}
\def\0{^{01}}
\def\1{^{10}}
\def\od{\sqrt{2}}
\def\aq{[\al(\phi)]_\e}
\def\hbl{\hat\beta_L}
\def\hbr{\hat\beta_R}
\def\hmu{\hat\mu}
\def\bi{\bibitem}
\def\teo{\noindent {\bf Theorem}}
\def\defi{\noindent {\bf Definition}}
\def\rem{\noindent {\bf Remark}}
\def\pro{\noindent {\bf Proof}}
\def\lem{\noindent {\bf Lemma}}
\def\saf{{\e\over 2}\vert \hat a\vert^2\la i\hal^\vee,\hat \phi\ra}
\begin{titlepage}
\begin{flushright}
{}~
IML 01-08\\
hep-th/0103118
\end{flushright}

\vspace{1cm}
\begin{center}
{\Huge \bf Quasitriangular WZW model}\\
[50pt]{\small
{\bf Ctirad Klim\v{c}\'{\i}k}
\\ ~~\\Institute de math\'ematiques de Luminy,
 \\163, Avenue de Luminy, 13288 Marseille, France}
\end{center}

\vspace{0.5 cm}
\centerline{\bf Abstract}
\vspace{0.5 cm}
A   dynamical system is canonically associated to every Drinfeld double
of any
 affine Kac-Moody group. The choice of the affine Lu-Weinstein-Soibelman
 double  gives
 a smooth one-parameter deformation of the standard
WZW model. In particular, the worldsheet
and the target of the classical version of the  deformed   theory  are the
 ordinary smooth
 manifolds.
The quasitriangular WZW model is exactly solvable and  it admits the chiral
 decomposition.
Its  classical action   is not invariant with respect to
the left and right action of the loop group, however it satisfies
the weaker condition of
the Poisson-Lie symmetry.
The  structure  of the deformed WZW model is characterized by several
 ordinary and  dynamical r-matrices with spectral parameter. They
describe the
 $q$-deformed current
algebras, they enter the definition of $q$-primary fields and
they characterize the
quasitriangular exchange (braiding)  relations. Remarkably,
the symplectic structure
of the deformed chiral WZW theory is cocharacterized by the same elliptic
   dynamical  $r$-matrix that appears   in the Bernard generalization of
the Knizhnik-Zamolodchikov  equation, with $q$ entering the modular
parameter
of the Jacobi theta functions. This reveals a tantalizing connection
between the classical $q$-deformed WZW model and the quantum standard
 WZW theory
on elliptic curves and opens the way for the systematic use of
 the dynamical  Hopf
algebroids in the rational  $q$-conformal field theory.
\end{titlepage}
\tableofcontents

\chapter{Introduction}
\noindent {\bf 1. Basic observation.} The WZW model \cite{Wi} is
 certainly one
of the most important models
of the two-dimensional (conformal) field theory. It is well-known
that many
 interesting
theories can be naturally obtained by its reductions, like e.g.
the coset
models \cite{GKO} or Toda theories \cite{FW}. Such fabrication of
 new structures from the
roof model was called "the WZW factory" in \cite{GawP}.

This paper is based on an   observation, that the roof of the WZW
 factory
is in fact two floors above the WZW model. In other words, we shall
argue that there is a master model from which the WZW model can be
 obtained
by two successive symplectic reductions. This master  model
 describes
the geodesical flow on the affine Kac-Moody group
 $\ti G=\br\times_{\hat S} \hat G$
 and its action reads
\be S(\ti g)=-{\k\over 4}\int d\tau
 \biggl(\ti g^{-1}{d\over d\tau}{\ti g},
\ti g^{-1}{d\over d\tau}{\ti g}\biggr)_{\ti\G}. \ee
Here  $\ti g(\tau)\in\ti G$, $\k$ is going to play the role
of the WZW level,
$(.,.)_{\ti\G}$ is the invariant inner product on
$\ti\G=Lie(\ti G)$, $\hat G=\widehat {LG_0}$ is the centrally
extended loop group   and
 $\times_{\hat S}$ means the semidirect product corresponding to
the loop group parameter shift automorphism $\si\to\si +s$.

  The reader may be surprised that
neither the world-sheet space derivative $\d_\si$ nor
the WZW term are present in the master action (1.1). We shall see,
 however,
that they are "born" under the process of the symplectic reduction.

The interest in lifting the
WZW model relies on the fact that the master model sitting two
 floors higher
has  extremely simple structure. Its phase space is the cotangent
 bundle
$T^*\ti G$ equipped with its canonical symplectic structure. We can
therefore easily construct deformations of the master model by using
the theory of various doubles (Manin, Drinfeld, Heisenberg)
of the group
$\ti G$. One simply replaces the cotangent bundle $T^*\ti G$ by
a chosen Drinfeld double $\ttD$ and the symplectic structure is then
canonically
given by the Semenov-Tian-Shansky two-form $\ti\omega$ on $\ttD$.
The left-right $\ti G$-symmetries of the master model (1.1) get thus
 deformed
to Poisson-Lie symmetries, and the  Hamiltonian charges
 (=Abelian moment maps
generating the standard symmetries) become non-Abelian
Poisson-Lie moment maps.
We then obtain the quasitriangular WZW theory
by performing the two step symplectic reduction of
the deformed master model.
\vskip1pc
\rem :
  {\small It is also interesting to note
 that the various symplectic reductions   can be applied also
in the Poisson-Lie case, or,
in other words, the whole WZW factory should survive  the
$q$-deformation.}

\vskip1pc
\noindent {\bf 2. Chiral decomposition.} It is well-known
that the standard
WZW model can be obtained by the appropriate glueing of two
identical copies
of a simpler dynamical system called the chiral WZW model \cite{FG}.
 The same
thing turns out to be  true also for the master model (1.1).
It can be glued up from two copies
of the chiral
geodesical model whose  (first order Hamiltonian) action
 is given by
\be \ti S_L(\ti k,\ti \phi)=\int d\tau
 [\la  \ti \phi,\ti k^{-1}{d\over d\tau}\ti k\ra +
{1\over 2\k}(\ti \phi,\ti \phi)_{\ti\G^*}],\ee
where  $\ti k(\tau)\in\ti G$ and $\ti \phi(\tau)\in\ti \A_+$.
Note that
   $\ti \A_+$ is the Weyl
alcove viewed as the subset of the dual of the Cartan subalgebra
 $\ti\T$ of $\ti\G$ .

We shall show that
   the  chiral WZW model   can
be also  obtained by a simple   two-step symplectic reduction from
the chiral master model
(1.2). In fact, the $\si$-shift   and the central
circle subgroups of $\ti G$ act in the standard
Hamiltonian way on the phase space
$\ti M_L\equiv\ti G\times \ti \A_+$. The
  reduction
is then induced by setting the $\si$-shift Hamiltonian  charge
 to $0$ and its central  circle fellow to $\k$.

It turns out that the master model (1.1) and the chiral geodesical
 model (1.2) have   natural deformations
based on the choice of an appropriate Drinfeld
 double $\ttD$ of the affine Kac-Moody group $\ti G$.
In particular,
 the  resulting deformed chiral geodesical model
is formulated on the same phase space $\ti M_L$  but now its action reads
\be \ti S_L^q(\ti k,\ti \phi)=\int
 [\theta^q +
{1\over 2\k}(\ti \phi,\ti \phi)_{\ti\G^*} d\tau].\ee
In order to explain the notation:
There exists a one-parameter family of embeddings of the  affine model
 space $\ti M_L$ into
the Drinfeld double $\ttD$. This parameter will
be  referred to as $q\equiv e^\e$. Then
$\theta^q$ is the solution\footnote{The classical action (1.3)
makes sense even if
this solution $\theta^q$ exists only locally on $\ti M_L$.}
 of the equation $d\theta^q=\Omega^q$, where $\Omega^q$
is the pull-back of the  Semenov-Tian-Shansky form  to the
$q$-embedded submanifold
$\ti M_L\hookrightarrow \ttD$.

There is the  crucial condition to be imposed
on $\ttD$, namely, the $\si$-shift   and the central
circle subgroups of $\ti G$ must still act  on $\ti M_L$ in the
standard
Hamiltonian way but now with respect
to the symplectic structure $\Omega^q$. Such good doubles
will be referred to as the
WZW doubles of $\ti G$. As in the non-deformed case,
the quasitriangular chiral WZW model will be  then  obtained
from the action (1.3)
by the $\k$-depending symplectic reduction based on setting
the corresponding $\si$-shift and central circle
Hamiltonian charges to $0$ and $\k$, respectively.

  Although the $\si$-shift and the central circle still act in
the standard Hamiltonian
way on $(\ti M_L,\Omega^q)$, this is non longer true for the
action of the remaining
loop group generators of $Lie(\ti G)$. Nevertheless, due
to the fact that $\Omega^q$
is the pull-back of the Semenov-Tian-Shansky form, the
remaining generators
act in the Poisson-Lie way. This means, in particular,
that the action (1.3) of
the deformed chiral geodesical  model is  Kac-Moody
Poisson-Lie symmetric.
Due to this property, the quantized quasitriangular
chiral WZW model  will enjoy
the $q$-Kac-Moody symmetry.
\vskip1pc
\noindent {\bf Remarks} : {\small i)  The deformation of
the master model (1)
exists for every Drinfeld double of $\ti G$. However, the
two-step symplectic
reduction can be performed only if    $\ttD$ is  the WZW
double  (see above).
We found with satisfaction that a nontrivial WZW double
$\ttD$ of $\ti G$
can be indeed constructed; it is in fact
nothing but the complexification $\ttD=\ti G^\bc$ of the
affine Kac-Moody group $\ti G$
equipped with certain invariant maximally noncompact inner
product on $\ttd=Lie(\ttD)$.
We shall   call $\ttD$ the "affine Lu-Weinstein-Soibelman double",
 since
it turns out to be the natural  affine generalization of the
standard Lu-Weinstein-Soibelman
double $D_0$ of the group $G_0$.

~~~~~~~ii)
The   Weyl alcove $\A_+$ is usualy viewed as the fundamental domain
of the action of the affine Weyl group  on the
Cartan subalgebra $\T$ of $\G_0\equiv Lie(G_0)$.   The affine
Weyl group
acts also on the Cartan subalgebra $\ti\T$ of
$\ti\G\equiv Lie(\ti G)$. The alcove $\ti \A_+$
is again the  fundamental domain of this action.
  It is also important
to note that the  second floor chiral  Hamiltonian
\be \ti H_L=-{1\over 2\k}(\ti\phi,\ti\phi)_{\ti\G^*}\ee
 does not depend on $q$. The
$q$-dependence of its descendant $H_L^{qWZ}$ will turn out to be the
  fruit of the reduction.
 The   reader will find more detailed
explanations of all that in the body of the paper.

 ~~~~~~~~~~~iii)
 The  $q$-deformations
of the  WZW model have been already studied in the
literature \cite{FG, BS,QWZW}.
However, in  those cases either the worldsheet or the
target of the $\sigma$-model was first {\it kinematically}
 deformed to become a lattice
or  a non-commutative manifold. Then a kind of a (discrete or
non-commutative)  dynamics was formulated
on this deformed background.
What we are doing  here is somewhat different;  we avoid any
 preliminary
kinematical deformation of the worldsheet or of the target.
The $q$-deformed
objects   are generated {\it dynamically}. This means that
they appear solely
as the result of  standard field theoretical quantization of
some (chiral) classical
theory  whose phase  space   $M_L^{WZ}$is topologically
{\it the same} as that
of the non-deformed chiral WZW theory. The  things that
get (smoothly) deformed are the symplectic
form  and the Hamiltonian function on this unchanged phase space. }

\vskip1pc
\noindent {\bf 3. Quasitriangular symplectic structure.} As already
stated in the remark iii) above,
the crucial result of the two-step symplectic reduction of (1.3)
is the fact that
the phase space of the quasitriangular chiral WZW model is
topologically {\it the same}
manifold as the phase space $M_L^{WZ}$ of the non-deformed
standard chiral WZW theory.
 Recall that points in $M_L^{WZ}$ are the maps $m:\br\to G_0$,
fulfilling the
monodromy condition
\be m(\si+2\pi)=m(\si)M.\ee
Here the monodromy\footnote{Sometimes people consider
\cite{Gaw,FG,Mad} the
bigger  chiral WZW phase
space in the sense that $M$ can be an arbitrary element of $G_0$.
 Such an enlargement
is useful for description of the (finite dimensional)
quantum group symmetries
of the standard   WZW model, however, it
is not necessary for recovering the full left-right WZW model by
appropriate glueing of two chiral models. The choice of the
maximal torus monodromy
is sufficient to do this job and we stick on it for
the rest of this paper.}
 $M=\exp{(-2\pi i a^\mu H^\mu)}$ sits in the
 fundamental Weyl alcove viewed as the subset
of the maximal torus $\bt$ of $G_0$. In what follows,
$a^\mu$ will be    coordinates on the alcove $\A_+$
corresponding to the choice of the orthonormal basis $H^\mu$ on
the Cartan subalgebra.

Although the phase space $M_L^{WZ}$ is the
same, the symplectic structure $\om_L^{qWZ}$   and the
 Hamiltonian $H_L^{qWZ}$
differ, however,  from their non-deformed WZW counterparts.
One of the  main results of this paper is
 the explicite description
of  the pair
$(\omega_L^{qWZ},H_L^{qWZ})$.
 Thus the symplectic structure corresponding to the two-form
$\om_L^{qWZ}$ is  fully characterized by the following
  Poisson bracket
\be  \{m(\si)\ptp m(\si') \bq =(m(\si)\otimes m(\si'))
 B_{\e}(a^\mu,\si-\si')+
 \e\hat r(\si-\si')(m(\si)\otimes m(\si')),\ee
where $B_{\e}(a^\mu, \si)$ is the so called quasitriangular braiding matrix
given by
 $$ B_{\e}(a^\mu,\si)=$$
\be =-{i\over \k}\rho({i\si\over 2\k\e},{i\pi\over \k\e})
H^\mu\otimes H^\mu
-{i\over \k}\sum_{\al\in\Phi}{\rn\over 2}\si_{ a^\mu\la
\al,H^\mu\ra}
( {i\si\over 2\k\e},{i\pi\over \k\e})E^\al\otimes E^{-\al} \ee
and $\hat r(\si)$  is defined as
\be  \hat r(\si)=
 r + C {\rm cotg}\jp \si . \ee
Here $r$ and $C$ are the ordinary (non-affine)  $r$-matrix and
Casimir elements, respectively,
given by
\be r=\sum_{\al\in\Phi_+}{i\vert \al \vert^2\over 2}(E^{-\al}
\otimes E^{\al}-E^\al\otimes E^{-\al});\ee
\be C=\sum_\mu H^\mu\otimes H^\mu+
\sum_{\al\in\Phi_+}{\vert \al \vert^2\over 2}(E^{-\al}
\otimes E^{\al}+E^\al\otimes E^{-\al}).\ee
The functions $\rho(z,\tau),\si_w(z,\tau)$ are defined as
 (cf. \cite{F,FW,EV})
\be \si_w(z,\tau)={\theta_1(w-z,\tau)\theta_1'(0,\tau)\over
\theta_1(w,\tau)\theta_1(z,\tau)},
\quad \rho(z,\tau)={\theta_1'(z,\tau)\over \theta_1(z,\tau)},\ee
where $\theta_1(z,\tau)$ is the Jacobi theta function
\be \theta_1(z,\tau)=-\sum_{j=-\infty}^{\infty}
e^{\pi i (j+\jp)^2\tau +2\pi i(j+\jp)(z+\jp)}.\ee
In (1.11), the apostrof ' means the derivative with respect
to the first argument $z$,
the argument $\tau$ (the modular parameter) is a nonzero
complex number such that Im$\tau>0$.

 \vskip1pc
\noindent {\bf Remark} :
{\small In Sections 5.1 and 5.2, we describe in detail the hard work
needed to arrive from the
 Semenov-Tian-shansky
form on the double $\ttD$ to the $q$WZW chiral Poisson bracket
$\{.,.\bq$. Inspite of the
intermediate complicated calculations, the resulting formula
(1.14) is simple and esthetically
appealing. Tantalizingly,  $B_\e(a^\mu,\si)$
 is the Felder-Wieczerkowski
   elliptic dynamical $r$-matrix that appears in the
Bernard generalization
of the Knizhnik-Zamolodchikov equation \cite{B,F,FW}.
 Apart from some deeper
sense lurking at us from this fact, there is yet another
 profitable circumstance:
the  corresponding quantum dynamical $R$-matrix
is known \cite{F}. No doubt,
 an important part of the structure of
a $q$-deformed conformal field theory will be then
 underlied by the concept of the
dynamical Hopf algebroid corresponding to $R$.}
 \vskip1pc

\noindent In the limit $\e\to 0$, Eq. (1.6) becomes
\be \{m(\si)\ptp m(\si') \}_{(q=1)WZ} =(m(\si)\otimes m(\si'))
 B_{0}(a^\mu,\si-\si') ,\ee
 where
\be B_0(a^\mu,\si)=
-{\pi\over\kappa}\biggl[\eta(\si)(H^\mu\ot H^\mu)-
i\sum_{\al\in\Phi}{\rn\over 2}{\exp{(i
\pi \eta(\si)\la \al,H^\mu\ra a^\mu)}\over
\sin{(\pi \la \al,H^\mu\ra a^\mu)}}
E^\al\ot E^{-\al}\biggr].\ee
Here  $\eta(\si)$ is the function defined by
\be \eta(\si)=2[{\si\over2\pi}]+1,\ee
where $[\si/2\pi]$ is the largest integer less than or equal
to ${\si\over 2\pi}$.
It is shown in Section 3.2, that the relation (1.13)
completely characterize
the symplectic
structure $\om_L^{WZ}$ of the standard non-deformed chiral WZW model.

 \vskip1pc

\noindent {\bf 4. Quasitriangular Hamiltonian.}
The Hamiltonian $H_L^{qWZ}$ descends (upon the reduction)
 from the second
floor master Hamiltonian (1.4).   We want to make
 explicit how $H_L^{qWZ}$ is defined as the
function on the phase space
$M_L^{WZ}$. For this purpose, it is more convenient to perform
the classical (inverse) vertex-IRF transformation defined by
\be k(\si)=m(\si)\exp{(ia^\mu H^\mu\si)}.\ee
Note that $k(\si)$ then becomes periodic, hence an element of the
loop group $G=LG_0$. Therefore, topologically, $M_L^{WZ}=LG_0\times \A_+$.

We shall first start with the case
$q=1$ that gives the standard chiral WZW Hamiltonian.
 Some basic knowledge
of the Poisson-Lie world is needed for  understanding the case
$q\neq 1$. The  interested reader may  consult Sections 4.1 where the
relevant Poisson-Lie  notions are explained.

The  standard chiral WZW Hamiltonian $H_L^{WZ}$ is usually
written in the monodromic
variables $m(\si)$ and it is given by the Sugawara formula
\be H_L^{WZ}(m)=
-{1\over 2\k}(\k\d_\si mm^{-1},\k\d_\si mm^{-1})_{\G^0}.\ee
The minus sign appears because in our conventions the form
$(.,.)_{\G_0}$
is negative definite.
In the variables $(k,a^\mu)$, it becomes
\be H_L^{WZ}(k,a^\mu)=-{1\over 2\k}(\phi,\phi)_{\G^*}-\la \phi,
k^{-1}\d_\si k\ra -
{\k\over 2}(k^{-1}\d_\si k,k^{-1}\d_\si k)_\G,\ee
where $(\hat\phi_\k)'=\phi$ (see   the meaning of this
notation in a while)
and $(.,.)_\G$ is the invariant scalar product on the
loop group Lie algebra $\G=L\G_0$.
As always in the paper $\la .,.\ra$ means the canonical
pairing between the elements
of mutually dual spaces.
The formula (1.18) can be rewritten in the following way
\be H_L^{WZ}(k,a^\mu)=
-{1\over 2\k}(\phi,\phi)_{\G^*}- ( \wti{Coad}_{\hat k}\hat\phi_\k)^0 .\ee
In turns out that   for generic $q$,   (1.19) generalizes to
\be H_L^{qWZ}(k,a^\mu)=-{1\over 2\k}(\phi,\phi)_{\G^*} -
  (\wti{Dres}_{\hat k}e^{\ti\Lm(\hat\phi_\k)})^0.\ee
Recall that $\hat\phi_\k$ is the function of $a^\mu$ hence the
Hamiltonians really
depend on the indicated variables.
\vskip1pc
\noindent {\bf Notations 1.1:}
\noindent i) Denote $\ti T^0\in \ti\G$ and $\ti T^\infty\in \ti\G$
the generators of
the $\si$-shift and
of the central circle, respectively. Then we have the linear space
 decomposition
$\ti\G=\br \ti T^0+\br \ti T^\infty+\G$ and its dual $\ti\G^*=
\br\ti t_0+\br \ti t_\infty
+\G^*$. Now we define
\be \ti x=\ti x^0\ti t_0+ \ti x^\infty \ti t_\infty +\ti x',
\quad \ti x\in\ti\G^*, \ti x'\in\G^*.\ee
In words: $\ti x^0$ is the $\si$-shift part of $\ti x$, $\ti x^\infty$
the central circle part
and $\ti x'$ the $\G^*$-part.
 The lifted alcove $\hat\phi_\k$ is then characterized by the relations :
\be (\hat\phi_\k)^0=0,\quad (\hat\phi_\k)^\infty=\k,\quad
(\hat\phi_\k)'=\phi. \ee
\vskip1pc
\noindent ii) Consider $\ti\B=Lie(\ti B)$ and $\B=Lie(B)$,
 where $\ti B,B$
are respectively the dual\footnote{The
existence and the properties of $\ti B,B$ are implied  by the
structure of the Drinfeld double
$\ttD$.} Poisson-Lie
groups of $\ti G,G$. There is the following unique
 decomposition of any element $\ti b\in\ti B$:
\be \ti b=\exp{( \ti b^0\ti\Lm(\ti t_0))}
\exp{(\ti b^\infty\ti\Lm(\ti t_\infty))}\ti b',
\quad \ti b\in\ti B,\ti b'\in B,\ee
where   $\ti b'$ is the Poisson-Lie analogue of $\ti x'$ and
the real numbers $\ti b^0,\ti b^\infty$
are the  analogues of $\ti x^0,\ti x^\infty$.
The map $\ti\Lm:\ti\G^*\to\ti\B$
is the identification map defined by the invariant bilinear
form $(.,.)_{\ttd}$
on the Drinfeld double $\ttD$ of $\ti G$. This form can be
 arbitrarily normalized. This
normalization parameter is actually the deformation parameter
 of our WZW story. We call
it either $\e$ or $q$, with $q=e^\e$. Note that $\ti\Lm$
then depends on $q$ which
gives the $q$-dependence of the quasitriangular Hamiltonian (1.20).

\vskip1pc
\noindent iii)  $\wti{Coad}$ means the $\ti G$-coadjoint
action on   $\hat\phi_\k$ viewed as the element of   $\ti\G^*$
and $\wti{Dres}$ means the $\ti G$-dressing action on
 $e^{\ti\Lm(\hat\phi_\k)}$
viewed as the element of $\ti B$.
\vskip1pc
\noindent iv)     $\hat G=\wh{LG_0}$ is the
principal $U(1)$-bundle over $G=LG_0$ with the projection $\pi$.
Then $\hat k$ is any
element of $\hat G$ such that $\pi(\hat k)=k$. Since
$\ti G=\br\times_{\hat S}\hat G$,
we can view $\hat k$ also as the element of $\ti G$
 and it is in this sense that
$\hat k$ appears in (1.19) and (1.20).
\vskip1pc
\noindent It is shown in section 5.2.7, that the
 $q\to 1$ limit of $H_L^{qWZ}$
gives indeed $H_L^{WZ}$. We have already learned that the symplectic
form $\om_L^{qWZ}$ has also the correct $q=1$ limit.
Thus we conclude that  the quasitriangular chiral WZW model
is indeed the smooth deformation of its standard counterpart.
\vskip1pc
\noindent {\bf 5. Quasitriangular classical action.}
We have just described explicitely the pair
$(\om_L^{qWZ},H_L^{qWZ})$.
Knowing these data, we can write down the following
classical action of the deformed
chiral WZW model
 \be  S_L^{qWZ}[\eta_L(\tau)]=
\int (\eta_L^*\theta_L^{qWZ}-H_L^{qWZ}(\eta_L )d\tau )\ee
Here $\eta_L(\tau)$ is a trajectory in the phase space $M_L^{WZ}$
parametrized by the ordinary
(continuous) time parameter $\tau$,  $\theta_L^{qWZ}$
is a $1$-form on the phase space called the symplectic potential
and $\eta_L^*\theta_L^{qWZ}$
is its pullback by the map $\eta_L$. The symplectic
form $\omega_L^{qWZ}$ on $M_L^{WZ}$ can be then written as
\be \omega_L^{qWZ}=d\theta_L^{qWZ}.\ee
Consider manifolds
$M_{L}^{WZ}=LG_0\times \A_+$ and  $M_{R}^{WZ}=LG_0\times \A_-$.
Here $\A_-= -\A_+$.
The full left-right quasitriangular WZW model has
the following classical action
\be S^{qWZ}[\eta_L,\eta_R,\lm_\mu]=
S_L^{qWZ}[\eta_L(\tau)]+S_R^{qWZ}[\eta_R(\tau)]  +
\int d\tau \lm_\mu(\tau)(
a_L^\mu(\tau)+a_R^\mu(\tau)).\ee
Here $\eta_L=(k_L,a^\mu_L)$, $\eta_R=(k_R,a^\mu_R)$ with
$k_{L,R}\in LG_0$.
The left and right chiral actions $S_L^{qWZ}(k_L,a_L^\mu)$
and $S_R^{qWZ}(k_R,a_R^\mu)$
 have exactly the same dependence
on their respective variables, but $a_L^\mu$'s run over
the positive standard
Weyl 	alcove $\A_+$
and $a_R^\mu$'s over the negative one $\A_-$. Finally,
the fields $\lm_\mu$ are the
Lagrange multipliers. We note, that in the limit $q\to 1$
the action (1.26) reduces
to the  classical action of the standard full left-right
WZW model written in the
form of Ref.\cite{CL}.
\vskip1pc
\rem : {\small
The variation of the action (1.24) does not depend on
the choice
of the symplectic potential $\theta_L^{qWZ}$
 but only on the pair  $(\omega_L^{qWZ},H_L^{qWZ})$.
  This explains why one can give the meaning to the
$classical$ action (1.24) also
in the case where $\omega_L^{qWZ}$ is not exact
(i.e. there is no
globally defined $\theta_L^{qWZ}$ such that
(1.25) is valid).}

\vskip1pc
\noindent {\bf 6. Quasitriangular current algebra.}
It is of the
crucial importance to understand the symmetries   of the models
(1.24) and (1.26).
We have learned already from the standard ($q=1$)
chiral WZW example, that
the canonical quantization of the model relies heavily on its
symmetry  structure.  In fact,
one needs the identification of suitable  observables
whose Poisson brackets get promoted to the quantum commutation
relations. In the $q=1$ case,
  such observables are the components of the Kac-Moody
current $j=\k \d_\si mm^{-1}$ who serve as the Hamiltonian
charges generating
the action of the  $G={LG_0}$ on the phase space
 $M_L^{WZ}=LG_0\times \A_+$.
The latter fact can be expressed pregnantly by the following
matrix Poisson brackets
\be \{k(\si)\ptp j(\si')\}_{WZ}=2\pi C\delta(\si-\si')(k(\si)\otimes 1),\ee
\be  \{j(\si)\ptp j(\si')\}_{WZ}=\pi\delta(\si-\si')[C,
j(\si)\otimes 1-1\otimes j(\si')]+2\pi\k C\d_\si\delta(\si-\si').\ee
  We note that the  quantum versions
  of  (1.27) and of (1.28) means respectively  that
$k$ is the Kac-Moody primary field and   that $j$ generates the action
of $ {L\G_0}$ on the quantum Hilbert space.

The present paper brings some clarification even of the symmetry
structure
of the standard   chiral   $q=1$ WZW model.  In fact, the second floor
master model (1.2) is strictly symmetric with respect to the left
$\ti G$ action.
The two-step symplectic reduction down to the chiral $WZW$ model
 reduces this
exact $\ti G$-symmetry to anomalous $LG_0$-symmetry (1.28),
 generated by the current $j$.

It turns out that the quasitriangular picture is exactly analoguous.
The deformed chiral
master model is strictly $\ti G$ Poisson-Lie symmetric with respect
 to the left action
of $\ti G$. This means that the chiral master Hamiltonian
$\ti H_L $ is strictly
$\ti G$-invariant
but $\ti G$-invariance of the symplectic form $\Omega^q$
is broken  in  certain  special (Poisson-Lie ) way.
Then also the $\ti G$-invariance of the
deformed chiral classical action (1.24) is broken in the special
 way dictated
by the Poisson-Lie symmetry (see \cite{AT} for a nice general
discussion of this issue).
 The two-step symplectic reduction then changes the strict $\ti G$
Poisson-Lie symmetry
into anomalous $LG_0$ Poisson-Lie symmetry, whose
 non-Abelian moment map will
satisfy the $q$-deformed version of the current algebra.
 The precise way how
the central anomaly manifests itself follows from the fundamental
Poisson bracket (1.6)   defining the chiral symplectic
form $\om_L^{WZ}$.

In order to find the quasitriangular analogue of the Kac-Moody
current $j\in\G=L\G_0$,
 we first write
$j$ in the variables $(k,a^\mu)$:
\be j =-\k a^\mu kT^\mu k^{-1}+\k\d_\si kk^{-1}.\ee
It then follows
\be (j,.)_\G= (\wti{Coad}_{\hat k}\hat\phi_\k)',\ee
where all necessary notations were already defined   after Eq. (1.20).
The $q$-Kac-Moody current $F(k,a^\mu)\in B$ is in turn  given by
\be F=(\wti{Dres}_{\hat k}e^{\ti\Lm(\hat\phi_\k)})',\ee
in full analogy with (1.30).
The Poisson brackets involving $F$ follow from the defining formula
 (1.6).  Their calculation
is somewhat involved but the result is  very simple. It reads
\be \{k\ptp F\bq=\hat r^\k(k\otimes F),\ee
\be \{F\ptp F\bq =[\hat r,F\otimes F],\ee
\be \{F^\dg\ptp F^\dg\bq =-[\hat r,F^\dg\otimes F^\dg],\ee
\be  \{  (F^\dg)^{-1}\ptp F\bq =\hat r^{2\k}( (F^\dg)^{-1}\otimes F)
-(  (F^\dg)^{-1}\otimes F)\hat r.\ee
The $r$-matrices $ \hat r^\k$ and
$\hat r^{2\k}$ are detailed
in Section 5.2.5 and
5.2.6.
The superscript $\k$ over $\hat r^\k$ indicates the presence
of the central extension.
In the limit $q\to 1$, the $q$-primary field condition
(1.32)  becomes (1.27) and  the $q$-current algebra (1.33)-(1.35)
yields (1.28).
\vskip1pc
\noindent {\bf Remarks:}
{\small i) We observe that the quantity
$\wti{Coad}_{\hat k}\hat\phi_\k\in \ti\G^*$
is {\it the} observable of the standard chiral WZW model.
Its $\G^*$-part
$(\wti{Coad}_{\hat k}\hat\phi_\k)'$ gives the
Kac-Moody current $j$, its central circle part
$(\wti{Coad}_{\hat k}\hat\phi_\k)^\infty$
 gives the level $\k$
 and its $\si$-shift part $(\wti{Coad}_{\hat k}\hat\phi_\k)^0$
contributes to
the Hamiltonian $H_L^{WZ}$.  The similar
observation can be made also
for the variable $\wti{Dres}_{\hat k}e^{\ti\Lm(\hat\phi_\k)}\in \ti B$
in the quasitriangular case.

~~~~~~~~~~~~~ii) It is worth stressing again that the
$q$-Kac-Moody brackets (1.33) -(1.35)
can be {\it derived} from the fundamental exchange relation
(1.6) containing
the elliptic dynamical $r$-matrix (1.7). This suggests, in
 particular, that there
might  be  a lurking $q$-current algebra also  in the
description of the standard WZW conformal
blocks on elliptic curves, Hitchin systems etc.}

\vskip1pc
\noindent It turns out convenient to introduce a new dynamical
variable defined by the relation
\be L=FF^\dg.\ee
The Poisson brackets (1.33) - (1.35) can be then equivalently rewritten
in terms of only one relation:
$$\{L(\si)\ptp L(\si')\}_{qWZ}=(L(\si)\ot L(\si'))\e\hat r(\si-\si')
+\e\hat r(\si-\si')(L(\si)\ot L(\si')) $$
 \be -(1\ot L(\si'))\e\hat r(\si-\si'+2i\e\k)(L(\si)\ot 1)
-(L(\si)\ot 1)\e\hat r(\si-\si'-2i\e\k)(1\ot L(\si')).\ee
This relation coincides with the definition of $q$-current
algebra introduced by Reshetikhin and Semenov-Tian-Shansky \cite{RST}.
They call $L(\si)$ the $q$-current. By abuse of notation we shall
refer to both quantities $F$ and $L$ as to  the $q$-currents.

Recall that the non-deformed current $j$ can be simply expressed
in terms of the primary field $m(\si)$:
\be j=\k\d_\si mm^{-1}.\ee
This relation can be called the classical Knizhnik-Zamolodchikov
equation \cite{KZ} since its quantum version becomes indeed the
KZ-equation written in the operatorial form \cite{FHT}. It turns
out, that the $q$-current $L(\si)$ can be also simply expressed
in terms of the $q$-primary field $m(\si)$:
\be L(\si)=m(\si+i\k\e)m^{-1}(\si-i\e\k).\ee
This nice relation can be interpreted as the classical $q$-KZ
equation. As expected, it is not differential but rather a difference
equation.
\vskip1pc

\noindent {\bf 7.   The plan of the paper:}   in Chapter 2 we
first explain the crucial notion of the central biextension
$\ti G$ of a Lie group $G$
and we derive explicit formulae for  adjoint and coadjoint
actions of $\ti G$.
 Then we detail our basic observation that the two-step
symplectic reduction
of the master model (1.1) on $\ti G$ gives the
standard WZW model on $G$.
We shall also see   that this construction can be performed for
any central biextension;
the affine Kac-Moody group being only the special case.
We are thus led to the
notion of the universal WZW model.

In Chapter 3, we study the case of the affine Kac-Moody group.
We show that the
master model on $\ti G$ can be decomposed in two copies of the
simpler chiral model.
Then we perform the two-step chiral symplectic reduction
to obtain the standard
chiral WZW model and we detail the symplectic structure of
the model in the
$(k,a^\mu)$ variables.

We devote Chapter 4 to the construction of the universal
quasitriangular model
based on any Drinfeld double $\ttD$ of the biextended group
$\ti G$. We first
review some basic notions of the theory of
the Poisson-Lie groups
 and then we
identify which  conditions    $\ttD$ must fulfil in order that
 the two-step
reduction could be performed. We call such good doubles $\ttD$
the WZW doubles of $\ti G$.
The rest of the chapter is devoted to the construction of the
 affine Lu-Weinstein-Soibelman
double $\ti G^\bc$ and to
the proving that it does fulfil the required conditions.
 The quasitriangular WZW model based on this particular double is
 the $q$-deformation of the standard loop group WZW model.

The core of the paper is Chapter 5. Starting from the affine
Lu-Weinstein-Soibelman
double, we first explain the construction of     the deformed
chiral geodesical model (1.3).
 Then we perform the
two step symplectic reduction down to the quasitriangular
chiral WZW model. We shall
make explicit the symplectic structure $\om_L^{qWZ}$ and
the Hamiltonian $H_L^{qWZ}$
of the model, we introduce
the $q$-current algebra and show how its commutation relations
follows from the symplectic
structure $\om_L^{qWZ}$. We also show that the model has the
correct $q\to 1$ limit
and finally we glue up  the two quasitriangular chiral WZW
models to obtain
the full left-right theory.

In  Chapter 6 we summarize the results and provide an outlook.
In particular, we outline
the further generalizations of the construction aiming to the
$q$-deformation of the
whole WZW factory, we draw the  plausible picture of
the quantization of the model and  we furnish also some
remarks on the role
of the Virasoro group in the $q$-deformed case.

Chapter 7 contains Appendices that  are of three types :
i) they provide more
background material for better understanding of the article;
 ii) they contain
the detailed technical proofs of some assertions in the text;
 iii) they give alternative
derivations of some results.
\chapter{Universal WZW model}
In this chapter, we shall be very general and we shall
work with an arbitrary Lie group $\ti G$  which is the central
biextension of a Lie group $G$. The loop group case leading to
the standard WZW model
will be only the special (though very important)
example of our construction. Indeed,
we shall see that the WZW-like symplectic structure is "universal";
it can be defined not only for the loop groups and it
does not depend on the detailed structure
of the group multiplication or of the central extension.

\section{Central  biextension}

Consider a central extension $\hat G$ of a Lie  group $G$
by the circle group $U(1)$. This means that there is
the  exact sequence of morphisms of groups:
\be 1\to U(1) \to \hat G \stackrel{\pi}{\to} G \to 1;\ee
where $U(1)$ is injected into the centre of $\hat G$. The
morphism from $\hat G$ to $G$ is denoted  as $\pi$. Note that
the circle fibration over the base $G$ can be topologically
nontrivial.
The  fundamental example of this exact sequence is the famous
central extension
of the loop groups $LG_0$.  It is reviewed in the appendix 7.1.
The reader is invited
to consult this appendix whenever he (or she) will need the
illustration of the
general scheme presented in this chapter.

It is clear that the exact sequence (2.1) of groups induces
the following
 exact sequence of their Lie algebras:
\be 0\to \br \to \hat\G\stackrel{\pi_*}{\to} \G\to 0.\ee
Here $\pi_*:\hat\G\to\G$ is the Lie algebra homomorphism
induced by the
group homomorphism $\pi$.
In general there need not exist a canonical map between
$\hat\G$ and $\G$
that would go in opposite direction than $\pi_*$. Suppose,
however, that we choose
\footnote{If $\hat G$ is constructed
from $G$ in a suitable way, the  existence of a natural
$\iota$ may be the consequence of
this construction; this happens in the case of the central
extensions of the
loop groups of simple compact Lie groups.}
such a map $\iota:\G\to\hG$ . We do not suppose,
however, that
$\iota$ is the homomorphism of Lie algebras! We just
claim that $\iota$ be
a linear injection of $\G$ into $\hG$ fulfilling the
following condition:
\be \pi_*(\iota(\xi)) = \xi\ee
for every $\xi\in\G$.

The existence of the map $\iota$  immediately implies,
that the structure
of the Lie algebra $\hG$ of $\hat G$ must be given by
the following
 commutator\footnote{We use the same symbol $[.,.]$ for
the commutators of different
Lie algebras. It should be clear which usage we have in
mind by realizing
to which Lie algebra the arguments of the  commutator belong.}
\be [\hat \xi,\hat \eta]=\iota([\pi_*\hat \xi,\pi_*\hat \eta])+
\rho(\pi_*\hat \xi,\pi_*\hat \eta)\hat T^\infty\ee
for some cocycle $\rho:\G\w\G\to\br$. Recall that the cocycle
condition means
\be \rho([\xi,\eta],\zeta)+\rho([\eta,\zeta],\xi)+
\rho([\zeta,\xi],\eta)=0,
\quad \xi,\eta,\zeta\in \G.\ee
The elements $\hat \xi,\hat \eta$ in (2.4) are from $\hG$,
the element $\hat T^\infty\in\hG$
 corresponds to the generator of $U(1)$ injected in $\hat G$
according to the exact sequence (2.1). In other words:
\be \pi_*\hat T^\infty=0.\ee
In particular, the relation (2.4) implies
\be [\iota(\xi),\iota(\eta)]=\iota([\xi,\eta])+
\rho(\xi,\eta)\tF.\ee
Suppose that there is a one-parameter subgroup $\hat S$
of automorphisms of $\hat G$ commuting with the central
circle action. It then gives
rise to the one-parameter subgroup $S$
of automorphisms of $G$. We denote as $\d$
the generator of $S$ and we define
\vskip1pc
\defi ~{\bf 2.1}: The group
$\ti G=\br\times_{\hat S} \hat G$ is
 the central  biextension  if
\be \rho(\xi,\eta)=(\xi,\d \eta)_{\G},\ee
where $(.,.)_{\G}$ is a symmetric non-degenerate invariant
 bilinear form on $\G=Lie(G)$,
such that
\be (\xi,\d \eta)_\G+(\d\xi,\eta)_\G=0.\ee
\vskip1pc
\noindent If $\ti G$
is the central biextension then there is the canonical
symmetric non-degenerate
 invariant bilinear form on $\ti\G=Lie(\ti G)$. It is given by
\be ((iX, \xi,ix),(iY, \eta,iy))_{\ti\G}=
(\xi,\eta)_\G-Xy-Yx.\ee

\noindent {\bf Conventions 2.2}: The generator of $\hat S$
in $\ti\G$ will
be denoted either as $\ti T^0$ or as  $(i,0,0)$. The elements
 $\iota(\xi)$ and $\tF$ of $\hat\G$
will be denoted either as $\ti\iota(\xi)$ and
$\ti T^{\infty}$  or as $(0,\xi,0)$
and $(0,0,i)$ when considered as the elements
of the Lie algebra $\ti\G$.

\vskip1pc

\rem : {\small The theory of the biextension was
developed by Medina and Revoy \cite{MR}
in their quest
of classifying the Lie algebras admitting the symmetric
non-degenerate
 invariant bilinear form. They used the terminology
"double extension".
We take the liberty of modifying this name to the
"biextension" since in our paper
the word double will often appear in the different
 sense. It should be also noted that
 the extensions
 considered here form only
the subclass of the Medina-Revoy extensions. For the
 physicists oriented survey of the
subject see also \cite{FFS}.}
\vskip1pc
\noindent It is not difficult to verify directly the
invariance of the bilinear form
$(.,.)_{\ti\G}$. For this, it is also useful to write
the commutator in $\ti\G$
in terms of the notation above:
\be  [(iX, \xi,ix),(iY, \eta,ix)]=(i0,[\xi,\eta]+
X\d \eta-Y\d \xi,i(\xi,\d\eta)_\G).\ee
\noindent The following theorem is of great importance
for our construction:
\vskip1pc
\teo ~{\bf 2.3}: Suppose that an element $\hat g\in\hat G$ is
viewed as the element of $\ti G$,
where $\ti G$ is the central biextension of $G$.
Then the adjoint action of $\hat g$
on the Lie algebra $\ti \G$ has the following
explicit form
\be \widetilde {Ad}_{\hat g}(iX,\xi,ix)=
(iX,Ad_g\xi -X\d gg^{-1},
ix-i(g^{-1}\d g,\xi)_{\G}+
\jp iX(g^{-1}\d g,g^{-1}\d g)_\G).\ee

\noindent {\bf Conventions 2.4}: Since the large part of our
technical work will consist
in "travelling" between the groups $G$,$\hat G$ and $\ti G$,
we should be very careful
in book-keeping  with respect to  which group the certain
operations are considered.
Thus, e.g. $\widetilde {Ad}$ means that the adjoint operation
is taken with respect to the
group $\ti G$ and
we shall always  use the convention that $g=\pi(\hat g)$, if
both $g$ and $\hat g$ appear
in the same formula. Moreover we denote
\be -\d gg^{-1}\equiv Ad_g\d -\d;\quad g^{-1}\d g\equiv
Ad_{g^{-1}}\d -\d.\ee
The $Ad$ operation in (2.13) is taken in the group $\br\times_S G$.
 We do not denote
this group by a special symbol because it will appear less
frequently than its
colleagues mentioned above. Finally,
it is clear that both $\d gg^{-1}$ and $g^{-1}\d g$ live in $\G$.
\vskip1pc
\rem : {In the context of the biextensions of the loop groups,
 $\d gg^{-1}$
equals to $\d_\si gg^{-1}$, where $\d_\si$ is the derivative
with respect
to the loop parameter.}
\vskip1pc
\noindent {\bf Proof of the theorem 2.3}:
First we   show that
\be \widetilde {Ad}_{\hat g}(0,\xi,ix)=(0,Ad_g\gamma,
ix-i(g^{-1}\d g,\xi)_{\G}).\ee
In this special case, we can forget about the group $\ti G$
and can  work only with $\hat G$. The reason is that $(0,\xi,x)$
is in $\hat\G$
and $\hat G$ is the subgroup of $\ti G$.
First of all we have
\be \pi_*(\widehat {Ad}_{\hat g}\iota(\xi))=Ad_g\xi,\ee
because $\pi$ is the homomorphism of groups. Moreover,
\be \widehat {Ad}_{\hat g}\tF=\tF,\ee
because $\tF$ is the generator of the central circle. Thus we
 conclude, that
\be \widehat {Ad}_{\hat g}(\iota(\xi)+x\tF)=\iota(Ad_g\xi)+
(x-F(g,\xi))\tF,\ee
where $F(g,\xi)$ is a function to be determined. First of
all, we know
$F(g,\xi)$ at the group origin $g=e$ (where it vanishes) and
 near the group origin
\be F(\chi,\xi)=(\d\chi,\xi)_\G,\quad \chi\in\G.\ee
Moreover, it is easy to check that
$ F(g,\xi)$ has to verify the following cocycle condition
\be F(g_1g_2,\xi)=F(g_2,\xi)+F(g_1,Ad_{g_2}\xi).\ee
Infinitesimally:
\be F(g_1\chi,\xi)=(\d\chi,\xi)_\G+F(g_1,[\chi,\xi]).\ee
Thus we obtain a first order differential equation with the
known initial condition.
One readily checks that
\be F(g,\xi)=(g^{-1}\d g,\xi)_{\G}\ee
is its solution.

The proof of the formula (2.12) with $X\neq 0$ then goes
like in \cite{PS}; i.e;
 we know that the r.h.s. of (2.12) must have the form
$(X,\dots,\dots)$ and
then one directly checks that the formula (2.12) is the
only possible one preserving
the invariance of the bilinear form $(.,.)_{\ti\G}$.

The theorem is proved.

\rightline{\#}
\noindent It will be also useful to have the explicit
 expressions for the invariant
bilinear form on the dual $\ti\G^*$ and for the coadjoint action
of $\hat g$ on $\ti\G^*$. They read, respectively
\be ((A,\al,a)^*,(B,\b,b)^*)_{\ti\G^*}=(\al,\b)_{\G^*}-
Ab-Ba;\ee
$$ \widetilde {Coad}_{\hat g}(C,\gamma,c)^*=$$\be =
(C+\la \gamma, g^{-1}\d g\ra+\jp c(g^{-1}\d g,g^{-1}\d g)_\G),
Coad_g\gamma +c
\U^{-1}(\d gg^{-1}),c)^*.\ee
We have also
\be \wh{Coad_{\hat g}}(\pi^*(\gamma)+c\tf)=\pi^*(Coad_g\gamma +c
\U^{-1}(\d gg^{-1})) +c\tf.\ee
\noindent {\bf Conventions 2.5} : The map $\U:\G^*\to\G$ is defined
 by the invariant
bilinear form $(.,.)_\G$.
The decomposition $\ti\G=\br\ti T^0+\ti\iota(\G)+\br\ti T^{\infty}$
induces the decomposition of the dual $\ti\G^*$, hence every
$\ti \al\in\ti\G^*$ can be cast
as\footnote{Note that there is no $i$ in the dual objects.
It is consistent to use
such a notation  because we shall never use in this paper the dual
of the {\it complexified} algebra $\ti\G^\bc$.}
\be \ti \al=(A,\al,a)^*,\ee
where $A$, $a$ are in $\br$ and $\al$ in $\G^*$.
The element $(1,0,0)^*$ will be sometimes denoted as $\ti t_0$,
 $(0,\al,0)^*$ as $\ti\pi^*(\al)$ and
$(0,0,1)^*$ as $\ti t_{\infty}$.

In a similar way the decomposition $\hat\G=\iota(\G)+\br\hat
T^{\infty}$
induces
\be \hat\al=\pi^*(\al)+a\tf, \quad \hat \al\in\hat\G^*,\quad
\al\in\G^*,\ee
where $\tf\in\hG^*$ is characterized by the property
\be \la \tf,\iota(\G)\ra=0,\quad \la \tf,\tF\ra=1.\ee
  Of course, $\pi^*:\G^*\to \hat \G^*$ is the map dual to $\pi_*:
\hG\to \G$ defined by the exact sequence (2.2). Note also that
by definition
\be \la \wti{Coad}_{\hat g}\ti\gamma, \ti\xi\ra=
\la \ti\gamma,\wti{Ad}_{\hat g^{-1}}\ti\xi\ra,\ee
where the pairing $\la .,.\ra$ is clearly defined as
\be \la (C,\ga,c)^*,(X,\xi,x)\ra=CX+cx+\la \ga,\xi\ra.\ee

\vskip1pc
\section{The symplectic reduction}
\subsection{Second floor master model on $\ti G$.}
As we have stated in the introduction, the classical action of
the geodesical
model on $\ti G$ reads
\be \ti S(\ti g)=-{\kappa\over 4}\int d\tau
(\ti g^{-1}{d\over d\tau}{\ti g},
\ti g^{-1}{d\over d\tau}{\ti g})_{\ti\G}, \ee
where  $\ti g(\tau)\in\ti G$ and $\kappa$ is a parameter
going to play the role
of the level of the WZW model. We first rewrite this
action in the first order
Hamiltonian form:
\be \ti S(\ti\beta_L,\ti g)= \int d\tau
 [\la \ti\beta_L,{d\over d\tau}\ti g \ti g^{-1}\ra +
{1\over \k}(\ti\beta_L,\ti\beta_L)_{\ti\G^*}].\ee
Here $\ti\beta_L\in\ti\G^*$ are the "momentum" and $\ti g$
the position
coordinates induced by the right trivialization of the
phase space $T^*\ti G$.
The symplectic potential on $T^*\ti G$ is
\be \ti\theta=\la \ti\beta_L,d\ti g \ti g^{-1}\ra\ee
in agreement with the general story explained after the
formula (1.3) in the Introduction.
The reader can find in Appendix 7.2 the  detailed account
of the canonical
symplectic structure on the
cotangent bundle of the group manifold. It is clear that we
 can obtain  (2.30) from
(2.31) by eliminating $\ti\beta_L$ via the field equations.

\subsection{The symplectic reduction to the
 first floor $\hat G$}

\noindent The field multiplet $\ti\beta_L\in\ti\G^*,
\ti g\in \ti G$ of (2.31) can be
 also parametrized as follows
\be \ti\beta_L=\wti{Coad}_u\ti\gamma_L,\quad \ti g =u\hat g u,\ee
where $\ti\gamma_L\in\ti\G^*$, $u=\exp{s\ti T^0}\in\ti G$
and $\hat g\in \hat G$.
Using the invariance of the form $(.,.)_{\ti\G^*}$, the
 action written
in the new variables becomes
 \be \ti S(\ti\gamma_L,\hat g, s)= \int d\tau
 [\la \ti\gamma_L,(\ti T^0+\hat g\ti T^0\hat g^{-1})\ra {ds\over d\tau}
+\la\ti\gamma_L,{d\over d\tau}\hat g \hat g^{-1}\ra +
{1\over \k}(\ti\gamma_L,\ti\gamma_L)_{\ti\G^*}].\ee
Following our list of conventions, we can represent $\ti\gamma_L$ as
\be \ti\gamma_L=(\ga_L^0,\gamma_L,\ga_L^\infty)^*.\ee
We can introduce once again a new way of parametrizing
the phase space $T^*\ti G$,
now by the set of coordinates $(\ga_L^s,\ga_L,
\ga_L^\infty,s,\hat g)$ where
\be \ga_L^s=\la \ti\gamma_L,(\ti T^0+\hat g\ti T^0\hat g^{-1})\ra=2\ga_L^0
 -\la \ga_L,\d gg^{-1}\ra+\ga_L^\infty
\jp (g^{-1}\d g,g^{-1}\d g)_\G.\ee
The action in these newest coordinates becomes
 $$ \ti S(\gamma_L^s,\ga_L,\ga_L^\infty,\hat g, s)= \int d\tau
 [ \gamma_L^s {ds\over d\tau}-{1\over \k}\ga_L^s\ga_L^\infty]$$
\be +\int d\tau [ \la\hat\gamma_L,{d\over d\tau}
\hat g \hat g^{-1}\ra  +
{1\over \k}(\gamma_L,\gamma_L)_{\G^*}-{\ga_L^\infty\over \k}
\la \ga_L,\d gg^{-1}\ra
+{1\over 2\k} (\ga_L^\infty )^2(g^{-1}\d g,g^{-1}\d g)_\G].\ee
 Here we have used the formula (2.12), the explicit form (2.22)
 of the scalar product
on $\ti\G^*$, and we have set \be \hat\ga_L=(0,\ga_L,\ga_L^\infty)^*.\ee
The reader should pay attention to the distribution of hats and
tildes. Of course,
we still use the convention that $g=\pi(\hat g)$, if both $g$
and $\hat g$ appear
in the same formula.

The terms containing (not containing) the level $\k$ encodes
 the Hamiltonian
(the symplectic potential $\ti\theta$)
in our new coordinates. It is clear that the coordinate
 $\ga_L^s$
Poisson-commute\footnote{Of course,
this can be seen also from the fact that $\ga_L^s$ is the
 moment map generating
the axial action $\ti g\to u\ti g u$, $\ti\ga_L\to Coad_u\ti\ga_L$,
 $u=\exp{s\ti T^0}$
 and the Hamiltonian
$\ti H=-{1\over 2\k}(\ti\ga_L,\ti\ga_L)_{\ti\G^*}$ is clearly
 invariant with respect to
this action.}
with the Hamiltonian since the latter does not contain the
variable $s$. We can therefore
consistently set $\ti\ga_L^s=0$ in the action (2.37). This
is the so-called symplectic
reduction of the dynamical system (2.31) with respect to the
moment map generating
the axial action of $\ti T^0$ on $T^*\ti G$. The resulting
reduced action reads
 $$\hat S(\hat\ga_L,\hat g)= $$
\be \int d\tau [ \la\hat\gamma_L,{d\over d\tau}\hat g \hat g^{-1}\ra  +
{1\over \k}(\gamma_L,\gamma_L)_{\G^*}-{\ga_L^\infty\over \k}
\la \ga_L,\d gg^{-1}\ra
+{1\over 2\k} (\ga_L^\infty)^2 (g^{-1}\d g,g^{-1}\d g)_\G].\ee
It can be rewritten
entirely from the point of view of the cotangent bundle
$T^*\hat G$. Indeed, we shall first note
that the symplectic potential $\hat \theta$  of the
reduced dynamical system (2.39)
is
\be \hat\theta= \la\hat\gamma_L,d\hat g \hat g^{-1}\ra \ee
which is the canonical symplectic potential on $T^*\hat G$
written in the right trivialization
coordinates $\hat\ga_L,\hat g$. The Hamiltonian $\hat H$ of
the reduced model can be also written
elegantly in terms of natural quantities related to $T^*\hat G$.
Indeed, it turns out that
\be \hat H=-{1\over 2\k}(\iota^*(\hat \ga_L),
\iota^*(\hat \ga_L))_{\G^*}-
{1\over 2\k}(\iota^*(\hat \ga_R),\iota^*(\hat \ga_R))_{\G^*},\ee
where $\iota^*:\hG^*\to\G^*$ is the map dual to the injection
$\iota:\G\to\hG$ and
$\hat\ga_R$ is the coordinate on $T^*\hat G$ given by the left
 trivialization. Note also
the minus signs reflecting the fact that the form $(.,.)_\G$
is negative definite.In other
words, $\hat \ga_L \hat g=\hat g\hat\ga_R$ or
$\hat \ga_L=\wh{Coad}_{\hat g}\hat\ga_R$.
In deriving  (2.41), we have used the formula (2.24)
 expressing the coadjoint action on $\hat\G^*$

We conclude that
 the reduced first floor action $\hat S$ can be written in a
completely left-right symmetric way as
$$\hat S=
 \jp\int d\tau [ \la\hat\gamma_L,
{d\over d\tau}\hat g \hat g^{-1}\ra
+\la\hat\gamma_R,\hat g^{-1}{d\over d\tau}\hat g \ra ]$$
\be +{1\over 2\k}\int d\tau [(\iota^*(\hat \ga_L),
\iota^*(\hat \ga_L))_{\G^*}+
(\iota^*(\hat \ga_R),\iota^*(\hat \ga_R))_{\G^*}].\ee
\rem : {\small The first floor universal WZW  action (2.42)
can be naturally written for
whatever central extension of the group $G$ admitting the
biinvariant metric. In other words,
even if the cocycle $\rho(x,y)$ does not have the form (2.8),
the action (2.42)
makes sense. Actually, we had begun to write this article from
the vantage point of the action
(2.42). This seemed  sufficient for our deformation programme
 since the symplectic structure of the model on $T^*\hat G$ is
already canonical. Thus we can
introduce the Heisenberg double, Semenov-Tian-Shansky form etc.
What happened was, however,
that we had canonically obtained  the symplectic structure of
the deformed model but
there was no clue how to single out the canonical choice of the
deformed Hamiltonian.
 In fact, it turned out that many Hamiltonians
have satisfied the basic condition of the Poisson-Lie symmetry
and have had correct limit
when the deformation parameter went to zero. It was the search
of the natural Hamiltonian
that finally opened our eyes and we realized that the model
(2.42) can be lifted to the
second floor $\ti G$ for the cocycles of the type (2.8).
On $\ti G$, there is the canonical
choice of the Hamiltonian even in the deformed case.
 Then the canonical Hamiltonian  on the first floor is the
one inherited from the master
model on $\ti G$.}

\subsection{The reduction to the ground floor $G$}

The second symplectic reduction is slightly more involved
since the central circle
bundle over $\hat G$ is nontrivial.  In the Appendix 7.3
the   reduction is performed by working directly with the
Poisson brackets
on $T^*\hat G$ and $T^*G$. Here we shall rather work with
the symplectic forms. This  form language
has the advantage of being briefer and for this reason we
expose it in the main body of the
paper. However, the dual (Poisson bracket) derivation has
the advantage of being
more transparently deformable to the case of the nontrivial
Drinfeld doubles.
 Anyway, we offer both derivations in this paper.

Recall that the canonical symplectic form $\hat\omega$ on
$T^*\hat G$ is given by $\hat\omega=
d\hat\theta$
where $\hat\theta$ is the symplectic potential:
\be \hat\theta = +\la \hat\gamma_L, d\hat g \hat g^{-1}\ra.\ee
In the same way, $\theta$ is the symplectic potential on $T^*G$:
\be \theta =+\la \gamma_L, dg g^{-1}\ra.\ee
Of course, we have
 trivialized the cotangent bundle $T^*G$ by the right-invariant
forms hence
every  point
$K\in T^*G$ can be  decomposed as $K=\gamma_L g$, where
$\gamma_L\in \G^*$ and   $g\in G$.
In this section 2.2.3  we shall never work with the opposite
 decomposition $K=g\ga_R$.

Consider an  extension $\pi_{ext}:T^*\hat G\to T^*G$ of the
map $\pi:\hat G\to G$ defined
by the exact sequence (2.1).  $\pi_{ext}$ can be easily
expressed in the right trivialization
as follows
\be \pi_{ext}(\hat\gamma_L\hat g)=\iota^*(\hat\gamma_L)\pi(\hat g).\ee
Recall that $\iota^*$ is the map dual to $\iota:\G\to\hG$.
Then there is a natural relation between the forms $\hat\theta$
and $\theta$ as the following
lemma states:
\vskip1pc
\lem ~{\bf 2.6}: It holds:
\be \hat\theta=  \ga_L^\infty (R^*_{\hat g^{-1}}\tf) +
\pi^*_{ext}\theta.\ee
\vskip1pc
\pro :
Let $T^i$ be a basis of the Lie algebra $\G$ and $t_i$ its
dual basis in $\G^*$.
We can then choose $\tF,\iota(T^i)$ as the basis in $\hG$.
Its dual is clearly $\tf,\pi^*(t_i)$,
since $\pi_*\circ\iota$ equals to the identity map $\G\to\G$ (cf. (2.3)).
The right invariant Maurer-Cartan form $\rho_{\hat G}=
d\hat g\hat g^{-1}$ can be clearly
written as
\be d\hat g\hat g^{-1}=
R^*_{\hat g^{-1}}\tf\otimes\tF+R^*_{\hat g^{-1}}\pi^*(t_i)
\otimes\iota(T^i),\ee
where $R^*$ is the pull-back map.
Then the symplectic potential $\hat\theta$ can be expressed as follows
\be \hat\theta =\ga_L^\infty (R^*_{\hat g^{-1}}\tf)+\la
 \hat\gamma_L,\iota(T^i) \ra
 (R^*_{\hat g^{-1}}\pi^*(t_i)).\ee
Now  we observe that:
\vskip1pc
\noindent 1) $\la \ga_L,T^i\ra$ is a function on $T^*G$.
If we calculate its $\pi_{ext}$-pullback
we obtain
\be \pi^*_{ext}\la\ga_L,T^i\ra= \la \iota^*(\hat\ga_L),T^i\ra=
\la \hat\ga_L,\iota(T^i)\ra.\ee

\noindent 2) Due to the fact that $\pi:\hat G\to G$ is the
 group homomorphism, we have
\be R^*_{\hat g^{-1}}\pi^*=\pi^*R^*_{\pi(\hat g)^{-1}}.\ee
Using 1) and 2) in (2.48) we infer  that
\be \hat\theta=\ga_L^\infty  (R^*_{\hat g^{-1}}\tf)+
\pi_{ext}^*\la\ga_L,T^i \ra
 \pi^*R^*_{\pi(\hat g)^{-1}}t_i= \ga_L^\infty
 (R^*_{\hat g^{-1}}\tf)+
\pi_{ext}^*\theta.\ee

\noindent The lemma is proved.

\rightline{\#}
\noindent By using the formula
\be d(d\hat g\hat g^{-1})=d\hat g\hat g^{-1}\w d\hat g\hat
g^{-1}\ee
  for the exterior derivative of the Maurer-Cartan form,
we can immediately calculate also the exterior derivatives
of its components. Thus we obtain,
in particular, that
\be d(R^*_{\hat g^{-1}}\tf)=
\jp\rho(T^i,T^j)(R^*_{\hat g^{-1}}\pi^*(t_i))\w
(R^*_{\hat g^{-1}}\pi^*(t_j))=\jp \pi^*_{ext}\rho(dgg^{-1}
\stackrel{\w}{,}dgg^{-1}),\ee
where $\rho(.,.)$ is the cocycle defining the central extension.
We can therefore express conveniently the symplectic form
$\hat\omega$ on $T^*\hat G$ as
\be \hat\omega=d\hat\theta= +d\ga_L^\infty\w(R^*_{\hat g^{-1}}\tf)
+\jp \ga_L^\infty\pi^*_{ext}\rho(dgg^{-1}\stackrel{\w}{,}dgg^{-1})+
\pi^*_{ext}d\theta.\ee
Now we can directly perform the symplectic reduction by setting
\be  \ga_L^\infty=\k.\ee
The restriction of the form $\hat\omega$ to the submanifold
determined by (2.55)
is clearly a $\pi^*_{ext}$-pullback of the two-form
$\omega_{red}$ living on the
manifold $T^*G$ and given by
\be \omega_{red}={\k\over 2}\rho(dgg^{-1}\stackrel{\w}{,}dgg^{-1})+
d\la \ga_L, dg g^{-1}\ra.\ee The form $\omega_{red}$ is the
reduced symplectic form as the
notation indicates.
\vskip1pc
\teo ~{\bf 2.7}: If $G$ is the loop group $LG_0$
then the  form $\omega_{red}$  on $T^*G$ is the  symplectic form
of the standard WZW model.

\vskip1pc
\pro : The loop group cocycle $\rho(\eta,\xi)$ reads
\be \rho(\eta,\xi)={1\over 2\pi}\int_{S^1}
(\eta,\d_\si\xi)_{\G_0^\bc}.\ee
The form $\om_{red}$ can be then rewritten as
\be \omega_{red}={\k\over 4\pi}\int_{S^1}(dgg^{-1}
\stackrel{\w}{,}\d_\si(dgg^{-1}))_{\G_0}+
{1\over 2\pi}d\int_{S^1}(J_L(\si),dgg^{-1})_{\G_0}.\ee
Here $J_L(\si)=\U(\ga_L)$, where $\U$ is the identification map
$\G^*\to\G$ induced
by $(.,.)_\G={1\over 2\pi}\int (.,.)_{\G_0}$.
It turns out that (2.58) is the standard  WZW symplectic form
of the reference  \cite{Mad}. (Actually, there is the difference  in the
 overall normalization factor $(-2\pi)$; if wished, this factor
  can be easily restored  in all our formulae. The reader should also note
that our bilinear form $(.,.)_{\G_0}$ is $-Tr$ of
 Ref.\cite{Mad}).

\noindent The theorem is proved.

\rightline{\#}
\noindent The Hamiltonian $\hat H$ of the first floor model on
$\hat G$ can be read off from
the formula (2.41). It clearly Poisson-commutes with the moment
map $\ti\ga_L^\infty$
since it is invariant with respect to the central circle action.
It thus descends to
the function on the ground-floor phase space $T^*G$ where it is
given by the formula
$$ H_{WZW}(\ga_L,g)=-{1\over \k}(\gamma_L,\gamma_L)_{\G^*}+
 \la \ga_L,\d gg^{-1}\ra
-{\k\over 2} (g^{-1}\d g,g^{-1}\d g)_\G=$$
\be  =-{1\over 2\k}(J_L,J_L)_{\G}-{1\over 2\k}(J_R,J_R)_{\G}, \ee
where
\be J_L=\U(\ga_L),\quad J_R(\si)=-Ad_{g^{-1}}J_L(\si)+\k g^{-1}
\d_\si g.\ee
 We immediately observe that our Hamiltonian $H_{WZW}$ coincides
(up the factor
$(2\pi)$ mentioned above) with the standard WZW Hamiltonian
of Ref. \cite{Mad}. Note that the symplectic form of Ref. \cite{Mad}
is the $(-2\pi)$-multiple of  our $\om_{red}$ and the
Hamiltonian \cite{Mad}
is $(2\pi)$-multiple of our$ H_{WZW}$. The discrepancy in the
 relative sign
is innocent. Indeed, if we change the sign of the symplectic
potential in (2.31) and integrate
away the momenta, we shall again obtain the same second order
action (2.30).
 Thus we have proved the following theorem
\vskip1pc
\teo ~{\bf 2.8}: The two step symplectic reduction
 of the master model (1.1)
 induced by equating
$\ga_L^s=0$, $\ga_L^\infty =\k$ yields the standard WZW model.
\vskip1pc
\rem : {\small We stress
 that the dynamics of the WZW model
is intrinsically left-right symmetric. The left-right asymmetry
in the Hamiltonian
(cf. (2.60)) is purely
coordinate effect which can be traced back to the asymmetric way
of performing the
symplectic reduction. Indeed, the choice of the right trivialization
of the
bundle in (2.32)
already breaks the symmetry. The left-right symmetric formalism
of Appendix 7.1
does not use the trivialization of the cotangent bundle.
 We can choose a diffeomorphism relating $\mk/U(1)$ and  $T^*G$
that does not break
the left-right symmetry. The reason  why we  are not making such a
  symmetric
choice   (but we prefer the   asymmetric one)  is simple:
It is because  we want to arrive at  the standard $T^*G$ presentation
of the WZW symplectic structure existing in the literature \cite{FG,Mad}.}
\vskip1pc
\noindent It is instructive to evaluate the Poisson bracket of
functions on $T^*G$
with respect to the reduced form $\om_{red}$.
 It is convenient to use a short-hand notation
$\la \ga_L,T^i\ra\equiv \ga_L^i$. Then the reduced form
$\omega_{red}$ can be written as
\be \omega_{red}=+d\ga_L^i\w R^*_{g^{-1}}t_i+
\jp(\ga_L^i f_i^{~mn}+\k\rho(T^m,T^n))R^*_{g^{-1}}t_m\w
R^*_{g^{-1}}t_n,\ee
 where $f_i^{~mn}$ are the structure constants of the Lie algebra $\G$.
This expression can be readily  inverted to give the
 corresponding Poisson tensor $\Pi_{red}$:
\be \Pi_{red}=\jp(\k\rho(T^i,T^j)+\ga_L^mf_m^{~ij})
{\d\over \d\ga_L^i}\w {\d\over \d\ga_L^j}-
{\d\over \d\ga_L^i}\w R_{g *}T^i.\ee
Since we have that $\la\nabla^L_G,\xi\ra=R_{g *}\xi$, for $\xi\in\G$,
we obtain from $\Pi_{red}$
the following WZW Poisson brackets (cf. (7.41) -(7.43)):
\be  \{\Phi_1(g),\Phi_2(g)\}_{red}=0;\ee
\be \{\Phi(g),\la \ga_L,\xi\ra\}_{red}=\dto\Phi(e^{s\xi}g)\equiv
\la\nabla^L_G\Phi,\xi\ra;\ee
\be \{\la\ga_L,\xi\ra,\la \ga_L,\eta\ra\}_{red}=\la\ga_L,
[\xi,\eta]\ra+\k\rho(\xi,\eta).\ee
From this we obtain for  the loop group case:
 $${1\over 2\pi}\{(T^\al,J_L(\si))_{\G_0},
(T^\beta,J_L(\si'))_{\G_0}\}_{red}=$$
\be =([T^\al,T^\beta],J_L(\si))_{\G_0}\delta(\si-\si')+
\k(T^\al,T^\beta)_{\G_0}\partial_\si
\delta(\si-\si');\ee
$${1\over 2\pi}\{(T^\al,J_R(\si))_{\G_0},
(T^\beta,J_R(\si'))_{\G_0}\}_{red}=$$
\be =([T^\al,T^\beta],J_R(\si))_{\G_0}\delta(\si-\si')-
\k(T^\al,T^\beta)_{\G_0}\partial_\si
\delta(\si-\si');\ee
\be {1\over 2\pi}\{g(\si),(T^\al,J_L(\si'))_{\G_0}\}_{red}=
T^\al g(\si)\delta(\si-\si');\ee
\be {1\over 2\pi}\{g(\si),(T^\al,J_R(\si'))_{\G_0}\}_{red}=
- g(\si)T^\al\delta(\si-\si');\ee
\be  {1\over 2\pi}\{(T^\al,J_L(\si))_{\G_0},
(T^\beta,J_R(\si'))_{\G_0}\}_{red}=0.\ee
Here $T^\al$ is some element of the Lie algebra $\G_0$,
$g(\si)$ is understood
as a matrix in some (typically fundamental) representation
and $\delta(\si-\si')$ is the
standard $\delta$-function given by
\be \delta(\si-\si')={1\over 2\pi}\Sigma_{n\in{\bf Z}}e^{in(\si-\si')}.\ee
Upon to the $(-2\pi)$  normalization (cf. the remark
above concerning the normalization of the symplectic form),
 our reduced  Poisson brackets (2.66)-(2.70)  coincide  with the Poisson
brackets (2.4) of the reference \cite{Mad} and thus
they define the standard WZW symplectic
 structure.
\vskip1pc
\noindent {\bf Remarks}:{\small  1) We should complete
the list of the Poisson brackets
(2.66)-(2.70)
by the following "trivial" bracket:
\be \{g(\si)\ptp g(\si')\}_{red}=0.\ee
It will turn out that in the quasitriangular generalization
of the WZW model such a bracket
will not vanish.

\phantom{\bf Remarks}~~~~~~~~~~~ 2)
It is important to note that the space derivate $\partial_\si$
 in the reduced Hamiltonian (2.59) was "born"
in the process of the symplectic reduction. So we observe
that field theoretic character
of the WZW model is in a sense the fruit of the central extension.}

\chapter{Chiral decomposition of the WZW model}
It exists a sort of the square root of
the dynamical structure of the standard WZW model. It is
called the chiral WZW model \cite{CGO,CL,FG}
and it describes the dynamics of left (or right) movers
independently. The full
WZW model is then obtained by the appropriate glueing of
the left and right chiral WZW
theories. The goal of this chapter is to
  present the derivation of the chiral WZW model starting from
the master model (1.1).  We shall first decompose (1.1) into the
chiral components (1.2) called the chiral master models  and then
we perform an appropriate  two step symplectic reduction of the
 latters. We shall
see that the result is indeed the standard chiral WZW model
\cite{CGO,CL,FG}.

 As it was often remarked \cite{Gaw,AT},
the analogue of the chiral decomposition exists already at
the level of finite-dimensional
Lie groups. We shall devote a section to the description
of  this finite-dimensional story in order to set the technical,
notational
a ideological background   for the more involved
infinite-dimensional case.

\section{Chiral geodesical model on $G_0$}

\subsection{Cartan decomposition}
The geodesical model can be naturally associated with every
Lie group possessing a biinvariant non-degenerate metric.
In other words, it is required
that an invariant symmetric non-degenerate $\br$-bilinear
form $(.,.)_{\G}$ exists on the
Lie algebra $\G$ of the group $G$. In this section, we are
going to study the case
of a simple compact connected and simply connected group $G_0$
 equipped with its standard
Kiling-Cartan form playing the role of $(.,.)_{\G_0}$.

In what follows, we shall introduce a map $\U_0$ that identifies
the dual $\G_0^*$ of $\G_0$
with $\G_0$ itself via the bilinear form $(.,.)_{\G_0}$.
 Thus
\be \la x^*,y\ra=(\U_0 (x^*),y)_{\G_0}, \qquad x^*\in\G_0^*,
\quad y\in \G_0.\ee
 Consider now a subspace $\U_0^{-1}(\T)$ of $\G_0^*$, where
$\T$ is the Lie algebra of
a chosen maximal torus $\bt$ in $G_0$.
We can view $\U_0^{-1}(\T)$ also  as the subspace of the
cotangent space at the unit element of
$G_0$ hence as the subgroup of $T^*G_0$. In the latter case
we shall denote $\U_0^{-1}(\T)$ as
$\A^0$ and we shall call it the  Cartan subgroup of $T^*G_0$.
This terminology is not standard but it is very
suitable for the purposes of this paper, in particular for
 the generalization to the loop group case.

Consider  now subgroups $Norm(\A^0)\subset G_0$
and $Cent(\A^0)\subset G_0$ given by
\be Norm(\A^0)=\{w\in G_0,  w\A^0 w^{-1}\in \A^0\};\ee
\be Cent(\A^0)=
\{w\in G_0,  w a w^{-1}= a\quad {\rm if}~~  a\in  \A^0\}.\ee
Here the group multiplication law is that of the cotangent bundle
 $T^*G_0$.
Clearly, $Cent(\A^0)$ is the normal subgroup of $Norm(\A^0)$.
\vskip1pc
\noindent{\bf Definition 3.1:} Weyl group $W$ is the factor group
$Norm(\A^0)/Cent(\A^0)$.
\vskip1pc
\rem : The group $Cent(\A^0)$ is the maximal torus $\bt$ of $G_0$.
\vskip1pc
\noindent The Weyl group acts  $\T$ or on $\A^0$. The fundamental
domains of this action on $\A^0$
are called (Weyl) chambers. One usually chooses one chamber which
is then called
the positive Weyl chamber and denoted as $\A^0_+$.

It is the  well-known fact that the $G_0$-adjoint orbit of
every  element of $\G_0$ intersects the Cartan subalgebra $\T$ of
$\G_0$ (the diagonalization
property in\cite{BT}).
This  fact, the trivializability of the cotangent bundle $T^*G_0$
 and the definition
of the Weyl chamber imply together
the following theorem:

\vskip1pc
\teo ~{\bf 3.2}: (Cartan decomposition) Every element $ K\in T^*G_0$ can be
decomposed as
\be  K= k_L \phi k_R^{-1}, \quad k_{L,R}\in G_0,\quad
\phi\in \A^0_+.\ee
The ambiguity of the decomposition is given by the simultaneous
right multiplication of $k_L$
and $k_R$  by the same element of $Cent(\A^0)=\bt$.
\vskip1pc
\pro : By the left trivialization, every element $K\in T^*G_0$
can be written as
$K=g_L\beta_R$, where $g_L\in G_0$ and $\beta_R\in \G_0^*$.
By diagonalization, $\beta_R$ can be
written as $\beta_R=k_R \phi k_R^{-1}$, for some $k_R\in G_0$
and $\phi\in \A^0$.
By writing $g_L$ as $k_L k_R^{-1}$ for certain $k_L\in G_0$
and by using the action of the Weyl group, we immediately arrive
 at the Cartan decomposition
formula (3.4).

The theorem is proved.

\rightline{\#}

\subsection{Standard chiral symplectic structure}
Recall from section 7.2 that
the symplectic potential $\theta$ on $T^*G_0$
can be   simply expressed in the right trivialization $K=\beta_L g_R$
as
\be \theta =\la \beta_L,\rho_{G_0}\ra\equiv\la \beta_L,
dg_R g_R^{-1}\ra.\ee
The dynamical system characterized by the symplectic form $d\theta$
and by the Hamiltonian
\be H(K)=-(\beta_L(K),\beta_L(K))_{\G_0^*}\ee
is called the standard geodesical model on $G_0$. Recall that
the form $(.,.)_{\G_0^*}$
is dual to $(.,.)_{\G_0}$. The latter form is defined by the
restriction of the
form the Killing-Cartan form $(.,.)_{\G_0^\bc}$ to the compact
real form $\G_0$.
As such, the form $(.,.)_{\G_0}$ is {\it negative} definite
which explains
 the minus sign
in the definition (3.6) of the Hamiltonian and also in the
second order action:
\be S((g)=-\js\int d\tau(g^{-1}{d\over d\tau}g,
g^{-1}{d\over d\tau}g)_{\G_0}.\ee
Now consider a manifold $G_0\times \A^0_+\times G_0$; we
shall denote its points as triples
 $(k_L,\phi,k_R)$.
The Cartan decomposition (3.4) then induces a natural map $\Xi$
from this manifold into the cotangent double $T^*G_0$. We can
then pull-back the polarization form
$\theta$
by the map $\Xi$. By noting that
\be\beta_L=k_L\phi k_L^{-1},\quad g_R=k_Lk_R^{-1},\ee
we obtain
\be \Xi^*\theta =\la Coad_{k_L}\phi,dk_L k_L^{-1}+k_L
dk_R^{-1}k_R k_L^{-1}\ra=
\la \phi,k_L^{-1}dk_L\ra
-\la \phi,k_R^{-1}dk_R\ra.\ee
Recall that $\phi$ is also the  element of $\G_0^*$ hence
the pairing in (3.9) makes sense.
We observe that the resulting form can be chirally decomposed
in the left and right parts
who talk to each other only via the variable $\phi$.
 We can make
the left and right form in (3.9) completely independent
by means of the following construction

Consider a manifold $M_L=G_0\times \A^0_+$. Its elements
are couples $(k_L,\phi_L)$ and it is clearly
a submanifold of $T^*G_0$. We can pullback the  symplectic
potential $\theta$ on $T^*G_0$ to $M_L$
by the map $(k_L,\phi_L)\to k_L\phi_L\in T^*G_0$, where the
multiplication is in the sense of the
group law in $T^*G_0$.
The result is clearly
\be \theta_L=\la \phi_L, k_L^{-1}dk_L\ra.\ee
We shall  prove soon  that the form $d\theta_L$ on
$G_0\times \A^0_+$ is nondegenerate,
hence it
defines the symplectic structure.

\vskip1pc
\defi~{\bf 3.3} : The manifold $M_L=G_0\times \A^0_+$ equipped with the
 symplectic form $d\theta_L$ is referred
to as the model space of the (simple  compact etc.) group $G_0$.
\vskip1pc

\noindent We have seen that we can obtain the symplectic structure
 on the model space by the
simple pullback
of the canonical symplectic form on $T^*G_0$. We can show with
the help
of the Cartan decomposition that a sort of the  "inverse"
procedure is also
possible. Indeed, consider a direct product $M_L\times M_R$
of two copies of the model space
$M_L= G_0 \times \A^0_+$  and $M_R=G_0\times \A^0_-$, where
$\A^0_-=-\A^0_+$. Equip the manifold
$M_L\times M_R$
with  a symplectic form
\be \omega_{L\times R}=d\theta_L +d\theta_R=
d\la \phi_L,k_L^{-1}dk_L\ra +
d\la \phi_R,k_R^{-1}dk_R\ra.\ee
The cotangent bundle $T^*G_0$ with its canonical symplectic
structure $\omega=d\theta$ can be
obtained by an appropriate  symplectic reduction of the
symplectic manifold
$(M_L\times M_R,\omega_{L\times R})$. Indeed, consider a
submanifold of $M_L\times M_R$
obtained by equating $\phi_L+\phi_R=0$. The form
$\omega_{L\times R}$ restricted to
this submanifold becomes just $\Xi^*\theta$.
It is clearly degenerate, since by construction of the map $\Xi$,
 its kernel is given by vector
fields generating
the simultaneous right action of the maximal torus $\bt$ on
 $k_L$ and $k_R$. By imposing
the equivalence $(k_L,k_R,\phi)\cong (k_L h,k_R h,\phi)$,
 where $h\in \bt$, we obtain the
 reduced
manifold. According to the Cartan decomposition theorem, the
latter is nothing but $T^*G_0$.

It turns out that the Hamiltonian (3.6) of the geodesical model
on $T^*G_0$ can be also
"descended" from a natural Hamiltonian on $M_L\times M_R$.
The latter is given by the following
formula
\be H_{L\times R}=H_L+H_R=-\jp(\phi_L,\phi_L)_{\G_0^*}-
\jp(\phi_R,\phi_R)_{\G_0^*}.\ee
Since  $H_{L\times R}$ restricted to $\phi_L+\phi_R=0$
is trivially invariant with respect to the maximal torus
action
 $(k_L,k_R,\phi)\cong (k_L h,k_R h,\phi)$, it defines certain
 Hamiltonian on the reduced manifold
$T^*G_0$. In order to show that   this is precisely the
 Hamiltonian of the geodesical
flow in (3.6), it is sufficient to note, that
$$(\beta_L(K),\beta_L(K))_{\G_0^*}=$$\be=
(\beta_L(k_L\phi k_R^{-1}),\beta_L(k_L\phi
k_R^{-1}))_{\G_0^*}
=(k_L\phi k_L^{-1},k_L\phi k_L^{-1})_{\G_0^*}=(\phi,\phi)_{\G_0^*}.\ee

\subsection{Dynamical r-matrix}
This section is devoted to the study of the chiral dynamical
system defined on the
model space $M_L$ and characterized by the symplectic potential $\theta_L$
and the Hamiltonian $H_L$. This system has been proposed in \cite{AS}
as the finite dimensional analogue of the chiral WZW model.
We have seen in the previous paragraph that the geodesical model
on $G_0$ admits the chiral decomposition in two chiral models.
 By this we mean that
 it can be defined by the symplectic reduction of the model on
$M_L\times M_R$, characterized by the symplectic form $\om_{L\times R}$
and by the Hamiltonian $H_{L\times R}$.

The chiral  dynamics can be derived from the following action principle
\be S=\int d\tau [\la \phi_L,k_{L}^{-1}\dot{k_L}\ra+\jp(\phi_L,
\phi_L)_{\G_0^*}],\ee
where the dot indicates the time derivative.
The equations of motion can be easily derived:
\be {d\over d\tau}(Coad_{k_L} \phi_L) =0,
\quad P_\T k_L^{-1}\dot{k_L} =-\U_0(\phi_L),\ee
where $P_\T$ denotes the orthogonal projection on
 $\T$ and $\U_0:\G_0^*\to\G_0$
is the map that identifies $\G_0^*$ with $\G_0$ via the  form
 $(.,.)_{\G_0}$.

From the first equation it follows
\be Coad_{k_L(\tau)} \phi_L(t)=Coad_{k_L(0)} \phi_L(0),\ee
or
\be \phi_L(\tau)=Coad_{k^{-1}_L(\tau)k_L(0)} \phi_L(0).\ee
This implies that  $k_L(\tau)^{-1}k_L(0)\in \bt$ and
$\phi_L(\tau)=\phi_L(0)$,
 where $\bt$ is the maximal torus of $G_0$.
From this and the other equation (3.15), we finally
obtain
\be k_L(\tau)=k_L(0)\exp{[-\U_0(\phi_L(0))\tau]}.\ee
The only thing that changes in the treatment of the right
model space $M_R$ is the
fact that $\phi_R(0)\in \A^0_-$. The solution
of the right dynamical system is
\be k_R(\tau)=k_R(0)\exp{[-\U_0(\phi_R(0))\tau]}.\ee
We glue the left and right system by identifying $\phi_L=-\phi_R$
which gives the standard
geodesical motion on the group manifold
\be k(\tau)=k_L(\tau)k_R^{-1}(\tau)=k_L(0)
\exp{[-2\U_0(\phi_L(0))\tau]}k_R^{-1}(0).\ee
The symplectic form $d\theta_L$ can be easily inverted to give
the Poisson bracket on the model space.
Although this calculation was already detailed in the
literature \cite{AT}, we shall repeat
it here as the simplest prototype of several similar but
technically more involved computations
that we shall be doing later on.

Recall that the Killing-Cartan form $(.,.)_{\G_0^\bc}$ on
$\G_0^\bc$ is normalized in such a way
that the square of the length of the longest root is
equal to two. We pick an orthonormal
basis $H^\mu\in i\T$ in the Cartan subalgebra
$\T^\bc$ of $\G_0^\bc$ with respect to
the Killing Cartan
form   $(.,.)_{\G_0^\bc}$. Note that the elements
$H^\mu$ are Hermitian hence they are not the
elements of the  Lie algebra $\T$ of the maximal torus $\bt$.
Consider the root space decomposition of $\G_0^\bc$:
\be \G_0^\bc=\T^\bc\bigoplus(\oplus_{\al\in\Phi}\bc E^\al),\ee
where $\al$ runs over the space $\Phi$ of all roots
$\al\in\T^{\bc *}$. The step generators $E^\al$
fulfil
\be [H^\mu,E^\al]=\al(H^\mu)E^\al,\quad(E^\al)^\dagger=
E^{-\al};\ee
\be \quad [E^\al,E^{-\al}]=\al^{\vee},\quad
[\al^\vee,E^{\pm\al}]=\pm 2E^{\pm\al},\quad
(E^\al,E^{-\al})_{\G_0^\bc}=
{2\over \vert \al\vert^2}.\ee
The element $\al^\vee\in i\T$ is called the coroot of
 the root $\al$.
Thus the basis of the {\it complex} vector space
$\G_0^\bc$ is $(H^\mu,E^\al)$, $\al\in\Phi$.
The corresponding dual basis of $(\G_0^\bc)^*$ will
be denoted as $(h_\mu,e_\al)$.

We want to invert the symplectic form on the model
space $M_L$ of the {\it compact real form}
$G_0$ of the simple complex group $G_0^\bc$. For
this we need a basis on the Lie algebra $\G_0$.
We can construct such a basis in a canonical way
from the basis $(H^\mu,E^\al)$ on $\G_0^\bc$ .
Set
\be T^\mu=iH^\mu,\quad B^\al=
{i\over \od}(E^\al+E^{-\al}),\quad C^\al=
{1\over \od}(E^\al-E^{-\al}).\ee
The set $(T^\mu, B^\al, C^\al)$,  is an orthogonal
basis of the {\it real} vector space $\G_0$
with respect to $(.,.)_{\G_0}$.
Note that $\al$ runs only over the positive roots in this context!
The dual basis of $\G_0^*$ will be denoted as $(t_\mu,b_\al,c_\al)$.
Using the relation
\be k^{-1}dk=L^*_{k^{-1}}t_\mu\otimes T^\mu+
\sum_{\al\in\Phi_+}(L^*_{k^{-1}}b_\al\otimes B^\al+
L^*_{k^{-1}}c_\al\otimes C^\al),\ee
 we can write down the symplectic form $d\theta_L$ on $M_L$ as
$$ d\theta_L=\la d\phi \stw k^{-1}dk\ra -
\la \phi,k^{-1}dk\w k^{-1}dk\ra=$$
\be = da^\mu\w L^*_{k^{-1}}t_\mu +\sum_{\al\in\Phi_+}\la
 \phi,i\al^\vee\ra L^*_{k^{-1}}b_\al\w
 L^*_{k^{-1}}c_\al.\ee
Here $a^\mu$'s are defined by the expansion $\phi=a^\mu t_\mu$.
In deriving (3.26), we have used the  commutation relation
\be [B^\al,C^\al]=-i\al^\vee.\ee
To make the formulas less cumbersome,
 we have also suppressed the index $L$ on $\phi_L$ and $k_L$.

It is now very easy to invert the symplectic form $d\theta$.
The corresponding Poisson tensor
reads
\be \Pi_L=  L_{k*}T^\mu \w {\partial\over \partial a^\mu}
-\sum_{\al\in\Phi_+}
{1\over \la \phi,i\al^\vee\ra} L_{k*}B^\al\w L_{k*}C^\al.\ee

It is useful to give an explicit formula for the Poisson
 brackets of functions
 that can be obtained
as matrix elements of representations of the group $G_0$.
Consider two finite-dimensional
representations $\rho_i:G_0\to EndV_0$, $i=1,2$. The matrix element of
 the representation can be obtained
as the function $\la w^*,\rho_i(k) v\ra$, where
 $v\in V_0$ and $w^*\in V^*_0$. The Poisson bracket of two such
functions then reads
$$ \{\la w_1^*,\rho_1(k) v_1\ra,\la w_2^*,\rho_2(k) v_2\ra\}_{M_L}
=$$\be=\la (w^*_1\otimes w^*_2),
(\rho_1(k)\otimes\rho_2(k))(\rho_1
\otimes\rho_2)(r_0(a))(v_1\otimes v_2)\ra,\ee
where
$$ r_0(a^\mu)=$$\be = \sum_{\al\in\Phi_+}
{-1\over \la \phi,i\al^\vee\ra}
(B^\al\otimes C^\al-C^\al\otimes B^\al)=
\sum_{\al\in\Phi_+}{i\vert \al\vert^2\over
2a^\mu\la \al,H^\mu\ra}
E^\al\otimes E^{-\al} .\ee
The last equality in (3.30) follows from (3.24) and from
 the well-known relations
\be \al^\vee={2\over \vert \al\vert^2}\la \al,H^\mu\ra
H^\mu,\quad {\rm or}\quad
i\al^\vee={2\over \vert \al\vert^2}\la \al,H^\mu\ra T^\mu.\ee
Note that $r_0(a^\mu)$ is an $a^\mu$-dependent element of
$\G_0\w\G_0$; it is called the
 dynamical $r$-matrix.
It is to be contrasted with the standard $r$-matrix (cf. (1.9))
which does not depend on $a^\mu$.
Both standard and dynamical $r$-matrices have to satisfy some
 consistency conditions
if  the Poisson brackets based on them  are to satisfy the Jacobi
 identities.
Those conditions
are called, respectively, the   Yang-Baxter and the dynamical
Yang-Baxter equations.
We do not worry about the Jacobi identity here because we know a
priori that the symplectic
form $d\theta_L$ is closed.

Physicists use the so called matrix Poisson brackets
(cf. e.g. \cite{Ba,Fad})
in order to make the expressions like (3.29)
more transparent. For simplicity, let us consider the case
where $\rho_1=\rho_2$ and choose some
basis of the  representation space $V_0$. Then the Poisson bracket
of two matrix valued functions
$A_{ij}$ and $B_{kl}$ is written as
\be \{A_{ij},B_{kl}\}\equiv \{A \stackrel{\otimes}{,} B\}_{ik,jl}.\ee
With such a notation, we can write the brackets  (3.29) of the
matrix elements in the following
matrix form:
\be \{\rho(k)\stackrel{\otimes}{,}\rho(k)\}_{M_L}=
(\rho(k)\otimes \rho(k))\rho(r_0(a^\mu)).\ee
Even more often, people use a notation where the dependence
on the representation $\rho$
is explicitely suppressed but tacitly assumed, i.e.
\be \{k\ptp k\}_{M_L}=(k\otimes k)r_0(a^\mu).\ee
The Poisson bracket between the $k$ and $a$ variable can also
be written in the matrix form
as follows
\be \{k,a^\mu\}_{M_L}= kT^\mu.\ee
 We complete this section with the commutation relation
of the moment maps
 generating the left action of
$G_0$ on $M_L=G_0\times \A^0_+$.
 The symplectic potential $\theta_L$
(hence the symplectic form $d\theta_L$)  is clearly
invariant with respect to the
left multiplication
by any $k_0\in G_0$. Consider the   infinitesimal vector
field $V=R_{k_L*}T$ on $M_L$ corresponding
to the left action of a generator $T\in\G_0$.
 As usual (cf. also (4.87)), the corresponding moment map
$\la M,T\ra$ is defined
 by the relation
\be  -i_V d\theta_L\equiv d\theta_L(.,V)=d\la M,T\ra.\ee
The invariance of the symplectic potential $\theta_L$ means
the vanishing of its Lie derivative
with respect to $V$. In other words:
\be (i_Vd+di_V)\theta_L=0.\ee
From this relation, it immediately follows that
\be \la M,T\ra =i_V\theta_L=\la \phi_L, L_{k_L^{-1}*}R_{k_L*}T\ra =
\la Coad_{k_L}\phi_L,T\ra=
\la\beta_L(k_L\phi_L),T\ra.\ee
 Recall that $\beta_L(K)$ is the $\G_0^*$-valued map defined
by the decomposition
$K=\beta_L(K) g_R(K)$.
From (3.36) , one can  immediately  infer:
\be \Pi_L(.,d\la M,T\ra)=R_{k*}T.\ee
Eqs. (3.38) and (3.39) imply that for any function
$f(k_L,\phi_L)$ on $M_L$ it holds
\be \{f(k_L,\Phi_L),\la \beta_L(k_L\phi_L),T\ra\}_{M_L}=
\la\nabla_{G_0}^L f,T\ra,\ee
where the differential operator $\nabla_{G_0}^L$ is
defined in (7.54).
In particular, by remarking that $\beta_L(k_L\phi_L)=
k_L\phi_Lk_L^{-1}$, we obtain
\be \{\la \beta_L(k_L\phi_L),x\ra,\la \beta_L(k_L\phi_L), y\ra\}_{M_L}=
\la \beta_L(k_L\phi_L),[x,y]\ra,\quad
x,y\in\G_0.\ee
Of course, the same relation can be directly obtained from the
 fact that $\beta_L(k_L\phi_L)$
is the moment map generating the left action of $\G_0$ on $M_L$.

\section{Chiral decomposition of the master model}
\subsection{Affine Cartan decomposition}

 Usually people derive the  standard WZW left-right decomposition
by using the
equations of motion of the full WZW theory \cite{FG}. The solutions
 of these equations of motion
split into left and right movers. Because the phase space of a
dynamical system can be
identified with its space of solutions, one infers that the phase
space itself can be
split in its chiral parts. The corresponding symplectic forms on
the chiral parts
have been derived by Gaw\c edzki \cite{Gaw}.

Here we shall show how the standard chiral WZW
dynamics \cite{Gaw} emerges from the   perspective of the master
model (1.1).
 We  does not start with the equations
of motion. Instead, we shall consider a Cartan-like decomposition
of the cotangent bundle
of the centrally biextended loop group $\ti G$. This will give the
 left-right splitting
without the use  of the field equations.

Denote $\T$ the Lie algebra of the maximal torus $\bt$ of the
simple compact group
 $G_0$. Clearly,
$\T$ can be interpreted also as the subalgebra of the loop group
algebra $L\G_0$ consisting
of the constant maps from $S^1$ into $\T$. In what follows,
we shall use the same notation
for $\T$ being the subalgebra of $\G_0$ or of $L\G_0$. Now we
consider following subalgebra
of $\ti\G$ :
\be \ti\T=\br\ti T^0+\ti\iota(\T)+\br \ti T^\infty.\ee
Their elements are triples $(iX,\xi_0,ix)$, $X,x\in\br$,
$\xi_0\in \G_0\subset L\G_0$
in the terminology of Section 2.1.
 The subalgebra $\ti\T$ can be  mapped
by $\ti\Upsilon^{-1}$ to the subgroup$\ti\U^{-1}(\ti\T)$ of
$T^*\ti G$, since $\ti\G^*$
can be identified with the cotangent space at the unit element
of the group $\ti G $.
Here, as usual, the identification map $\ti\U:\ti\G^*\to\ti\G$ is induced
by the invariant bilinear
form  (2.10) on $\ti\G$.

\vskip1pc
\noindent{\bf Definition 3.4:} $\ti\U^{-1}(\ti\T)\equiv\ti \A$
is called the  Cartan subgroup
of the cotangent space $T^*_{\ti e}\ti G$ at the unit element
$\ti e$ of $\ti G$.
In the terminology  of Section 2.1, their elements are triples
$\ti\phi =(a^0,\phi,a^\infty)^*$,
where $\phi\in\ti\U^{-1}(\T)$, $a^0,a^\infty\in\br$.
Of course, $\ti A$  can be also interpreted as the subalgebra
of the Lie algebra
$\ttd$ of the group $\ttD=T^*\ti G$.  We shall also define two
subalgebras of $\ti \A$
denoted $\hat\A$ and $\A$, the former is spanned by elements
having $a^0=0$ and the
latter by those having $a^0=a^\infty=0$.
\vskip1pc
\noindent Consider  now subgroups $Norm(\ti A)\subset\hat G$
and $Cent(\ti A)\subset \hat G$ given by
\be Norm(\ti \A)=\{\hat w\in\hat G, \wti{Coad}_{\hat w}\ti\A\in\ti\A\};\ee
\be Cent(\ti \A)=
\{\hat w\in\hat G, \wti{Coad}_{\hat w}\ti\phi=\ti \phi
 \quad {\rm if}~~ \ti \phi\in \ti \A\}.\ee
Clearly, $Cent(\ti \A)$ is a normal subgroup of $Norm(\ti \A)$.
\vskip1pc
\noindent{\bf Definition 3.5:} Affine Weyl group $\ti W$ is the factor group
$Norm(\ti \A)/Cent(\ti \A)$.
\vskip1pc
\rem : We shall see soon that the group $Cent(\ti \A)$ is nothing
but the direct product
$\bt\times U(1)$, where $\bt$ is the maximal torus of $G_0$ and
$U(1)$ is the central circle subgroup of $\hat G=\wh{LG_0}$.It
 is important to realize
in this context that the circle bundle over $\bt\subset LG_0$
is trivial hence $\bt$ {\it can}
be embedded in $\hat G$.
\vskip1pc

\noindent The coadjoint action of $\hat G$ on $\ti \A$  is given
by the formula (2.23):
$$ \widetilde {Coad}_{\hat g}\ti\phi =
\widetilde {Coad}_{\hat g}(a^0, \phi, a^\infty)^*=$$\be =
( a^0+\la \phi, g^{-1}\d g\ra+\jp  a^\infty(g^{-1}
\d g,g^{-1}\d g)_\G,Coad_g\phi + a^\infty
\U^{-1}(\d gg^{-1}), a^\infty)^*
.\ee
 Consider now an element $h(\sigma)=e^{iv\si}$
from the coroot group $Hom(U(1),\bt)$. We have
$$ \widetilde {Coad}_{\hat h}(a^0, \phi,a^\infty)^*=$$\be =
(a^0+\la \phi,  iv\ra+\jp a^\infty( iv,iv)_\G,\phi +a^\infty
\U^{-1}( iv),a^\infty)^*
.\ee
Here $\hat h$ is some $\pi^{-1}$-lift of $h(\si)$ into $\hat G$
and $v\in \H(=i\T)$ is an
$\si$-independent element
  called the coroot
which corresponds to the element $h(\si)$ in

\noindent $Hom(U(1),\bt)$.
This correspondence is clearly one-to-one and for this reason
the coroot group $Hom(U(1),\bt)$
is often viewed as the coroot lattice in $i\T$ or in $\T$.

The inspection of the formulae (3.45) and (3.46) tells us what
is the affine Weyl group.
 It is   the semidirect product
of the standard Weyl group of $G_0$ (which can be also naturally
 embedded
in $\hat G$) and of the coroot group $Hom(U(1),\bt)$.
By construction, the affine Weyl group acts on $\ti \A$.
Elements of $\ti W$ can be represented
by the elements of $\hat G$. For the elements of the ordinary Weyl
 group $W$, standard representation
by the elements  $G_0$ can be chosen. For the elements of the
coroot lattice, the representation is
evident since every element of $Hom(U(1), \bt)$ can be viewed
by definition as the element of the
loop group $L\bt$.
By the way, the fact that $Cent(\hat A)=
\bt\times U(1)$ follows directly from the formulae (3.45) and (3.46).

We immediately realize from (3.46), that the affine Weyl group
acts not only on $\ti \A$,
but also on $\hat\A$ and $\A$. However, in the latter case the
action depends on $A^\infty$
as on the parameter.
  The fundamental domains of this action on
$\A$
are called alcoves. Consider the decomposition of the positive
Weyl chamber into alcoves.
The alcove attached to the origin (zero) of this positive Weyl
chamber
is referred to as the fundamental alcove $\A^{a^\infty}_+$. It
clearly depends on $a^\infty$.
Now an element from $\ti \A$ of the form
$(a^0, \phi ,a^\infty)^*$ is said to be in
$\ti \A_+$, if $\phi$ is in the fundamental alcove $\A^{a^\infty}_+$.
  Hoping not to create too
 much confusion,
we shall call $\ti \A_+$  the fundamental alcove, too.

\vskip1pc
\noindent After this preliminary discussion we can now state
the important theorem.
\vskip1pc
\noindent {\bf Theorem 3.6:} Every element $\ti K\in T^*\ti G$ can
be decomposed
as
\be \ti K=\ti k_L\ti \phi \ti k_R^{-1}, \quad
\ti k_{L,R}\in\ti G,\quad \ti \phi\in\ti \A_+.\ee
The ambiguity of the decomposition is given by the simultaneous
 right multiplication of $\ti k_L$
and $\ti k_R$  by the same element of $Cent(\ti A)\times
\br_S=\exp{\ti\T}$.
\vskip1pc
\pro :  Obviously, we must  prove that every element of $\ti \G^*$
can be connected to some
element in $\ti A_+$ by the coadjoint action of $\ti G$.
In other words, for every
$(C,\gamma,c)^*\in
\ti\G^*$ it exists
$\ti k_L\in\ti G$ and $(a^0, \phi ,a^\infty)^*\in
\ti A_+$ such that
\be\widetilde {Coad}_{\ti k_L}(a^0, \phi ,a^\infty)^*=
(C,\gamma,c)^*.\ee
In fact, it turns out that already  $\hat k_L\in\hat G$
does the job. Indeed, we have
$$=\widetilde {Coad}_{\hat k_L}(a^0, \phi ,a^\infty)^*=$$
\be
( a^0+\la \phi, k_L^{-1}\d k_L\ra+\jp  a^\infty(k_L^{-1}
\d k_L,k_L^{-1}\d k_L)_\G,
Coad_{k_L}\phi + a^\infty
\U^{-1}(\d k_L k_L^{-1}), a^\infty)^*,\ee
where we remind our convention $k_L=\pi(\hat k_L)$ if both
$\hat k_L$ and $k_L$ are present in the
same formula.
 Thus we have to show that every $\ga$ can be written
as
\be
\U(\ga) =k_L\U(\phi)k_L^{-1}+a^\infty\partial_\si k_L k_L^{-1},\ee
where $\phi$ is in the  fundamental alcove $\A^{a^\infty}_+$. Define
\be V(\si)=\stackrel{\leftarrow}{P}\exp{\int_0^\si {\U(\ga)(\si)
\over a^\infty}d\si}.\ee
The monodromy $V(2\pi)$ is the element of the compact group $G_0$,
 hence it can be diagonalized
\cite{BT2,FG} as
\be V(2\pi)=k_0e^{2\pi\rho} k_0^{-1},\ee
where $k_0$ is in $G_0$ and $\rho$ in the alcove $\U(\A^1_+)$.
Define now $k_L$ and $\phi$  as follows
\be k_L(\si)=V(\si) k_0\exp{(-{\rho \si })},
\quad \phi=a^\infty\U^{-1}(\rho).\ee
It can be easily check that the pair $(k_L,\phi)$ defined
in this way satisfies
(3.50).
Thus we see that the elements $\ti k_L$, $\ti k_R$ and $\ti \phi$
from the statement of the theorem
always exist, moreover, $\ti\phi$ is unique. It then easily
follows that ambiguity of the choice
of $\ti k_L$ and  $\ti k_R$ is given by the simultaneous
right multiplication by an element
from $ \exp{\ti\T}$.

 The theorem is proved.

\rightline{\#}

\subsection{Affine model space}
We wish to construct the phase space of the loop group chiral WZW model.
 The first part
of the exposition of this section will follow the spirit of the
Section 3.1. Indeed,
we shall equip the  affine model space $\ti M_L=\ti G\times \ti A_+$
 with a symplectic
structure by taking the pull back of the canonical symplectic form on
 the cotangent
bundle $T^*\ti G$ of the {\it centrally biextended} loop group. We
shall
also write down the natural Hamiltonian thus constructing the chiral
 geodesical
model on the affine Kac-Moody group $\ti G$. Then
 we shall make  steps
which are not rooted in Section 3.1; namely, we perform the symplectic
 reduction
of that chiral master model (down to the chiral WZW model).

Recall from section 7.2  that
the symplectic potential $\ti \theta$ on $T^*\ti G$
can be   simply expressed in the right trivialization
$\ti K=\ti\beta_L \ti g$
as
\be \ti \theta =\la \ti \beta_L,d\ti g \ti g^{-1}\ra.\ee
The dynamical system characterized by the symplactic form
$d\ti\theta$ and by the
Hamiltonian
\be\ti H(\ti K)=
-{1\over\k}(\ti\beta_L(\ti K),\ti\beta_L(\ti K))_{\ti\G^*}\ee
is nothing but the master model (1.1).

Consider the  affine model space  $\ti M_L=\ti G\times \ti \A_+$.
Its elements are couples
 $(\ti k_L,\ti \phi_L)$ and it is clearly
the
 submanifold of $T^*\ti G$. We can pullback the  symplectic
potential $\ti\theta$ on $T^*\ti G$
to $\ti M_L$
by the map $(\ti k_L,\ti \phi_L)\to \ti k_L\ti \phi_L\in T^*\ti G$,
where the group multiplication
law is considered in the sense of $T^*\ti G$.
The result is clearly
\be \ti\theta_L=\la \ti \phi_L, \ti k_L^{-1}d\ti k_L\ra.\ee
 The form $d\ti\theta_L$ on $\ti G\times \ti \A_+$ wil turn out to
be non-nondegenerate,
hence it defines a symplectic structure.

We have seen that we can obtain the symplectic structure on the
affine model space by the
simple pullback
of the canonical symplectic form on $T^*\ti G$. We can show with the help
of the affine Cartan decomposition that a sort of
the "inverse" procedure is also
possible. Indeed, consider the  direct product $\ti M_L\times
\ti M_R$ of two copies of the model
 space
$\ti M_L= \ti G \times \ti \A_+$  and $\ti M_L=\ti G\times \ti \A_-$,
where $\ti \A_-=-\ti \A_+$.
Equip the manifold $\ti M_L\times \ti M_R$
with  a symplectic form
\be \ti\omega_{L\times R}=d\ti\theta_L +d\ti\theta_R=
d\la \ti \phi_L,\ti k_L^{-1}d\ti k_L\ra +
d\la \ti \phi_R,\ti k_R^{-1}d\ti k_R\ra.\ee
The cotangent bundle $T^*\ti G$ with its canonical symplectic structure
$\ti\omega=d\ti\theta$ can be
obtained by the appropriate  symplectic reduction of the symplectic
manifold
$(\ti M_L\times \ti  M_R,\ti\omega_{L\times R})$ induced by
equating $\ti \phi_L+\ti \phi_R=0$.
The argument proving this statement is step by step identical
to the finite dimensional
argument of Section 3.2 and we shall not repeat it here.

We have just shown that the symplectic structure of the geodesical
model on $\ti G$
can be obtained by the symplectic reduction of the product of two
 affine model
spaces. It turns out that also the Hamiltonian on $T^*\ti G$
 can be descended
from a Hamiltonian on $\ti M_L\times \ti M_R$. The latter is given by
 \be \ti H_{L\times R}=\ti H_L+\ti H_R=
-{1\over 2\k}(\ti \phi_L,\ti \phi_L)_{\ti\G^*}-
{1\over 2\k}(\ti \phi_R,\ti \phi_R)_{\ti\G^*}.\ee
Here $(.,.)_{\ti\G^*}$ is   the form (2.22).

The symplectic reduction from $\ti M_L\times \ti M_R$ to $T^*\ti G$
 is governed
by the  moment maps $\ti \phi_L+\ti \phi_R$. It generates the
 simultaneous right action
of the group $\exp{\ti\T}\equiv \bt\times U(1)\times \br_S$ on
$\ti k_L$ and $\ti k_R$.
The Hamiltonian $\ti H_{L\times R}$
is invariant with respect to this action hence it defines the
Hamiltonian $\ti H$ on
$T^*\ti G$ given by
$$ \ti H=-{1\over 2\k}(\wti{Coad}_{\ti k_L}\ti \phi_L,
\wti{Coad}_{\ti k_L}\ti \phi_L)_{\ti\G^*}-
{1\over 2\k}(\wti{Coad}_{\ti k_L}\ti \phi_L,
\wti{Coad}_{\ti k_L}\ti \phi_L)_{\ti\G^*}=$$
\be =-{1\over \k}(\ti\beta_L(\ti k_L\ti\phi\ti k_R^{-1}),
\ti\beta_L(\ti k_L\ti\phi\ti k_R^{-1}))_{\ti\G^*}
=-{1\over \k}(\ti \b_L(\ti K),\ti \b_L(\ti K))_{\ti\G^*}.\ee
Thus we have recovered the Hamiltonian $\ti H$ given by the formula (3.55).

In what follows, we are therefore going to study the chiral model
 on $\ti M_L$
given by the action
\be \ti S_L(\ti k_L,\ti \phi_L)=\int d\tau[\la
\ti \phi_L,\ti k_L^{-1}{d\over d\tau}\ti k_L\ra
+{1\over 2\k}(\ti \phi_L,\ti \phi_L)_{\ti\G^*}].\ee
So far we have learned that the master model (1.1) on $\ti G$
 admits the chiral
decomposition into two chiral models (3.60). By this we mean
that it can be defined
by the symplectic reduction of the model defined on
$\ti M_L\times \ti M_R$
and  characterized by the symplectic form $\ti\om_{L\times R}$
and by the Hamiltonian
$\ti H_{L\times R}$. Combining this fact  with the results of
Section 2.2, we  learn
  that the standard WZW model can be produced  by glueing
the two models (3.60) and then performing the symplectic reduction.
 Next we shall
show that we arrive at the same result (WZW model) if we first perform
a simple symplectic reduction at the chiral level (3.60)    and then glue
two such reduced (chiral WZW) models.

\subsection{Chiral reduction to the first floor}
In order to perform the first step of the reduction, we
should evaluate the standard
(Abelian) moment maps generating the left action of
$\ti G$ on $\ti M_L$.
The symplectic potential $\ti\theta_L$
(hence the symplectic form $d\ti\theta_L$)  is clearly
invariant with respect to the
left multiplication
by any $\ti k_0\in \ti G$. Consider an  infinitesimal vector field
 $\ti V=R_{\ti k_L*}\ti T$ on
$\ti M_L$ corresponding
to the left action of a generator $\ti T\in\ti G$. As usual,
the corresponding moment map
$\la \ti M_, \ti T\ra $ is defined
 by the relation
\be  -i_{\ti V} d\ti\theta_L\equiv d\ti\theta_L(.,\ti V)=
d\la \ti M, \ti T\ra.\ee
The invariance of the symplectic potential $\ti\theta_L$
means the vanishing of its Lie derivative
with respect to $\ti V$. In other words:
\be (i_{\ti V}d+di_{\ti V})\ti\theta_L=0.\ee
From this relation, it immediately follows that
\be  \la \ti M, \ti T\ra =i_{\ti V}\ti\theta_L=
\la \ti \phi_L, L_{\ti k_L^{-1}*}R_{\ti k_L*}\ti T\ra =
\la \wti{Coad}_{\ti k_L}\ti \phi_L,\ti T\ra=
\la\ti \beta_L(\ti k_L\ti \phi_L),\ti T\ra.\ee
Since  $\ti \beta_L(\ti k_L\ti \phi_L)=
\wti{Coad}_{\ti k_L}\ti \phi_L$ is the moment map of the
standard Hamiltonian left action of $\ti G$ on $\ti M_L$, we
infer immediately the following
Poisson brackets of its coefficient functions on $\ti M_L$:
\be \{\la \ti\beta_L(\ti k_L\ti \phi_L),\ti x\ra,\la
\ti\beta_L(\ti k\ti \phi_L),
\ti y\ra\}_{\ti M_L}=
\la \ti\beta_L(\ti k\ti \phi_L),[\ti x,\ti y]\ra,\quad
\ti x,\ti y\in\ti\G.\ee
The particular case of these Poisson brackets will play the
important role in what follows:
 $$ \{\la \ti\beta_L(\ti k_L\ti \phi_L),(0,\xi,0)\ra,\la
\ti\beta_L(\ti k\ti \phi_L),
(0,\eta,0)\ra\}_{\ti M_L}$$
\be =\la \ti\beta_L(\ti k_L\ti \phi_L), (0,[\xi,\eta],0)\ra +
 a_L^\infty\rho(\xi,\eta),\quad
\xi,\eta\in\G.\ee
Here  $\rho$ is the loop group cocycle (7.16) and we have used
the fact that
\be \la \ti\beta_L(\ti k_L\ti a_L),\ti T^\infty\ra = a_L^\infty.\ee
$a_L^\infty$ is defined by the decomposition (for the notation
 cf. Section 3.2.1)
\be \ti \phi_L=(a^0_L,\phi_L,a^\infty_L)^*.\ee
Note also that $\ti k_L\ti\phi_L$ is the product in the sense of
$T^*\ti G$.

Now we parametrize $\ti k_L=u\hat k_L$, $u=\exp{s\ti T^0}$ and
rewrite the action (3.60)  as
\be \ti S_L(s,\hat k_L,\ti \phi_L)=\int d\tau[
\la \wti{Coad}_{\hat k_L}\ti \phi_L,
\ti T^0 \ra {ds\over d\tau}+\la \hat \phi_L,
\hat k_L^{-1}{d\over d\tau}\hat k_L\ra  +
{1\over 2\k}(\ti \phi_L,\ti \phi_L)_{\ti\G^*}].\ee
Of course, $\la \wti{Coad}_{\hat k_L}\ti \phi_L,
\ti T^0 \ra\equiv \ti\b_L^0$ is the moment map generating the
 infinitesimal left action of
$\ti T^0$ and we put
\be \hat \phi_L= (0,\phi_L,a_L^\infty)^*.\ee
Now we introduce the set of coordinates $(s,\ti\b_L^0,\hat k_L,
\hat \phi_L)$ and, using
the formula (2.23), we calculate
\be \ti\b_L^0=\la \wti{Coad}_{\hat k_L}\ti \phi_L,
\ti T^0 \ra=a_L^0+\la \phi_L, k_L^{-1}\d k_L\ra+
\jp a_L^\infty(k_L^{-1}\d k_L,k_L^{-1}\d k_L)_\G.\ee
As usual, here $k_L=\pi(\hat k_L)$.  The action (3.6.8)
finally becomes
$$\ti S_L(s,\ti\b_L^0,\hat k_L,\hat a_L)=\int d\tau[
 \ti\b_L^0{ds\over d\tau}-{1\over \k}\ti\b_L^0 a_L^\infty+$$
\be +\la \hat \phi_L,\hat k_L^{-1}{d\over d\tau}\hat k_L\ra  +
{1\over 2\k}( \phi_L, \phi_L)_{\G^*} +{a_L^\infty\over \k}\la
\phi_L, k_L^{-1}\d k_L\ra +
{(a_L^\infty)^2\over 2\k}(k_L^{-1}\d k_L,k_L^{-1}\d k_L)_\G.]\ee
 The Hamiltonian $\ti H_L=
-{1\over 2\k}(\ti \phi_L,\ti \phi_L)_{\ti\G^*}$ is obviously
invariant with respect to the action of $\ti T^0$ on $\ti M_L$,
 hence it Poisson-commutes
with the  moment map $\ti\b_L^0$ and we can consistently set
$\ti\b_L^0$ to zero in (3.71).
This constitutes the first step of the symplectic reduction.

\subsection{Ground floor: standard chiral WZW model}

Recall that the first step of the  symplectic reduction from
the chiral master model (3.60)
gave the result
$$\hat S_L(\hat k_L,\hat \phi_L)=
\int d\tau[\la \hat \phi_L,\hat k_L^{-1}{d\over d\tau}\hat k_L\ra +$$
\be
+{1\over 2\k}( \phi_L, \phi_L)_{\G^*} +{a_L^\infty\over \k}\la
\phi_L, k_L^{-1}\d k_L\ra +
{(a_L^\infty)^2\over 2\k}(k_L^{-1}\d k_L,k_L^{-1}\d k_L)_\G.]\ee
This first floor chiral theory is formulated on the phase space
$\hat M_L=\hat G\times
\hat \A_+$ that
we shall call the reduced affine model space.  Recall that the
elements of the
alcove $\hat\A_+$ have the form $(0,\phi_L,a^\infty)$, where
$\phi_L\in \A^{a^\infty}_+$.
The Hamiltonian $\hat H_L$ is given by the collection of terms
in (3.72) depending on $\k$.
On the other hand, the symplectic potential is independent on $\k$.

 In order to perform the second step of the
symplectic reduction, it will be convenient to express the
symplectic form
\be  \hat\omega_L=d\hat\theta_L = d\la \hat \phi_L,
\hat k_L^{-1}d \hat k_L\ra \ee
 in some basis of the Lie algebra $\hat\G=\wh{L\G_0}$.
The convenient basis can be obtained by injecting a basis of
 $\G=L\G_0$ into $\hG$ by the map
$\iota$ and adding the generator $\tF$. The basis of $L\G_0$,
in turn,  can be naturally
constructed from the canonical Cartan-Weyl basis of the
 complexified Lie algebra $L\G_0^\bc$.
The step generators of the latter are of the form
\be E^\al e^{in\si}\equiv E^\al_n, \quad n\in\bz,
\qquad H^\mu e^{in\si}\equiv H^\mu_n, \quad n\in\bz, n\neq 0,\ee
where $E^\al,H^\mu$ is the basis of $\G_0^\bc$ (cf. Section 3.1.3)
and $\si$ is the loop parameter.
 In what follows, we shall
often denote   a generic element of the set (3.74) as $E^{\hat\al}$,
where $\hal\in\hat\Phi$ stands
for the corresponding labels $(\al,n)$ or $(\mu,n\neq 0)$.
If $\hal$
is such that $\al,\mu$ are   arbitrary and $n>0$, or $\al\in\Phi_+$
 and $n=0$,
we say that $\hal\in\hat\Phi_+$. The basis of the Lie algebra
${L\G_0}$
can be then chosen as $(T^\mu,B^{\hal},C^{\hal})$,
 $\hat\al\in\hat\Phi_+$ where
\be T^\mu=iH^\mu,\quad B^{\hal}={i\over \od}(E^{\hal}+E^{-\hal}),\quad
C^{\hal}={1\over \od}(E^{\hal}-E^{-\hal}).\ee
Here by $-\hal$ we mean $(-\al,-n)$ for $\hal=(\al,n)$ and
 $(\mu,-n)$ for $\hal=(\mu,n)$.
 It turns out that  this basis is orthogonal with respect
to the form $(.,.)_\G$ defined in (7.1). The dual basis to (3.75)
will be denoted as
$t_\mu,b_{\hal},c_{\hal}$, $\hal\in\hat\Phi_+$.

Now we can finally write down the basis of $\hG=\wh{L\G_0}$
alluded above; it reads
\be \tF,\iota(T^\mu),\iota(B^{\hal}),\iota(C^{\hal}),
\quad \hal\in\hat\Phi_+.\ee
The dual basis is
\be \tf,\pi^*(t^\mu),\pi^*(b^{\hal}),\pi^*(c^{\hal}),
\quad \hal\in\hat\Phi_+,\ee
where the map $\pi^*:\G^*\to\hG^*$  is induced by the exact
sequence (2.2).

By using the general formula (7.14) and the explicit form (7.16)
of the cocycle, there is no problem to write down all commutation
 relations among the generators of
the basis (3.76). Here we shall write down only the commutators
relevant for further discussion,
or, in other words, only the commutators which are not
annihilated by all elements of
$Span(\tf,\pi^*(t_\mu))$.
Thus we have for every $\hal\in\hat\Phi_+$
\be [\iota(B^{\hal}),\iota(C^{\hal})]= -i\hal^\vee,\ee
where $\hal^\vee $ is the so-called affine coroot. It is given
explicitely
as follows
\be -i\hal^\vee= \iota(-i\al^\vee)-{2n\over \vert \al\vert^2}\tF,
\quad \hal=(\al,n),\qquad
-i\hal^\vee=-n\tF,\quad \hal=(\mu,n).\ee
Now we are ready to study the symplectic form $\hat\omega_L=
d\hat\theta_L$ on the
reduced   model space $\hat M_L$. In what follows, we shall
suppress the subscript $L$
 on the coordinates $(\hat k_L,\hat \phi_L)$ of the model space.
First of all, $\hat\omega_L$  can be written as
$$ \hat\omega_L=
\la d\hat \phi\stw \hat k^{-1}d\hat k\ra-\la \hat \phi,
\hat k^{-1}d\hat k\w
\hat k^{-1}d\hat k\ra =$$
$$ = da^\infty\w L^*_{\hat k^{-1}}\tf+d(a^\infty a^\mu)\w
L^*_{\hat k^{-1}}\pi^*(t_\mu)
 +$$\be +\sum_{\hal\in\hat\Phi_+}
\la \hat \phi,i\hal^\vee\ra L^*_{\hat k^{-1}}\pi^*(b_{\hal})\w
L^*_{\hat k^{-1}}\pi^*(c_{\hal}).\ee
Here we have set
\be \hat \phi=a^\infty \tf+ a^\infty a^\mu \pi^*(t_\mu)
 \ee
and used
$$ \hat k^{-1}d\hat k= L^*_{\hat k^{-1}}\tf\ot \tF+L^*_{\hat k^{-1}}
\pi^*(
t_\mu)\ot \iota(T^\mu)+$$\be + L^*_{\hat k^{-1}}\pi^*(b_{\hal})\ot
\iota(B^{\hal})+
L^*_{\hat k^{-1}}\pi^*(c_{\hal})\ot \iota(C^{\hal}).\ee
  Note  that the normalization is chosen in the way that
 $a^\mu$'s parametrize  the alcove
$\A^1_+$, i.e.
 $a^\mu t_\mu\in \A^1_+$. We have
\be \la \hat \phi ,i\hal^\vee\ra={2 a^\infty\over
\vert \al\vert^2}(n  +
\la\al,H^\mu\ra   a^\mu),
\quad \hal=(\al,n);
\ee
\be \la \hat \phi,i\hal^\vee\ra = n  a^\infty, \quad \hal=(\mu,n).\ee
This follows from the   well-known fact
\be \al^\vee={2\over \rn}\la \al,H^\mu\ra H^\mu.\ee
\vskip1pc
\teo ~{\bf 3.7}: The symplectic reduction of $\hat\omega_L$, induced by
setting the moment map
$ \la\hat\beta_L(\hat k\hat \phi),\tF\ra$ equal to some real
 number $\kappa$, gives
the chiral WZW symplectic form $\omega^{WZ}_L$ on the WZW model space
$M^{WZ}_L=G\times\A^1_+=LG_0\times\A^1_+$.
\vskip1pc
\pro :
First we observe that $ \la\hat\beta_L(\hat k\hat \phi),
\tF\ra =a^\infty$.
The form $\hat\omega_L$ restricted
to the surface $ a^\infty=\kappa$   becomes
$$\hat\omega_L|_{\hat a^\infty=\kappa}=
\k d a^\mu\w \pi^* L^*_{\pi(\hat k)^{-1}}
t_\mu+$$ $$ +\k \sum_{ \hal=(\al,n)\in\hat\Phi_+}{2\over
\vert \al\vert^2}(n  +\la\al,H^\mu\ra
a^\mu) \pi^* L^*_{\pi(\hat k)^{-1}}
b_{\hal}\w
\pi^* L^*_{\pi(\hat k)^{-1}}c_{\hal}+$$
\be +\k\sum_{\hal= (\mu,n)\in\hat\Phi_+}n
  \pi^* L^*_{\pi(\hat k)^{-1}}
b_{\hal}\w
\pi^* L^*_{\pi(\hat k)^{-1}}c_{\hal}.
\ee
In deriving this formula, we have used (3.83) and (3.84) and
also the fact that $\pi$ is the
group homomorphism, which implies that
$\pi^*L^*_{\pi(\hat k)^{-1}}=L^*_{\hat k^{-1}}\pi^*$.

Since we know that the moment map
$\la\hat\beta_L(\hat k\hat a),\tF\ra=  a^\infty$ generates
the central circle action,
we conclude immediately that the kernel of the form $\hat\omega_L$
restricted to $  a^\infty=\kappa$ is spanned by the
vectors $L_{\hat k*}\tF$.  This can be seen also directly from the
formula (3.86) since
the central circle does not act on the coordinates
 $a^\mu, a^\infty$ of the
reduced affine model space
$\hat M_L$. The restricted form (3.86) is therefore pullback
 of some two-form $\om'_L$ on
the manifold $M_L=G\times\A^1_+$ by the map
$\pi:(\hat k, a^\mu)\to(\pi(\hat k),  a^\mu)$.
It remains to find  this two-form $\om'_L$ on $M_L$.
The first term in (3.86) can be rewritten as
\be \k d a^\mu\w \pi^* L^*_{\pi(\hat k)^{-1}}t_\mu
 =\k\pi^*\biggl(d  a^\mu \w \la t_\mu,k^{-1}dk\ra\biggr) .\ee
Then we have
$$ \k \sum_{ \hal=(\al,n)\in\hat\Phi_+}{2\over \vert \al\vert^2}
\la\al,H^\mu\ra
 a^\mu \pi^* L^*_{\pi(\hat k)^{-1}}
b_{\hal}\w
\pi^* L^*_{\pi(\hat k)^{-1}}c_{\hal}=$$
\be =-\k\pi^*\biggl (\la  a^\mu t_\mu,  k^{-1}d k\w
 k^{-1}d k\ra\biggr ).\ee
Here we have used the commutation relations in the Lie algebra $L\G_0$:
\be [B^{\hal},C^{\hal}] =-i\al^\vee, \quad \hal=(\al,n);\ee
\be [B^{\hal},C^{\hal}]=0, \quad \hal=(\mu,n).\ee
By using the same commutation relations and the cocycle formula
 (7.16), we directly find
that the remaining term proportional to $\kappa$ is in fact equal to
$-{\kappa\over 2}\pi^* (k^{-1}d k\stw \d_\si(k^{-1}d k))_\G$.
Putting all together
\be  \om'_L=\k d  a^\mu \w \la t_\mu,k^{-1}dk\ra -\k\la  a^\mu t_\mu,
  k^{-1}d k\w
 k^{-1}d k\ra
 -{\kappa\over 2} (k^{-1}d k\w \d_\si(k^{-1}d k))_\G.\ee
Now we make a comparison with the formula  (4.5) of  \cite{FG}
to conclude that,
up to the $(2\pi)$ normalization (cf. Section 2.2.3),
our $\om'_L$ is indeed the symplectic form $\om^{WZ}_L$ of the
 chiral WZW model.

The theorem is proved.

\rightline{\#}
\vskip1pc
\noindent We conclude this paragraph by writing the formula for
the (doubly) reduced
Hamiltonian   on the WZW model space $M^{WZ}_L$. It  can be read off
from the formula (3.72) :
\be H^{WZ}_L =-{1\over 2\k}( \phi_L, \phi_L)_{\G^*} -
\la \phi_L, k_L^{-1}\d k_L\ra
-{\k\over 2}(k_L^{-1}\d k_L,k_L^{-1}\d k_L)_\G,\ee
where $\phi_L =\k a^\mu t_\mu$. This  coincides with the
Sugawara Hamiltonian
of the chiral WZW model as we shall see in Section 3.2.6. Having
obtained the correct
 symplectic form and Hamiltonian, we have indeed produced
 the standard chiral WZW theory   by the two-step
chiral symplectic reduction from the chiral master model (1.2).
The fact that
the full left-right WZW model can be obtained by glueing
two chiral WZW theories was explained e.g. in \cite{FG} and we
shall not repeat
this argument here.
\subsection{Affine dynamical $r$-matrix}
Our next task is to prepare the land for the
quasitriangular generalization
described later on in this paper. For this
we have to invert the chiral WZW symplectic form $\omega^{WZ}_L$.
We shall write it as follows
\be  \omega^{WZ}_L=
\k da^\mu\w  L^*_{ k^{-1}}t_\mu
   +\sum_{\hal\in\hat\Phi_+} \la \hat \phi_\k,i\hat{\al}^\vee\ra
 L^*_{ k^{-1}}
b_{\hal}\w
 L^*_{ k^{-1}}c_{\hal},\ee
 where \be \hat\phi_\k=\k\tf + \k a^\mu\pi^*(t_\mu).\ee
From here we find immediately the corresponding Poisson bivector
$\Pi^{WZ}_L$:
\be \Pi^{WZ}_L(k, a^\mu)= -{\partial\over \k\partial
 a^\mu}\w L_{k*}T^\mu
 -\sum_{\hal\in\hat\Phi_+}
{1\over \la \hat \phi_\k,i\hat{\al}^\vee\ra}
 L_{k*}B^{\hal}\w L_{k*}C^{\hal}.\ee
Let us calculate the Poisson bracket of certain special
 functions of the
variables $(k, a^\mu)\in G\times \A^1_+$. These functions are simply the
matrix elements in some representation of $LG_0$. The bivector formula
(3.95) then immediately implies  the following Poisson brackets
\be \{k\ptp k\}_{WZ}= (k\otimes k)\hat r_0(\hat\phi_\k),\ee
\be \{k,a^\mu\}_{WZ}=kT^\mu,\quad \{a^\mu,a^\nu\}_{WZ}=0,\ee
where
\be \hat r_0(\hat\phi_\k)=
\sum_{\hal\in\hat \Phi}
{i\over \la \hat \phi_\k,i\hat{\al}^\vee\ra }E^{\hal}\otimes E^{-\hal}.
 \ee
The brackets (3.96) and (3.97) characterize completely the Poisson
structure on the WZW model
space $M^{WZ}_L$.
Note also that   the summation in (3.98) is not restricted only
 to the positive roots.

The fundamental braiding relation (3.96) can be rewritten
in the representation
corresponding to the pointwise action of the loop group element
 on the $G_0$-representation
space $V_0$. In other words, consider a function on $LG_0$
  of the form
$\rho_{ij}^{\si'}(k)$, where $\rho$ is a matrix representation
 of the group $G_0$
and $\si'$ is some point on the loop. In  words, this function
is defined as follows:
take an element $(k,a^\mu)\in M^{WZ}_L$, forget about $a^\mu$,
consider the
 element $k(\si')$
of the group $G_0$ obtained by evaluation of $k$ at the point
$\si'$ and finally take
the matrix element ${ij}$ of the element $k(\si')$ in the
$G_0$-representation $\rho$.
If we recall the definition (3.75) of $B^{\hal}$ and $C^{\hal}$
in terms of $E^{\hal}$ and
the definition (3.74) of $E^{\hal}$ in terms of  $E^\al,H^\mu$
and  $e^{in\si}$,
we can directly
derive from (3.98)
\be\{k(\si)\ptp k(\si')\}_{WZ}=
(k(\si)\ot k(\si'))\hat r_0(\hat\phi_\k,\si-\si'),\ee
where the affine dynamical $r$-matrix is denoted as
$\hat r_0(\hat \phi_\k,\si-\si')$
and defined as
$$  \hat r_0(\hat \phi_\k,\si-\si')=i\sum_{\al\in\Phi,n\in\bz}
{\rn\over 2} {1\over \k(n +\la\al,H^\mu\ra
a^\mu)}E^{\al}\ot E^{-\al} \exp{in(\si-\si')}$$
\be
 +i\sum_{\mu,n\in\bz,n\neq 0}{1\over n \kappa}H^\mu\ot H^\mu
 \exp{in(\si-\si')}.\ee
It is important to note that the summation goes over all
roots $\al\in\Phi$, not only over
the positive ones. From (3.97), we can   also derive  the
following  bracket
\be \{ k(\si),a^\mu\}_{WZ}=  {1\over\k}k(\si)T^\mu.\ee
 It is simple  to sum up the Fourier series in (3.100). The result is
$$ \hat r_0(\hat \phi_\k,\si-\si')={1\over \k}{\rm Per}
(\si-\si'-\pi)\sum_\mu
H^\mu\ot H^\mu+$$
\be +\sum_{\al\in\Phi}
{\rn\over 2\k}{2\pi\over e^{-2\pi ia^\mu\la \al,H^\mu\ra}-1}
{\rm Per}(e^{-ia^\mu\la \al,H^\mu\ra(\si-\si')})E^{\al}\ot E^{-\al},\ee
where the notation Per($f(\si)$) means the function of $\si$ periodic
with the period $2\pi$ and defined as $f(\si)$ for $\si\in [0,2\pi]$.

\subsection{Vertex-IRF transformation and braiding relation}
The formula of the type (3.96) appears in the WZW literature
\cite {AS,Fad,Bl,CGO,CG} under the
name of the exchange (braiding) relation.
The reader might have noticed however, that our formula (3.99)
does not at all resemble e.g.
the braiding relation (26,31) of the reference \cite{CGO}. The
reason is that we have used
different coordinates on the WZW model space $M^{WZ}_L$.
In order to establish
the equivalence of our approach with that  of \cite{CGO},
 we must perform the so-called
classical vertex-IRF transformation (the terminology is
borrowed from \cite{FG}).

Consider a map $\si\to m(\si)\in G_0$ defined as
\be m(\si)=k(\si)\exp{(a^\mu\U(t_\mu)\si)  },\ee
where $(k,a^\mu)$ are the old coordinates of the WZW model
 space $M^{WZ}_L$.
Now we introduce the  new set of "monodromic" coordinates
$m(\si)$. The name
is motivated by the fact that $m(\si)$ is no longer a
single-valued function but it
developes a monodromy upon going around the circle of sigmas.
This monodromy is encoded
in  the variables $a^\mu$, hence $m(\si)$ encodes the
information about both $k$ and
$a^\mu$.

We wish to calculate the exchange relation (3.99)
in terms of the variables $m$.
 We shall use the following obvious matrix relation
$$\{AB\ptp CD\}=(A\ot 1)\{B\ptp C\}(1\ot D) + (A\ot C)\{B\ptp D\}+$$
\be
+\{A\ptp C\}(B\ot D)+ (1\ot C)\{A\ptp D\}(B\ot 1) \ee
and write
$$\{m(\si)\ptp m(\si')\}_{WZ}=\{k(\si)e^{a^\mu\U(t_\mu)\si}
 \ptp
 k(\si')e^{a^\mu\U(t_\mu)\si'} \}_{WZ}=$$
$$(m(\si)\ot m(\si'))\biggl[-{\si -\si'\over \kappa}
(H^\mu\ot H^\mu)+$$ $$ +
(e^{- a^\mu\U(t_\mu)\si} \ot e^{-a^\mu\U(t_\mu)\si'})
\hat r_0(\hat\phi_\k,\si-\si')(e^{ a^\mu\U(t_\mu)\si}
\ot e^{a^\mu\U(t_\mu)\si'})
 \biggr]\equiv $$
\be \equiv (m(\si)\ot m(\si'))B_0(\hat \phi_\k,\si-\si').\ee
We shall call $B_0(\hat \phi_\k,\si-\si')$ the quasiclassical
braiding matrix.  It is important to note
that the argument $\si$ of the braiding matrix $B_0(\hat\phi_\k,\si)$
 is the element of
$\br$ and not of $S^1$.  This is related to the fact that
the monodromic coordinate $m(\si)$
on $M_L$ is multi-valued from the point of view of $S^1$.
Considering $\si$ as
an element of $\br$ makes the quantity $m(\si)$ single-valued
and the Poisson bracket (3.105)
well-defined.

 Now combining (3.102) and (3.105), we immediately arrive
at (cf. \cite{CGO})
\be B_0(\hat \phi_\k,\si)=
-{\pi\over\kappa}\biggl[\eta(\si)(H^\mu\ot H^\mu)
-i\sum_\al{\rn\over 2}{\exp{(i
\pi \eta(\si)\la \al,H^\mu\ra a^\mu)}\over
\sin{(\pi \la \al,H^\mu\ra a^\mu)}}
E^\al\ot E^{-\al}\biggr].\ee
Here $\eta(\si)$ is the function defined by
\be \eta(\si)=2[{\si\over2\pi}]+1,\ee
where $[\si/2\pi]$ is the largest integer less than or equal
to ${\si\over 2\pi}$.

 It turns out
that important dynamical variables  can be particularly simply
expressed in terms of the new variables $m(\si)$. Before
 showing this, some discussion is needed
about the  moment maps generating the left $\hat G$-action on $\hat M_L$
and about their behaviour under the symplectic reduction leading
from the reduced affine model space $\hat M_L$
to the WZW model space $M^{WZ}_L$.

We already know that  the moment maps $\hat \beta_L(\hat k\phi)=
\wh{Coad}_{\hat k}\hat\phi$
generate
(via the Poisson bracket
on $\hat M_L$) the left action of the group $\hat G$ on $\hat M_L$.
In particular, the moment $\la \hat\beta_L,\tF\ra$ generates
the central circle action on $\hat M_L=\hat G\times \hat \A_+$.
This means that
\be \{\la \hat\b_L,\tF\ra,\la \hat\b_L,\iota(x)\ra\}_{\hat M_L}=0,
\quad x\in\G.\ee
It then follows that
\be \la \hat \beta_L(\hat k\hat \phi ),\iota(x)\ra=
\la\hat \beta_L(e^{s\tF}\hat k\hat \phi),\iota(x)\ra,\quad x\in\G;\ee
or, in other words,  $\la\hat\beta_L,\iota(x)\ra$ is invariant function
with respect to the central
circle action. As such, it  gives rise to some function on the space
of the central circle
orbits located at the submanifold  $ a^\infty=\kappa$ of the affine
 model space $\hat M_L$.
The latter space of orbits is nothing but the reduced model space
$M^{WZ}_L$ hence we conclude
that $\la\hat\beta_L,\iota(x)\ra$  can be interpreted as the honest
function on $M^{WZ}_L$.
 We denote it as $j_L^x(k,a^\mu)$.

Actually, the functions $j_L^x(k,a^\mu)$  are the "important dynamical
variables"
mentioned above. In fact they are nothing but the generators of
the chiral current algebra.
To see this, we calculate their Poisson brackets $\{j^x_L,j^y_L\}_{WZ}$.
 The computation
follows the general procedure of the  symplectic
reduction at the level of the Poisson brackets as described in the
Appendix 7.3.

Consider  a pair of functions $\phi_i$, $i=1,2$ on $M_L^{WZ}$. We
wish to calculate their
reduced Poisson bracket $\{\phi_1,\phi_2\}_{WZ}$.   In our
particular situation, the procedure works as follows: define two functions
$\hat\phi_i$
on $\hat M_L$ that fulfil
\be \hat\phi_i(\hat k,a^\mu,a^\infty=\k)=\phi_i(\pi(\hat k),a^\mu),
\quad \hat k\in\widehat{LG_0}.\ee
Calculate then the   Poisson bracket
$\{\hat\phi_1,\hat\phi_2\}_{\hat M_L}$ on $\hat M_L$.
It verifies
\be \{a^\infty,\{\hat\phi_1,\hat\phi_2\}_{\hat M_L}\}_{\hat M_L}=0\ee
as the simple consequence of the Jacobi identity and the central circle
invariance of
$\hat\phi_i$. This  means that it exists a function on $M_L^{WZ}$
denoted suggestively as $\{\phi_1,\phi_2\}_{WZ}$ which verifies
\be \{\hat\phi_1,\hat\phi_2\}_{\hat M_L}(\hat k,a^\mu,a^\infty=\k)
=\{\phi_1,\phi_2\}_{WZ}(\pi(\hat k),a^\mu).\ee
Needless to say, the function $\{\phi_1,\phi_2\}_{WZ}$ is the seeken
reduced Poisson bracket.

Now the function $\la \hat\b_L,\iota(x)\ra$ on $\hat M_L$ plays the
role of $\hat\phi_1$
for the function $\phi_1=j_L^x$ on $M_L^{WZ}$. But we know from (3.64)
and (2.7) that
\be \{\la \hat\b_L,\iota(x)\ra, \la \hat\b_L,\iota(y)\ra\}_{\hat M_L}
=\la \hat\b_L,\iota([x,y])\ra +\la \hat\b_L,\tF\ra \rho(x,y).\ee
Using the fact that $\la \hat\b_L,\tF\ra =a^\infty$ and the relation
(3.112), we obtain
immediately \be \{j^x_L,j^y_L\}_{WZ}=j_L^{[x,y]}+\k\rho(x,y).\ee
 This is nothing but the basic relation defining the chiral current
algebra.

Let as calculate the currents $j_L^x$ as explicit functions of $k,a^\mu$.
By using the formulas (3.38)  and (2.24), we infer
$$ \la \hat\beta_L(\hat k\hat\phi),\iota(x)\ra=
\la \wh{Coad}_{\hat k}(a^\infty \tf +a^\infty a^\mu\pi^*(t_\mu)),
\iota(x)\ra=$$
\be \la \pi^*(Coad_k(a^\infty a^\mu t_\mu)+a^\infty \U^{-1}
(\d_\si k k^{-1})),\iota(x)\ra,\ee
where $k=\pi(\hat k)$. Thus for the currents we obtain
\be j_L^x(k,a^\mu)=\k (a^\mu k\U(t_\mu)k^{-1} +\d_\si kk^{-1},x)_\G.\ee
This expression look quite complicated but it drastically simplifies in the
monodromic variables $m(\si)$:
\be j_L^x=
 \k(\d_\si mm^{-1},x)_\G.\ee
Moreover, also the Hamiltonian $H_L^{WZ}$ given by (3.92)  simplifies
 considerably:
\be H^{WZ}_L=-{1\over 2\k}(\k\d_\si mm^{-1},\k\d_\si mm^{-1})_{\G}.\ee
This  is the Sugawara formula expressing the chiral  WZW Hamiltonian
 solely in terms
of the Kac-Moody currents. The reader should not be confused by the
minus sign.
It is related to the negative definiteness of our bilinear form $(.,.)_\G$.

The monodromic variables are generally used in the study of the
standard WZW model.
It turns out, however, that viewing things from the points of view
of the variables
$k,a^\mu$ will be more insightful for seeking the quasitriangular
generalization
of the story.

Another
important Poisson bracket is  the following one
\be \{m(\si),(\kappa\d_\si m(\si')m^{-1}(\si'),x_0)_{\G_0}\}_{WZ}=x_0
m(\si)\delta(\si-\si'), 
\quad x_0\in\G_0.\ee
 It is the direct consequence of the  braiding relation (3.105)
and  it expresses the simple fact, that
even after the symplectic reduction the current $\k\d_\si mm^{-1}$
 continues to generate
the left action of the centrally extended loop group on the WZW
model space $M^{WZ}_L$.
 However,
the action of the generator of  the central circle is now trivial.
Upon the
quantization, the bracket (3.119)
   means
that  $m(\si)$ is the Kac-Moody primary field.

 Finally, as the illustration of the suitability
of the monodromic variables, we give   the explicit expression
 for the solution of the classical
field equations of the chiral WZW model.
 It turns out, that the time evolution of the coordinates $m(\si)$
is given by the following simple formula
\be [m(\si)](\tau)=m(\si-\tau).\ee
This fact can be derived  from  the braiding relation (3.105) and from
the Sugawara
formula (3.118) (because those two ingredients anyway characterize
completely
 the structure of the standard chiral WZW model). Here we shall
offer another derivation
based on the solution of the second floor chiral master  model (1.2).
Indeed,
varying the action (1.2), we find in full analogy with the
finite-dimensional
calculation of Section 3.1.3 that
 the classical solution of the second floor chiral model
\be \ti k(\tau)=\ti k_0
\exp{\biggl({-\ti\U(\ti \phi_0)\over\k}\tau\biggr)}.\ee
 Since we have performed the reduction with respect to the
{\it left} action of
the group $\br_S$, we have to cast $\ti k(\tau)$ as
\be \ti k(\tau)=e^{s(\tau)\ti T^0}  \hat k(\tau),\ee
and suppress $e^{s(\tau)\ti T^0}\in \br_S$  . This corresponds
to the first step of the reduction.
The result is
\be \hat k(\tau) =\biggl(e^{-{a^\infty\tau\over\k}\ti T^0}\hat k_0
e^{{a^\infty\tau\over\k}\ti T^0}\biggr)
e^{{1\over\k}(a^0\tF-a^\infty a^\mu \U(t_\mu))\tau},
  \ee
The reduction $a^\infty=\k$ to the ground floor gives
\be k(\tau)=\pi(\hat k(\tau))=k_0(\si-\tau)
e^{-a^\mu \U(t_\mu )\tau},\ee
where $\pi(\hat k_0)\equiv k_0$. Combining the definition (3.103) of
 the monodromic
variables with the last formula (3.124), we arrive directly at the
desired relation (3.120).

\chapter{Universal  quasitriangular WZW model}

In two preceding sections, we have described two important
dynamical systems. The first
one - the geodesical
model - was constructed for any Lie group $G$ possessing an
invariant symmetric non-degenerate
bilinear form on the Lie algebra $\G$ of $G$. The construction
of the second one - the WZW model -
necessitated moreover an existence of the central biextension
$\ti G$ of $G$.
The symplectic structures of both these models have been either
identified to
or derived from  the canonical symplectic
structure of the cotangent bundle $T^*G$ (for the geodesical model)
 and of $T^*\ti G$
(for the WZW model).

In this section, we are going to show that one can generalize both
 geodesical
and WZW models mentioned above. A recipe how   to do this consists
in a crucial
observation that $T^*G$ is the so called Heisenberg double of the
group $G$. The latter
is a certain group equipped with an additional structure that we
 shall describe in what
follows. There may exist many different Heisenberg doubles of a
given group $G$; for us it
is important that the geodesical model and the WZW one can be
constructed by using only
those properties of $T^*G$ that are shared also by all  others
Heisenberg doubles of $G$.
In particular, given a nontrivial Heisenberg double  of the
centrally biextended group $\ti G$,
we can define an associated WZW-like model.
We shall refer to the  latter as to the
universal  quasitriangular WZW model.

In order to keep this paper as selfcontained as possible, we are
going to give here
a  quick review  of  theory of Poisson-Lie
groups and of related various doubles of Lie groups.  The reader may
find somewhat inconvenient to burden the text also with demonstrations
 of the relevant
  standard  propositions. However, these demonstrations involve many
important facts, technical skills and computational tools that may
facilitate the understanding
of the present article.

\section{Poisson-Lie primer}
\subsection{The Drinfeld double}

A Poisson bracket on a Lie group manifold $G$ that is compatible with
the group  multiplication law is called the {\bf Poisson-Lie} bracket.
 Denote $\Delta:Fun(G)
\to Fun(G)\otimes Fun(G)$ the standard coproduct defined as
\be (\Delta F)(g,g')=F(gg'),\quad g,g'\in G,\quad F\in Fun(G).\ee
Then the condition of compatibility reads
\be \{\Delta F_1,\Delta F_2\}_{G\times G}=\Delta \{F_1,F_2\}_G.\ee
Here $\{.,.\}_{G\times G}$ is the direct product Poisson bracket on
$G\times G$
characterized by the condition
$$\{F_1(x)G_1(y),F_2(x)G_2(y)\}_{G\times G=}$$
\be=\{F_1(x),F_2(x)\}_G G_1(y)G_2(y)
+F_1(x)F_2(x)\{G_1(y),G_2(y)\}_G,\ee
where $x$ and $y$ are coordinates on the first and second copy of
 $G$ respectively.  Note that upon the quantization of the algebra of
functions on the group
manifold, we obtain the so called quantum group. The quantum version
of the Poisson-Lie
condition (4.2) then becomes the usual statement  in the theory
of Hopf algebras saying that  the coproduct is the algebra homomorphism.

On a given group $G$ one may have several inequivalent Poisson-Lie
brackets. In this paper,
we shall always be concerned with one privileged way of constructing
the Poisson-Lie
structures on $G$ that uses the concept of the Heisenberg  double of
$G$. Before
defining this notion, we have to recall respectively the definitions
of the
Manin and Drinfeld doubles.

So
the {\bf Manin} double of a Lie group $G$ is any Lie group $D$
 whose dimension is twice as big as that of $G$ and that fulfils
two other conditions:

\noindent 1) The double $D$ contains $G$ as its subgroup;

\noindent 2) The Lie algebra $\D$ of the double $D$ is equipped with
an invariant  symmetric
nondegenerate bilinear form $(.,.)_\D$ such that the Lie algebra
$\G$ of $G$ is isotropic;
i.e. $(\G,\G)_\D =0$.

Suppose now that the same group $D$ equipped with the form
$(.,.)_\D$ on $\D$
is the Manin double of two  Lie groups $G$ and $B$. We say that
$D$ is the {\bf Drinfeld} double of $G$ (and of $B$), if the Lie algebras
$\G$ and $\B$ of $G$ and $B$ linearly generate all the Lie algebra
$\D$ of $D$.
In other words:
\be \D=\G +\B,\quad (\G,\G)_\D=0,\quad (\B,\B)_\D=0.\ee
It is very important to realize that $\G+\B$ means here the direct
sum of two
vector spaces and not the direct sum of two Lie algebras.
The Lie bracket on $\D$ can be conveniently encoded in terms of the
Lie brackets
on $\G$ and $\B$ and of the form $(.,.)_\D$ as follows
\be [X+\al,Y+\beta]_\D=[X,Y]_\G+[\al,\beta]_\B + Coad_X\beta-Coad_Y\al
+Coad_\al Y-Coad_\beta X.\ee
Here $X,Y\in\G$ and $\al,\beta\in\B$.
As far as the coadjoint action is concerned, the elements
 $\al,\beta$ are viewed as the elements of $\G^*$ by the prescription
\be \la \al,X\ra =(\al,X)_\D,\ee
and similarly $X,Y$ are viewed as the elements of $\B^*$ under the
coadjoint action of $\B$.

\subsection{The Heisenberg double}

The group multiplication law in the Drinfeld double $D$ induces
smooth maps
${\cal M}_L:G\times B\to D$ and ${\cal M}_R:B\times G\to D$ given by
\be {\cal M}_L(g,b)=gb,\quad {\cal M}_R(b,g)=bg,
\quad g\in G,\quad b\in B.\ee
The  crucial fact underlying  all this article can be formulated
as the following theorem
\vskip1pc
\noindent {\bf Theorem 4.1} :   If the maps ${\cal M}_{L,R}$
defined above are bijective then
$D$ is a symplectic manifold and
the following expression defines the   symplectic form
 \be \omega=\jp(b_L^*\lambda_B\stackrel{\w}{,}g_R^*\rho_G)_\D+
\jp(b_R^*\rho_B\stackrel{\w}{,}g_L^*\lm_G)_\D.
\ee
The corresponding Poisson bracket reads
\be \{\phi,\psi\}_D=
\jp(\nl \phi,\R^*\nl \psi)_{\D^*}+\jp(\nr \phi,\R^*\nr \psi)_{\D^*}.\ee

\vskip1pc
\noindent {\bf Remark}: {\small The Poisson bracket (4.9) was
introduced by Semenov-Tian-Shansky
in \cite{ST2}. If
the maps ${\cal M}_{L,R}$ are  not bijective then (4.9) still defines
a Poisson bracket and
 the symplectic leaves of the corresponding Poisson structure
were described in \cite{AM}.
The bijectiveness of ${\cal M}_{L,R}$
is often referred to as the property that the group $D$ is
smoothly globally decomposable as
$D=GB$ and $D=BG$.}
\vskip1pc
\noindent Let us explain the symbols appearing in (4.8):
The  maps $b_L:D\to B$ and $g_R:D\to G$ are induced by the
decomposition $D=BG$
and $g_L:D\to G$ and $b_R:D\to B$     by  $D=GB$. The expression
$\lm_G$ ($\rho_G$) denotes
the left(right)invariant $\G$-valued Maurer-Cartan form on the
group $G$. Recall that
\be \lm_G(X_g)=L_{g^{-1}*}X_g,\quad \rho_G(X_g)=R_{g^{-1}*}X_g,
\quad X_g\in T_gG.\ee
Note that the forms $\lm_G$ and $\rho_G$ are often   written also as
\be \lm_G=g^{-1}dg,\quad \rho_G=dgg^{-1}.\ee
The notation used in (4.9) is as follows:
$(.,.)_{\D^*}$  is  the  bilinear form on the dual of $\D$ induced
by the (nondegenerate) bilinear form $(.,.)_\D$ and $\R^*:\D^*\to\D^*$
is the map dual to $\R:\D\to \D$. The latter is  given by
\be \R=Pr_\B-Pr_\G,\ee
where $Pr_\G$ ($Pr_\B$) is the projector on $\G$ ($\B$) with the
kernel $\B$ ($\G$).
Clearly, the decomposition $\D=\G+\B$ induces the corresponding
 decomposition
of the dual $\D^*=\G^*+\B^*$ and
\be \R^*=Pr_{\B^*}-Pr_{\G^*}.\ee
 Recall also the definitions (cf. (7.26),(7.27)
of the differential operators $\nl$,$\nr$  on $D$:
\be \nl:Fun(D)\to Fun(D)\otimes \D^*;
\qquad \nr:Fun(D)\to Fun(D)\otimes \D^*;\ee
\be \la \nl \phi,\al\ra (K)=\dto\phi(e^{s\al}K),\qquad
\la \nr \phi,\al\ra (K)=\dto\phi(K e^{s\al}).\ee
Here $\al\in\D$, $K\in D$ and $\phi\in Fun(D)$.

 It is useful to write the bracket (4.9)
in some basis $T^i,t_i; i=1,\dots, dimG$ of $\D$ where $T^i$'s
form the basis
of $\G$ and $t_i$'s the corresponding dual basis of $\B$. We obtain
\bea\{\phi,\psi\}_D=\jp\la\nl \phi, T^i\ra\la \nl \psi,t_i\ra
 -\jp\la \nl \phi,t_i\ra\la\nl \psi, T^i\ra
 \cr
 + \jp\la\nr \phi, T^i\ra\la \nr \psi,t_i\ra -
\jp\la \nr \phi,t_i\ra\la\nr \psi, T^i\ra,\eea
where the standard Einstein summation convention is used.
 By the duality of the basis $t_i$ with respect to the basis $T^i$
we mean that the following relation holds:
\be (t_i,T^j)_\D=\delta_i^j.\ee

\noindent
In order to prove the theorem 4.1, we need to handle in an efficient
way the Semenov-Tian-Shansky
symplectic form $\om$ given by (4.8). We  shall first prove the
following lemma that will
be used also for the proof of the theorem 4.6 of the next section.
\vskip1pc
\lem ~{\bf 4.2}: Consider a point $K\in D$ and four linear subspaces of the
tangent space $T_K D$ defined as
$S_L=L_{K*}\G$, $S_R=R_{K*}\G$, $\ti S_L=L_{K*}\B$ and $\ti S_R=
R_{K*}\B$. Let $\Pi_{L\ti R}$
be a projector on $\ti S_R$ with a kernel $S_L$ and $\Pi_{\ti L R}$
a projector on $S_R$
with a kernel $\ti S_L$. Then
\be \om(t,u)=(t,(\Pi_{\ti L R} -\Pi_{L\ti R})u)_\D,\ee
where  $t,u$ are arbitrary two vectors in the tangent space $T_K D$
 at the point $K\in D$
and the metric $(.,.)_\D$ at the point $K$ is defined by the right
or left transport
of the bilinear form $(.,.)_\D$ defined at the unit element $E\in D$.

\vskip1pc
\pro :
First we rewrite (4.8) as
\be \om=\jp(b_L^*\rho_B\stw \rho_D)_\D+\jp( b_R^*\lm_B\stw \lm_D)_\D,\ee
where $\lambda_D$($\rho_D$) is the left (right)
 invariant Maurer-Cartan form on the double $D$. In order to see that
(4.19) is correct, we write
$$ \jp(dKK^{-1}\stw db_Lb_L^{-1})_\D=$$
 \be =\jp(db_Lb_L^{-1}+b_Ldg_Rg_R^{-1}b_L^{-1}\stw db_Lb_L^{-1})_\D =
\jp (g_R^*\rho_G\stackrel{\w}{,}b_L^*\lambda_B)_\D\ee
and
$$ \jp(K^{-1}dK\stw b_R^{-1}db_R)_\D=$$\be=
\jp(b_R^{-1}g_L^{-1}dg_L b_R+b_R^{-1}db_R\stw b_R^{-1}db_R)_\D=
\jp(g_L^*\lm_G\stackrel{\w}{,}b_R^*\rho_B)_\D.
\ee
Here we used $K=b_L(K)g_R(K)$ in the first relation (4.20) and
$K=g_L(K)b_R(K)$
 in the second one (4.21).

\noindent Take a vector $v\in S_L\subset T_KD$ and calculate
the expression
\be \la b_L^*\rho_B,v\ra= R_{b_L^{-1}*}b_{L*}v=0.\ee
The vanishing of this expression
follows from the fact that $b_L(Ke^{sT})=b_L(K)$ for every $T\in \G$.
Now consider another vector $w\in \ti S_R\subset T_KD$. We have
\be \la b_L^*\rho_B,w\ra= R_{b_L^{-1}*}b_{L*}w=
R_{b_L^{-1}*}R_{g_R^{-1}*}w=R_{K^{-1}*}w.\ee
This follows from the fact that $b_L(e^{st}K)=e^{st}b_L(K)=
e^{st}Kg_R^{-1}(K)$ for every
$t\in \B$. We thus obtain  for an arbitrary vector $u\in T_KD$ that
\be \la b_L^*\rho_B,u\ra=
\la b_L^*\rho_B,(\Pi_{\ti RL}+\Pi_{L\ti R})u\ra=
R_{b_L^{-1}*}b_{L*}\Pi_{L\ti R}u=R_{K^{-1}*}\Pi_{L\ti R}u.\ee

\noindent Much in the same way as above we derive
\be \la b_R^*\lm_B,u\ra = L_{K^{-1}*}\Pi_{R\ti L}u.\ee
Combining the formulas (4.19), (4.24) and (4.25), we arrive at
$$ -2\om(t,u)=(R_{K^{-1}*}t,R_{K^{-1}*}\Pi_{L\ti R}u)_\D+
(L_{K^{-1}*}t,L_{K^{-1}*}\Pi_{R \ti L}u)_\D$$
$$ -(R_{K^{-1}*}u,R_{K^{-1}*}\Pi_{L\ti R}t)_\D-(L_{K^{-1}*}u,
L_{K^{-1}*}\Pi_{R \ti L}t)_\D=$$
\be =(t,\Pi_{L\ti R}u)_\D+(t,\Pi_{R \ti L}u)_\D
 -(u,\Pi_{L\ti R}t)_\D-(u,\Pi_{R \ti L}t)_\D. \ee
Now we use the obvious fact
$$ \Pi_{L\ti R}+\Pi_{\ti R L}=1,\quad \Pi_{R \ti L}+\Pi_{\ti L R}=1$$
(here the first index of the projector  denotes its kernel and the
second its image)
and  obtain    three relations
\be (t,\Pi_{R\ti L}u)_\D=(t,u)-(t,\Pi_{\ti LR}u)_\D;\ee
\be -(u,\Pi_{L\ti R} t)_\D=-(\Pi_{\ti RL} u,\Pi_{L\ti R}t)_\D=
-(\Pi_{\ti RL} u, t)_\D=
-(u,t)+(\Pi_{L\ti R}u,t)_\D;\ee
\be -(u,\Pi_{R\ti L}t)_\D=-(\Pi_{\ti LR}u,\Pi_{R\ti L}t)_\D=
-(\Pi_{\ti LR}u,t)_\D.\ee
Inserting (4.27),(4.28) and (4.29) into (4.26), we obtain
\be  \om(t,u)=(t,(\Pi_{\ti L R}-\Pi_{L\ti R})u)_\D.\ee
The lemma is proved.

\rightline{\#}

\vskip1pc
\noindent {\bf Proof of the theorem 4.1}: The strategy of
the proof is as follows:
Since the antisymmetry of the bracket (4.9) is obvious from (4.16),
we   have to prove
only the validity of the Jacobi identity, in order to show
that (4.9) is really the Poisson bracket. Then we shall prove the
 non-degeneracy
of this Poisson structure by showing
 that the form $\om$ is dual (inverse) to the Poisson bivector
corresponding to the bracket (4.9).

 The Poisson bracket (4.9)  on $D$ can be rewritten in terms of a certain
bivector (antisymmetric two-tensor)  $\alpha$ such that the
following relation holds
\be \{\phi,\psi\}_D=  \al(d\phi,d\psi),
 \quad \phi,\psi\in Fun(D).\ee
We can easily identify $\al$ by noting that (4.16) can be rewritten as
\be \{\phi,\psi\}_D=\la r,\nl\phi\otimes\nl\psi\ra+\la r,
\nr\phi\otimes\nr\psi\ra,\ee
where $r\in\Lambda^2\D$ is the so called classical $r$-matrix.
It is clearly given by
\be r=\jp(T^i\otimes t_i-t_i\otimes T^i) .\ee
From this we have
\be \al_K=(L_{K*}+R_{K*})r,\quad K\in D.\ee
The   Jacobi identity for the Poisson bracket is equivalent  to the
following condition for the bivector $\al$
\be [\al,\al]_S=0.\ee
Here  $[.,.]_S$ is the Schouten bracket of the multivectors \cite{AKM}.
 Recall
its main properties
\be [\al,\beta]_S=-(-)^{(\vert\al\vert -1)
(\vert\beta\vert -1)}[\beta,\al]_S,\ee
\be [\al,\beta\w\gamma]_S=[\al,\beta]_S\w\gamma+(-1)^{(\vert \al\vert-1)
\vert\beta\vert}\beta
\w[\al,\gamma]_S.\ee
Moreover, for any vector field $X$, the bracket $[X,\al]_S$ is just
the Lie derivative.

Let as calculate $[\al,\al]_S$ for $\al$ given by (4.34). Since
 right-invariant
 vector fields on the Lie group manifolds commute with
left-invariant ones, the result
reads
\be [\al,\al]_S=[L_{K*}r,L_{K*}r]_S+[R_{K*}r,R_{K*}r]_S =
(L_{K*}-R_{K*})[r,r]_S.\ee
Here we use the same symbol $[.,.]_S$ also for the Schouten
bracket based on the Lie
algebra commutator. Thus we see that the Jacobi identity is
fulfilled iff
$[r,r]_S$ is the invariant element of $\Lambda^3\D$. Actually,
using (4.33), the direct calculation
gives
\be [r,r]_S=([t_i,t_l],T^k)_\D t_k\w T^i\w T^l+([T^i,T^l],t_k)_\D
T^k\w t_i\w t_l.\ee
 Using
the bracket (4.5),
the $\D$-invariance of this  expression can be checked by the
direct calculation.
But another way to see it is to realize that the r.h.s. of (4.39)
coincides with the invariant
Cartan (or WZW) element of $\Lambda^3\D$ canonically associated
to the  invariant bilinear
form $(.,.)_\D$ (cf. \cite{AKM}).

It is well-known that   the bivector
   $\al$ satisfying (4.35) defines a symplectic structure iff it
 exists a dual (or inverse) 2-form
$\omega\in\Lambda^2T^*D$. The latter is then automatically
 closed $(d\omega=0)$, as the consequence of (4.35).  The duality
means that
\be\al(.,\omega(.,u))=u\ee for any vector field $u\in TD$.
Let us prove (4.40) for $\al$ given by (4.34) and $\om$ by (4.18).

First we note that the element $t_i\otimes T^i+T^i\otimes t_i\in
\D\otimes\D$ is invariant
since it corresponds to the invariant bilinear form $(.,.)_\D$.
 Due to this fact the expression
for $\al$ can be rewritten as
\be \al=L_{K*}(T^i\otimes t_i)-R_{K*}(t_i\otimes T^i).\ee
Thus we have
\be\al(.,\omega(.,u))=R_{K*}t_i(R_{K*}T^i, (\Pi_{L\ti R}-
\Pi_{\ti LR})u)_\D
-L_{K*}T^i(L_{K*}t_i, (\Pi_{L\ti R}-\Pi_{\ti LR})u)_\D.\ee
This can be rewritten as
\be \al(.,\omega(.,u))=(\Pi_{R\ti R}-\Pi_{\ti LL}) (\Pi_{L\ti R}-
\Pi_{\ti LR})u.
 \ee
Recall that the first subscript of the projector stands for the kernel
 and the second
for the image (cf. Proof of Lemma 4.2). Now we have
$$ (\Pi_{R\ti R}-\Pi_{\ti LL}) (\Pi_{L\ti R}-\Pi_{\ti LR})=
\Pi_{R\ti R}\Pi_{L\ti R}- \Pi_{\ti LL}\Pi_{L\ti R} +
\Pi_{\ti L L}\Pi_{\ti L R}=$$
\be =\Pi_{L\ti R} - \Pi_{\ti LL}\Pi_{L\ti R} +\Pi_{\ti L L}=
 (\Pi_{L\ti R}+\Pi_{\ti RL})+(\Pi_{\ti LL}-\Pi_{\ti R L} -
\Pi_{\ti L L}\Pi_{ L \ti R})=1+0.\ee
Combining (4.43) and (4.44), we arrive finally at
\be \al(.,\omega(.,u))= u.\ee
The theorem is proved.

\rightline{\#}

\rem : {\small In order to show that $d\om =0$,
we can also use the Polyakov-Wiegmann formula \cite{PW} applied
for $K=b_L(K)g_R(K)$
and $K=g_L(K)b_R(K)$ respectively:
$$ {1\over 3}(\rho_D\stackrel{\w}{,}\rho_D\w\rho_D)_\D=
d(g_R^*\rho_G\stackrel{\w}{,}b_L^*\lambda_B)_\D+$$
\be +{1\over 3}(b_L^*\rho_B\stackrel{\w}{,}b_L^*\rho_B\w
b_L^*\rho_B)_\D +
{1\over 3}(g_R^*\rho_G\stackrel{\w}{,}g_R^*\rho_G\w g_R^*\rho_G)_\D;
\ee
$$ {1\over 3}(\rho_D\stackrel{\w}{,}\rho_D\w\rho_D)_\D=
d(b_R^*\rho_B\stackrel{\w}{,}g_L^*\lambda_G)_\D+$$
\be +{1\over 3}(b_R^*\rho_R\stackrel{\w}{,}b_R^*\rho_B\w
 b_R*\rho_B)_\D +
{1\over 3}(g_L^*\rho_G\stackrel{\w}{,}g_L^*\rho_G\w g_L^*\rho_G)_\D.
\ee
Note that the last two terms in (4.46) and also in (4.47) vanish
because of the isotropy
of the Lie algebras $\G$ and $\B$ with respect to the bilinear
form $(.,.)_\D$.
Using (4.46),(4.47) and the definition (4.8) of the Semenov-Tian-Shansky
form $\om$, we arrive at
$$ -d\om=\jp d(g_R^*\rho_G\stackrel{\w}{,}b_L^*\lambda_B)_\D+
\jp d(g_L^*\lm_G\stackrel{\w}{,}b_R^*\rho_B)_\D=$$
\be ={1\over 6}(\rho_D\stackrel{\w}{,}\rho_D\w\rho_D)_\D
-{1\over 6}(\rho_D\stackrel{\w}{,}\rho_D\w\rho_D)_\D =0.\ee}
 We note also that  physicists write the Polyakov-Wiegmann formula (4.46)
 as follows
$$ {1\over 3}(dKK^{-1}\stackrel{\w}{,}dKK^{-1}\w dKK^{-1})_\D=
d(dg_Rg_R^{-1}\stackrel{\w}{,}b_L^{-1}db_L)_\D+$$
\be +{1\over 3}(db_Lb_L^{-1}\stackrel{\w}{,}db_Lb_L^{-1}\w
 db_Lb_L^{-1})_\D +
{1\over 3}(dg_Rg_R^{-1}\stackrel{\w}{,}dg_Rg_R^{-1}\w dg_Rg_R^{-1})_\D.
\ee
\vskip1pc
\noindent {\bf Definition 4.3:} The {\bf Heisenberg} double is the
 Drinfeld double equipped
with the Poisson structure (4.9).
\vskip1pc
\noindent In what follows we shall always suppose that the multiplication
maps ${\cal M}_{L,R}$ are
 bijective hence for us the
Heisenberg
double will always be the {\it symplectic} manifold.

The Poisson bracket on the Heisenberg double of $G$  has the
following  crucial property:
\vskip1pc
\noindent {\bf Proposition 4.4:} (Semenov-Tian-Shansky \cite{ST2}):
The algebra $Fun(D)^B$
of right  $B$-invariant  functions on the Heisenberg double $D$
is a Lie subalgebra
in the Poisson algebra (4.9)  of all functions $Fun(D)$. The same
is true for the algebra
$\ ^B Fun(D)$ of left $B$-invariant functions  and also for the
algebras $Fun(D)^G$
and $\ ^G Fun(D)$.
\vskip1pc
\noindent {\bf Proof:}
Since the role of the groups $G$ and $B$
is completely symmetric in $D$, we shall restrict our attention
 only to the cases $Fun(D)^G$
and $\ ^G Fun(D)$.
 Let $\rho,\eta\in Fun(D)^G$,
which means that
\be \la\nr\rho,T^i\ra=\la\nr\eta,T^i\ra=0.\ee
 From (4.16) we see immediately that
\be \{\rho,\eta\}_D=\jp\la \nl \rho,T^i\ra\la\nl \eta, t_i\ra -
\jp\la \nl \rho,t_i\ra\la\nl \eta, T^i\ra=\jp(\nl \rho,\R^*\nl
\eta)_{\D^*}.\ee
Now having in mind that the left and right derivatives $\nl$ and
 $\nr$ commute with each
other, we obtain
\be  \la\nr\{\rho,\eta\}_D,T^i\ra=0.\ee
The case $\ ^G Fun(D)$ can be treated in the same way. The
 proposition is proved.

\rightline{\#}

\noindent This proposition permits to construct the
Poisson-Lie brackets on $B$ (and
 on $G$). Indeed,  both  $Fun(D)^G$
and $\ ^G Fun(D)$ are naturally isomorphic to $Fun(B)$
(due to the existence of the global decomposition $D=GB=BG$) and
 the following
proposition holds
\vskip1pc
\noindent {\bf Proposition 4.5:} The Poisson brackets on $Fun(B)$
induced from the
Semenov-Tian-Shansky bracket on $Fun(D)^G$
and $\ ^G Fun(D)$ coincide up to sign and they verify the
Poisson-Lie condition
(4.2).
\vskip1pc
\noindent {\bf Proof}:
 Take  two functions $\Phi$ and $\Psi$ in $Fun(B)$ and calculate
their Poisson bracket
$\{\Phi,\Psi\}_B^R$
induced from $Fun(D)^G$. We have by definition
$$ b_L^*\{\Phi,\Psi\}_B^R=\{b_L^*\Phi,b_L^*\Psi\}_D=$$
\be =\jp\la \nabla^L_D (b_L^*\Phi),T^i\ra\la\nabla^L_D (b_L^*\Psi),
t_i\ra -
\jp\la \nabla^L_D (b_L^*\Phi),t_i\ra\la\nabla^L_D (b_L^*\Psi), T^i\ra.\ee
The superscript $R$ over the Poisson bracket on $B$ indicates that
the latter
originates from $Fun(D)^G$. Now it is obvious that
\be \la\nabla^L_D (b_L^*\Psi), t_i\ra= b_L^*\la \nabla^L_B \Psi,t_i\ra,\ee
where the subscripts $D$ and $B$ indicates the group on which
the differential
operators live. It is less straightforward to evaluate
$\la \nabla^L_D (b_L^*\Phi),T^i\ra$.

We shall proceed as follows: we note that a map
$b_L\circ {\cal M}_L:G\times B\to B$ given
by
\be g\times b\to b_L(gb)\equiv Dres_g b\ee
defines a left  action of the group $G$ on the manifold $B$. It is
 easy to see this since
the following relation holds
\be b_L(g_1g_2b)=b_L(g_1b_L(g_2b)g_R(g_2b))=b_L(g_1b_L(g_2b)),
\quad g_1,g_2\in G,\quad b\in B.\ee
Now by definition
\be \la \nabla^L_D (b_L^*\Phi),T^i\ra(K)=\dto\Phi(b_L
(e^{sT^i}b_L(K))),\quad K\in D.\ee
Looking at the relations (4.56) and (4.57), we see that
there exists a vector field
(a differential operator)
$\nabla^i_B$ acting on $Fun(B)$ such that
\be \la \nabla^L_D (b_L^*\Phi),T^i\ra=b_L^*(\nabla^i_B\Phi).\ee
Such operator $\nabla^i_B$ can be certainly expressed as a linear
 combination
of the operators $\la\nabla^L_B ,t_j\ra$. In other words, there
exists a matrix valued function
$\Pi_R^{ij}(b)$ on $B$ such that
\be \nabla^i_B\Phi=\Pi_R^{ij}(b)\la \nabla^L_B \Phi,t_j\ra.\ee
Hence from (4.53), we obtain for our (right) Poisson bracket the
following expression
\be \{\Phi,\Psi\}_B^R=\jp\la \nabla^L_B
\Phi,t_i\ra(\Pi_R^{ij}(b)-\Pi_R^{ji}(b))
\la \nabla^L_B \Psi,t_j\ra .\ee
Proceeding in the same way, we define the left bracket
 \be b_R^*\{\Phi,\Psi\}_B^L=\{b_R^*\Phi,b_R^*\Psi\}_D\ee
and we have
\be \{\Phi,\Psi\}_B^L=\jp\la \nabla^R_B \Phi,t_i\ra(\Pi_L^{ij}(b)-
\Pi_L^{ji}(b))
\la \nabla^R_B \Psi,t_j\ra \ee
for certain matrix valued function $\Pi_L^{ij}(b)$ on $B$.
In order to show that the left Poisson bracket (4.60) is equal to
the minus right one (4.62),
we have to know more about the matrices $\Pi_{L,R}^{ij}$. Introduce first
the following matrices
(cf. \cite{KS1})
\be  A_i^{~j}(K)=(K^{-1} t_i  K,T^j)_\D,
\quad  B^{ij}(K)=(K^{-1}T^i K,T^j)_\D,
\quad K\in D.\ee
Now calculate
\be T^i b=bb^{-1}T^ib=b(B^{ij}(b)t_j+A_j^{~i}(b^{-1})T^j).\ee
Since the differential operator $\la \nabla^R_B,t_j\ra $ corresponds
to the vector field $L_{b*}t_j=bt_j$, we see from (4.64) that the
 operator $\nabla^i_B$ can
be expressed as
\be \nabla^i_B =B^{ij}(b)\la \nabla^R_B,t_j\ra=
B^{ik}(b) A_k^{~j}(b^{-1})\la\nabla^L_B,t_j\ra.\ee
Hence
\be
\Pi_R^{ij}(b)= -B^{ik}(b) A_k^{~j}(b^{-1}).\ee
On the other hand, we have
\be bT^i=bT^ib^{-1}b=(B^{ij}(b^{-1})t_j+A_j^{~i}(b) T^j)b\ee
and from this we arrive at
\be \Pi_L^{ij}(b)=-B^{ik}(b^{-1})A_k^{~j}(b)=-B^{ki}(b)A_k^{~j}(b).\ee
Let us now show that the tensors $\Pi_L^{ij}$ and $\Pi_R^{ij}$ are
antisymmetric.
We have
$$ -\Pi_L^{ij}(b)=B^{ki}(b)A_k^j(b)=(b^{-1}T^k b,T^i)_\D (b^{-1} t_k
 b,T^j)_\D=$$
\be (Ad_bT^i,T^k)_\D(t_k, Ad_bT^j)=-(Ad_bT^i,t_k)_\D (T^k, Ad_bT^j)=
\Pi_L^{ji}(b)\ee
and similarly for $\Pi_R^{ij}$. Thus the expressions (4.60) and (4.62)
simplify as follows
\be \{\Phi,\Psi\}_B^R=\la \nabla^L_B \Phi,t_i\ra \Pi_R^{ij}(b)
\la \nabla^L_B \Psi,t_j.\ra\ee
\be \{\Phi,\Psi\}_B^L=\la \nabla^R_B \Phi,t_i\ra \Pi_L^{ij}(b)
\la \nabla^R_B \Psi,t_j.\ra\ee
Using the obvious relation
\be \la \nabla^L_B, t_i\ra=A_i^j(b)\la \nabla^R_B, t_j\ra\ee
we can rewrite the  right bracket as
\be \{\Phi,\Psi\}_B^R=\la\nabla^R_B \Phi,t_m\ra A_i^{~m}(b)\Pi_R^{ij}(b)
A_j^{~l}(b)\la\nabla^R_B \Psi,t_l\ra.\ee
Now we have
\be A_i^{~m}(b)\Pi_R^{ij}(b)
A_j^{~l}(b)=-A_i^{~m}(b) B^{ik}(b) A_k^{~j}(b^{-1})
A_j^{~l}(b)=-A_i^{~m}(b) B^{il}(b)=\Pi_L^{lm}(b).\ee
From (4.70), (4.71)  and (4.74), we conclude that
\be \{\Phi,\Psi\}_B^R= -\{\Phi,\Psi\}_B^L .\ee
It remains to show that the Poisson bracket $\{.,.\}_B^R$ verifies
 the Poisson-Lie
condition (4.2). Recall that a Poisson bracket on any manifold can
 be equivalently
described by a Poisson bivector (=antisymmetric
tensor) $\alpha\in\Lambda^2TB$; the Poisson-Lie bracket on $B$ in
terms of $\alpha$ is given by
\be \{\phi,\psi\}_B=\la \alpha,d\phi\otimes d\psi\ra, \quad \phi,
\psi\in Fun(B).\ee
  The Poisson-Lie condition (4.2)   can be directly rewritten
as
\be \al_{ab}=L_{a*}\al_b+R_{b*}\al_a,\quad a,b\in B,\ee
where $\al_b\in \Lambda^2T_bB$.
Since the bivector bundle $\Lambda^2TB$ on any group manifold
 is trivializable by the right-invariant
vector fields, we loose no information about the Poisson-Lie
structure $\alpha$
if we trade it for another object, namely a map
$\Pi:G\to\Lambda^2\G$ defined as follows
\be \Pi_R(b)=R_{b^{-1}*}\alpha_b.\ee
Now $\Pi_R(b)$ can be expressed in the basis $t_i$ as
\be \Pi_R(b)=\Pi_R^{ij}(b) t_i\otimes t_j.\ee
In fact, the  matrix  $\Pi_R^{ij}$ just introduced is the
same  as the one in (4.59) (the notation is thus consistent!)
 because the
operator $\la \nl,t_i\ra$ corresponds
to the vector field defined in each point $b\in B$ as $R_{b*}t_i$.
The condition (4.77) translates in terms of $\Pi_R(b)$ as
\be \Pi_R(ab)=\Pi_R(a)+Ad_a\Pi_R(b).\ee
One can directly check that  the expression (4.66) fulfils (4.80),
by using  the definition (4.63)  of the matrices $A_i^{~j}(b)$
and $B^{ij}(b)$.

 The theorem is proved.

\rightline{\#}
\subsection{Lu-Weinstein-Soibelman double}
 We are now going to present two important examples  of the
Heisenberg doubles. We shall present a third important (loop group)
example in section 4.4.

1)
The cotangent bundle $T^*G$ of any Lie group is its Heisenberg double.
 The bilinear
form $(.,.)_\D$ is defined as in (7.22).
The role of the group $B$ is played by the subgroup of all elements
 $K$ of $T^*G$  for which
$P_K=e$. Looking at the $T^*G$ multiplication law (7.18), we immediatelly
discover that $B$ is
an Abelian group, in fact, it is nothing but the dual vector space
 space $\G^*$ of the Lie algebra $\G$ of $G$. The global
 decomposability $D=GB=BG$ follows
from the fact that the cotangent bundle of any Lie group is
trivializable.
The Poisson-Lie bracket on $G$, induced  from the bracket on
 the Heisenberg double $T^*G$,
identically
vanishes. This follows from (7.44). On the other hand, the
Poisson-Lie bracket on $B$ is
nontrivial. From (7.53), we see that it is in fact the
Kirillov-Kostant Poisson bracket on $\G^*$.

2) Consider now any finite-dimensional   simple compact
 connected Lie group $G_0$.
For its  Heisenberg double $D$ we take simply its
complexification (viewed
as the $real$ group)
$G_0^{\bf C}$ of $G_0$. So, for example,
the
double of $SU(2)$ is  $SL(2,{\bf C})$. The invariant non-degenerate
form $( .,.)_\D$ on the Lie algebra $\D=\G_0^\bc$ of $D=G_0^\bc$ is
given by
\be ( x,y)_\D={1\over \e}{\rm Im}(x,y)_{\G_0^\bc},\ee
or, in other words, it is just the imaginary part of
the Killing-Cartan form $(.,.)_{\G_0^\bc}$
divided by a real parameter $\e$. We shall see later that $\e$ plays
the role of a deformation parameter. Since $G_0$ is the compact real
 form of $G_0^\bc$,
clearly the imaginary part of $(x,y)_{\G_0^\bc}$ vanishes
if $x,y\in\G_0$. Hence,
$G_0$ is indeed isotropically embedded in $G_0^\bc$.  The
double $G_0^\bc$ equipped with the metric (4.81) is usually
referred to  as the double of Lu\& Weinstein \cite{LW} and of
Soibelman \cite{S}.

It turns out that $G_0^\bc$ is indeed the Drinfeld double, because
 $D=G_0^\bc$ is at the same time the double
of its another  subgroup  $B_0$ which coincides with
the so called $AN$ group in the Iwasawa decomposition of $G_0^\bc$:
\be G_0^\bc=G_0AN=ANG_0.\ee
For the groups $SL(n,{\bf C})$, the group $AN$ can be identified
 with the
upper triangular matrices of determinant $1$
and with positive real numbers on the diagonal.
In general, the elements of $AN$ can be uniquely represented by means
of the exponential map as follows
\be \ti g={\rm e}^{\phi}{\rm exp}[\Si_{\al>0}v_\al E^\al]\equiv
 {\rm e}^{\phi}n.\ee
Here $\al$'s denote the roots of $\G_0^\bc$, $v_\al$ are complex numbers,
 $E^\al$ are the
step operators
and $\phi$ is an Hermitian element\footnote{Recall that the
Hermitian element of  any complex simple Lie algebra $\Gc$
is an eigenvector of the
involution which defines the compact real form $\G$;
the corresponding eigenvalue
is $(-1)$. This involution originates from the group involution
$g\to (g^{-1})^\dagger$.
 The anti-Hermitian elements that span
 the compact real form are eigenvectors
of the same involution with the eigenvalue equal to $1$. For elements
of $sl(n,{\rm C})$ Lie algebra, the Hermitian element is
indeed a Hermitian
matrix in the standard sense.}
 of the Cartan subalgebra
of $\G_0^\bc$. Loosely said, $A$ is the "noncompact part" of
the complex maximal
torus of $G_0^\bc$. The isotropy of the Lie algebra $\B_0$ of
$B_0=AN$
follows from (4.81); the fact that $\G_0$ and $\B_0$ generate together
the Lie algebra $\D$ of the whole double  is evident from (4.82).
The Iwasawa decomposition itself is the global decomposition
$D=G_0B_0=B_0G_0$ needed
for ensuring that the  Semenov-Tian-Shansky Poisson bracket (4.9)
 does indeed define the
(everywhere non-degenerate) symplectic structure on $D$.

\subsection{Non-Abelian moment maps}
The concept of a Poisson-Lie symmetry \cite{ST2} is the
 generalization of the traditional Hamiltonian
symmetry of a dynamical system defined by a symplectic manifold
 and a Hamiltonian.
Here we shall partially follow the exposition of the papers
\cite{FG,AT} and \cite{KS8}.

First we need to recall the definition of the  {\bf dressing action}
of $G$ on its dual
Poisson-Lie group $B$.
An element $g\in G$ acts on an element $b\in B$ to give
\be Dres_gb\equiv b_L(gb),\ee
where the multiplication $gb$ is taken in the Drinfeld double $D$
and the map $b_L:D\to B$ is induced by the decomposition $D=BG$.
It follows from (4.56)  that this is really the group action, i.e.
\be Dres_eb=b, \quad Dres_{(g_1g_2)}b=Dres_{g_1}(Dres_{g_2} b).\ee
Suppose now that there is a symplectic form $\om$ on a manifold $P$
and there is a left action
of a Poisson-Lie group $G$ on $P$, infinitesimally generated by
a section $v$ of the bundle $TP\otimes \G^*=TP\otimes \B$. The vector
field
corresponding to the action of a generator $T\in \G$ is then
$\la v,T\ra\in TP$. Recall
that  $B$
is the dual Poisson-Lie group  and $\B$ is its Lie algebra;
the dual space $\G^*$
is identified with $\B$ via the invariant bilinear form $(.,.)_\D$
on the Drinfeld double $\D=\G+\B$ (cf. section 4.1.1 and 4.1.2).

Moreover, suppose that there is a $G$-equivariant   map $M:P\to B$,
where $G$ acts on $P$
as above and acts on $B$ via the dressing action.
 Finally,  such a map $M$ is called the non-Abelian
moment map if it holds
\be -i_v\om\equiv \om(.,v)= M^*\rho_B.\ee
Said in words: the contraction of the symplectic form $\om$ by the
 section $v\in TP\otimes \B$
is equal to the pull-back of the right-invariant Maurer-Cartan form
$\rho_B$ on $B$
by the map $M$. The condition (4.86) is often written as
\be \om(.,v)=dMM^{-1}.\ee
\vskip1pc
\teo {\bf 4.6:} Let the manifold $P$ be the Heisenberg double
$D$ of $G$ and
$\om$ is
the Semenov-Tian-Shansky
symplectic form (4.8). Then

\noindent 1) The map $b_L:D\to B$ induced by the decomposition
 $D=BG$ is the
non-Abelian moment map $M$ of the standard left action
 $G\times D\to D$ given by the group
multiplication law in $D$, i.e. $(g,K)\to gK$, $g\in G,K\in D$.

\noindent 2) The map $b_R^{-1}:D\to B$  induced by the
decomposition $D=GB$ is the
non-Abelian moment map $M$ of the  left action $G\times D\to D$
given by  $(g,K)\to Kg^{-1}$,
$g\in G,K\in D$.
\vskip1pc

\vskip1pc
\noindent {\bf Proof}:
The first part 1) is slightly easier: We have to show, that for a
 generator $T\in \G$, it holds
\be \om(u,R_{K*}T)=(T, \la b_L^*\rho_B,u\ra )_\D,\ee
where $K\in D$ and $u\in T_KD$.
We have
\be \om(u,R_{K*}T)=(R_{K*}T,(\Pi_{L\ti R}-\Pi_{\ti L R})u)_\D=
(R_{K*}T,\Pi_{L\ti R}u)_\D=(T, \la b_L^*\rho_B,u\ra )_\D.\ee
Here the first equality follows from the Lemma 4.2, the second
 from the isotropy
of  the space $S_R$ with respect to $(.,.)_\D$ and the third one
from the relation (4.24).

For the second part 2), consider again the generator $T\in \G$.
We want to show, that
\be \om(L_{K*}T,u)=(T,\la d(b_R^{-1})(b_R^{-1})^{-1},u\ra)_\D.\ee
The last relation
can be rewritten as
\be \om(u,L_{K*}T)=(T,\la b_R^{-1}db_R,u\ra)_\D=
(T,\la b_R^*\lm_B,u\ra)_\D.\ee
Let us prove (4.91). We have
$$ \om(u,L_{K*}T )=(L_{K*}T,(\Pi_{L\ti R}-\Pi_{\ti L R})u)_\D=$$
$$=(L_{K*}T,(\Pi_{L\ti R}+\Pi_{\ti R L})u)_\D-(L_{K*}T,(\Pi_{\ti L R}+
\Pi_{R\ti L})u)_\D
+(L_{K*}T,\Pi_{R\ti L}u)_\D=$$
$$=(L_{K*}T,u)_\D-(L_{K*}T,u)_\D
+(L_{K*}T,\Pi_{R\ti L}u)_\D=$$\be =(L_{K*}T,\Pi_{R\ti L}u)_\D=
(T,\la b_R^*\lm_B,u\ra)_\D.\ee
The last equality in (4.92) follows from (4.é(). The $G$-equivariance
of $M$ is obvious
in the first case 1); in the second it follows from the relation
\be b_R^{-1}(K)=b_L(K^{-1}),  \quad K\in D.\ee

\noindent The theorem is proved.

\rightline{\#}

\noindent We end up this section with a definition:
\vskip1pc
\defi {\bf 4.7:} A dynamical system characterized by a symplectic
 manifold $P$, a symplectic
form $\om$ and a Hamiltonian $H$ is \underline{Poisson-Lie symmetric}
with respect
to a left action
of a Poisson-Lie group $G$, if the Hamiltonian is $G$-invariant
and the $G$-action
on $P$ is generated by the non-Abelian moment map $M:P\to B$
fulfilling the condition (4.86).
\vskip1pc
\rem: {\small The classical action of any dynamical system can
be written in the
standard way
\be S=\int (\theta -H dt),\ee
where $d\theta$ is the  symplectic form  and $H$ the Hamiltonian.
The variation of the
action of a general  Poisson-Lie symmetric system was calculated
in \cite{AT} with the
result
\be \delta_T S=-\int \epsilon_i A^i.\ee
Here $A^i$ is the set of functions on the phase space $P$,
satisfying the non-Abelian
zero-curvature condition
\be dA^i=\tilde f^i_{~~kl}A^k A^l;\ee
  $\epsilon_i$ are the coefficients of the generator $T\in\G$ in
some basis $T^i$ of $\G$
and $\tilde f^i_{~~kl}$ are the structure constants of $\B$ in the
dual basis $t_i$.

If $A^i=dX^i$ for some collection of functions $X^i$ on $P$,
then $\tilde f^i_{~~kl}$
obviously vanishes and the integrand on the r.h.s. of (4.95)
is the total derivative. The action $S$ is then strictly symmetric
 with respect to the
$G$-action and $X^i$ are nothing but the standard Hamiltonian
charges generating the
symmetry. However, if $\tilde f^i_{~~kl}$ do not vanish, the action
$S$ is not symmetric
in the strict sense of this word. However, it is by definition
Poisson-Lie symmetric
since its variation has the special form encoded in (4.95) and (4.96).}

\section{Quasitriangular geodesical model}
\defi {\bf 4.8:}  Consider now a Lie group $G$ and let  $D$  be some of its
Heisenberg doubles.
Choose an appropriate $G$-biinvariant Hamiltonian function $H(K)$,
 $K\in D$.
The  dynamical
system defined by the Semenov-Tian-Shansky symplectic structure on $D$
and by the biinvariant Hamiltonian $H(K)$ will be  called the
quasitriangular geodesical model.

\vskip1pc
\noindent   We know from the
results of the previous section,  that  the quasitriangular geodesical
model
is distinguished
by two independent Poisson-Lie symmetries given by the left
multiplication $k_{L}K$,
$k_{L}\in G$ or the right multiplication (but the left action!)
 $Kk_{R}^{-1}$, $k_{R}\in G$.
Even without knowing anything more about the Hamiltonian $H(K)$,
its $G$-biinvariance
entails immediately an important information about the  trajectories
of this dynamical
  system (we shall call such a trajectory a quasitriangular geodesics).
Indeed:
 Let $K(t)$ be the  quasitriangular geodesics. Then
\be b_L(K(t))=b_{L}(K(0))=b_L^0,\quad b_R(K(t))=b_R(K(0))=b_R^0.\ee
In other words, $b_L^0$ and $b_R^0$ are nonlinear constant of motions.
 We may say loosely
that the nonlinear momentum $b_L$ (or  $b_R$)  is constant, and the
 quasitriangular
geodesical motion is therefore "nonlinearly free" in the nonlinear
coordinate $g_R$ (or $g_L$).

The independence of $b_L$ and $b_R$ on time is the consequence of the
$G$-biinvariance of $H(K)$
and of the following
theorem:
\vskip1pc
\noindent {\bf Theorem 4.9}: The Semenov-Tian-Shansky Poisson bracket
of a $G$-leftinvariant function on $D$ with a $G$-rightinvariant
function on $D$
always vanishes.
\vskip1pc
\noindent {\bf Proof}:
Due to the existence of the global decompositions $D=GB=BG$, each
$G$-rightinvariant
function on $D$ is a $b_L$-pullback of some function $\Phi\in Fun(B)$
and similarly
each  $G$-leftinvariant
function on $D$ is a $b_R$-pullback of some function $\Psi\in Fun(B)$.
Using the
formula (4.41), the
Semenov-Tian-Shansky bracket (4.9) of such two functions then becomes
\be \{b_L^*\Phi,b_R^*\Psi\}_D=
\jp\la \nabla^R_D( b_L^*\Phi),T^i\ra\la\nabla^R_D (b_R^*\Psi), t_i\ra
-\jp \la\nabla^L_D (b_L^*\Phi), t_i\ra
\la \nabla^L_D (b_R^*\Psi),T^i\ra .\ee
Now the left (right) $G$-invariance of $b_R^*\Psi$ ( $b_L^*\Phi$) means,
 respectively,
\be \la \nabla^L_D (b_R^*\Psi),T^i\ra (K) =0,\ee
\be \la \nabla^R_D (b_L^*\Phi),T^i\ra (K) =0.\ee
Thus \be \{b_L^*\Phi,b_R^*\Psi\}_D=0.\ee
The theorem is proved.

\rightline{\#}

\noindent As an example, we can take  for $G$  the simple compact
connected group $G_0$ and its Heisenberg double is
 the Lu-Weinstein-Soibelman  double $D_0=G_0^\bc$ of
the example 2) of the previous paragraph. Now we look for a biinvariant
 Hamiltonian.
Recall that  in the case of the standard geodesical model
 the choice of the biinvariant Hamiltonian
  was canonical.
 It turns out that in the $D_0=G_0^\bc$ case, there
is also the canonical choice. Up to a normalization
(to be fixed later),
it is given by
\be H=(ln(b_L^\dg b_L),ln(b_L^\dg b_L))_{\G_0^\bc}=
(ln(b_R^\dg b_R),ln(b_R^\dg b_R))_{\G_0^\bc},\ee
where $\dg$ is the Hermitian conjugation defined in the section 4.1.3.
We shall return to the quasitriangular geodesical model on $G_0^\bc$
later when we shall study
its chiral decomposition. It will be on this occasion when we shall
give the exact solution
of the model.

\section{WZW Drinfeld doubles of $\ti G$}
In this section, $\ti G$ will denote the central biextension of the
group $G$.
Recall that in the non-deformed case, we have performed the two-step
symplectic
reduction starting from the  geodesical model (1.1) on $\ti G$ with its
canonical symplectic structure
on $T^*\ti G$   and arriving at the WZW model on $G$ with
its (non-canonical) WZW symplectic form on $T^*G$.
In principle, we can construct the deformation of the master model (1.1)
 for whatever
Drinfeld double of $\ti G$. However, if we want to make also the two-step
symplectic reduction,
the double must be of the special form described in the following
definition:

\vskip1pc
\noindent {\bf Definition 4.10:}  We shall say
that a double $\ti{\ti D}(=\ti G\ti B=\ti B\ti G)$ of the central
 biextension $\ti G$
is of the WZW type (or simply is the WZW double), iff
 there exist a double $\hhd=(\hat G\hat B=\hat B\hat G)$ of $\hat G$
and a
 double $D(=GB=BG)$ of $G$ such that
\vskip1pc
\noindent i) The group $\ti B$ is isomorphic to the direct product of
the real line with
$\hat B$, i.e. $\ti B=\br\times \hat B$.

\noindent ii) The group $\hat B$  is isomorphic to the semi-direct
product of the real
line with $B$, i.e. $\hat B=\br\times_{Q}B$, where $Q$ is some one
 parameter group
of automorphisms of $B$.
\vskip1pc
\teo {\bf 4.11:} The properties 1)-2)  of the definition 4.10  are
satisfied by the
triple of Drinfeld doubles  $\ttD=T^*\ti G$, $\hhD=T^*\hat G$ and
$D=T^*G$.
\vskip1pc
\pro :
The dual
group $\ti B$ is clearly $\ti G^*$ viewed as the Abelian group (linear
 space).
Their elements are $(A,\al,a)^*$ (cf. Conventions 2.5).  The subgroups
formed by the elements
of the forms $(0,\al,a)^*$   and   $(0,\al,0)^*$ are $\hat B$ and $B$,
 respectively.
Since $\ti B$ is Abelian, the theorem is proved.

\rightline{\#}

 \noindent In order to give the definition of the universal
 quasitriangular WZW model, it is useful
to set appropriate conventions:

\vskip1pc
\noindent {\bf Conventions 4.12:} Consider the group $\ti B=\br\times
\hat B=\br\times(\br\times_Q B)$
from the definition 4.10. The symbol $\times$ without (with)
the subscript means the direct (semidirect)
product of groups. This decomposition induces several natural maps; our
 conventions
will give them names.
First of all, from $\ti B=\br\times \hat B$ we obtain two natural maps
 $m^0:\ti B\to\br$
and $\hat m:\ti B\to\hat B$.  Then from $\hat B=\br\times_Q B$ we
 produce $m^\infty:\hat B\to \br$
and two maps $m_{L,R}:\hat B\to B$ given by two possible
decompositions  $\hat B=B\br =\br B$.
Recall also that  the notation $\ti b_{L,R}$
is induced by the two canonical decompositions $\ttD=\ti G\ti B=
\ti B\ti G$.

\vskip1pc
\defi {\bf 4.13:} The universal quasitriangular WZW model is the rule that
associates a  dynamical system  to every WZW Drinfeld double $\ttD$ (of
 the central biextension $\ti G$) and to every $\ti G$-biinvariant
function $\ti H(\ti K)$,
$\ti K\in\ttD$.
This system is obtained from the quasitriangular geodesical
model on $\ttD$ by the two-step  symplectic reduction induced by setting
\be m^0(\ti b_L(\ti K))+m^0(\ti b_R(\ti K))=0;\ee
\be (m^\infty \circ\hat m)(\ti b_L(\ti K))+(m^\infty \circ\hat m)
(\ti b_R(\ti K))=2\k.
\ee
We shall call this dynamical system the $(\ttD,\ti H)$-quasitriangular
 WZW model.
\vskip1pc
\noindent {\bf Remarks}:

 {\small \noindent 1) For a general WZW double $\ttD$,
 we do not have a natural choice of Hamiltonian $\ti H$. However, two
 important WZW doubles
of the affine Kac-Moody group $\ti G$ permit to choose the Hamiltonian
in the canonical
way. The first one is the cotangent bundle $T^*\ti G$ which leads to the
 definition
of the standard loop group WZW model described in Chapters 2 and 3.
 The second one is
the  affine Lu-Weinstein-Soibelman double which will be introduced
in the
next Section and  which will lead to the main result
of this paper: the construction of the loop group quasitriangular
 WZW model.

\noindent 2) It is
important to note that the moment maps appearing in the  relations (4.103)
 and (4.104)
are ordinary functions.   They  generate respectively
the axial
actions (i.e. $\ti K\to u\ti K u$) of the group $\br_S$ generated by
$\ti T^0$ and
of the central circle $S^1$ generated by $\tF$.  For a  non-WZW double
 of $\ti G$
such axial action would be only of the Poisson-Lie type and, generically,
 we would not
be able to disentangle the moment maps of
the $\ti T^0$ and $\tF$ symmetries from the moment maps of the other
$\ti G$-symmetries.
The choice of the WZW double insures that these particular symmetries
are Hamiltonian in the
standard sense of this word hence the symplectic reduction can be
performed.

\noindent 3)  We shall describe the details of the symplectic reduction
for the affine Lu-Weinstein-Soibelman double
 of the affine Kac-Moody group.  There is no interest to list here
explicitly
the reduction for whatever WZW double
since the corresponding formulas are anyway too general to be
 illuminating.}

\section{Affine Lu-Weinstein-Soibelman double}
Now we are advancing to our most important example of the triple
 $(\ttD,\hhD,D)$ of the Drinfeld
doubles of the groups $(\ti G,\hat G,G)$ (cf. Definition 4.10).   $G$
 will be  the loop group $G=LG_0$, where
$G_0$ is a simple compact connected and simply connected Lie
 group\footnote{Note that $G$ is then
connected.}; $\hat G=\wh{LG_0}$ will be its standard central
 extension described in Appendix 7.1
and $\ti G$ will be the affine Kac-Moody group $\br\times_{\hat S}
\wh{LG_0}$.
 Although the construction
that we are going to present here is apparently original, we choose
the name
"affine Lu-Weinstein-Soibelman double" because the resulting
double $\ttD$ has
many features similar as the finite-dimensional
 Lu-Weinstein-Soibelman double $D_0=G_0^\bc$.
\subsection{The double $D=LG_0^\bc$}
 For the Heisenberg double $D$ of $G$, we take the loop group
$LG_0^\bc$ consisting
of smooth maps from the circle $S^1$ into $G_0^\bc$.
It will be often convenient to view the loop group $LG_0^\bc$
as a group of holomorphic
maps from the $\si$-Riemann sphere without poles into the
complex group $G_0^\bc$. Clearly, the
loop circle $S^1$ is identified with the equator. The $\si$-Riemann
sphere is in fact the ordinary Riemann sphere but  since in this
 section
we shall encounter the notion
of the  Riemann sphere in several different contexts, we shall
use the labels to distinguish them.

The complex Lie algebra $\G^\bc=\wh{L\G_0^\bc}$  is equipped with
an invariant
nondegenerate bilinear form  given by
\be (x,y)_{\G^\bc}={1\over 2\pi}\int d\si ( x(\si), y(\si))_{\G_0^\bc},\ee
where the elements $x,y\in\G^\bc$
are smooth maps from $S^1$ into $\G_0^\bc$.
Recall that $D$ is the group $\wh{LG_0^\bc}$ viewed as the real group.
 The invariant
nondegenerate bilinear form on $\D=Lie(D)$ is then defined as
\be  (x,y)_\D={1\over \e} Im (x,y)_{\G^\bc}, \ee
 where $\e$ is  a real positive parameter. Note the full analogy with
 the finite-dimensional
definition (4.81).
The Lie algebra $\G=L\G_0$ is isotropic with respect
to $(.,.)_\D$ since it is "pointwise" isotropic with respect to
$Im(.,.)_{\G_0^\bc}$.
Thus we see that
$D$ is indeed the Manin double of $G$. In order to show that it is
also the Drinfeld double,
we need the complementary (or dual) group $B$. In fact, $B$ is the
group $L_+G_0^\bc$ consisting
of loops in $G_0^\bc$ that  are boundary values of holomorphic maps
from the unit disc
into $G_0^\bc$. In other words, we may view $L_+G_0^\bc$
as the group of holomorphic maps from the $\si$-Riemann sphere
without the north pole
into the complex group $G_0^\bc$.
We require moreover, that the value  of this holomorphic map
at the origin of the disc
(=the south pole
of the $\si$-Riemann sphere)
is an element of $B_0=AN\in G_0^\bc$.

As we have already said, the isotropy of the Lie algebra $L\G_0$
of $G=LG_0$ with respect
to the bilinear form (4.106)
is obvious. But also the Lie algebra $\B=L_+\G_0^C$ is isotropic
with respect to
$(.,.)_\D$. Indeed, the expression to be integrated in (4.106) has only the
non-negative Fourier
modes. The integral of all strictly positive modes then vanishes.
The zero mode does not contribute
either since
\be Im(Lie(AN),Lie(AN))_{\G_0}=0\ee
as in the example 2) of Section 4.13.
Finally, the existence of the global decomposition $D=LG_0^\bc=
(LG_0)(L_+G_0^\bc)=GB=BG$
was proved in \cite{PS}.

\subsection{The double $\hhD=\br\times_Q \rlgc$}

Now we are going to construct the   double $\hhD$ of the
centrally extended loop group $\hat G$.

 Consider the  group $DG_0^\bc$ of smooth maps from the unit disc
into $G_0^\bc$ with
the usual pointwise multiplication. We can now define an extended
 group $\ ^\br\widehat {DG_0^\bc}$ whose
elements are pairs $(\bar l,\lm)$ where $\bar l\in DG_0^\bc$ and
$\lm\in U(1)$ and whose multiplication
law reads
\be (\bar l_1,\lm_1)(\bar l_2,\lm_2)=(\bar l_1\bar l_2,\lm_1\lm_2
\exp{[2\pi i\beta_\br (\bar l_1,
\bar l_2)]}).\ee
Here  $\beta_\br$ is a real valued 2-cocycle on $DG_0^\bc$ given by
\be \beta_\br(\bl_1,\bl_2)={1\over 8\pi^2}\int_{Disc} Re(\bl_1^{-1}
d\bl_1\stackrel{\w}{,}d\bl_2
\bl_2^{-1})_{\G_0^\bc},\ee
where $(.,.)_{\G_0^\bc}$ is again the (standardly normalized)
 Killing-Cartan form on $\G_0^\bc$. It is crucial
to note the presence of the real part symbol in the definition of the
 2-cocycle. This
real part is also reflected by the (left) superscript $\br$ in the
symbol  $\rdgc$.

Consider now a subgroup $\d G^\bc$ of $DG_0^\bc$ consisting of all
smooth maps
from the $Disc$ into $G_0^\bc$ such that their value  at every point
of the boundary $\d D=S^1$
is the unit element $e_0$ of $G_0^\bc$. Any $\bl\in\d G^\bc$ can be
thought of as a map
$\bl:S^2\to G_0^\bc$ by identifying the boundary $S^1$ of $Disc$ with
the north pole of $S^2$.
The Riemann sphere thus obtained will be called the $D$-Riemann sphere.
It turns out that there is a homomorphism
$\Th^\bc_\br:\d G^\bc\to\rdgc$  defined by
\be \Th^\bc_\br(\bl)=(\bl,\exp{[-2\pi iC^\bc_\br(\bl)]}),\ee
where
\be C^\bc_\br (\bl)={1\over 24\pi^2}\int_{Ball} Re(d\bl\bl^{-1}\stw
d\bl\bl^{-1}\w d\bl\bl^{-1}])_{\G_0^\bc}.\ee
Here $Ball$ is the unit ball whose boundary is the $D$-Riemann sphere
and we have extended
the map $\bl:S^2\to G_0^\bc$ to a map $\bl: Ball\to G_0^\bc$. The
proof of the fact that
$\exp{[-2\pi iC^\bc_\br(\bl)]}$ does not depend on the extension
of $\bl$ to $Ball$ reduces to the
same proof as for the compact group $G_0$ since $G_0^\bc$ has the
same homotopies as $G_0$.
 The demonstration
that $\Th^\bc_\br$ is indeed a homomorphism and the fact that the
image $\Th^\bc_\br(\d G^\bc)$
 is  the normal subgroup
in $\rdgc$ follows again from  the Polyakov-Wiegmann formula \cite{PW}
 (see Appendix 7.1)
which asserts that
\be C^\bc_\br(\bl_1\bl_2)=C^\bc_\br(\bl_1)+C^\bc_\br(\bl_2)-
\beta_\br(\bl_1,\bl_2).\ee
The subgroup $\hat D\equiv \rlgc$ of the double $\hhD$ is now
defined as the factor group
$\rdgc/\Th^\bc_\br(\d G^\bc)$. This group is a (nontrivial) circle
bundle over the
base space $LG_0^\bc=D$. The projection $\Pi_0$ is
 $(\bl,\lm)\to \bl\vert_{S^1}$ and the center of $\hat D$ is
represented by the
  $\Th$-equivalence classes  represented by $(1,\lm)\in
\rdgc$.
The projection homomorphism from $\rdgc$ onto $\hat D$ will be
referred to as
$\wp_\bc$.

In order to construct the  Drinfeld
double $\hhD$, we need a one-parameter
group  $\br_{\hat Q}$
of automorphisms of $\hat D$. For this, it is convenient first to
define a one-parameter
group of automorphisms $\br_{\bar Q}$ of $\rdgc$.
 If $(\bl(z,r),\lm)$ is an element in $\rdgc$
the action of an element $w\in\br_{\bar Q}$   reads
\be ^q(\bl(z,r),\lm)=(\bl(qz,r),\lm),\ee
where $q=e^w$.
Recall, that we view the loop group $LG_0^\bc$ as the group of
 holomorphic maps from the
Riemann sphere without poles into the complex group $G_0^\bc$.
The standard polar coordinates
$(\si,r)$ of the {\it Disc} thus get traded for $(z,r)$. We
can view  the disc as the intersection
of the equatorial plane  with the interior of the $\si$-Riemann
sphere. We stress however that
this $\si$-Riemann sphere is not the same as the $D$-sphere used
for the definition of the term
$\exp{[-2\pi iC^\bc_\br(\bl)]}$ for elements $\bl\in\d G^\bc$.

We have to prove that
\be ^q((\bl_1,\lm_1)(\bl_2,\lm_2))=\ ^q(\bl_1,\lm_1)^q(\bl_2,\lm_2),\ee
where the product is considered in $\rdgc$.
This in turn amounts to show that
\be \beta_\br(\bl_1(z,r),\bl_2(z,r))=\beta_\br (\bl_1(qz,r),
\bl_2(qz,r)).\ee
Recalling the definition of the cocycle $\beta_\br$, we can
rewrite (4.109)
as
$$ \beta_\br(\bl_1,\bl_2)=$$
\be ={1\over 8\pi^2}\int_0^1 dr \int_{\vert z \vert =1} dz
(Re(\bl_1^{-1}\d_r \bl_1,\d_z \bl_2 \bl_2^{-1})_{\G_0^\bc}-
Re(\bl_1^{-1}\d_z \bl_1,\d_r \bl_2
 \bl_2^{-1})_{\G_0^\bc}).\ee
Here the integration over $\si$ is replaced by the contour
integration along the equator
 $\vert z \vert=1$
on the $\si$-Riemann sphere. It is straightforward to show that
$\beta_\br (\bl_1(qz,r),\bl_2(qz,r))$
is given by the same integral as (4.116) but along the new contour
$\vert z\vert =q$.
Since the integrand is everywhere holomorphic function between
the two contours, they can
be deformed one to the other and we conclude that (4.115) holds.

Now we have to show that the group action (4.113) survives the
factorization by the group
 $\delta G^\bc$.
In other words, the   automorphisms ${\bar Q}$ of $\rdgc$
descends to  automorphisms $\hat Q$ of the factor group $\hat D=\rlgc$,
or still rephrased differently:
$\bar Q$ acts on the  $\d G^\bc$ classes in $\rdgc$.
This amounts to show that
\be C^\bc_\br(\bl(z,r))= C^\bc_\br(\bl(qz,r)).\ee
Looking at the formula (4.111),
the integration goes now over three variables parametrizing the
interior of the $D$-Riemann sphere.  We can again change the
$\si$-integration
to the contour integration on the $\si$-Riemann sphere and
then we prove (4.117) by the
identical contour
deformation argument as  above.

Thus our group $\hhD$ is defined as the group of couples $(X,q)$,
 $X\in\rlgc$,
$q\in\br^+$ with the following composition law
\be (X_1,q_1)(X_2,q_2)=(X_1 \ ^{q_1}X_2,q_1q_2).\ee
Now we have to prove several lemmas in order
to show that $\hhD$ is indeed the Drinfeld double of $\hat G$.

\vskip1pc
\noindent {\bf Lemma 4.14}: The  subgroup $\Pi_0^{-1}(LG_0)\subset\hat D$
is isomorphic
to $\widehat{LG_0}=\hat G$.

\vskip1pc
\noindent {\bf Proof}:
Consider an element $\hat l\in \Pi_0^{-1}(LG_0)\subset\rlgc$. It can
be lifted by the "map" $\wp_\bc^{-1}$ (of course non-uniquely) to
some element
$\bl \in \hdg\subset\rdgc$. We stress that the element $\bl$
can be chosen in $\hdg$. The element $\bl$ can be then projected by
the map $\wp$
(not $\wp_\bc$ !) to some element of $\widehat{LG_0}$. We now show
that this element of
$\widehat{LG_0}$ does not depend on the choice $\bl\in\hdg$, which
 means that
we have constructed certain
 map
$\mu: \Pi_0^{-1}(LG_0)\to\widehat{LG_0}$. Indeed, if we have two
 elements
 $\bl_1,\bl_2\in \widehat{DG_0}$
such that $\wp_\bc(\bl_1)=\wp_\bc(\bl_2)=\hat l$ then it certainly
exists an element
 $F\in\d G(\subset
\d G^\bc)$ such that $\bl_1=F\bl_2$. This means, in other words,
that $\wp(\bl_1)=\wp(\bl_2)$
hence $\mu$ is a well-defined map.

Let us show that $\mu$ is a homomorphism. First of all the unit
element in $\Pi_0^{-1}$
can be lifted directly to the unit element of $\widehat{DG_0}$ and
mapped by $\wp$
to the unit element of $\widehat{LG_0}$ (since $\wp$ is  a homomorphism).
 Then we want
to prove that
$$\mu(\hat l_1\hat l_2)=\mu(\hat l_1)\mu(\hat l_2), \quad
\hat l_1,\hat l_2\in
\Pi_0^{-1}(LG_0).$$
But if $\bl_1,\bl_2\in\hdg$ are such that $\wp_\bc(\bl_i)=\hat l_i;
i=1,2$ then
 $\wp_\bc(\bl_1\bl_2)=\hat l_1\hat l_2$
since $\wp_\bc$ is the homomorphism. Thus we have
$$\mu(\hat l_1\hat l_2)=\wp(\bl_1\bl_2)=\wp(\bl_1)\wp(\bl_2)=
\mu(\hat l_1)\mu(\hat l_2)$$
because also $\wp$ is  the  homomorphism.

Injectivity: if we take again $\hat l_1,\hat l_2\in \Pi_0^{-1}(LG_0)$
 such that
$\Pi_0(\hat l_1)\neq\Pi_0(\hat l_2)$
then clearly $\pi(\mu(\hat l_1))\neq \pi(\mu(\hat l_2))$ hence
$\mu(\hat l_1)\neq\mu(\hat l_2)$
(recall
that $\pi$ denotes  the homomorphism from $\widehat{LG_0}$ to $LG_0$
defined by the exact sequence
(2.1)). If $\Pi_0(\hat l_1)=\Pi_0(\hat l_2)$ and $\hat l_1\neq \hat l_2$
then $\hat l_1 =Y\hat l_2$,
 where $Y$ is
a non-unit
 central circle element in $\rlgc$.
  Then also $\bl_{1,2}\in\hdg$ can be chosen to be connected
by the same  non-unit central circle element viewed as the element of
 $\hdg$ and
  $\wp(\bl_1)=\mu(\hat l_1)
=Y\mu(\hat l_2)$, where now the same $Y$ is viewed as the element of
 $\hlg$.

Since the surjectivity is evident, the lemma is proved.

\rightline{\#}

\vskip1pc
\noindent {\bf Lemma 4.15}: The  group $B=L_+G_0^\bc$ can be
homomorphically
injected into
$\hat D=\rlgc$. Moreover, the image of this injection is preserved by
the automorphisms
$\hat Q$.

\vskip1pc
\noindent {\bf Proof}:
Consider an element $b\in\l+$. By definition, it is the    boundary
value
 of the holomorphic map $\bar b$ from the unit disc
into $G_0^\bc$. Consider now the map $\bar\nu:\l+\to\rdgc$ defined by
\be \bar\nu(b)=(\bar b,1).\ee
The crucial thing is that the map $\bar\nu$ is the homomorphism of
groups. This follows
from the fact that the cocycle $\beta_\br(\bar b_1,\bar b_2)$ vanishes
if $\bar b_i's$
are the holomorphic maps. Indeed,
by using the contour representation (4.116) of this
cocycle, we see immediately that the contour can be contracted to the
 origin of the unit disc
without encountering any singularity
 because the integrated function is everywhere holomorphic.

Consider now the map $\nu:\l+\to\hat D=\rlgc$ defined by
\be \nu=\wp_\bc\circ\bar\nu.\ee
First of all, $\nu$ is the group homomorphism being the composition of
two homomorphisms.
Moreover,  it holds
 \be (\Pi_0\circ\wp_\bc\circ\bar\nu)(b)=b\in LG_0^\bc.\ee
From this it follows that $\nu$ is the injection. The invariance of
$\nu(B)$ under
the action of $\hat Q$ is obvious.

The lemma is proved.

\rightline{\#}
\vskip1pc
\noindent {\bf Remark}: {\small The two preceding lemmas say, in other
 words, that both $\hat G=
\widehat{LG_0}$
and $\hat B=\br\times_Q\l+  $ are  subgroups
of $\hhD$ which  is one of the basic properties of the Heisenberg double
 of $\hat G$ and of
$ \hat B$.}
\vskip1pc
\noindent {\bf Lemma 4.16}: The group $\hhD$ can be globally decomposed as
 $\hhD=\hat G\hat B=
\hat B\hat G$ where $\hat G=\hlg$ and $\hat B=\br\times_Q\l+$.
\vskip1pc
\noindent {\bf Proof}:
Of course, in the sense of the two lemmas above,
here $\Pi_0^{-1}(LG_0) \subset \hat D$ is viewed as $\hat G$ and
$\br\times_Q\nu(\l+)$ as $\hat B$.
Take any element $\hat {\hat K}$ in the extended double $\hhD$. Since
$\hhD=\br\times_Q\hat D$,
$\hat {\hat K}$ can be uniquely decomposed as
\be \hat {\hat  K}=(\hat K,1)(1,q),\ee
where $\hat K\in\hat D$ and $q=e^w$, $w\in\br_Q$.

Now we have to prove that $\hat K$ can
be uniquely decomposed as
\be \hat K=\hat a\nu(b),\quad
\hat a\in\hat G=\Pi_0^{-1}(LG_0), \quad b\in B=\l+.\ee
Consider first the element $\Pi_0(\hat K)\in \lgc$. It can be uniquely
decomposed as
\be \Pi_0(\hat K)=ab,\quad a\in LG_0,\quad b\in\l+.\ee
Now it is certainly true that $\hat K$ can be found among the elements
 of the circle fiber
  $\Pi^{-1}_0(a)\nu(b)$ above $ab$. The existence of $\hat a$ and $b$
from (4.123) then
follows immediately.

It remains to be proved that the decomposition (4.123) is unique. So if
\be \hat K=\hat a\nu(b)=\hat a'\nu(b'), \quad a,a'\in\hat G,
\quad b,b'\in B,\ee
then
\be \Pi_0(\hat a\nu(b))=\Pi_0(\hat a)\Pi_0(\nu(b))=\Pi_0(\hat a)b=
\Pi_0(\hat a')b'.\ee
The second equality here follows from (4.121).
It is now clear that $b=b'$ because they are both in $\l+$ and the
decomposition
of the element $\Pi_0(\hat a\nu(b))\in\lgc$ into elements of $LG_0$
and $\l+$ is unique.
We conclude that $b=b'$, hence $\hat a=\hat a'$.

The lemma is proved.

\rightline{\#}

\vskip1pc
\noindent {\bf Lemma 4.17:}
The group $\hhD=\br\times_Q \rlgc$   is indeed the common  Heisenberg
double of the groups $\hat{LG_0}$ and $\br\times_Q\l+$,
if we define the invariant bilinear form on its Lie algebra
$\hhd$ as
\be  (\hat\iota(x),\hat\iota(y))_{\hhd}=
{1\over 2\pi\e}\int d\si Im(  x(\si), y(\si))_{\G_0^\bc}=
(x,y)_\D,\ee
\be (\hat T^\infty,\hat\iota(\D))_{\hhd}=
(\hat t^1_\infty,\hat\iota(\D))_{\hhd}=0;
\quad (\hat T^\infty,\hat t^1_\infty)_{\hhd}={1\over \e},\ee
Here $\hat t^1_\infty$ is the generator of $\br_Q$; $\tF$ that of
the central $U(1)$
 and $\hat\iota:\D\to\hhd$ is the natural extension of
 the injection $\iota:\G\to\hat\G$. The injection $\hat\iota$ acting
on $\B$
is given by  the derivation  map $\nu_*$.
\vskip1pc
\noindent {\bf Proof:} We know already that $\hhD$ contains both
$\hat G=\hat {LG_0}$
and $\hat B=\br\times_Q \l+$ as its subgroups and that the
global decomposition $\hat D=\hat G\hat B=\hat B\hat G$ takes place.
 The isotropy
of the corresponding Lie subalgebras $\hat\G$ and $\hat\B$ follows
from the isotropy
of $\G$ and $\B$ with respect to the form $(.,.)_\D$. Indeed,  for
instance,
$\hat \G=\iota(\G)+Span(\tF)$ then for  $\xi,\eta\in \G$
we have
$$(\iota(\xi),\iota(\eta))_{\hhd}=
(\xi,\eta)_\D=0.$$
Adding to this the fact (4.128) that $(\tF,\iota(\G))_{\hhd}=0$ we obtain
$(\hat \G,\hat\G)_{\hhd}=0$.

The bilinear form $(.,.)_{\hhd}$ is also symmetric (if we complete
 appropriately the
part (4.128) of its definition)  since the form $(.,.)_\D$ is symmetric.
The non-degeneracy of the form $(.,.)_{\hhd}$ also follows immediately
from the defining
relations (4.127), (4.128) and from the fact that $(.,.)_\D$ is
non-degenerate.

The remaining thing to show is the invariance of the form $(.,.)_{\hhd}$.
 This can
be derived by the direct calculation from the Lie bracket in $\hhd$:
$$[ w_1\tF +\xi^c_1(\si)+\vr_1\tf^1, w_2\tF +\xi^c_2(\si)+\vr_2\tf^1]=$$
$$= +{1\over 2\pi}\int d\si Re(\xi^c_1,\d_\si \xi^c_2)
\tF+ [\xi^c_1,\xi^c_2](\si)-\vr_1 i \d_\si \xi^c_2(\si) +\vr_2 i\d_\si
\xi^c_1(\si).$$
Here   $w_i,\vr_i$ are real numbers and $\xi^c_i(\si)$
  are from $L\G_0^\bc$.

The lemma is proved.

\rightline{\#}
\vskip1pc
\rem : {\small We observe that $\hat B=\br\times_Q B$
as required by the Definition 4.10.}

\subsection{The double $\ttD=\br^2\times_{S,Q}\hlgc$}
As the title of this subsection suggests, the double $\ttD$ of the
affine Kac-Moody
group $\ti G$ will be the semidirect product of the plane $\br^2$
 with certain group $\hlgc$.
In order
to construct $\hlgc$, we shall  proceed in close analogy with the
 previous section
and our intermediate explanations will be therefore much briefer.

 Consider the  group $DG_0^\bc$ of smooth maps from the unit disc
into $G_0^\bc$ with
the usual pointwise multiplication. We can now define an extended
group
$ \widehat {DG_0^\bc}$ whose
elements are pairs $(\bar l,v)$ where $\bar l\in DG_0^\bc$ and
$v\in \bc^\times$ and
 whose multiplication
law reads
\be (\bar l_1,v_1)(\bar l_2,v_2)=(\bar l_1\bar l_2,v_1v_2 \exp{[2\pi
i\beta (\bar l_1,
\bar l_2)]}).\ee
Here  $\bc^\times$ is the complex plane without the origin viewed
as the Abelian  multiplicative group of complex
numbers and  $\beta$ is a complex-valued 2-cocycle on $DG_0^\bc$
given by
\be \beta(\bl_1,\bl_2)={1\over 8\pi^2}\int_{Disc} (\bl_1^{-1}d\bl_1
\stackrel{\w}{,}d\bl_2
\bl_2^{-1})_{\G_0^\bc}.\ee
The remaining notations are the same as in the previous section. It
 is crucial
to note the absence of the real part symbol in the definition of
the 2-cocycle.

Consider now the  subgroup $\d G^\bc$ of $DG_0^\bc$.
 It turns out that there is a homomorphism
$\Th^\bc:\d G^\bc\to\hdgc$  defined by
\be \Th^\bc(\bl)=(\bl,\exp{[-2\pi iC^\bc(\bl)]}),\ee
where
\be C^\bc (\bl)={1\over 24\pi^2}\int_{Ball} (d\bl\bl^{-1}\stw
d\bl\bl^{-1}\w d\bl\bl^{-1}])_{\G_0^\bc}.\ee
 The group $\hlgc$ is
  now defined as the factor group
$\hdgc/\Th^\bc(\d G^\bc)$. This group is a (nontrivial) $\bc^\times$
 bundle over the
base space $LG_0^\bc=D$. The projection $\Pi^\bc_0$ is
 $(\bl,v)\to \bl\vert_{S^1}$ and the center of $\hlgc$ is represented
 by the
  $\Th^\bc$-equivalence classes  represented by $(1,v)\in
\hdgc$.
The projection homomorphism from $\hdgc$ onto $\hlgc$ will be referred
to as
$\hat\wp_\bc$.

In order to construct the  Drinfeld
double $\ttD$, we need  two commuting one-parameter
groups
of automorphisms of $\hlgc$. We shall denote them as
    $\br_{\hat Q}$ and $\br_{\hat S}$.
To define them, associate first to every complex number $w+is$ an
automorphism
of $\hdgc$ given by
\be ^q(\bl(z,r),v)=(\bl(qz,r),v),\ee
where $q=\exp{(w+is)}$ and $\bl(z,r)\in\dgc$.
Recall, that we view the loop group $LG_0^\bc$ as the group of
holomorphic maps from the
$\si$-Riemann sphere without poles into
 the complex group $G_0^\bc$.
Similarly as  in Section 4.4.2, we can prove  that
\be ^q((\bl_1,v_1)(\bl_2,v_2))=\ ^q(\bl_1,v_1)^q(\bl_2,v_2),\ee
 and  that the group action (4.133) survives the factorization by the
 group
 $\d G^\bc$.

Thus our group $\ttD$ is defined as the group of couples $(X,q)$,
 $X\in\hlgc$,
$q\in\bc^\times$ with the following composition law
\be (X_1,q_1)(X_2,q_2)=(X_1 \ ^{q_1}X_2,q_1q_2).\ee
Now we have to prove several lemmas in order
to show that $\ttD$ is indeed the Drinfeld double of $\ti G$.

\vskip1pc
\lem {\bf 4.18:} The group $\ti G=\br\times_S\hat G$ is the
 subgroup of $\ttD$.
\vskip1pc
\pro : First we  prove that $\hat G=\hlg$ is the subset of
$\hlgc$. Consider the set
\be {\cal S}=\{\hat l\in \hlgc, \exists \bl\in \widehat{DG_0}
\subset\hdgc, \quad \hat\wp_\bc(\bl)
=\hat l\}.\ee
We are going to show that the set ${\cal S}$ is the subgroup of $\hlgc$
 isomorphic to $\hlg$.
The isomorphism $\mu:{\cal S}\to \hlg$ is defined as follows:
\be \mu(\hat l)=\wp(\bl).\ee
Recall that $\wp:\hdg\to\hlg$ is the map associating to every element
of $\hdg$ its
$\Th$-class (cf. Section 7.1).  It is immediate to check that the
definition of $\mu$
 does not depend
on the choice of the representative $\bl\in\hdg$.

If $\bl_1,\bl_2$ are the respective representatives of
$\hat l_1,\hat l_2$, then obviously
$\bl_1\bl_2$ can be chosen as the representative of $\hat l_1\hat l_2$.
 Since
$\bl_1\bl_2\in\hdg$, it follows that the set ${\cal S}$ is the subgroup
 of $\hlgc$.
 Moreover one has
\be \wp(\bl_1\bl_2)=\wp(\bl_1)\wp(\bl_2),\ee
hence $\mu$ is the group homomorphism.  It remains to show the
 injectivity and surjectivity
of $\mu$.

If $\hat l_1\neq\hat l_2$ and
$\Pi_0^\bc(\hat l_1)\neq\Pi_0^\bc(\hat l_2)$ then obviously
$\wp(\bl_1)\neq\wp (\bl_2)$.  If $\hat l_1\neq\hat l_2$ and
 $\Pi_0^\bc(\hat l_1)=\Pi_0^\bc(\hat l_2)$ then we can choose
$\bl_1,\bl_2$ in such a way
that
\be \bl_1=\bl_2 (1,e^{i\phi}), \quad e^{i\phi}\neq 1.\ee
Then obviously $\wp(\bl_1)\neq\wp (\bl_2)$ and the injectivity follows.
 The surjectivity
is also clear since the central circle acts freely on ${\cal S}$.

The lemma is proved.

\rightline {\#}
\lem {\bf 4.19:} The  group $B=L_+G_0^\bc$ can be homomorphically injected
into
$\hlgc$. Moreover, the image of this injection is preserved by the
 automorhisms
$\hat Q$.

\vskip1pc
\noindent {\bf Proof}:
Consider an element $b\in\l+$. By definition, it is the
boundary value
 of the holomorphic map $\bar b$ from the unit disc
into $G_0^\bc$. Consider now the map $\bar\nu:\l+\to\hdgc$ defined by
\be \bar\nu(b)=(\bar b,1).\ee
As in the proof of  Lemma 4.15, it can be shown that
 the map $\bar\nu$ is the homomorphism of groups.

Consider now the map $\hat\nu:\l+\to \hlgc$ defined by
\be \hat\nu=\hat\wp_\bc\circ\bar\nu.\ee
First of all, $\hat\nu$ is the group homomorphism being the
 composition of two homomorphisms.
Moreover,  it holds
 \be (\Pi_0^\bc\circ\wp_\bc\circ\bar\nu)(b)=b\in LG_0^\bc.\ee
From this it follows that $\hat\nu$ is the injection. The invariance
of $\hat\nu(B)$ under
the action of $\hat Q$ is obvious.

The lemma is proved.

\rightline{\#}
\vskip1pc
\rem :  {\small The two preceding lemmas say, in other words, that
 both $\ti G=\br\times_S\hat G$
 and $\ti B =(\br\times_Q\l+)\times \br  $ are  subgroups
of $\ttD$ which  is one of the basic properties of the Heisenberg
 double of $\ti G$ and of
$ \ti B$.  Of course, the direct product factor  $\br$ here is the
central line subgroup
of $\hlgc$ corresponding to $\Th^\bc$-classes in $\hdgc$ of the
form $(1,e^t), t\in\br$.}
\vskip1pc

\vskip1pc
\noindent {\bf Lemma 4.20}: The group $\ttD$ can be globally decomposed
as $\ttD=\ti G\ti B=
\ti B\ti G$ where $\ti G=\br\times_S\hlg$ and $\ti B=
 (\br\times_Q\l+)\times \br$.
\vskip1pc
\noindent {\bf Proof}:
Of course, it is sufficient to prove that $\hlgc$ can be decomposed
as $\hlgc=\hat G(\hat\nu(B)\times\br)$, where $\br$ stands for
 the central line in the
sense of the remark above.
Thus we are going  to prove that $\hat K\in\hlgc$ can
be uniquely decomposed as
\be \hat K=\hat a\hat\nu(b)e^t,\quad \hat a\in\hat G, \quad
   b \in B= L_+G_0^\bc,
\quad t\in \br. \ee
Consider first the element $\Pi_0^\bc(\hat K)\in \lgc$. It can be
uniquely
decomposed as
\be \Pi_0^\bc(\hat K)=ab,\quad a\in LG_0,\quad b\in\l+.\ee
Now it is certainly true that $\hat K$ can be found among the elements
of the $\bc^\times$ fiber
  $(\Pi^\bc_0)^{-1}(a)\hat\nu(b)$ above $ab$. The existence of
$\hat a$,$b$ and $t$
from (4.143) then
follows immediately.

It remains to be proved that the decomposition (4.143) is unique.
The required argument
is very similar to that of the proof of Lemma 4.16 and we shall not
repeat it here.

The lemma is proved.

\rightline{\#}

\noindent There exists the pregnant way of presenting the commutator
in the Lie algebra $\ttd$.
 It reads
$$[(X^c,\xi^c(\si), x^c),(Y^c,\eta^c(\si),y^c)]=$$\be =
(0,  [\xi^c,\eta^c] -iX^c \d_\si \eta^c +iY^c\d_\si \xi^c,
{i\over 2\pi}\int  (\xi^c,\d_\si \eta^c)_{\G_0^\bc}).\ee
Here $X^c,Y^c,x^c,y^c$ are {\it complex} numbers and
$\xi^c,\eta^c\in L\G_0^\bc$.
The identification of various generators is as follows:
$\ti T^0=(i,0,0)$
corresponds
to the automorphisms $S$ and $\ti t^1_\infty=(1,0,0)$ to the
automorphisms $Q$. Moreover,
$\ti T^\infty =(0,0,i)$ is to be identified with the central circle
generator and
$\ti t_0^1=(0,0,1)$ with the central line generator corresponding to
the group $\br$
in the decomposition $\ti B=\br\times \hat B$.
We stress that we view $\ttD$ as the real Lie algebra, nevertheless
we observe
  that the commutator just defined is complex bilinear hence $\ttd$
  possesses
the natural complex structure. Moreover,
it is instructive to compare this formula with the commutator (2.11)
of the central biextension
algebra $\ti\G$. We observe immediately that $\ti\G$
is the real form of $\ttd$ viewed as the complex Lie algebra.
In other words, the affine Lu-Weinstein-Soibelman double of $\ttD$
is nothing but the complexification   $\ti\G^\bc$, in full analogy with
 the state
of matters for the finite dimensional ordinary  Lu-Weinstein-Soibelman
double.  We can define
the following invariant non-degenerate
bilinear form on $\ti\G^\bc$
 $$ ((X^c,\xi^c, x^c),(Y^c,\eta^c,y^c))_{\ti\G^\bc}=$$
\be (\xi^c,\eta^c)_{\G^\bc}+X^c y^c+Y^c x^c,
\ee
where the form $(.,.)_{\G^\bc}$ was defined in (4.105).
The invariant non-degenerate
bilinear form on the double $\ttd$  is then defined as
\be ((X^c,\xi^c, x^c),(Y^c,\eta^c,y^c))_{\ttd} =
 {1\over\e}Im((X^c,\xi^c, x^c),(Y^c,\eta^c,y^c))_{\ti\G^\bc}\ee
Note that the form (4.147)  is  the analogue
of the forms (4.106) and (4.81).

\vskip1pc
\noindent {\bf Lemma 4.21:}
The group $\ttD$  is indeed the common  Heisenberg
double of the groups $\ti G$ and $\ti B$.

\vskip1pc
\noindent {\bf Proof:}
The parts of this proposition were already proved in the preceding
 lemmas. The only
thing that remains is to prove the isotropy of the Lie algebras $\ti\G$
and
$\ti\B$ with respect to the bilinear form (4.147).

We start with $\ti\G$. In the parametrization of $\ttd$ given by (4.145),
we have
$\ti\G= Span (iR,\G,ir)$, $R,r\in\br$ and from (4.146) and
(4.147) it follows
 $(\ti\G,\ti\G)_{\ttd}=0$;
On the other hand, $\ti\B=Span(R,\B,r)$, $R,r\in\br$ and again
 $(\ti\B,\ti\B)_{\ttd}=0$.

The lemma is proved.

\rightline{\#}

 \vskip1pc
\noindent {\bf Conclusion}:  We have shown that $\ti B$ is
  the direct product
 of the central line $\br$  with $\hat B$, i.e.
 $\ti B=\br\times \hat B$
and $\hat B=\br\times_Q B$. Thus the requirements of
the definition 4.10 of the WZW double
$\ttD$ are  indeed verified for the affine Lu-Weinstein-Soibelman double
and we
can safely perform the symplectic reduction leading to the
quasitriangular WZW model.
\vskip1pc

\chapter{Loop group quasitriangular WZW model}
We are arriving at the core of this paper. We shall consider
 the loop group quasitriangular
WZW model corresponding to the affine Lu-Weinstein-Soibelman
double $\ttD$ introduced
in the previous chapter.  The definition of this model involves
certain symplectic reduction
 (cf. Definition 4.13 of Section 4.3), which is the quasitriangular
analogue
of the full left-right reduction described in Sections 2.2.2 and 2.2.3.
However,  in the case of
 the standard loop group WZW model
described in Chapters 2 and 3, we have used also the alternative
approach by  performing
 the reduction
 at the {\it chiral} level.  The full left-right WZW model was
then obtained by glueing
two copies of the reduced chiral model.
Although both approaches gave the same result,
 the chiral reduction was much simpler from both conceptual and
technical points of view.

Here we shall take the advantage of the fact that in the loop group
case one can also construct   a {\underline {chiral}} second
floor  \underline {quasitriangular}
  master model. Therefore we can
perform the symplectic reduction  directly  at the chiral level. As
 in the standard case,
the full left-right quasitriangular WZW  model will be then obtained
by an
appropriate glueing of
the reduced chiral copies.  It can be shown that this gives the same
result as the reduction   of the full left-right master model on $\ttD$
in the spirit
of the Definition 4.13.

 Our strategy here will be therefore as  follows:
first we introduce the quasitriangular chiral master  $\ti G$-model
and perform
the quasitriangular analogue of the two-step chiral symplectic
reduction of Chapter
3. In this way, we construct the main result of our paper
which is the chiral quasitriangular WZW model.
 Moreover, we shall
give the very explicit description of its  symplectic structure and
of its Hamiltonian
  making thus
evident  that we have really to do with
 the one-parameter deformation  of the standard chiral WZW model
 described
in Section 3.2.4. We  shall finally glue up two copies of the
chiral model to obtain the full left-rigtht quasitriangular WZW theory.

\section{ Quasitriangular chiral geodesical model}
We shall first illustrate the idea of the  quasitriangular chiral
decomposition in the (finite-dimensional) case of the geodesical
model on
the \LWD \  $G_0^\bc$.
 This section  should serve as the ideological and technical reference
for the more complicated infinite-dimensional structures described later.
\subsection{Chiral splitting of the Semenov-Tian-Shansky form}
It is the well-known fact \cite{Zel} that every element $K$ of a simple
 complex connected
and simply connected group $G_0^\bc$ can be decomposed as
\be K=k_Lak_R^{-1}, \quad k_{L,R}\in G_0, \quad  a\in A_+=
\exp{\Lambda_0(\A^0_+)}.\ee
Here $G$ is the compact real form of $G_0^\bc$ and
$\A^0_+\subset\U_0^{-1}(\T)\subset\G_0^*$ is the positive Weyl chamber
introduced in Section 3.1.1. Moreover,
$\Lambda_0$ is the identification map $\Lambda_0: \G_0^*\to\B_0$ defined
 as
\be (\Lambda_0(x^*),y)_\D=\la x^*,y\ra,\quad x^*\in\G_0^*, y\in\G_0.\ee
Note that $\Lambda_0$ depends on the parameter $\e$ (cf. (4.81)), since
the form $(.,.)_\D$
does, too. It is easy to see that $A_+=\exp{\Lambda_0(\A^0_+)}$ is the
 subset of the group $A$
appearing in  the Iwasawa decomposition $G_0^\bc=G_0AN$. This subset
 {\it does not}
depend on $\e$, however, since the Weyl chamber is invariant under
the scaling.

The decomposition (5.1) is called the Cartan
one and its  ambiguity
is given by the simultaneous right multiplication of $k_L$ and $k_R$ by
 the same
element of $\bt$. Note that the same thing was true for the Cartan
decomposition of $T^*G_0$
(cf. Theorem 3.2).

Consider the
 manifold $G_0\times A_+\times G_0$; we shall denote its points as
triples $(k_L,a,k_R)$.
The Cartan decomposition (5.1) then induces a natural map $\Xi$
from this manifold into the complex Heisenberg double $D=G_0^\bc$.
 We can then pull-back the Semenov-Tian-Shansky symplectic form $\om$
by the map $\Xi$. The following lemma is of a crucial  importance for
the success of all our
programme:

\vskip1pc
\lem  ~{\bf 5.1:} Consider maps $\Xi_{L,R}: G_0\times A_+\times G_0\to D$
defined as  $\Xi_L(k_L,a,k_R)=k_La$
and $\Xi_R(k_L,a,k_R)=ak_R^{-1}$. Then
\be \Xi^*\om=\Xi_L^*\om+\Xi_R^*\om.\ee
\vskip1pc

\rem : {\small The proposition of the lemma 5.1 can be restated intuitively
 as follows:

\noindent The pullback form $\Xi^*\om$  can be chirally decomposed on
the left and right part
who talk to each other only via the variable $a$.}

\vskip1pc

\pro :

First we write the Semenov-Tian-Shansky form $\om$ as follows
\be \om=-\jp( g_R^{-1}dg_R\stw K^{-1}dK)_\D -\jp ( dg_Lg_L^{-1}\stw
 dKK^{-1})_\D.\ee
Recall (cf. (4.11)) that $K^{-1}dK$ denotes the left invariant
Maurer-Cartan form on the
Heisenberg double $D$ and $g_R^{-1}dg_R$ is $g_R^*\lm_{G_0}$,
where $g_R$ and $g_L$ are induced
by the decomposition
\be K=b_L(K)g_R(K)=g_L(K)b_R(K).\ee We can easily recover the original
formula (4.8) for $\om$ by using the isotropy of $\G_0$ with respect to
$(.,.)_\D$ and by  noting that
\be  K^{-1}dK=g_R^{-1}(b_L^{-1}db_L)g_R+g_R^{-1}dg_R,\quad dKK^{-1}=
dg_Lg_L^{-1}+g_L(db_R b_R^{-1})
g_L^{-1}.\ee
Consider now another  decomposition of $K\in D$:
\be K=p_Lk=kp_R,\ee
where
\be k=k_Lk_R^{-1}, \quad p_L=k_Lak_L^{-1},\quad p_R=k_Rak_R^{-1}.\ee
Clearly, $k_{L,R}$ and $a$ come from the Cartan decomposition (5.1).
Although they are
not given unambiguously, this ambiguity disappears in (5.8);
in other words: $k$ and $p_{L,R}$
are uniquely fixed by  $K\in D$. We can successively rewrite the
form $\om$ as follows
$$-2\om=( g_R^{-1}dg_R\stw K^{-1}dK)_\D + ( dg_Lg_L^{-1}\stw
dKK^{-1})_\D=$$
$$=( g_R^{-1}dg_R\stw k^{-1}p_L^{-1}dp_L k)_\D + ( dg_Lg_L^{-1}\stw
kdp_Rp_R^{-1}k^{-1})_\D=$$
$$=( kg_R^{-1}d(g_Rk^{-1})\stw p_L^{-1}dp_L )_\D +
( d(k^{-1}g_L)g_L^{-1}k\stw dp_Rp_R^{-1})_\D+$$
\be +( dkk^{-1}\stw p_L^{-1}dp_L)_\D +( k^{-1}dk\stw dp_R
 p_R^{-1})_\D.\ee
In deriving this equality, we have  used the isotropy of $\G_0$ with
respect to $(.,.)_\D$.
Now  we see from (5.5) and (5.7) that
\be g_Rk^{-1}=b_L^{-1}p_L,\quad k^{-1}g_L=p_Rb_R^{-1}.\ee
This permits  to write
$$-2\om= ( dp_Lp_L^{-1}\stw db_Lb_L^{-1})_\D +( p_R^{-1}
dp_R\stw b_R^{-1}db_R)_\D+$$
\be +( dkk^{-1}\stw p_L^{-1}dp_L)_\D +( k^{-1}dk\stw dp_R
p_R^{-1})_\D.\ee
 Now there are four terms on the r.h.s. of (5.11).  We insert
the expressions (5.8) into the last
two of them. The result reads
$$\Xi^*\om=\jp ( db_Lb_L^{-1}\stw dp_Lp_L^{-1})_\D +\jp( b_R^{-1}
db_R\stw  p_R^{-1}dp_R)_\D+$$
$$ \jp ((daa^{-1}+a^{-1}da)\stw k_L^{-1}dk_L)_\D-\jp( (daa^{-1}+
a^{-1}da)\stw k_R^{-1}dk_R)_\D+$$
\be+  \jp (k_L^{-1}dk_L\stw ak_L^{-1}dk_L a^{-1})_\D-
\jp (k_R^{-1}dk_R\stw ak_R^{-1}dk_R a^{-1})_\D.\ee
Now observe that
\be b_L(k_Lak_R^{-1})=b_L(k_La), \quad b_R(k_Lak_R^{-1})=
b_R(ak_R^{-1}).\ee
This means that quantities bearing the index $L$ $(R)$ do
 not depend on $k_R$ ($k_L$) hence
the proposition of the lemma follows.

\rightline{\#}

\noindent Consider now the model spaces $M_L=G_0\times \A^0_+$
and $M_R=G_0\times \A^0_-$, where
$\A^0_-=-\A^0_+$.

The symplectic form $\om_L$ on $M_L$ is defined as
$$ \om_L=\ti\Xi^*_L\om=$$\be =
\jp  (db_Lb_L^{-1}\stw dp_Lp_L^{-1})_\D +
(daa^{-1}\stw k_L^{-1}dk_L)_\D
+ \jp (k_L^{-1}dk_L\stw ak_L^{-1}dk_L a^{-1})_\D.\ee
The map  $\ti\Xi_L:M_L\to D$ is given by $\ti\Xi(k_L,\phi_L)=k_La_L$
for $(k_L,\phi_L)\in M_L$,
where
\be a_L=\exp{\Lambda_0(\phi_L)}\ee
 and $p_L$ and $b_L$ are defined as before.

Now a subtlety: we define the symplectic form $\om_R$ on $M_R$ by
exactly the same formula, i.e,
\be \om_R=\ti\Xi^*_R\om,\ee
where $\ti\Xi_R(k_R,\phi_R)=k_Ra_R$, where $ a_R=exp{\Lambda_0(\phi_R)}$.
 Such a definition may look surprising
because the right part of the form $\Xi^*\om$ in (5.12) was obtained by
 pulling back by the map
$a_Lk_R^{-1}$ rather than by $k_R a_R$. We shall see in a while
 that this gives the same thing,
however.
The advantage of our definition is evident: it allows as to study
only  the case $M_L$ since  the symplectic structure on $M_R$ is the same
(up to the change in the domain of the variable $\phi$:
$\A^0_+\to \A^0_-$).

Consider the manifold $M_L\times M_R$ equipped with the symplectic form
\be \om_{L\times R}=\om_L +\om_R.\ee
\vskip1pc
\lem ~{\bf 5.2:} The submanifold of $M_L\times M_R$, given by equating
$a_La_R=1$, is naturally
diffeomorphic to $G_0\times A_+\times G_0$ and the form
$\om_{L\times R}$ restricted
to this
 submanifold is nothing but $\Xi^*\om$ given by the equation (5.12).
\vskip1pc
\pro :
The left part of $\Xi^*\om$ (cf. (5.12)) coincides with $\om_L$ by
definition. Let us show that the
right part gives $\om_R$. We have from (5.14) and (5.16)
$$\om_R =\jp  (db_L(k_Ra_R)b_L^{-1}(k_Ra_R)\stw  dp_Rp_R^{-1})_\D +$$
\be +(da_Ra_R^{-1}\stw k_R^{-1}dk_R)_\D
+ \jp (k_R^{-1}dk_R\stw a_R k_R^{-1}dk_R a_R^{-1})_\D,\ee
where $p_R=k_Ra_Rk_R^{-1}$. Now we have to set $a_R=a_L^{-1}\equiv
 a^{-1}$ and insert in (5.18).
We obtain
$$\om_R =\jp  (db_L(k_Ra^{-1})b_L^{-1}(k_Ra^{-1})\stw
d(k_Ra^{-1}k_R^{-1})k_Rak_R^{-1})_\D +$$
\be +(k_R^{-1}dk_R\stw daa^{-1})_\D
-\jp (k_R^{-1}dk_R\stw ak_R^{-1}dk_R a^{-1})_\D.\ee
Now we see that the second and third term on the r.h.s. of (5.19)
have their counterparts
in the right part of $\Xi^*\om$. It remains to show that
$$ ( d(k_Ra^{-1}k_R^{-1})k_Rak_R^{-1}
\stw db_L(k_Ra^{-1})b_L^{-1}(k_Ra^{-1}))_\D =$$\be =
((k_R a^{-1} k_R^{-1})d(k_Rak_R^{-1})\stw b_R^{-1}(ak_R^{-1})
db_R(ak_R^{-1}))_\D.\ee
But this equality follows from the following obvious relation
\be b_L(k_Ra^{-1})=b_R^{-1}(ak_R^{-1}).\ee
\noindent The lemma is proved.

\rightline{\#}

\noindent {\bf Corollary 5.3}: The symplectic reduction of
the form $\om_{L\times R}$ by the relation  $a_La_R=1$ is the
 Semenov-Tian-Shansky
form $\om$ on $G_0^\bc$.

\vskip1pc

\pro : The form $\om_{L\times R}$ restricted on $a_La_R=1$
coincides with $\Xi^*\om$
given by (5.12) and it is
clearly degenerate along orbits
of the maximal torus $\bt$ acting as $(a,k_L,k_R)\to (a,k_Lh,k_Rh)$,
 $h\in\bt$. The orbit
space is nothing but $G_0^\bc$.

\rightline{\#}

\noindent So far we have shown that the symplectic structure
$\om$ on $D=G_0^\bc$ naturally originates from $\om_{L\times R}$
under the symplectic reduction induced by setting $a_La_R=1$.
But  also the   Hamiltonian $H (K)$ introduced in Section 4.2
can be  "descended" from some  Hamiltonian on $M_L\times M_R$. Indeed,
the latter is defined
as follows
\be H_{L\times R}=H_{L}+H_{R}=-\jp(\phi_L,\phi_L)_{\G_0^*}-
\jp(\phi_R,\phi_R)_{\G_0^*}=\jp a^\mu_L a^\mu_L + \jp a^\mu_R a^\mu_R.\ee
Recall that
\be a_L=\exp{(a^\mu_L \Lambda_0(t_\mu))}, \quad  a_R=\exp{(a^\mu_R
\Lambda_0(t_\mu))}.\ee
Clearly, $\Lambda_0(t_\mu)$'s are in $Lie(A)\subset \B_0$, they
fulfil $(\Lambda_0(t_\mu),T^\nu)_\D=
\la t_\mu,T^\mu\ra= \delta_\mu^\nu$
and $T^\mu\in\G_0$ were defined in (3.24). By the abuse of notation, we
shall often write
$\Lambda_0(t_i)=t_i$, i.e; the identification map will be tacitly
 assumed (cf. (4.6)).
However, in the case where a confusion can arise we shall use the
 symbol $\Lambda_0$
explicitely.

Note that the Hamiltonians $H_{L,R}$ on $M_{L,R}$ coincide
with the Hamiltonians of the standard geodesical model (3.12).
It is the symplectic form $\om_L$, defined by (5.14), that differs
from $d\theta_L$
given by (3.11). We shall see, however, that for $\e\to 0$ it holds
$\om_L\to d\theta_L$.
The Hamiltonians do
 not depend on $k_L,k_R$, respectively, hence they trivially survive
the symplectic reduction and give  the Hamiltonian (4.102) on $D=G_0^\bc$.
\vskip1pc
\defi ~{\bf 5.4:}  The chiral quasitriangular geodesical model
is the dynamical
system whose phase space
is $M_L$ parametrized by couples $(k\in G_0,\phi\in \A^0_+)$,
whose Hamiltonian is
\be H_{L}=-\jp(\phi,\phi)_{\G_0*}=\jp  a^\mu a^\mu\ee
 and whose  symplectic form is
$$ \om_L=$$\be =\jp (db_L(ka)b_L^{-1}(ka)\stw dpp^{-1})_\D +
(daa^{-1}\stw k^{-1}dk)_\D
+ \jp (k^{-1}dk\stw a(k^{-1}dk) a^{-1})_\D,\ee
where $p=kak^{-1}$ and  $a=\exp{\Lambda_0(\phi)}=
\exp{(a^\mu \Lambda_0(t_\mu)}$.
\vskip1pc
 We have learned in this section that the quasitriangular
 geodesical model
formulated on the Lu-Weinstein-Soibelman double $G_0^\bc$
admits the chiral decomposition
into two chiral models defined above.
By this we mean that it  can be defined
by the symplectic reduction of the model defined on $M_L\times M_R$
and characterized
by the symplectic form $\om_{L\times R}$ and the Hamiltonian
$H_{L\times R}$.
\vskip1pc
 \rem : {\small The chiral quasitriangular geodesical model was
apparently first proposed in \cite{AT}.
 In what follows, we shall invert the symplectic form $\om_L$ in
the way which
technically  differs from that of \cite{AT}.  Its spirit is  the same,
however, in that we use
the Poisson-Lie symmetry of $\om_L$. }
\vskip1pc
\subsection{The power of the Poisson-Lie symmetry}
In the paragraphs 5.1.2 and 5.1.3, we shall often parametrize
the points of $M_L$
by couples $(k,a)$, where $k\in G_0$ and $a=e^{\Lambda_0(\phi)}\in A_+$.
\vskip1pc
\lem ~{\bf 5.5:}  The chiral quasitriangular geodesical model is Poisson-Lie
symmetric
(cf. Definition 4.7) with respect to the
left action $(k,a)\to (k_0k,a)$ of the group $G_0$ on the
 model space $M_L$. The corresponding
non-Abelian moment map $M:M_L\to B$ is given by
$M(k,a)=b_L(ka)=Dres_ka$.
\vskip1pc
\pro: Consider a point $(k,a)$ in $M_L$. This point can be
mapped into $D$ as $\ti\Xi(k,a)=ka$
under
the embedding $M_L\hookrightarrow D$.
Multiplication of $(k,a)$ on the left by an infinitesimal
generator $T\in \G_0$ gives the vector
$v_T=(Tk,a)$ hence
the vector $\ti\Xi_*v_T= Tka=R_{(ka)*}T\in T_{(ka)}D$.
We want to show that
\be \om_L(.,v_T)=(T, dMM^{-1})_\D.\ee
Since the form $\om_L$ is the pullback
$\ti\Xi^*\om$, we infer
\be \om_L(u,v_T)=\om(\ti\Xi_*u,\ti\Xi_*v_T)=
\om(\ti\Xi_*u,R_{(ka)*}T),\ee
where $u$ is an arbitrary vector at the point $(k,a)\in M_L$.
Now  from (4.88), we conclude
\be \om(\ti\Xi_*u,R_{(ka)*}T) =(T,\la b_L^*\rho_B,\ti\Xi_* u\ra)_\D=
(T,\la M^*\rho_B,u\ra)_\D
=(T,\la dMM^{-1},u\ra)_\D,\ee
because $\ti\Xi^*b_L=M$.
 The Hamiltonian  $H_L$ does not depend on $k$,
it is therefore clearly invariant with respect to $(k,a)\to(k_0k,a)$.

The lemma is proved.

\rightline{\#}
\noindent
The existence of the non-Abelian moment map $M:M_L\to B$ entails
 certain differential
condition on the Poisson bivector $\Si_L$, corresponding to the
form $\om_L$. Recall first, that
\be \Si_L(.,\om_L(.,u))=u,\ee
for arbitrary vector $u$ at arbitrary point of $M_L$. It then follows
from (5.26), that
\be \Si_L(.,dMM^{-1})=v,\ee
where $v\in TM_L\otimes \G_0^*(\equiv TM_L\otimes\B_0)$ generates the
left action of
 $G_0$ on the model space $M_L$.

Calculate now the Lie derivative  $\L_v\Si_L$ with respect to $v$.
 The result is
$$ \L_v\Si_L=\L_v\om_L^{-1}=-\Si_L \L_v\om_L\Si_L=
-\Si_L di_v\om_L\Si_L=$$ \be
+\Si_L d(dMM^{-1})\Si_L=\Si_L (dMM^{-1}\w dMM^{-1})\Si_L=-v\w v.\ee
The last relation can be also written in some basis $T^i$ of $\G_0$:
\be \L_{v^i}\Si_L=-\jp \ti f^i_{~jk} v^j\w v^k,\ee
where
\be \ti f^i_{~jk}=(T^i,[t_j,t_k])_\D, \quad (t_i,T^j)_\D=\delta_i^j,
\quad v^i=(v,T^i)_\D.\ee
This is the differential condition that the Poisson tensor $\Si_L$
 must obey.

Note first that there is a particular solution  of the
differential condition (5.32). It  reads
\be \Si_L^{part}(k,a)=-\jp\Pi^R_{ij}(k)R_{k*}T^i\w R_{k*}T^j,\ee
  where the matrix $\Pi^R_{ij}(k)$ is given  by
\be \Pi^R_{ij}(k)=-B_{ik}(k)A^k_{~j}(k^{-1}),\ee
\be B_{ik}(k)=(k^{-1}t_ik,t_k)_\D,\quad A^k_{~j}(k)=
(k^{-1}T^kk,t_j)_\D.\ee
 Now we replace the role of the groups $G$ and $B$ in the Proposition
4.5 in order to realize that,  up to the  sign minus,
 the r.h.s. of (5.35) is the
Poisson-Lie bivector on the group manifold $G_0$ (here viewed as
the bivector on $G_0\times A_+$).
If we regard the condition (5.32) as the differential equation for
 the unknown bivector $\Si_L$,
we see immediately that its general solution can be given as the
sum of the particular
solution (5.34) and any solution of the homogeneous equation. But
the latter is nothing but
every  $G_0$-right-invariant bivector because $v^i=R_{k*}T^i$.
 We can therefore write the following ansatz for the seeken
Poisson bivector $\Si^L$
$$ \Si_L(k,a) =-\jp\Pi^R_{ij}(k)R_{k*}T^i\w R_{k*}T^j+$$
\be +\Si^0_{ij}(a)L_{k*}T^i\w L_{k*}T^j+  \si_{i}^{~\mu}
(a)L_{k*}T^i\w
{\partial\over \partial a^\mu}
+s^{\mu\nu}(a){\partial\over \partial a^\mu}\w{\partial\over
\partial a^\nu}\ee
Our task is to find the coefficient functions  $\Si^0_{ij}(a)$,
$\si_i^\mu(a)$
and $s^{\mu\nu}(a)$.

Thus we observe the power of the Poisson-Lie symmetry:
 the Poisson tensor  $\Si_L$
on the model space $M_L$ is completely determined by its value at
the points $(e,a)\in M_L$,
in other words: at the unit element  $e$ of $G_0$.

\subsection{Deformed dynamical $r$-matrix}
Here we shall evaluate the unknown functions $\Si^0_{ij}(a)$,
$\si_{i}^{~\mu} (a)$
and  $s^{\mu\nu}(a)$. For this, we first calculate the matrix
of the symplectic form
$\om_L$ at points $(e,a)$ (in the basis $R_{a*}T^i,R_{a*}t_\mu$ of
$T_{(e,a)}M_L$)
and then we invert it to
obtain the  unknown coefficient functions of the Poisson bracket.

The basis $T^i$ of $\G_0$ was canonically chosen in (3.24), i.e.
 $T^i=(T^\mu,B^\al,C^\al)$,
where
$$ T^\mu=iH^\mu,\quad B^\al={i\over\sqrt{2}}(E^\al+E^{-\al}),\quad
 C^\al={1\over \sqrt{2}}
(E^\al -E^{-\al}),\eqno(3.24).$$ Recall then
that
$a$ is an element of the group $A$; its Lie algebra $Lie(A)$ is
 generated by the generators
$t_\mu=\e H^\mu$ which are dual to $iH^\mu$ with respect to the
bilinear form $(.,.)_\D$
given by the formula (4.81). The full dual basis reads
\be t_i=(t_\mu,b_\al,c_\al)=(\e H^\mu,
{\e\vert \al\vert^2\over \od}E^\al, -i
{\e\vert \al\vert^2\over \od}E^\al)\ee
and it satisfies the basic condition $(t_i,T^j)_\D=\delta_i^j$.
\vskip1pc
\rem : {\small Note that the form $(.,.)_\D$ given by (4.81) depends
on $\e$ therefore
also the dual generators $t_i$ are $\e$-dependent. We shall
 occasionally stress this dependance
by writing  $(.,.)_\e$ and  $t_i^\e$. }
\vskip1pc
We decompose the elements of the basis of $T_{(e,a)}M_L$ into several
  parts:
the $\al$-part generated by
$R_{a*}B^\al,R_{a*}C^\al$, $\al\in\Phi_+$ and the
 $\mu$-part generated by $R_{a^*}T^\mu,R_{a^*}t_\mu$.
Now we use the formula (4.18)
$$\om(t,u)=(t,(\Pi_{\ti L R}-\Pi_{L\ti R})u)_\D.$$
We immediately observe that
\be \om_L(\al,\mu)=0,\quad \om_L(\al,\beta)=0,\quad \al,\beta\in\Phi_+,
\quad \al\neq\beta.\ee
This is because we are at the point $(e,a)\in M_L(\subset D)$.
It is hence sufficient to
invert the matrix $\om_L$ for the $\mu$-sector and for every
$\alpha$-sector separately.
We have
\be \Pi_{L\ti R}R_{a*}T^\mu=0,\quad \Pi_{\ti LR}R_{a*}T^\mu=
R_{a*}T^\mu.\ee
From this we obtain
\be \om_L(R_{a*}t_\nu, R_{a*}T^\mu)=\delta^\mu_\nu.\ee
One has
\be R_{a*}C^\al =( R_{a*}C^\al, L_{a*}t_i)_\D  L_{a*}T^i+(R_{a*}C^\al,
 L_{a*}T^i)_\D L_{a*}t_i\ee
and from (5.38)
\be L_{a*}b_\al=e^{a^\mu\la\al,t_\mu\ra}R_{a*}b_\al,\quad L_{a*}c_\al
=e^{a^\mu\la\al,t_\mu\ra}
R_{a*}c_\al.\ee
Then we have
\be  \Pi_{L\ti R}R_{a*}C^\al=e^{a^\mu\la \al,t_\mu\ra} (L_{a*}B^\al,
R_{a*}C^\al)_\D R_{a*}b_\al
+e^{a^\mu\la \al,t_\mu\ra} (L_{a*}C^\al,R_{a*}C^\al)_\D R_{a*}c_\al;\ee
\be   \Pi_{\ti LR}R_{a*}C^\al=R_{a*}C^\al.\ee
It follows that
\be  \om_L(R_{a*}B^\al, R_{a*}C^\al)=-e^{a^\mu\la \al,t_\mu\ra}
 (L_{a*}B^\al,R_{a*}C^\al)_\D={1\over
\e\vert \al\vert^2}(e^{2a^\mu\la \al,t_\mu\ra}-1).\ee
Our conclusion is  that
\be s^{\mu\nu}=0, \quad \si_\al^\mu=0,\quad \si_\mu^\nu=\delta_\mu^\nu,
\quad  \Si^0_{\mu\nu}=
\Si^0_{\al\mu}=\Si^0_{\al\beta}=0\ee
and
$$ \Si_L(k,a) =-\jp\Pi^R_{ij}(k)R_{k*}T^i\w R_{k*}T^j+$$
\be +\sum_{\al\in\Phi_+}{\e\vert\al \vert^2\over (1-e^{2\e a^\mu\la\al,
H^\mu\ra} )}L_{k*}B^\al\w
L_{k*}C^\al  +   L_{k*}T^\mu\w
{\partial\over \partial a^\mu}.
 \ee
Recall that in the formulas above, we have set $a=e^{a^\mu t_\mu}$ and
$t_\mu=\e H^\mu$.

Denote $\{.,.\}_{qM_L}$ the Poisson bracket corresponding to the
bivector $\Sigma_L$.
  Now we wish to calculate  the   bracket of the type
 $\{k\ptp k\}_{qM_L}$;
according to (5.48), it can be decomposed in two parts
\be  \{k\ptp k\}_{qM_L}= \{k\ptp k\}_{\Si^0}-\{k\ptp k\}^R_{G_0},\ee
where, of course,
$k$ is  understood in some  representation $\rho$, the bracket
$\{k\ptp k\}_{\Si^0}$
is associated to the bivector on the second line of (5.48)
 and the bracket  $\{k\ptp k\}^R_{G_0}$ is the standard
Poisson-Lie bracket (4.70)  on the group $G_0$. Let us calculate
the latter more explicitely :
\be  \{k\ptp k\}^R_{G_0}=(T^ik\otimes T^jk) \Pi_{ij}^R(k)=
-(T^ik\otimes T^jk)(k^{-1}t_ik,t_l)_\D
(kT^lk^{-1},t_j)_\D.\ee
From the isotropy of $\G_0$, it follows
$$ (kT^lk^{-1},t_j)_\D T^j=kT^lk^{-1}.$$
Inserting this back into (5.50), we obtain
\be  \{k\ptp k\}^R_{G_0} =-(T^i k\otimes kT^l)(k^{-1}t_ik,t_l)_\D.\ee
We insert in (5.51)  another obvious identity:
$$k^{-1}t_ik=(k^{-1}t_ik,T^l)_\D t_l+(k^{-1}t_ik,t_l)_\D T^l$$
to obtain finally
\be  \{k\ptp k\}_{G_0}^R=-
T^ik\otimes t_i k +kT^i\otimes kt_i =[(k\otimes k),(T^i\otimes t_i)].\ee
We recall that $D=G^\bc_0$ was viewed as the {\it real} group. Among the
representations
of $\D$, we can therefore consider those originating from the complex
representations
of $\G^\bc_0$.
From now on we restrict our attention to the faithful representations
 of this type.
The representatives of the elements $B^\al, C^\al, b_\al, c_\al$ are
then obtained
from the representatives of $E^\al$ by using the formulae (3.24)
and (5.38).
In such representations and
using the  canonical choice of the basis (3.24) and (5.38),
we can calculate
\be r_+\equiv T^i\otimes t_i=
i\e (H^\mu\otimes H^\mu +\sum_{\al>0}
{\vert\al\vert^2}E^{-\al}\ot E^{\al})=i\e C +\e r.\ee
Here
\be C\equiv H^\mu\otimes H^\mu +\sum_{\al\in\Phi_+}
{\vert\al\vert^2\over 2}(E^\al\ot E^{-\al}+
E^{-\al}\ot E^{\al})\ee
is the Casimir element and
\be r\equiv
\sum_{\al\in\Phi_+}{i\vert\al\vert^2\over 2}
(E^{-\al}\ot E^\al-E^\al\ot E^{-\al})\ee
is the so-called classical $r$-matrix. We note that the Casimir
element commutes with the diagonal
elements like $(k\ot k)$. Hence, we can rewrite (5.52) as
\be  -\{k\ptp k\}^R_{G_0}=\e[r,(k\otimes k)].\ee
Now we use (5.56) and (5.48) to write down the  Poisson bracket:
\be  \{k\ptp k\}_{M_L}=(k\ot k)r'_\e(a^\mu) -\{k\ptp k\}^R_{G_0}=
(k\ot k)r'_\e(a^\mu)+
\e[r,(k\otimes k)],\ee
where
\be r'_\e(a^\mu)=\sum_{\al\in\Phi_+}{\e\vert\al \vert^2\over
(1-e^{2\e a^\mu\la\al,H^\mu\ra })}
(B^\al\ot C^\al-C^\al
\ot B^\al).\ee
The Poisson bracket (5.57) can be rewritten as
\be \{k\ptp k\}_{qM_L}=(k\otimes k) r_{\e}(a^\mu)+\e r(k\otimes k),\ee
where $r_{\e}(a^\mu)$ is the so called canonical dynamical $r$-matrix
associated to a
 simple Lie algebra
( see e.g. \cite{Xu}).
It is given by
\be r_\e(a^\mu)=r_\e'(a^\mu)-\e r=i\e \sum_{\al\in\Phi}{\rn\over 2}
{\rm coth}(\e a^\mu\la \al,
H^\mu\ra)
  E^\al\ot E^{-\al}.\ee
 It is interesting to observe that the deformed braiding relation
(5.59) involves
two canonical $r$-matrices: the standard one and the dynamical one.
The description of the full  Poisson bivector $\Sigma_L$
 on $M_L$ is then completed by the following bracket
\be \quad \{k,a^\mu\}_{qM_L}=kT^\mu .\ee
It is important to calculate the limit $\e\to 0$ (or, equivalently,
$q\to 1$)
of the dynamical $r$-matrix $r_\e(a^\mu)$
and of the Poisson brackets $\{k\ptp k\}_{qM_L}$, $\{k,a^\mu\}_{qM_L}$.
 Recall the explicit expression for the dynamical $r$-matrix
$r_0(a^\mu)$ obtained in (3.30):
\be  r_0(a^\mu) = \sum_{\al\in\Phi_+}
{i\vert \al\vert^2\over 2a^\mu\la \al,H^\mu\ra}E^\al\ot
E^{-\al}.\ee
 We observe immediately that $\displaystyle\lim_{\e\to 0}r_\e(a^\mu)=
 r_0(a^\mu)$.
  Looking at (3.34), (3.35), (5.59) and  (5.61), we conclude that
$lim_{q\to 1}\{.,.\}_{qM_L}=\{.,.\}_{M_L}$. In other words: the
 symplectic
structure of the chiral quasitriangular geodesical model is the smooth
$q$-deformation of the symplectic
structure of the standard chiral geodesical model. The same conclusion
 can be obviously
obtained also by studying directly the $\e\to 0$ limit of the
 bivector (5.48).

Our next task is to calculate the Poisson bracket
$\{b_L(ka)\ptp b_L(ka)\}_{qM_L}$ of the
 non-Abelian
moment maps $b_L(ka)$. The simplest way to do it is to use Lemma 5.5 and
 realize that for any function
$f(k,a)$ on $M_L$ it holds
\be ( T^i,\{f(k,a),b_L(ka)\}_{qM_L}b_L^{-1}(ka))_\D=
\la\nabla^L_{G_0} f,T^i\ra\ee
or, equivalently,
\be \{f(k,a),b_L(ka)\}_{qM_L}=\la\nabla^L_{G_0} f,T^i\ra t_ib_L(ka).\ee
In particular, we have
\be \{k\ptp b_L(ka)\}_{qM_L}=(T^i\otimes t_i) (k\otimes b_L(ka)) =
r_+(k\otimes b_L(ka)).\ee
Replacing $f(k,a)$ by the matrix valued function $b_L(ka)$, we obtain
$$\{b_L(ka)\ptp b_L(ka)\}_{M_L}=b_L(T^ika)\otimes t_ib_L(ka)=$$
$$=(b_L^{-1}T^ib_L,T^j)_\D b_Lt_j\otimes t_ib_L=$$
\be (T^i\otimes t_i)(b_L\otimes b_L)-(b_L\otimes b_L)(T^i\otimes t_i),\ee
where the last equality follows from
\be b_L^{-1}T^ib_L=( b_L^{-1}T^ib_L, T^j)_\D t_j+
(b_L^{-1}T^ib_L,t_j)_\D T^j.\ee
Using (5.53), we can finally write
\be \{b_L(ka)\ptp b_L(ka)\}_{qM_L}=\e[r,b_L(ka)\ot b_L(ka)].\ee
In the case of the compact group $G_0$, the Poisson  brackets of the
type  $\{k\ptp k\}$ determines completely
the Poisson tensor. In the case of the group $AN$ this is no
longer true, because (5.68)
  computes
only the Poisson brackets of the holomorphic functions of the
 variables $v_\al$ in the formula (4.83).
It turns out that knowing two other matrix Poisson brackets of the form
$\{b_L^\dg(ka)\ptp b_L^\dg(ka)\}_{qM_L}$ and
 $\{(b_L^\dg(ka))^{-1}\ptp b_L(ka)\}_{qM_L}$
 is already sufficient. The former bracket can be calculated
similarly as before with
the result
\be \{b_L^\dg(ka)\ptp b_L^\dg(ka)\}_{qM_L}=-\e[r,b_L^\dg(ka)\ot
 b_L^\dg(ka)].\ee The calculation
of the latter goes as follows
 $$\{(b_L^\dg(ka))^{-1}\ptp b_L(ka)\}_{qM_L}=-(b_L^{-1}T^ib_L,T^j)_\D
(b_L^\dg)^{-1}t^\dg_j\ot
t_i b_L=$$
$$=T^i(b_L^\dg)^{-1}\ot t_ib_L-(b_L^{-1}T^ib_L,t_j)_\D (b_L^\dg)^{-1}T^j
\ot t_ib_L=$$
\be =(T^i\otimes t_i)((b_L^\dg)^{-1}\otimes b_L)-((b_L^\dg)^{-1}
\otimes b_L)(T^i\otimes t_i)=
[r_+, ((b_L^\dg)^{-1}\otimes b_L)].\ee
In the derivation of the relation above, we have used the following
 formula
\be -b_L^\dg T^i (b_L^\dg)^{-1}=( b_L^{-1}T^ib_L, T^j)_\D t_j^\dg-
(b_L^{-1}T^ib_L,t_j)_\D T^j,\ee
which is the consequence of (5.67).
The Poisson brackets (5.68), (5.69)  and (5.70) constitute the
 quasitriangular
 generalization
of the commutation relation (3.41) between the coefficients of the
 standard Abelian
moment map $M(k,a)=\beta_L(ka)$  (cf. Section 3.1.3). To  see this
we compute the
$\e\to 0$ limit of (5.68), (5.69) and (5.70). This calculation requires
 some notational care
(cf.  Remark after Eq. (5.23)).  Until the end of this paragraph, we shall
make the notational
 distinction
between $\G_0^*$ and $\B_0$; i.e.   $t_i$'s will be always the
elements of $\G_0^*$
dual to $T^i\in\G_0$ with respect to the pairing $\la .,.\ra$ and
 $t_i^\e=\Lambda_0(t_i)$ the elements of $\B_0$ dual to $T^i\in\G_0$
with respect to the pairing
$(.,.)_\e$.

We start from the formula
\be b(ka)=Dres_k(a)= 1+ a^\mu C_\mu^{~j}(k)t^\e_j+ O(\e^2)
 \ee
written is some matrix representation of $G_0^\bc$.
Here $C_i^{~j}(k)$ is the matrix defined by
\be Coad_kt_i=C_i^{~j}(k)t_j.\ee
The formula (5.72)  can be inferred from
\be a=\exp{(a^\mu t_\mu^\e)}=\exp{(a^\mu \e H^\mu)}=1+ a^\mu t_\mu^\e +
O(\e^2)\ee
and from the relation (4.5) rewritten as
\be [T^i,t_j^\e]= f^{kj}_{~~i}t^\e_k+([t_i^\e,t_k^\e],T^j)_\e T^k.\ee
Here $f^{kj}_{~~i}$ are the structure constants of $\G_0$
defined by the relation $[T^i,T^j]=f^{ij}_{~~l}T^l$.
Now recall the formula (3.38) for the Abelian moment maps
$\la M(k,a^\mu),T^j\ra$
studied in Section 3.1.3. It can be rewritten as
\be \la M(k,a^\mu),T^j\ra =\la \beta_L(k\phi),T^j\ra=
a^\mu C_\mu^{~j}(k).\ee
Note that here the multiplication $k\phi$ is in the sense of $T^*G_0$.
We conclude that the $\e$-expansion (5.72) can be rewritten as
\be b(ka)=Dres_k(a)= 1+\e  \la M(k,a^\mu),T^j\ra t^1_j+ O(\e^2).
 \ee
Inserting this expansion into the formulae (5.68), (5.69) and (5.70)
 and using
 the explicit
form (5.38) of the dual basis $t_j^1$ is the chosen representation, we
 obtain (in the lowest
nontrivial order  $\propto\e^2$ ) the following formula
 \be \{\la M,T^i\ra,\la M,T^j\ra\} =f^{ij}_{~~l}\la M,T^l\ra .\ee
 The result (5.78)  coincides with the standard  formula (3.41) of
Section 3.1.3.

\subsection{The classical solution}
It is very easy to solve classically the quasitriangular chiral
geodesical model.
 It is enough to use
the Hamiltonian (5.22)
and the Poisson bracket (5.61) to conclude that the quasitriangular
 geodesics
satisfy the equation
\be {d\over d\tau}a^\mu_L=\{a^\mu_L,H_L\}_{qM_L}=0,
\quad {d\over d\tau}k_L=\{k_L,H_L\}_{qM_L}=  k_LT^\mu a_L^\mu.\ee
The  general solution of these equations has the form
\be k_L(\tau)=k_L(0)\exp{(a_L^\mu T^\mu)\tau},\quad a^\mu_L(\tau)=
a^\mu_L(0).\ee
By comparing with (3.18), we see that the deformation has not changed
 the classical
solutions of the geodesical model!  So what got deformed after all?
It is in fact the
symplectic structure on the space of solutions (phase space) that
 got deformed. This
means that  the natural dynamical variables of the group
theoretical origin will have modified Poisson bracket and, upon
the quantization,
modified commutation relations. For instance, the correlation
functions  (in the field theoretical applications) will change.

\section{Quasitriangular chiral WZW model}
We proceed in full analogy with the previous Section 5.1, where the
chiral   model
was constructed from the full  geodesical  model.
Thus the  deformed chiral master model  is going to live on the same
 affine
model space $\ti M_L$
as the non-deformed one (3.60).
Its symplectic structure $\ti \om_L^q$ will be obtained by embedding
 the affine
model space $\ti M_L$ (defined in Section 3.2.2) into the double $\ttD$
and by
pulling back the Semenov-Tian-Shansky
form from $\ttD$  to $\ti M_L$ by the embedding map. The chiral
 Hamiltonian on $\ti M_L$
will be the same as in the standard non-deformed master model (3.58).
Then the
two step symplectic reduction of this quasitriangular chiral master
model will be
  the deformed chiral WZW theory.

\subsection{Quasitriangular chiral  master model}
Consider the affine Lu-Weinstein-Soibelman double $\ttD$ introduced in
 section 4.4.
There is the  distinguished subgroup $\ti A\subset \ttD$ that we
 shall call the Cartan subgroup.
It is defined as
\be \ti A = \br_Q\times \nu(A) \times\br_l.\ee
Here the first copy $\br_Q$ corresponds to the automorphisms $Q$
obtained by
exponentiating the generator
$\ti t^1_\infty=(1,0,0)$ and the second copy $\br_l$ to
the (central) line  generated by
 $\ti t^1_0=(0,0,1)$.

Recall that   $\nu$ is the injective homomorphism sending the group
$B=L_+G_0^\bc$ into $\ttD$ (cf. Section 4.4.2). Moreover, we have
 already encountered
the  group $A$ in section 4.1.3. We have $A\subset B_0\subset B$ hence
 the notation $\nu(A)$
makes sense. Note that the group $B_0=AN$ is the zero mode subgroup
of $B=L_+G_0^\bc$, hence
the automorphisms from $Q$ do not act on $A$. From this  fact we
 conclude that
the Cartan subgroup $\ti A\subset\ttD$ is commutative and the direct
products in (5.81) make sense.

Recall that the Cartan subgroup $\ti \A\subset T^*\ti G$ was defined
 as $\ti\U^{-1}(\ti\T)$,
where $\ti\T=\T\oplus \br (i,0,0)\oplus \br (0,0,i)$ and $\T=Lie(\bt)$.
Consider the identification map $\ti\Lambda:\ti\G^*\to\ti\B=Lie(\ti B)$
defined by
\be ( \ti\Lm(\ti x^*),\ti y)_{\ttd}=\la \ti x^*,\ti y\ra,
\quad \ti x^*\in\ti\G^*,\ti y\in\ti\G. \ee
Note that  $\ti\Lm$ depends on $\e$ because $(.,.)_{\ttd}$ does
 (cf. (4.147)).
  We can naturally define the fundamental Weyl
alcove  $\ti A_+$ in the Cartan subgroup $\ti A\subset \ttD$ as
\be \ti A_+=\exp{\ti\Lm(\ti \A_+)},\ee
where $\ti \A_+$ is the fundamental Weyl alcove in the Cartan
subgroup $\ti A\subset T^*\hat G$
as explained in Section 3.2.1.

Recall that the Semenov-Tian-Shansky form $\ti\om$ on the
Heisenberg double $\ttD$ is
given
by the formula
\be \ti\omega=\jp(\ti b_L^*\lambda_{\ti B}
\stackrel{\w}{,}\ti g_R^*\rho_{\ti G})_{\ttD}+
\jp(\ti b_R^*\rho_{\ti B}\stackrel{\w}{,}\ti g_L^*\lm_{\ti G})_{\ttD}.
\ee
 Here the  maps $\ti b_L:\ttD\to \ti B$ and $\ti g_R:\ttD\to \ti G$ are
 induced by the
decomposition $\ttD=\ti B\ti G$
and $\ti g_L:\ttD\to \ti G$ and $\ti b_R:\ttD\to \ti B$     by
 $\ttD=\ti G\ti B$.
The expression $\lm_{\ti G}$ ($\rho_{\ti G}$) denotes
the left(right) invariant $\ti\G$-valued Maurer-Cartan form on
 the group $\ti G$.

Consider now the affine model space $\ti M_L =\ti G\times \ti \A_+$.
 The Semenov-Tian-shansky form $\ti\om$ can be pulled back to  the
 space $\ti M_L$
by the map $\ti\Xi:\ti M_L\to\ttD$ defined by
 $\ti\Xi(\ti k,\ti a)=\ti k\ti a\in\ttD$,
where $\ti k\in\ti G$, $\ti a=\exp{\ti\Lm(\ti\phi)}$
 and $\ti\phi\in\ti\A_+$.
The pull-back form $\ti\om_L\equiv\ti\Xi^*\ti\om$ defines
the chiral symplectic structure on $\ti M_L$.
Its explicit form can be  found by reconducting step by step
the finite
dimensional calculation of Section 5.1.1.
 The result is
$$ \ti\om_L\equiv\ti\Xi^*\ti\om =\jp (
d\ti b_L(\ti k\ti a)\ti b_L^{-1}(\ti k\ti a)\stw
d\ti p\ti p^{-1})_{\ttD}+$$\be +
(d\ti a\ti a^{-1}\stw \ti k^{-1}d\ti k
)_{\ttD}
+ \jp (\ti k^{-1}d\ti k\stw \ti a(\ti k^{-1}d\ti k)
 \ti a^{-1})_{\ttD},\ee
where $\ti p=\ti k\ti a\ti k^{-1}$.
This leads to the following definition
\vskip1pc
\defi {\bf 5.6:} The quasitriangular chiral master model is the
 dynamical system on the  phase space
$\ti M_L$,
 whose symplectic structure is given by (5.85) and whose
 Hamiltonian is
\be \ti H_L=-{1\over 2\k}(\ti\phi,\ti\phi)_{\ti\G}.\ee
\vskip1pc
\noindent Note that the Hamiltonian (5.86)  is the same as
that of the standard chiral master model (3.58), however,
the symplectic structure is different.  It is also crucial
to remark that the model  (5.85-86) is Poisson-Lie symmetric
in the sense of Definition 4.7. The proof of this fact is
very similar  as that of the analogous
finite-dimensional result  expressed in Lemma 5.5. In particular,
 the non-Abelian moment map
corresponding to the left $\ti G$-action on $\ti M_l$ is given
 by $\ti b_L(\ti k\ti a)$.

\subsection{Chiral symplectic reduction: the first step}
\noindent Now we are going to perform the symplectic reduction
of the
chiral master  model  introduced
in the previous subsection.
 As we know, we can
do it in the language of the symplectic forms (like in Section 2.2.3)
 or using the Poisson
bracket formalism (as in Section 7.3). Here we choose the first
 (second) possibility for the
first (second) step of the reduction.

We first  need some preliminary description of the objects with
which we are going to work.
 The Cartan subgroup $\hat A$
of the first floor double
$\hhD$  is defined as
\be \hat A=\br_Q\times\nu(A),\ee
with the same notation as in (5.81). Then define the identification
map $\hat\Lm:\hat\G^*\to\hat \B =Lie(\hat B)$
as follows
\be ( \hat\Lm(\hat x^*),\hat y)_{\hhd}=\la \hat x^*,\hat y\ra,\quad
\hat x^*\in\hat\G^*,\hat y\in\hat\G. \ee

Recall also the definition of the alcove $\hat\A_+$:
$$\hat \A_+=\{\hat\phi\in\ti\A_+, \hat\phi =(0,\phi,a^\infty)^*\}.$$
The hat-alcove $\hat A_+$ in $\hhD$ is then set to be
\be \hat A_+=\exp{\hat\Lm(\hat\A_+)}.\ee

Consider now the reduced model space $\hat M_L =\hat G\times \hat \A_+$.
 The Semenov-Tian-shansky form $\hat\om$ on $\hhD$
 can be pulled back to the reduced model space $\hat M_L$
by the map $\hat\Xi:\hat M_L\to\hhD$ defined by $\hat\Xi(\hat k,\hat a)=
\hat k\hat a\in\hhD$,
where $\hat k\in\hat G$, $\hat a=\exp{\hat\Lm(\hat\phi)}$ and
$\hat\phi\in\hat\A_+$.
The pull-back form $\hat\om_L\equiv\hat\Xi^*\hat\om$ defines the
symplectic structure on $\hat M_L$:
 $$ \hat\om_L\equiv\hat\Xi^*\hat\om =\jp (
d\hat b_L(\hat k\hat a)\hat b_L^{-1}(\hat k\hat a)\stw
 d\hat p\hat p^{-1})_{\hhd}+$$\be +
(d\hat a\hat a^{-1}\stw \hat k^{-1}d\hat k
)_{\hhd}
+ \jp (\hat k^{-1}d\hat k\stw \hat a(\hat k^{-1}d\hat k)
 \hat a^{-1})_{\hhd},\ee
where $\hat p=\hat k\hat a\hat k^{-1}$.

We note that due to the fact that $\ti G=\br\times_S\hat G$,
 the affine  model space
$\ti M_L$ is diffeomorphic to $\hat M_L\times \br_S\times\br_l$.
The  natural additive
coordinates
on $\br_l$ and $\br_S$ can be denoted as $a^0$ and $s$, respectively.
Thus any element $(\ti k,\ti \phi)$ of $\ti M_L$ can be naturally
represented by the quadruple
$(\hat k,\hat \phi,a^0,s)$, where $\ti k=e^{s\ti T^0}\hat k$, and
 $\ti a=\hat a e^{a^0\ti\U(\ti t_0)}$.

In order to perform the symplectic reduction, it is useful to
change the  coordinates on $\ti M_L$. For this,
recall that
the maps $m^0:\ti B\to\br$ and $\hat m:\ti B\to\hat B$
(cf. Conventions 4.12) were
 induced by the decomposition $\ti B=\hat B\times\br_l$. We shall write
 occasionaly $m^à(\ti b)=\ti b^0$
nad $\hat m(\ti b)=\ti b'$. We define also the maps $m^0_L:\ttD\to\br$
and $\hat m_L:\ttD\to \hat B$
as $m^0_L=m^0\circ\ti b_L$ and $\hat m_L=\hat m\circ \ti b_L$.
The essence of the first step of the
symplectic reduction is contained in the following proposition:
\vskip1pc
\teo {\bf 5.7:} 1) It holds
\be m_L^0(\hat k,\hat \phi,a^0,s)=a^0+ (\wti{Dres}_{\hat k}\hat a)^0,
\quad \hat a=\exp{\hat\Lm(\hat\phi)};\ee
  \vskip1pc
\noindent 2)
In the coordinates $(\hat k,\hat \phi,m_L^0,s)$ on $\ti M_L$,
 the symplectic form $\ti\om_L$   can be written as
\be \ti\om_L=-ds\w dm^0_L+\hat\om_L,\ee
where $\hat\om_L$ is the symplectic form on $\hat M_L$  defined in (5.90).
\vskip1pc
 \noindent The proof of this theorem will necessitate the
 following lemma:
\vskip1pc
\lem {\bf 5.8:} It  holds
\be \hat m_L(\wti{Ad}_{\hat k}\hat a)=\hat b_L(\wh{Ad}_{\hat k}\hat a),
\quad
\ti g_R(\wti{Ad}_{\hat k}\hat a)=\hat g_R(\wh{Ad}_{\hat k}\hat a).\ee
\vskip1pc
\noindent This lemma is proved in the Appendix 7.4
\vskip1pc

\noindent {\bf Proof of the Theorem 5.7:} 1) We have
\be \ti b_L(e^{s\ti T^0}\hat k \hat a e^{a^0\ti\U(\ti t_0)})=
e^{a^0\ti\U(\ti t_0)}\ti b_L(e^{s\ti T^0}\hat k \hat a).\ee
because  the line $e^{a^0\ti\U(\ti  t_0)}$
is central in $\ttD$.  We stress that the multiplication
 $\hat k\hat a$ in (5.94) is taken in $\ttD$.
Then we use the fact that the automorphism $S$   preserves both
subgroups $\ti G$ and $\ti B$.
This permits to write
$$ \ti b_L(e^{s\ti T^0}\hat k\hat a)=$$
\be \ti b_L(e^{s\ti T^0}\hat k\hat a e^{-s\ti T^0})=
 e^{s\ti T^0}\ti b_L(\hat k\hat a) e^{-s\ti T^0}=
(\exp{(m^0_L(\hat k\hat a)\ti\U(\ti t_0))})e^{s\ti T^0}
\hat m_L(\hat k\hat a)  e^{-s\ti T^0}, \ee
where the maps $m^0,\hat m$ were defined in Conventions 4.12.
Since $S$ preserve also $\hat B$, it follows from (5.94) and (5.95)
that
\be m_L^0(\hat k,\hat \phi,a^0,s)=a^0+ m^0_L(\hat k\hat a)
\equiv a^0+(\wti{Dres}_{\hat k}\hat a)^0;\ee
   \vskip1pc
\noindent 2) We have needed the statement 1) in order to show
that the coordinate
$a^0$ can be traded for the new coordinate $m^0_L$. So from
now until the end of the proof
we shall work with the variables $(\hat k,\hat\phi, m_L^0,s)$
on $\ti M_L$.The quasitriangular chiral
master model is Poisson-Lie symmetric, hence we
 know that  the group $\ti G$ acts on the affine model space
$\ti M_L$ in the Poisson-Lie
way. Recall that this follows from the fact that $\ti M_L$ can be
 viewed as the
submanifold of  $\ttD$ preserved by the $\ti G$-action on $\ttD$.
The latter
is of the Poisson-Lie type by construction and it has the non-Abelian
moment map
given by the factorization map $\ti b_L:\ttD\to\ti B$.
The restriction of the map
$\ti b_L$ to the affine model space $\ti M_L$ is the Poisson-Lie
 moment map on $\ti M_L$.

The relation (4.87) proved in Section  5.1.2 implies the following formula
\be \ti\om_L(., v_{\ti T^0})=(\ti T^0,d\ti b_L\ti b_L^{-1})_{\ttd}.\ee
Here on the l.h.s., $v_{\ti T^0}={\d\over \d s}$
 is viewed as the vector field on $\ti M_L$ corresponding
to the left action of the generator $\ti T^0$ and on the r.h.s.,
 $\ti T^0$ is viewed
simply as the element of $\ti\G$.

Now recall the definition of the map $m^0:\ti B\to\br$.
It is induced by the
decomposition $\ti B=\hat B\times\br$. Since $\br$ commutes
 with $\hat B$, it follows
that
\be \ti\om_L(.,v_{\ti T^0})=(\ti T^0,d\ti b_L\ti b_L^{-1})_{\ttd}=
(\ti T^0, \ti\Lm(\ti t_0))_{\ttd} dm^0_L=dm^0_L.\ee
In other words, the structure of our WZW double $\ttD$ insures,
that the $\ti T^0$-action on
$\ti M_L$ is Hamiltonian in the standard (Abelian) sense.

It follows immediately from the formula (5.92), that
\be \ti\om_L=-ds\w dm^0_L+ \ti\Omega(\hat k,\hat \phi, m_L^0)\ee
where $\ti\Omega$ does not contain $ds$. Moreover, $\ti\Omega$
neither depends on $s$, because
$\ti\om_L$ is invariant with respect to the Hamiltonian vector
field $\d/\d s$.
In fact, it is evident that $\ti\Omega$ is nothing but the
 pull-back of $\ti\om_L$
on the surface $s=0$ in $\ti M_L$. We can therefore write
$\ti\Omega=\ti\om_L(s=0)$.
Now we have from (5.85)
$$ \ti\om_L (s=0) =
\jp (d\ti b_L(\hat k\ti a)\ti b_L^{-1}(\hat k\ti a)
\stw d\ti p_0\ti p_0^{-1})_{\ttd}+$$\be +(d\ti a\ti a^{-1}
 \stw \hat k^{-1}d\hat k)_{\ttd}
+ \jp (\hat k^{-1}d\hat k\stw \ti a(\hat k^{-1}d\hat k)
\ti a^{-1})_{\ttd},\ee
where $\ti p_0=\wti{Ad}_{\hat k}\ti a$ and $\ti a =
\hat a \exp{(a^0(m_L^0,\hat k,\hat a)\ti\U(\ti t_0))}$.
The reader should pay attention to the distribution of
the hats and tildes and to the
fact that the group multiplication in this formula is
 taken in $\ttD$.
Now we study (5.100) term by term:
\be   (d\ti a\ti a^{-1}\stw \hat k^{-1}d\hat k)_{\ttd}=
(d\hat a\hat a^{-1}\stw \hat k^{-1}d\hat k)_{\hhd};\ee
this follows from the the fact that $(\hat k^{-1}d\hat k,
\ti\U(\ti t_0))_{\ttD}=0$. Then we have
\be(\hat k^{-1}d\hat k\stw \ti a(\hat k^{-1}d\hat k)
\ti a^{-1})_{\ttd}
=(\hat k^{-1}d\hat k\stw \wti{Ad}_{\hat a}
(\hat k^{-1}d\hat k)  )_{\ttd},\ee
because $\exp{(a^0\ti\U(\ti t_0))}$ is in the center of $\ttD$.
 Moreover, one checks directly that
\be (\hat k^{-1}d\hat k\stw \wti{Ad}_{\hat a}
(\hat k^{-1}d\hat k)  )_{\ttd}=
(\hat k^{-1}d\hat k\stw \wh{Ad}_{\hat a}
(\hat k^{-1}d\hat k)  )_{\hhd}.\ee
It is first convenient to rewrite the remaining term as
$$ \jp (d\ti b_L(\hat k\ti a)\ti b_L^{-1}(\hat k\ti a)
\stw d\ti p_0\ti p_0^{-1} )_{\ttd}=$$
\be = \jp( \ti b_L^{-1}(\wti{Ad}_{\hat k}\ti a)
d\ti b_L(\wti{Ad}_{\hat k}\ti a)\stw
d\ti g_R(\wti{Ad}_{\hat k}\ti a)
\ti  g_R^{-1}(\wti{Ad}_{\hat k}\ti a))_{\ttd}.\ee
We know from Lemma 5.8 that $\ti  g_R(\wti{Ad}_{\hat k}
\ti a)\in\hat G$, hence it follows
$$ \jp (d\ti b_L(\hat k\ti a)\ti b_L^{-1}(\hat k\ti a)
\stw d\ti p_0\ti p_0^{-1} )_{\ttd} =$$
 \be = \jp( \hat m_L^{-1}(\wti{Ad}_{\hat k}\hat a)
 d\hat m_L(\wti{Ad}_{\hat k}\hat a)\stw
d\ti g_R(\wti{Ad}_{\hat k}\hat a)
\ti  g_R^{-1}(\wti{Ad}_{\hat k}\hat a))_{\ttd},\ee
where $\hat m_L=\hat m\circ \ti b_L$.
We use again the Lemma 5.8 to rewrite finally (5.105) into
$$ \jp (d\ti b_L(\hat k\ti a)\ti b_L^{-1}(\hat k\ti a)
\stw  d\ti p_0\ti p_0^{-1})_{\ttd}=$$
\be = \jp(\hat b_L^{-1}(\hat{Ad}_{\hat k}\hat a)
d\hat b_L(\hat{Ad}_{\hat k}\hat a)\stw
d\hat g_R(\hat{Ad}_{\hat k}\hat a)
\hat  g_R^{-1}(\hat{Ad}_{\hat k}\hat a) )_{\hhd}.\ee
Collecting (5.101), (5.103) and (5.106), we conlude
$$ \ti\om_L (s=0)=\hat\om_L$$
 The theorem is proved.

\rightline{\#}
\noindent The first step of the symplectic reduction now consists
in setting $m_L^0=0$. This is consistent,
since the Hamiltonian
$\ti H_L=-{1\over 2\k}(\ti\phi,\ti\phi)_{\ti\G^*}$ Poisson commutes
with  $m_L^0$ because it is obviously
invariant with respect to the left $\ti G$-action on $\ti M_L$.

From the Theorem 5.7 one concludes immediately that the reduced
 dynamical
system has $\hat M_L$ for its phase space,  its  Hamiltonian reads
\be \hat H_L=-{1\over 2\k} (\phi,\phi)_{\G^*}-{a^\infty\over\k}
 (\wti{Dres}_{\hat k}\hat a)^0,\ee
where $\hat a=e^{\hat\Lm(\hat\phi)}$, $\hat\phi= a^\infty \tf
+\pi^*(\phi)$
  and its
 symplectic form is
$$\hat\om_L=\jp ( d\hat b_L(\hat k\hat a)\hat b_L^{-1}(\hat k\hat a)
\stw d\hat p\hat p^{-1})_{\hhD}+$$\be +(d\hat a\hat a^{-1}\stw
\hat k^{-1}d\hat k
)_{\hhD}
+ \jp (\hat k^{-1}d\hat k\stw \hat a(\hat k^{-1}d\hat k)
\hat a^{-1})_{\hhD},\ee
where $\hat p=\hat k\hat a\hat k^{-1}$.

 \subsection{Chiral symplectic reduction: the second step}

In order to perform the second step of the reduction, we have to
find the standard Hamiltonian
charge generating the central circle action on $\hat M_L$ . Its
 existence is guaranteed because we know that
the affine Lu-Weinstein-Soibelman double $\ttD$ is of the WZW type
(cf. Definition 4.10).
We find this charge from the fundamental relation (4.87)
\be \hat\om_L(.,v_{\hat T^\infty})=
(\tF, d\hat b_L\hat b_L^{-1})_{\hhd}=dm_L^\infty(\hat k\hat a),\ee
which holds because $\hat \om_L$ was pulled back from the Drinfeld
double $\hhD$. Here
$v_{\hat T^\infty}$ is the vector field on $\hat M_L$ corresponding
to the left
action of $\hat T^\infty$ and $m_L^\infty =m^\infty\circ\hat b_L$.
Recall that $m^\infty:\hat B\to \br$ was
defined by the decomposition $\hat B=\br\times_Q B$.  It is easy to
evaluate $m^\infty_L$,
in fact, it holds $m_L^\infty (\hat k\hat a)=a^\infty$.

Fix now the submanifold of $\hat M_L$ defined
by $a^\infty =\k$ and consider the space of the central
circle orbits on this submanifold. This space of orbits is nothing
 but the WZW model space
$M_L^{WZ}=G\times\A_+^1$, where
$\A_+^1$ is the standard Weyl alcove.
Before giving the explicit description of the (doubly) reduced
symplectic structure on $M_L^{WZ}$,
we first write the reduced Hamiltonian $H_L^{qWZ}$
on $M_L^{WZ}$.
The consistency requires that the first floor Hamiltonian
$\hat H_L(\hat k\hat a)$
be invariant with respect to the central circle action. But this
is evident since
\be \wti{Dres}_{\hat k}\hat a=\ti b(\hat k\hat a)=
\ti b(\wti{Ad}_{\hat k}\hat a).\ee
Thus we obtain
\be H_L^{qWZ}(k,a^\mu)=-{1\over 2\k}
 (\phi,\phi)_{\G^*}-(\wti{Dres}_{\hat k}e^{\hat\Lm(\hat \phi_\k)})^0,\ee
where $\hat \phi_\k= \k\tf +\pi^*(\phi)$,
 $\phi =\k a^\mu t_\mu$ and $a^\mu t_\mu\in\A_+^1$.
The Hamiltonian $H_L^{qWZ}$ depends on $\e$ due
to the $\e$-dependence of $\hat\Lm$ and of the map
$m^0:\ti B\to\br$. We shall see in Section 5.2.7 that
in the $\e\to 0$ (or
$q\to 1$) limit
 the formula (5.111) yields the standard chiral Sugawara
Hamiltonian $H_L^{WZ}$ given by (3.118).

Let us now study the quasitriangular symplectic structure
on $M_L^{WZ}$.
As we have already said, it is convenient to perform the
second step of the
symplectic reduction by working with the Poisson brackets.
This means  that we shall first need  to invert the symplectic
form $\hat\om_L$
to obtain the corresponding Poisson bivector $\hat\Si_L$
 on $\hat M_L$. Our strategy for
 inverting  $\hat\om_L$ will rely on using its Poisson-Lie symmetry.
Indeed, the story  of the finite-dimensional Section 5.1.2 can
 be  directly used also in our affine situation
if we replace $G_0$ and $\A^0_+$ of section 5.1.2 by our
 $\hat G$ and $\hat \A_+$, respectively.
Thus we have
\be \hat\om_L(.,v_{\hat T})=(\hat T,
 d\hat b_L(\hat k\hat a)\hat b_L^{-1}(\hat k\hat a))_{\hhD},\ee
where $\hat T$ is any generator of $\hG$ and $v_{\hat T}$ is
 the vector field on $\hat M_L$
corresponding to the left action of $\hat T$ on $\hat M_L$.
Following the reasoning after
the proof of Lemma 5.5,
we conclude that the Poisson bivector $\hat\Si_L$ on
 $\hat M_L$ fulfils
\be \L_{\hat v}\hat\Si_L=-\hat v\w \hat v,\ee
where $\hat v\in T\hat M_L\ot\hG^*=T\hat M_L\ot\B$
generates the left action of $\hat G$ on $\hat M_L$
and $\L_{\hat v}$ is the Lie derivative. From (5.113), it
 follows (cf. (5.37)) that $\hat\Si_L$
must be of the form
$$ \hat\Si_L(\hat k,\hat \phi) =
-\jp(\Pi^R_{\hat G})_{ij}(\hat k)R_{\hat k*}\hat T^i\w
R_{\hat k*}\hat T^j+$$
\be +\hat \Si^0_{ij}(\hat \phi)L_{\hat k*}\hat T^i\w
 L_{\hat k*}\hat T^j+
\hat\si_{i}^{~\hat\mu} (\hat \phi)L_{\hat k*}\hat T^i\w
{\partial\over \partial   \phi^{\hat\mu}}
+\hat s^{\hat\mu\hat\nu}(\hat \phi){\partial\over \partial
  \phi^{\hat\mu}}\w
{\partial\over \partial   \phi^{\hat\nu}}.\ee
Here $\hat \phi=\phi^\infty \tf + \phi^\mu\pi^*(t_\mu)\equiv
 a^\infty\tf +a^\infty a^\mu\pi^*(t_\mu)$,
and the notation $  \phi^{\hat\mu}$ means that we consider
at the same time the coordinates $  \phi^\mu$ and $\phi^\infty$;
 in other words,
the subscript $\hat\mu$ runs over $\mu$ and over $\infty$.
The expression $\jp(\Pi^R_{\hat G})_{ij}(\hat k)R_{\hat k*}\hat
T^i\w
R_{\hat k*}\hat T^j$ is the Poisson-Lie bivector on the group
$\hat G$ written is some
basis $\hat T^i$ of $\hG$. Recall that the Poisson-Lie bracket
 on $\hat G$ is entirely
specified by the structure of the affine Lu-Weinstein-Soibelman
double $\hhD$, as it was
explained in section 4.1.2.

Thus we again observe the power of the Poisson-Lie symmetry:
 the Poisson tensor  $\hat\Si_L$
on the affine model space $\hat M_L$ is completely determined by
its value at the points
$(\hat e,\hat \phi)\in \hat M_L$,
in other words: at the unit element  $\hat e$ of $\hat G$.

Our next task is to find the coefficient functions
 $\hat \Si^0_{ij}(\hat \phi)$,
 $\hat\si_i^{\hat\mu}(\hat\phi)$
and $\hat s^{\hat\mu\hat\nu}(\hat \phi)$.
 For this, we first calculate the matrix of the symplectic form
$\hat\om_L$ at points $(\hat e,\hat \phi)$. We do it in  the basis
$R_{\hat a*}\hat T^i,R_{\hat a*}
\hat\Lm(\hat t_{\hat\mu})$ of the $\hat\Xi$-pushed-forward tangent
space
 $T_{(\hat e,\hat \phi)}\hat M_L$; recall that
 $\hat a=\exp{\hat\Lm(\hat\phi)}$ and we set
$\hat t_{\hat\mu}=(\tf,\pi^*(t_\mu))$.
 We then invert this matrix to
obtain the  unknown coefficient functions of the Poisson bivector.

The convenient  basis of $\hG=\widehat{L\G_0}$  reads
\be \hat T^i=\tF,\iota(T^\mu),\iota(B^{\hal}),\iota(C^{\hal}),\quad
\hal\in\hat\Phi_+.\ee
Recall that
\be T^\mu=iH^\mu,\quad B^{\hal}={i\over \od}(E^{\hal}+E^{-\hal}),\quad
C^{\hal}={1\over \od}(E^{\hal}-E^{-\hal}),\ee
where
\be E^{\hal}=E^\al e^{in\si}, \quad \hal=(\al,n);\qquad
 E^{\hal}=H^\mu e^{in\si},
\quad \hal=(\mu,n\neq 0).\ee
The complet list of the  conventions and notations concerning this basis
can be found in Section 3.2.4.
The corresponding dual basis of $\hG^*$ is
\be \hat t_i=\tf,\pi^*(t_\mu),\pi^*(b_{\hal}),\pi^*(c_{\hal}),
\quad \hal\in\hat\Phi_+.\ee
From this we obtain the dual basis
  of $\hat \B$; it reads
\be \hat \Lm(\hat t_i)=\e\tf^1,\e\nu_*(H^\mu),\e\nu_*({\vert
\hal\vert^2\over \od}E^{\hal}),
\e\nu_*( -i
{\vert \hal\vert^2\over \od}E^{\hal} ),\quad \hal\in\hat\Phi_+,\ee
where $\tf^1=-i\d_\si$ and $\nu_*:\B\to\hat\B$ were
 defined in Section 4.4.2. Moreover,
 $\vert \hal\vert^2=\vert \al\vert^2$ for $\hal=(\al,n)$ and $\vert
\hal\vert^2=2$
for $\hal=(\mu,n)$.
 Of course, $\hat \B$
is the Lie algebra of $\hat B\subset\hhd$ and $\B=Lie(B)$ where
$L_+G_0^\bc=B\subset D=LG_0^\bc$.
The dual basis depends on $\e$
and satisfies the basic condition $(\hat\Lm(\hat t_i),\hat T^j)_{\hhd}=
\delta_i^j$.

 Recall
that
$\hat a=\exp{\hat\Lm(\hat\phi)} $ is the element of the alcove
$\hat A_+\subset \hat A$; the
Lie algebra $\hat A$
 is generated by
 the generators $\hat\Lm(\hat t_{\hat\mu})=\hat\Lm(\tf,\pi^*(t_\mu))=
(\e\nu_*(H^\mu),-i\e\d_\si).$
We decompose the elements of the basis of $T_{\hat a}\hat\Xi(\hat M_L)$
 into two  parts:
the $\hal$-part generated by
$R_{\hat a*}\iota(B^{\hal}),R_{\hat a*}\iota(C^{\hal})$,
 $\hal\in\hat\Phi_+$ and the $\hat\mu$-part
generated by $R_{\hat a^*}\hat T^{\hat\mu},R_{\hat a^*}
\hat\Lm(\hat t_{\hat\mu})$.
 Here by $\hat T^{\hmu}$
we mean either $\iota(T^\mu)$ or $\tF$.
Now we use the formula (4.18) from Section 4.1.2, expressing the
 Semenov-Tian-Shansky form on $\hhD$.
$$\hat\om(t,u)=(t,(\Pi_{\ti LR}-\Pi_{L\ti R})u)_{\hhd}.$$
We immediately observe that
\be \hat\om_L(\hal,\hmu)=0,\quad \hat\om_L(\hal,\hat\beta)=0,
\quad \hal,\hat\beta\in\hat\Phi_+,\quad
\hal\neq\hat\beta.\ee
This is because we are at the point $(\hat e,\hat a)\in \hat M_L(
\subset \hhD)$.
It is hence sufficient to
invert the matrix $\hat\om_L$ for the $\hat\mu$-sector and for
every $\hal$-sector separately.
  We have
\be \Pi_{L\ti R}R_{\hat a*}\hat T^{\hmu}=0,
\quad \Pi_{\ti LR}R_{\hat a*}\hat T^{\hmu}=
R_{\hat a*}\hat T^{\hmu}.\ee
From this we obtain
\be \om_L(R_{\hat a*}\hat\Lm(\hat t_{\hat\nu}),
R_{\hat a*}\hat T^{\hmu})=\delta^{\hmu}_{\hat\nu}.\ee
One has \be R_{\hat a*}\iota(C^{\hal})=(R_{\hat a*}\iota(C^{\hal}),
L_{\hat a*}\hat\Lm(\hat t_i))_{\hhd}
L_{\hat a*} \hat T^i+(R_{\hat a*}\iota(C^{\hal}),L_{\hat a*}
 \hat T^i)_{\hhd} L_{\hat a*}\hat\Lm(\hat t_i)\ee
and
\be L_{\hat a*}\hat\Lm(b_{\hal})=e^{\saf}
 R_{\hat a*}\hat\Lm(b_{\hal}),\quad
 L_{\hat a*}\hat\Lm(c_{\hal})=e^{\saf}
 R_{\hat a*}\hat\Lm(c_{\hal}).\ee
Then we have for every $\hal\in\hat\Phi_+$
$$  \Pi_{L\ti R}R_{\hat a*}\iota(C^{\hal})=$$
\be e^{\saf}\biggl[ (L_{\hat a*}\iota(B^{\hal}),R_{\hat a*}
\iota(C^{\hal}))_{\hhd} R_{\hat a*}\hat\Lm(b_{\hal})
+(L_{\hat a*}\iota(C^{\hal}),R_{\hat a*}\iota(C^{\hal}))_{\hhd}
 R_{\hat a*}
\hat\Lm(c_{\hal})\biggr];\ee
\be   \Pi_{\ti LR}R_{\hat a*}\iota(C^{\hal})=
R_{\hat a*}\iota(C^{\hal}).\ee
It follows that
$$  \hat\om_L(R_{\hat a*}\iota(B^{\hal}),
 R_{\hat a*}\iota(C^{\hal}))=$$
\be =-e^{\saf}
(L_{\hat a*}\iota(B^{\hal}),R_{\hat a*}\iota(C^{\hal}))_{\hhd}=
{1\over
\e\vert \hal\vert^2}(e^{2\saf}-1).\ee
Recall from Section 3.2.4, that  for
$\hal=(\al,n)$:
\be\saf = \e a^\infty(a^\mu\la\al,H^\mu\ra   +n)\ee
and for $\hal=(\mu,n)$
\be\saf=\e a^\infty n.\ee
Our conclusion is  that
\be \hat s^{\hmu\hat\nu}=0, \quad \hat\si_{\hal}^{\hmu}=0,\quad
\hat\si_{\hmu}^{\hat\nu}
=\delta_{\hmu}^{\hat\nu},\quad  \hat\Si^0_{\hmu\hat\nu}=
\hat\Si^0_{\hal\hmu}=\hat\Si^0_{\hal\hat\beta}=0\ee
and the explicit formula for the Poisson bivector $\hat\Si_L$ on
 $\hat M_L$ reads
$$ \hat\Si_L(\hat k,\hat \phi^{\hat\mu}) =
-\jp(\Pi^R_{\hat G})_{ij}(\hat k)R_{\hat k*}\hat T^i\w
R_{\hat k*}\hat T^j+$$
\be +\sum_{\hal\in\hat\Phi_+}{\e\vert\hal \vert^2\over
 (1-e^{2\saf})}L_{\hat k*}\iota(B^{\hal})\w
L_{\hat k*}\iota(C^{\hal})+
L_{\hat k*}\hat T^{\hmu}\w
{\partial\over \partial \hat \phi^{\hat\mu}}.\ee
So far we have inverted the form $\hat\om_L$,
now we are ready to perform the second step of the
symplectic reduction induced by
setting $a^\infty=\k$.
Consider  a pair of functions $\phi_i$, $i=1,2$ on
$M_L^{WZ}$. We wish to calculate their
reduced Poisson bracket $\{\phi_1,\phi_2\}_{qWZ}$.
The general procedure of the symplectic
reduction at the level of the Poisson brackets is
described in the Appendix 7.3. In our
particular situation, it works as follows: define
two functions $\hat\phi_i$
on $\hat M_L$ as
\be \hat\phi_i(\hat k,a^\mu,a^\infty=
\k)\equiv\phi_i(\pi(\hat k),a^\mu),
\quad \hat k\in\widehat{LG_0}.\ee
Calculate then the   quasitriangular Poisson bracket
 $\{\hat\phi_1,\hat\phi_2\}_{q\hat M_L}$ on $\hat M_L$.
It verifies
\be \{a^\infty,\{\hat\phi_1,\hat\phi_2\}_{q\hat M_L}\}_{q\hat M_L}=0\ee
as the simple consequence of the Jacobi identity and the central circle
 invariance of
$\hat\phi_i$. This  means that it exists a function on $M_L^{WZ}$
denoted suggestively as $\{\phi_1,\phi_2\bq$ which verifies
\be \{\hat\phi_1,\hat\phi_2\}_{q\hat M_L}(\hat k,a^\mu,a^\infty=\k)=
\{\phi_1,\phi_2\bq(\pi(\hat k),a^\mu).\ee
Needless to say, the function $\{\phi_1,\phi_2\bq$ is the seeken
 reduced Poisson bracket.
This method we now apply for the functions $\phi_i$ of
 particular form.
\subsection{Deformed affine dynamical $r$-matrix}

Our next task will consist in computing the reduced
Poisson bracket
$ \{k\ptp k\bq$. The reader who still did not get used to the
notation for the matrix Poisson bracket should again  consult
Sections 3.1.3 and 3.2.6.
We have
 \be \{k \ptp k \bq =\{k \ptp k \}_{0}- \{k\ptp k\}_G,\ee
where the first   bracket correspond (modulo the reduction) to
the   bivector on the second line
  on the r.h.s. of (5.131), and the second bracket to that on the
first line.
 Of course, the relation (5.135) is the affine analogue
of (5.49). Also the way of the evaluating of the two terms in (5.135)
follows
in spirit the calculation of Section 5.1.3. For doing it, we first
 introduce some necessary notions:

Define the identification map $\Lm:\G^*\to\B=Lie(B)$ as usual
\be (\Lm(x^*),y)_\D=\la x^*,y\ra,\quad x^*\in\G^*, y\in\G.\ee
We can directly check from the definition (4.127),
 (4.128) of $(.,.)_{\hhd}$
 that
\be \hat\Lm\circ\pi^*=\nu_*\circ\Lm.\ee
Now using the relations (4.63) and (4.66),
we first calculate the relevant  Poisson-Lie bracket on
 $\hat G=\widehat{LG_0}$ :
$$ \{\pi(\hat k)\ptp \pi(\hat k)\}^R_{\hat G}=
 -\biggl[\pi(\iota(T^i)\hat k)\otimes
\pi(\iota(T^j)\hat k)\biggr]\times$$ $$\times
\biggl[(\hat k^{-1}(\hat\Lm\circ\pi^*)(t_i)\hat k,
(\hat\Lm\circ\pi^*)(t_l))_{\hhd}
(\hat k^{-1}(\hat\Lm\circ\pi^*)(t_j) \hat k, \iota(T^l))_{\hhd}+$$
 $$
(\hat k^{-1} (\hat\Lm\circ\pi^*)(t_i)\hat k,\hat\Lm(\tf))_{\hhd}
(\hat k^{-1} (\hat\Lm\circ\pi^*)(t_j)\hat k,\tF)_{\hhd}\biggr]=$$
\be -\biggl[T^i\pi(\hat k)\otimes
T^j\pi( \hat k)\biggr] \times (\pi(\hat k)^{-1} \Lm(t_i)\pi(\hat k),
 \Lm(t_l))_{\D}
(\pi(\hat k)^{-1} \Lm(t_j)\pi(\hat k), T^l)_{\D}.\ee
 Now we use the formulas (5.88) and (5.137) to calculate the
 $\{.,.\}_G$-contribution to the
Poisson bracket $\{.,.\bq$ according to the decomposition (5.135).
The result is
$$ \{k\ptp k\}^R_G=$$ \be = -(T^i  k\otimes
T^j  k) \times (  k^{-1}\Lm( t_i ) k, \Lm(t_l))_{\D}
(  k^{-1}\Lm( t_j)  k, T^l)_{\D}=[(k\otimes k),(T^i\otimes \Lm(t_i))],\ee
which is by the way  nothing but the Poisson-Lie bracket on $G=LG_0$
induced by the double $D=LG_0^\bc$.
The calculation giving this result follows step-by-step the
 computation leading
from (5.50) to (5.52).

Although the continuation of this affine story
 is very similar to the finite
dimensional case described in section 5.1.3., some additional care
 is needed
in the affine case because of the infinite number of the elements of
 the basis
 $(T^i,\Lm(t_i))$ of $\D=L\G_0$. Indeed, we must give the meaning to
the series
$T^i\otimes \Lm(t_i)$.
Recall that the basis is given by (cf. (5.116) and (5.119)).
\be  T^i=T^\mu,B^{\hal}, C^{\hal},\quad \Lm(t_i)=\e H^\mu,
 {\e\vert \hal\vert^2\over \od}E^{\hal},
 -i{\e\vert \hal\vert^2\over \od}E^{\hal},
\quad \hal\in\hat\Phi_+.\ee
 A simple computation then shows that
\be T^i\otimes \Lm(t_i)= i\e H^\mu\otimes H^\mu+i\e\sum_{\hal\in\hat\Phi}
\vert \hal\vert^2E^{-\hal}\otimes E^{\hal}.\ee
We wish to calculate this expression in the evaluation representation
of $L\G_0$. Recall that its  representation space $LV_0$ is given
by square-integrable maps from the loop circle into the representation
space $V_0$ of some finite-dimensional (typically irreducible)
representation
of $\G_0$. This means (e.g. for the affine root $\hal=(\al,n)$) that
$E^{\hal}$ is to be viewed as $E^\al e^{in\si}$, where  $E^\al\in End(V_0)$
and $e^{in\si}$ is "the multiplication by function" operator in $End(LV_0)$.

Among the summation over all affine roots, we can consider the subsummation
over $\hal=(\al,n)$, where $\al$ is kept fixed and $n$ acquires
whatever integer
value. Then it is easy to see that in the evaluation
representation the Fourier series over $n$ diverges. This divergence shows
that we have to care about the analytic aspect
of working with the infinite dimensional symplectic manifolds.
 In other words, we have
to give a meaning to the divergent Fourier series (5.141).
The reader should
understand that the prescription associating a well-defined
function of $\si$
to the series (5.141) is the part of the definition of our chiral
 quasitriangular
WZW model. Indeed, the resulting functions appear in the definition
of the symplectic structure of the model.

Of course, our prescription for summing the divergent series
must fulfil some consistency conditions.
Among them there is the most important one: the Poisson bracket (5.139)
must fulfil the Jacobi identity. On the top of this, we shall require
that the resulting function of $\si$ be meromorphic. In this way we shall
find ourselves in the standard world of the $r$-matrices.

It turns
out that the prescription fulfilling both conditions exists: it is called
the Abel-Poisson summation method and is based on the following observation:
The series
$\Sigma_{n>0}(a_n \cos{n\si}+b_n\sin{n\si})$ becomes (uniformly) convergent
if we replace $(a_n,b_n)$ by $(r^na_n,r^nb_n)$, where $0\leq r<1$.
Its sum we denote $S_r(\si)$. If the limit lim$\ _{r\to 1}S_r(\si)$ exists,
it is called the Abel-Poisson sum of the original series. For example,
we have
(cf. p. 83 of \cite{PS})
\be \sum_{n>0}(e^{in\si}-e^{-in\si})=i{\rm cotg}\jp\si\ee
in the Abel-Poisson sense. It is indeed this formula, which permits to
compute $T^i\otimes \Lm(t_i)$ in the evaluation representation:
 \be T^i\otimes \Lm(t_i)\equiv \e\hat r(\si-\si') = \e r+\e
C{\rm  cotg}\jp(\si-\si') .\ee
   Recall that
\be r=\sum_{\al\in\Phi_+}{i\vert \al \vert^2\over 2}(E^{-\al}\otimes
 E^{\al}-E^\al\otimes E^{-\al});\ee
\be C=\sum_\mu H^\mu\otimes H^\mu+
\sum_{\al\in\Phi_+}{\vert \al \vert^2\over 2}(E^{-\al}\otimes E^{\al}+
E^\al\otimes E^{-\al}).\ee
 Inserting (5.143) into the formula (5.139), we obtain
\be -\{k(\si)\ptp k(\si')\}^R_G= \e[\hat r,(k(\si)\otimes k(\si'))].\ee
This Poisson bracket obeys indeed the Jacobi identity, since it can be
easily
checked that the $r$-matrix $\hat r(\si-\si')$ satisfies
the ordinary Yang-Baxter
equation with  spectral parameter, i.e.
\be [\hat r^{12}(\si_1-\si_2), \hat r^{13}(\si_1-\si_3)
+  \hat r^{23}(\si_2-\si_3)]+
[\hat r^{13}(\si_1-\si_3), \hat r^{23}(\si_2-\si_3)]=0.\ee
The calculation of the bracket $\{k\ptp k\}_0$ in (5.135)  is
even more straightforward. We use (5.131)
and (5.134) to arrive at
\be \{k\ptp k\}_0=(k\otimes k)\hat r'_\e(\hat\phi_\k),\ee
where
\be \hat r'_\e(\hat\phi_\k)=
\sum_{\hal\in\hat\Phi_+}{\e\vert\hal \vert^2\over
(1-e^{{\e\over 2}\vert \hal\vert^2\la i\hal^\vee,\hat\phi_\k\ra})}
 B^{\hal}\w
 C^{\hal}. \ee
Recall also that $\hat\phi_\k=\k\tf+\k a^\mu\pi^*(t_\mu)$.
Putting together (5.146) and (5.149), we arrive at the most important
 formula of this paper:
\be \{k\ptp k \bq =(k\otimes k)\hat r_{\e}(\hat\phi_\k)+\e
\hat r(k\otimes k)\ee
where $\hat r$ is the standard affine $r$-matrix (5.143) and
\be \hat r_\e(\hat\phi_\k)=
i\e\sum_{\hal\in\hat \Phi}{\vert \hal \vert^2\over 2}
{\rm coth}({\e\over 2}\vert \hal\vert^2\la i\hal^\vee,\hat\phi_\k\ra)
 E^{\hal}\otimes E^{-\hal}.\ee
Note that here the summation goes over all roots.
It is useful to note, that $\hat r_{\e} $ is nothing but the
direct affinization
of the formula (5.60) for $r_{\e}$.  We shall refer to
 $\hat r_{\e}(\hat\phi_\k)$ as to the deformed
affine dynamical $r$-matrix.  We shall see in a moment that
(5.151) in the evaluation representation will indeed fulfil
the dynamical Yang-Baxter equation with spectral parameter.

It is insightful to visualise the $\si$-dependence in (5.151).
Recalling the explicit expressions (3.83), (3.84)
for $\la i\hal^\vee,\hat\phi_\k\ra$, we can write
$$\hat r_\e(\hat\phi_\k)(\si-\si')=-i\e\sum_{\mu,n\neq 0}
{\rm coth}(-\e\k n)(H^\mu\ot H^\mu)e^{in(\si-\si')}$$ \be
-i\e\sum_{\al\in\Phi,n\in\bz}{\rn\over 2}{\rm coth}(-\e\k a^\mu
\la \al,H^\mu\ra -\e\k n)(E^\al\ot E^{-\al})e^{in(\si-\si')}.\ee
Of course, the summation is to be taken in the Abel-Poisson sense.

We use the following classical formulae \cite{WHW}
\be\si_{-y}(z,\tau)=\pi({\rm cotg}\pi z+{\rm cotg}\pi y)
+ 4\pi \Sigma_{m,n>0}e^{2\pi i\tau mn}\sin{2\pi(mz+ny)};\ee
\be \rho(z,\tau)=\pi {\rm cotg}\pi z +
4\pi\Sigma_{n>0}{e^{2\pi in\tau}\sin{2\pi nz}\over 1-e^{2\pi in\tau}},\ee
where  the functions $\rho(z,\tau),\si_w(z,\tau)$ are defined as
(cf. \cite{F,FW,EV})
\be \si_w(z,\tau)={\theta_1(w-z,\tau)\theta_1'(0,\tau)\over
\theta_1(w,\tau)\theta_1(z,\tau)},
\quad \rho(z,\tau)={\theta_1'(z,\tau)\over \theta_1(z,\tau)}.\ee
Note that  $\theta_1(z,\tau)$ is the Jacobi theta
function\footnote{We have
$\theta_1(z,\tau)=\vartheta_1(\pi z,\tau)$ with
$\vartheta_1$ in \cite{WHW}.}
\be \theta_1(z,\tau)=-\sum_{j=-\infty}^{\infty}e^{\pi i (j+\jp)^2\tau
+2\pi i(j+\jp)(z+\jp)},\ee
 the prime ' means the derivative with respect to the
 first argument $z$ and
the argument $\tau$ (the modular parameter )
is a nonzero complex number such that Im $\tau>0$.

  Now  we  can  sum up the   Fourier series (5.152) by using the classical
 formulae  (5.153), (5.154) and the relation (5.142). First we
 obtain\footnote{The formulae
(5.157) and (5.158)   appear
also in \cite{EV} but with several misprinted signs.  Those wrong signs
turn out to be innocent in the context of \cite{EV} since they
conspire to give an $r$-matrix which also fulfils the dynamical
Yang-Baxter equations (5.163). In fact, the correct and wrong $r$-matrices
differ by the gauge transformation of type 4.
 (cf. Section 4.2. of \cite{EV}).}
 \be \sum_{n\in{\bf Z}}e^{2\pi i zn}(1+{\rm coth}(a+i\pi n\tau))=
-{i\over \pi}\si_{ia\over \pi}(z,\tau),\ee
and
\be 1+ \sum_{n\in{\bf Z}\backslash 0}e^{2\pi i zn}
(1+{\rm coth}(i\pi n\tau))=
-{i\over \pi}\rho (z,\tau).\ee
Now by using (5.157) and (5.158), we finally arrive at
\be \{k(\si)\ptp k(\si') \bq =(k(\si)\otimes k(\si'))
\hat r_{\e}(a^\mu,\si-\si')+
 \e\hat r(\si-\si')(k(\si)\otimes k(\si')),\ee
where $\hat r_{\e} (a^\mu,\si)$ is the Felder-Wieczerkowski
\cite{FW} elliptic dynamical $r$-matrix
given by
 $$ \hat r_{\e}(a^\mu,\si)=$$
\be =-{\e\over \pi}\rho({\si\over 2\pi},{i\k\e\over \pi})
H^\mu\otimes H^\mu
-{\e\over \pi}\sum_{\al\in\Phi}{\rn\over 2}\si_{{\e\k a^\mu\la
\al,H^\mu\ra\over \pi i}}
({\si\over 2\pi},{i\k\e\over \pi})E^\al\otimes E^{-\al}.\ee
The (quasitriangular) braiding relation (5.159) plays the same role
in the quasitriangular
chiral WZW model as the   braiding relation (3.99) in the standard
 chiral WZW theory.
The description of the bracket $\{.,.\bq$ on $M_L^{WZ}$ is then
completed by the
following formula, that can be easily derived from (5.131) and (5.134):
\be \{k,a^\mu\bq={1\over\k} kT^\mu,\quad \{a^\mu,a^\nu\bq =0. \ee
 It remains to verify that our Abel-Poisson summation method did indeed
give the consistent result. First of all, both $r$-matrices appearing
in (5.159) are clearly meromorphic functions of $\si$. The Jacobi identity
for the Poisson brackets (5.159) and (5.161) then requires (5.147) and
\be [\hat r_{\e},1\otimes T^\mu +T^\mu\otimes 1]\equiv
[\hat r_{\e}^{12},(T^\mu)^1 +(T^\mu)^2]=0;\ee
$$ [\hat r_{\e}^{12}(\si_1-\si_2), \hat r_{\e}^{13}(\si_1-\si_3)
+  \hat r_{\e}^{23}(\si_2-\si_3)]+
[\hat r_{\e}^{13}(\si_1-\si_3), \hat r_{\e}^{23}(\si_2-\si_3)]+$$
\be +{1\over\k}( {\d\over\partial a^\mu}\hat r_{\e}^{12})(T^\mu)^3+
{1\over\k}({\d\over\partial a^\mu}\hat r_{\e}^{23})(T^\mu)^1+
{1\over\k}( {\d\over\partial a^\mu}\hat r_{\e}^{31})(T^\mu)^2=0.\ee
The relation (5.163) is called the dynamical Yang-Baxter equation
with spectral parameter \cite{F,EV}.
It is straightforward to check, that the elliptic $r$-matrix
$\hat r_{\e}(a^\mu,\si)$ does verify both conditions (5.162) and (5.163).

 It is instructive to rewrite the defining $q$WZW Poisson brackets
(5.159) and (5.161) in terms of the monodromic variables $m(\si)$
defined by the relation (3.103). The result is
\be  \{m(\si)\ptp m(\si') \bq =(m(\si)\otimes m(\si'))
 B_{\e}(a^\mu,\si-\si')+
 \e\hat r(\si-\si')(m(\si)\otimes m(\si')),\ee
where $B_{\e}(a^\mu, \si)$ is the quasitriangular braiding matrix
generalizing the matrix $B_0(a^\mu,\si)$ defined in (3.105) and (3.106).
We find it by generalizing the computation (3.105). The result is
 $$ B_{\e}(a^\mu,\si)=$$
\be =-{i\over \k}\rho({i\si\over 2\k\e},{i\pi\over \k\e})
H^\mu\otimes H^\mu
-{i\over \k}\sum_{\al\in\Phi}{\rn\over 2}\si_{ a^\mu\la
\al,H^\mu\ra}
( {i\si\over 2\k\e},{i\pi\over \k\e})E^\al\otimes E^{-\al}.\ee
We observe that the quasitriangular braiding matrix is again given
by the Felder $r$-matrix but with different modular parameter of the
elliptic functions. Indeed, we have derived (5.165) by using the following
modular identities
\be \si_{-\tau y}(z,\tau)=-{1\over \tau}
e^{-2\pi i yz}\si_y(-{z\over \tau},-{1\over \tau});\ee
\be \rho(z,\tau)=-{2\pi i z\over \tau} -{1\over \tau}
\rho(-{z\over\tau},-{1\over\tau}).\ee
We can conclude this section by saying that the vertex-IRF transformation
can be in a sense
interpreted as the modular transformation in the deformation parameter.

\subsection{$q$-Kac-Moody  primary fields}
We return for a moment to the symplectic structure of the standard
(nondeformed) chiral
WZW model and we note that the equations (3.117) and (3.119) imply
\be \{m(\si,j_L^x \}_{WZ}=x(\si)m(\si).\ee
Here $m(\si)$ was defined in (3.103),  $x(\si)$ is some element
of $L\G_0$ and
$j_L^x$
is the corresponding component of the   Kac-Moody current.

The Poisson bracket (5.168) plays a very important role in the quantum
theory where it becomes
the commutator of two quantum fields $m$ and $j_L$. The former plays
the role
of the vertex operator and the latter is the Kac-Moody current.
The quantized bracket (5.168) then expresses the fact that $m$ is the
 Kac-Moody primary field ,
or, in other words, the Kac-Moody tensor operator.

It will be
convenient to rewrite (5.168) by using the inverse vertex-IRF
 transformation (cf. Section 3.2.6).
Let $j_L$ be such $\G$-valued function on $M_L^{WZ}$ that
$j_L^x=(j_L,x)_{\G}$. Then the bracket
(5.168) can be rewritten in the matrix form as follows

\be \{k(\si)\ptp j_L(\si')\}_{WZ}=
2\pi C\delta(\si-\si')(k(\si)\otimes 1),\ee
where
\be k(\si)=m(\si)\exp{(-{a_\mu  \U(t_\mu)\si})}\ee
and $C$ is the  Casimir element defined in (5.145).

 We wish to find the quasitriangular generalization of the
relation (5.169). The quantity
$k(\si)$ keeps its meaning also in the deformed case; however the
standard moment map $\hat\beta_L(\hat k\hat a)\in\hat\G^*$ defined
on $\hat M_L$   is to be
replaced by the non-Abelian Poisson-Lie moment map
$\hat b_L(\hat k\hat a)\in\hat B$.
Note  that the group multiplication $\hat k\hat a$
(in the argument) then  takes place in the
Drinfeld double  $\hhD$ and $\hat a=\exp{\hat\Lambda(\hat \phi)}$,
where $\hat\phi=a^\infty\tf+a^\infty a^\mu\pi^*(t_\mu)$ (cf. (5.114)).
 What is the analogue of $j_L$?
In fact, $j_L$ can be written
as \be \pi^* j_L=\U\circ\iota^*\circ\hat\beta_L.\ee
Note that $\iota^*$ is the map from $\hat\G^*\to\G^*$ dual to
 $\iota:\G\to\hat\G$.
 But  we do not
have a canonical map from $\hat B$ into $B$ because $\hat B$
is the non-Abelian group.
There
are however two natural possibilities exploiting the maps $m_{L,R}$
defined in Section 4.3, Conventions 4.12. Both are equally good to
 work with and we shall
 choose $m_R$.
The deformed case analogue of $\iota^*\circ\hat\beta_L$ will
 be therefore the
map \be F_L'=m_R\circ \hat b_L.\ee
It is easy to see that similarly (as in the non-deformed case),
$F_L'(\hat k\hat a)$
is the invariant $B$-valued function on $\hat M_L$ with respect
to the central circle action. Restriction of this invariant function
on the surface $a^\infty=\k$
can be interpreted as the $B$-valued function on the quasitriangular
 model space $M_L^{WZ}$
which will be denoted as $F_L(k,a^\mu)$. The quasitriangular
 analogue of $\{k\ptp j_L\}_{WZ}$
will be now the Poisson bracket $\{k\ptp F_L\bq$. We want to
 calculate
this bracket explicitely.

By definition, we have
\be F_L'(\hat k,a^\mu,a^\infty=\k)=F_L(\pi(\hat k),a^\mu).\ee
We note, moreover, that (cf. Section 5.2.3)
\be a^\infty=m_L^\infty(\hat k\hat a) =(m^\infty\circ\hat b_L)
(\hat k\hat a),\ee where the map
$m^\infty:\hat B\to \br=\exp{Span(\hat\Lm(\tf))}$ was introduced
 in Conventions 4.12 of  Section 4.3.

The Poisson bivector  on $\hat M_L$ was denoted as
$\hat\Si_L$. Because $\hat b_L(\hat k\hat a)$ is the non-Abelian
 moment map, we have the
relation (cf. (5.30))
\be \hat\Si_L(.,d\hat b_L \hat b_L^{-1})=\hat\Lm(\tf)\otimes
\nabla^L_{\tF}+
(\nu_*\circ\Lm)(t_i)\otimes  \nabla^L_{\iota(T^i)},\ee
where $(\tF,\iota(T^i))$ is the basis of $\hG=\widehat{L\G_0}$
and $(\hat\Lm(\tf),(\nu_*\circ\Lm)(t_i))$
 is the
basis of $\hat\B$ in the sense of the double $\hhD$. Recall also
that $\hat\Lm\circ\pi^*=\nu_*\circ\Lm$.
The vector fields $\nabla^L$ correspond
to the left action of the group $\hat G=\widehat{LG_0}$ on the model
space $\hat M_L=\hat G
\times \hat A_+$. Because
\be \hat b_L=\exp{(m_L^\infty\hat\Lm(\tf))}F_L'=
\exp{(a^\infty\hat\Lm(\tf))}F_L' ,\ee we infer
$$ \hat\Si_L(., da^\infty \hat\Lm(\tf)+e^{a^\infty\hat\Lm(\tf)}
 d F_L' (F_L')^{-1}e^{-a^\infty\hat\Lm(\tf)}
 )
=$$ \be =\hat\Lm(\tf)\otimes \nabla^L_{\tF}+
(\nu_* \circ\Lm)(t_i)\otimes  \nabla^L_{\iota(T^i)}.\ee
From this relation, we obtain readily
\be \{\pi(\hat k)\ptp F_L'\}_{\hat M_L}= T^i\pi(\hat k)\otimes
 e^{-a^\infty\hat\Lm(\tf)}
 (\nu_*\circ\Lm)(t_i)e^{a^\infty\hat\Lm(\tf)} F_L'.\ee
By using the relation (5.134) and the fact that $a^\infty =\k$,
 the symplectic reduction is trivially performed to give
\be \{k\ptp F_L\bq =
 ( T^i\otimes \ ^{-\k}\Lm(t_i) )(k\otimes  F_L)\equiv \e\hat r^\k
 (k\otimes F_L).\ee
The notation  means
\be \ ^{-\k}\Lm(t_i)\equiv e^{-\k(-i\e\d_\si)} \Lm(t_i)
e^{\k(-i\e\d_\si)} .\ee
Thus we observe the presence of yet another $r$-matrix in our game.
It is instructive to give explicit formulas for the elements
$\ ^{-\k}\Lm(t_i)$:
\be{\ ^{-\k}\Lm(t_\mu)}=\Lm(t_\mu),\quad \ ^{-\k}\Lm(b_{\hal})=
e^{-\e\k n}\Lm(b_{\hal}),\quad
 \ ^{-\k}\Lm(c_{\hal})=e^{-\e\k n}\Lm(c_{\hal}),\ee
where $n$ is taken from $\hal =(\al,n)$ or from $\hal=(\mu,n)$.
The $\si$ dependence of the matrix $\hat r^\k$ is therefore as
follows
\be \hat r^\k(\si -\si')=\hat r(\si -\si' -i\e\k)=
r+C{\rm cotg}\jp(\si-\si'-i\e\k).\ee
The formula (5.179) is the principal result of this section. It is
the quasitriangular
generalization of the standard  primary field condition (5.169).
Upon the quantization, the relation
should express the crucial property   that the primary field
should be  the
tensor operator with respect to the $q$-current algebra.
\vskip1pc
\rem : {\small The characterization of the tensor operators of
certain quantum groups by means of
suitable $r$-matrices was discussed in \cite{Byts,BS}. Our
results fit in spirit
in the framework of those references.}
\subsection{$q$-deformed current algebra}
Recall the basic relation (3.114) defining the standard chiral
current algebra:
\be \{j_L^x,j_L^y\}_{WZ}=j_L^{[x,y]} +\kappa \rho(x,y).\ee
 Here  $x,y\in\G=L\G_0$. Recall that
\be j_L =\kappa\d_\si mm^{-1}=\k
  a^\mu k\U(t_\mu) k^{-1}+\kappa\d_\si kk^{-1}  .\ee
  The basic relation (5.183) can be cast in the
following matrix form
\be \{j_L(\si)\ptp j_L(\si')\}_{WZ}=\pi\delta(\si-\si')[C, j_L(\si)\otimes
 1-1\otimes j_L(\si')]+\k  2\pi C\d_\si\delta(\si-\si'),\ee
where    $C$ is the Casimir element defined in (5.145).

Our goal is to calculate the quasitriangular analogue of
the current commutator (5.185).
For this, we have to evaluate the Poisson brackets of the
$q$-currents $\{F_L\ptp F_L\bq$,
$\{F_L^\dg\ptp F_L^\dg\bq $ and
$\{(F_L^{\dg} )^{-1}\ptp F_L\bq$. The reason why the knowledge
of the first  bracket only
is not sufficient is the same as in the finite case (cf. the
 text between (5.68)
and (5.69)). The calculation is  similar as the one performed in
 the previous section. We start with the basic relation (5.175)
$$ \hat\Si_L(.,d\hat b_L \hat b_L^{-1})=\hat\Lm(\tf)\otimes
\nabla^L_{\tF}+
(\nu_*\circ \Lm)(t_i)\otimes  \nabla^L_{\iota(T^i)}.$$
Suppose that $f(\hat k)$ is some function invariant with respect
to the central
circle action. Then we obtain readily (cf. (5.178))
\be \{f(\hat k), F'_L\}_{\hat M_L}=
\nabla^L_{\iota(T^i)}f(\hat k)\times
( e^{-a^\infty\hat\Lm(\tf)}(\nu_*\circ\Lm)(t_i)
 e^{a^\infty\hat\Lm(\tf)} F_L').\ee
Now we take for $f$ the function $F_L'$ itself.
Since $F_L'=m_R\circ \hat b_L$, we can
easily calculate the derivative
$$ \nabla^L_{\iota(T^i)}F_L'=$$\be=(\hat b_L^{-1}
\iota(T^i)\hat b_L,\iota(T^p))_{\hhd}
F_L'(\nu_*\circ\Lm)(t_p) .\ee
Thus we obtain
$$ \{F_L'\ptp F_L'\}_{\hat M_L}=$$
\be = F_L' (\nu_*\circ\Lm)(t_p)
\otimes( e^{-a^\infty\hat\Lm(\tf)}(\nu_*\circ\Lm)(t_i)
 e^{a^\infty\hat\Lm(\tf)})
  F_L'
\times  ({F_L'}^{-1}(\ ^{-a^\infty}T^i) F_L' ,T^p)_\D.
\ee
Now we are ready to get down to the Poisson bracket
 $\{.,.\bq$ on $M_L^{WZ}$. We obtain
\be \{F_L\ptp F_L\bq=F_L \Lm(t_p)\otimes \ ^{-\k}\Lm(t_i)
 F_L\times
({F_L}^{-1}(\ ^{-\k}T^i)  F_L ,T^p)_\D.\ee
Now we use the obvious relation
\be {F_L}^{-1}(\ ^{-\k}T^i)  F_L=({F_L}^{-1}(\ ^{-\k}T^i)
 F_L,T_p)_\D \Lm(t_p)+
({F_L}^{-1}(\ ^{-\k}T^i)  F_L,\Lm(t_p))_\D T_p\ee
to rewrite (5.189) in the  form
$$ \{F_L\ptp F_L\bq=$$ $$=( \ ^{-\k} T^i\otimes
 \ ^{-\k}\Lm(t_i))(F_L\otimes  F_L) -
F_L T_p\otimes \ ^{-\k}\Lm(t_i) F_L\times
 ({F_L}^{-1}(\ ^{-\k}T^i)  F_L,\Lm(t_p))_\D=$$\be =
( \ ^{-\k} T^i\otimes \ ^{-\k}\Lm(t_i))(F_L\otimes
 F_L) -
(F_L\otimes F_L)( T^i \otimes  \Lm(t_i)).\ee
By using the concrete properties of the basis
 $T^i,\Lm(t_i)$ given by (5.140),
it can be easily checked that (5.191) can be rewritten as
\be \{F_L\ptp F_L\bq =\e[\hat r,F_L\otimes F_L].\ee
Note that this formula is fully analoguous to the
 relation (5.68) holding in the
finite dimensional (non-affine) case.

The case  $\{F_L^\dg\ptp F_L^\dg\bq$ can be calculated
similarly to yield
\be \{F_L^\dg\ptp F_L^\dg\bq =-\e[\hat r,F_L^\dg\otimes F_L^\dg].\ee
It remains to calculate $\{(F_L^\dg)^{-1}\ptp F_L\bq $. We
 proceed as before to arrive
at the following counterpart of the relation (5.188):
$$ \{(F_L'^\dg)^{-1}\ptp F_L'\}_{\hat M_L}=
 -({F_L'}^{-1}(\ ^{-a^\infty}T^i) F_L' ,T^p)_\D\times$$
\be \times (F_L'^\dg)^{-1} \nu_*((\Lm(t_p))^{\dg})\otimes
( e^{-a^\infty\hat\Lm(\tf)}(\nu_*\circ\Lm)(t_i)
  e^{a^\infty\hat\Lm(\tf)})
  F_L'
 \ee
Getting down to the Poisson bracket $\{.,.\bq$ on $M_L^{WZ}$,
 we obtain
$$ \{ (F_L^\dg)^{-1}\ptp F_L\bq=
-(F_L^\dg)^{-1} (\Lm(t_p))^\dg\otimes\ ^{-\k}\Lm(t_i) F_L \times
(\ ^{-\k}T^i , F_L T^p {F_L}^{-1})_\D=$$ \be=- (F_L^\dg)^{-1}
 (\Lm(t_p))^\dg\otimes F_LT^p+
 (F_L^\dg)^{-1} (\Lm(t_p))^\dg\otimes \ ^{-\k}T^i F_L\times
(\ ^{-\k}\Lm(t_i) , F_L T^p {F_L}^{-1})_\D. \ee
Now we use two obvious relations
\be  (F_L^{-1}\ ^{-\k}\Lm(t^i) F_L)^\dg=
(F_L^{-1}\ ^{-\k}\Lm(t^i) F_L,T^p)_\D (\Lm(t_p))^\dg,\quad
(\ ^{-\k}\Lm(t_i))^\dg=\ ^{\k}((\Lm(t_i))^\dg)\ee
to rewrite (5.195) in the  form
$$ \{ (F_L^\dg)^{-1}\ptp F_L\bq=$$ \be=-(  (F_L^\dg)^{-1}\ot F_L)
( (\Lm( t_p))^\dg\otimes T^p)
+\ ^{\k}((\Lm(t_p))^\dg)\otimes \ ^{-\k}T^p)((F_L^\dg)^{-1}\ot F_L).\ee
Using the explicit form
of the base $(T^i,\Lm(t_i))$, we find easily   that
 $\e\hat r=(\Lm( t_p))^\dg\otimes T^p$.
Thus we obtain the final form of the third defining
 Poisson bracket of the
$q$-current algebra:
 \be \{ (F_L^\dg)^{-1}\ptp F_L\bq=
 \e\hat r^{2\k}((F_L^\dg)^{-1}\ot F_L)
-((F_L^\dg)^{-1}\ot F_L)\e\hat r,\ee
where
 \be \hat r^{2\k}(\si-\si')=\hat r(\si -\si'-2i\e\k).\ee

\rem : {\small
The Poisson bracket of the type (5.198) resembles the  brackets
arising in the description
of the
structure of the so called twisted Heisenberg double of the
 reference \cite{ST}.
Although we do not see here any direct connection of \cite{ST} with (5.198)
(since
$M_L^{WZ}$ does not have even the structure of the double),
we believe, nevertheless, that
there is a deeper reason why similar formulas appear here and
in \cite{ST}. Most probably,
the double $D=LG_0^\bc$ equipped with the $qWZW$ bracket reduced
from $\hhD$ could be a sort of the
real form of the twisted Heisenberg double of \cite{ST}.
More precisely: there exists a method of generating a new
(non-twisted)
Heisenberg double $D_{real}$ from an
old one $D_{complex}=GB$ if the latter is equipped with an involution.
Indeed, $D_{real}$ is simply
the subgroup of $D_{complex}$ consisting of the elements which are
 stable under the involution.
There is  a condition to fulfil, however, that the restriction
of the bilinear form $(.,.)_\D$ on $Lie(D_{real})$  must be
non-degenerate.  Then $D_{real}$ has
 canonically
the structure of the Heisenberg double. It can be decomposed as
 $D_0=G_0B_0$ where $G_0,B_0$
play the role of the mutually dual isotropic Poisson-Lie groups
 and they themselves consist
of the elements of $G$ and $B$ stable under the involution.
 We conjecture that the
similar realification can be made also for the twisted doubles
and that among the class
of the twisted Heisenberg doubles of $LG_0$ we can find such
$D_{complex}$ that the derived
twisted double
$D_{real}=LG_0^\bc$
(given by  a suitable involution to be specified) is canonically
equipped with
 the $qWZW$ symplectic structure.  However, the fact whether the
 conjecture is true or false
does not have direct implications for our $qWZW$  story and we
 shall not further
study
possible connections of our formalism with the theory of the
twisted Heisenberg doubles.}
 \vskip1pc
\noindent  There already exists in the literature
(cf. \cite{RST}) the concept of the $q$-deformed current algebra. We now
show that our construction fits in this framework due to  Reshetikhin
and Semenov-Tian-Shansky \cite{RST}. First we consider the
 (matrix) observable
\be L=F_LF_L^\dg.\ee The defining commutations relations (5.192),
 (5.193) and (5.198)
can be then equivalently written in terms of one relation only:
$$\{L(\si)\ptp L(\si')\}_{qWZ}=(L(\si)\ot L(\si'))\e\hat r(\si-\si')
+\e\hat r(\si-\si')(L(\si)\ot L(\si')) $$
 \be -(1\ot L(\si'))\e\hat r(\si-\si'+2i\e\k)(L(\si)\ot 1)
-(L(\si)\ot 1)\e\hat r(\si-\si'-2i\e\k)(1\ot L(\si')).\ee
Our  formula (5.201) now exactly coincides with the defining formula
of $q$-current algebra in the version of Reshetikhin and
Semenov-Tian-Shansky
\cite{RST}.

Recall that in the undeformed WZW model, the current $j_L(\si)$ can be
written in terms of the primary field $m(\si)$ as follows
\be j_L=\k\d_\si mm^{-1}.\ee
This relation can be called the classical Knizhnik-Zamolodchikov equation
\cite{KZ}
since its quantum analogue is nothing but the standard KZ-equation written
in the operatorial form \cite{FHT}. Recall that here we have used the
monodromic variables $$m(\si)=k(\si)\exp{(-a^\mu T^\mu\si)}$$
for the description of the phase space $M_L^{WZ}$
of the undeformed (and also deformed) chiral WZW model.
The quasitriangular
analogue of (5.202) can be easily derived from (5.200) and from (5.172)
rewritten
as
\be F_L=b_L(k(\si+i\e\k)e^{\e\k a^\mu H^\mu}).\ee
The result is simple and esthetically appealing:
\be L(\si)=m(\si+i\e\k)m^{-1}(\si-i\e\k).\ee
This is the classical version of the $q$-KZ equation; as expected, it is
not differential but rather a difference equation.

\vskip1pc
\rem : {\small  It is an instructive exercise to calculate the Poisson
bracket $\{L(\si)\ptp L(\si')\}_{qWZ}$ starting with the representation
(5.204) and using the formula    (5.164). In order to arrive
at the formula (5.201), one needs to know the (quasi)periodic behaviour
of the involved elliptic functions:
\be \si_w(z+\tau,\tau)=\si_w(z,\tau)e^{2\pi iw},\qquad
\si_w(z+1,\tau)=\si_w(z,\tau); \ee
\be \rho(z+\tau,\tau)=\rho(z,\tau)-2\pi i,\qquad \rho(z+1,\tau)=
\rho(z,\tau).\ee}
\subsection{The limit $q\to 1$}

The symplectic structure of the quasitriangular chiral WZW model
 is fully described\footnote{The description in the nonmonodromic
variables $k(\si),a^\mu$ is given by the formulae (5.159) and (5.161).}
 by the
fundamental quasitriangular braiding relation (5.153)
$$  \{m(\si)\ptp m(\si') \bq =(m(\si)\otimes m(\si'))
 B_{\e}(a^\mu,\si-\si')+
 \e\hat r(\si-\si')(m(\si)\otimes m(\si')),\eqno(5.164)$$
where $B_{\e}(a^\mu, \si)$ is the quasitriangular braiding matrix
given by
 $$ B_{\e}(a^\mu,\si)=$$
$$ =-{i\over \k}\rho({i\si\over 2\k\e},{i\pi\over \k\e})
H^\mu\otimes H^\mu
-{i\over \k}\sum_{\al\in\Phi}{\rn\over 2}\si_{ a^\mu\la
\al,H^\mu\ra}
( {i\si\over 2\k\e},{i\pi\over \k\e})E^\al\otimes E^{-\al}.\eqno(5.165)$$
We wish to show that  in the limit $\e\to 0$, Eq. (5.164) gives
$$  \{m(\si)\ptp m(\si') \}_{(q=1)WZ} =(m(\si)\otimes m(\si'))
 B_{0}(a^\mu,\si-\si') ,\eqno(3.105)$$
 where
$$ B_0(a^\mu,\si)=
-{\pi\over\kappa}\biggl[\eta(\si)(H^\mu\ot H^\mu)-i\sum_\al{\rn\over 2}
{\exp{(i
\pi \eta(\si)\la \al,H^\mu\ra a^\mu)}\over
\sin{(\pi \la \al,H^\mu\ra a^\mu)}}
E^\al\ot E^{-\al}\biggr].\eqno(3.106) $$
Recall that $\eta(\si)$ is the function defined by
\be \eta(\si)=2[{\si\over2\pi}]+1,\ee
where $[\si/2\pi]$ is the largest integer less than or equal
to ${\si\over 2\pi}$.

Now we observe that   the
  term containing $\hat r$ in (5.164)
disappears in
the limit $\e\to 0$. Hence it is enough to show
\be lim_{\e\to 0}\si_{a^\mu\la \al,H^\mu\ra}({i\si\over 2\k\e},
{i\pi\over \k\e})=
-\pi {\exp{(i
\pi \eta(\si)\la \al,H^\mu\ra a^\mu)}\over
\sin{(\pi \la \al,H^\mu\ra a^\mu)}},\ee
\be lim_{\e\to 0}\rho({i\si\over 2\k\e},{i\pi\over \k\e})=
 -i\pi\eta(\si).\ee
Using the formulae (5.205) and (5.206) and the fact
$\eta(\si+2\pi)=\eta(\si) +2$
we conclude that it is enough to establish the limits (5.208), (5.209) for
$\si\in[-\pi,\pi]$. Then we can represent the elliptic
functions on the l.h.s. by the series (5.153) and (5.154) .
For $\si\in[-\pi,\pi]$, the contributions of the sums over $m,n>0$
  disappear in the $\e\to 0$ limit
and we can consider only the "cotangents". This gives immediately
(5.208) and (5.209).
The correct $\e\to 0$ limit  of the fundamental braiding relation
 is therefore
established.

Our next task is to establish the $\e\to 0$ limit of the
quasitriangular Hamiltonian
$$ H^{qWZ}_{L}=
-{1\over 2\k}(\phi,\phi)_{\G^*} -(\wti{Dres}_{\hat k}
e^{\hat\Lm(\hat\phi_\k)})^0.  $$
 We remind that $(\ti b)^0\equiv m^0(\ti b)$ and that
 the map $m^0:\ti B\to \br$ was defined in Conventions 4.12 of
Section 4.3 as
\be\hat m(\ti b) \exp{(m^0(\ti b)\ti\Lm(\ti t_0))}=\ti b,\ee
where $\hat m(\ti b)\in \hat B$. This decomposition is unambiguous
thus defining
the function $m^0$. Moreover, since the generator $\ti\Lm(\ti t_0)$
itself depends on $\e$
also $m^0$ does which we may indicate by the subscript $m^0_\e$.

Now we use the same reasoning as in the finite dimensional case
of Section 5.1.3 showing that
in the $\e\to 0$ limit the dressing action becomes the coadjoint
one. Thus we have
$$ lim_{\e\to 0}(m^0_\e\circ \widetilde{Dres}_{\hat k})
(e^{\hat\Lambda_\e(\hat\phi_\k)})=
\la \wti{Coad}_{\hat k}\hat\phi_\k,\ti T^0\ra=$$
\be  =\la  \phi,k^{-1}\d_\si k\ra
+\jp \k (k^{-1}\d_\si k,k^{-1}\d_\si k)_\G ,\ee
where $k$ stands for $\pi(\hat k)$ for brevity and we indicated by
the subscript that $\hat\Lambda$ depends
on $\e$.
From the relation (5.211), we then obtain immediately
\be lim_{\e\to 0}H_L^{qWZ}=
-{1\over 2\k} (\phi,\phi)_{\G^*}- \la  \phi,k^{-1}\d_\si k\ra
- {\k\over 2} (k^{-1}\d_\si k,k^{-1}\d_\si k)_\G \equiv H_L^{WZ}.\ee
Thus the standard chiral Hamiltonian (3.92) is indeed recovered in
the limit $\e\to 0$.
We conclude that the quasitriangular chiral WZW model gives in
the limit $\e\to 0$ (or $q\to 1$)
the standard chiral WZW theory.

We know that  the fundamental
bracket   (5.164) implies  the $q$-current algebra bracket (5.201).
Our next task is to show that in the limit $\e\to 0$ (or,
equivalently, $q\to 1$),
we recover  from (5.179) the
 standard Kac-Moody primary field
condition (5.169) and from (5.201)  the
ordinary current algebra bracket (5.185).

First we rewrite (5.179) in the equivalent way using the
$q$-current $L(\si)$
defined in (5.200). We obtain
$$ \{k(\si)\ptp L(\si')\}_{qWZ}=$$
\be =\e \hat r(\si-\si'-i\e\k)(k(\si)\ot L(\si'))-
(1\ot L(\si'))\e \hat r(\si-\si'+i\e\k)
(k(\si)\ot 1).\ee
From the classical $q$-KZ equation (5.204), we derive
\be L(\si)= 1+2i\e\k \d_\si mm^{-1}+O(\e^2)=1+2i\e j_L(\si)+O(\e^2).\ee
Inserting this into (5.213) and using (5.182), we obtain in the lowest
order in $\e$ the desired relation (5.169)
$$ \{k(\si)\ptp j_L(\si')\}_{WZ}=2\pi C\delta(\si-\si')(k(\si)\otimes 1).$$
Here
the  $\delta $-function was produced as the following limit $\e\to 0^+$
\be 4\pi i\delta(\si-\si')={\rm cotg}\jp(\si-\si'-i0^+)-
{\rm cotg}\jp(\si-\si'+i0^+).\ee
Now we establish the   $q\to 1$ limit of the $q$-current algebra
$$\{L(\si)\ptp L(\si')\}_{qWZ}=(L(\si)\ot L(\si'))\e\hat r(\si-\si')
+\e\hat r(\si-\si')(L(\si)\ot L(\si')) $$
 $$ -(1\ot L(\si'))\e\hat r(\si-\si'+2i\e\k)(L(\si)\ot 1)
-(L(\si)\ot 1)\e\hat r(\si-\si'-2i\e\k)(1\ot L(\si')).\eqno(5.201)$$
Inserting the $\e$-expansion (5.214) into (5.201), we obtain in the
lowest order
$\e^2$ the correct result
$$ \{j_L(\si)\ptp j_L(\si')\}_{WZ}=\pi\delta(\si-\si')[C, j_L(\si)\otimes
 1-1\otimes j_L(\si')]+  2\pi\k C\d_\si\delta(\si-\si').\eqno(5.185)$$
Here we have needed three formulae:
\be 8\pi i \d_\si\delta(\si-\si')={1\over {\rm sin}^2\jp(\si-\si'+i0^+)}
-{1\over {\rm sin}^2\jp(\si-\si'-i0^+)};\ee
\be 2\pi i\delta(\si-\si')={\rm cotg}\jp(\si-\si'-i0^+)-
{\rm cotg}\jp(\si-\si').\ee
\be 2\pi i\delta(\si-\si')={\rm cotg}\jp(\si-\si')-
{\rm cotg}\jp(\si-\si'+i0^+).\ee
The relation (5.216) can be obtained by deriving (5.215) and the remaining
equalities (5.217) and (5.218) can be proved by using the
Plemelj-Sokhotsky formula.

\subsection{Quasitriangular exact solution}
The simplest way for describing the
classical solutions of the quasitriangular chiral WZW model
consists in using the
monodromic variables $m(\si)$ (cf. (3.103)) on the phase space $M_L^{WZ}$.
In turns out that
the time evolution is the same as in the non-deformed case, i.e.
\be [m(\si)](\tau)=m(\si-\tau).\ee
In order to prove that we have to combine the arguments of Sections
 3.1.3  and 3.2.6.
First of all, the quasitriangular chiral master model on $\ti G$
can be solved in the same way
as its  finite-dimensional counterpart (3.14) on $G_0$, with the result
 (in $\ti k,\ti \phi$ variables
on the affine model space $\ti M_L$)
\be \ti k(\tau)=\ti k_0 \exp{({-\ti\U(\ti\phi_0)\over\k}\tau)},
\quad \ti\phi(\tau)=\ti\phi_0.\ee
Here the multiplication is taken in the sense of the group $\ti G$.

Now this solution can be projected  on the (doubly) reduced phase
$M_L^{WZ}$ following
step-by-step the argumentation between (3.121) and (3.124). The result
 is then (5.219).

\vskip1pc
\rem : {\small It may seem astonishing that the set of classical
 solutions does not
change under the $q$-deformation. We have met the finite dimensional
version of this phenomenon
already in Section 5.1.4. The solution of the puzzle is the same.
It is the symplectic
structure on the phase space that gets deformed in such the way
that the
$G$-action ceases to be Hamiltonian but becomes Poisson-Lie. This
means that the
natural dynamical variables of the group theoretical origin will
have modified Poisson
brackets and, upon the quantization, modified commutation relations.
For instance,
the field theoretical  correlation functions change. }

\subsection{The left-right glueing}
Consider the topological direct product $M_L^{WZ}\times M_R^{WZ}$ where
$M_{L}^{WZ}=LG_0\times \A^1_+$ and  $M_{R}^{WZ}=LG_0\times \A^1_-$.
 Recall that $\A^1_-\equiv -\A^1_+$.
The product symplectic structure on $M_L^{WZ}\times M_R^{WZ}$ is
 given by the symplectic
form
\be \om_{L\times R}^{qWZ}=\om_L^{qWZ}+\om_R^{qWZ}.\ee
The symplectic form $\om_R^{qWZ}$ differs from $\om_L^{qWZ}$ only by
the domain of definition
of the variables $a^\mu$.  The Hamiltonian on $M_L^{WZ}\times M_R^{WZ}$
is given by
\be  H_{L\times R}^{qWZ}(k_L,k_R,a^\mu_L,a^\mu_R)=
H_L^{qWZ}(k_L,a^\mu_L)+H_R^{qWZ}(k_R,a_R^\mu).\ee
Now  we perform a symplectic reduction by setting
\be a_L^\mu+a_R^\mu=0.\ee
We learn from the Poisson brackets (5.161) that the quantities
 $a_L^\mu+a_R^\mu$ are the moment
maps generating the simultaneous right action of the Cartan
generators $T^\mu$ on
$M_L^{WZ}$ and $M_R^{WZ}$. The reduction makes sense only if
the Hamiltonian $ H_{L\times R}^{qWZ}$
is invariant with respect to this $\bt$-action. But this is
the case as we see from (1.20)
and from the following chain of formulas
$$  \wti{Dres}_{(\hat ke^{\eta_\nu T^\nu})}
(e^{\hat\Lambda_\e(\hat\phi_\k)})\equiv
\ti b_L(\hat ke^{\eta_\nu T^\nu} e^{\hat\Lambda_\e
(\hat\phi_\k)}) =$$ \be =
\ti b_L(\wti{Ad}_{(\hat ke^{\eta_\nu T^\nu})}
 e^{\hat\Lambda_\e(\hat\phi_\k)})=
\ti b_L(\wti{Ad}_{\hat k }  e^{\hat\Lambda_\e(\hat\phi_\k)})
\equiv\wti{Dres}_{\hat k }
(e^{\hat\Lambda_\e(\hat\phi_\k)}).\ee
Recall in this respect that (5.224) makes sense since $\bt$ {\it can}
 be embedded as the
subgroup into $\wh{LG_0}$.

We conclude that the reduced symplectic form   and
Hamiltonian   live on the
 reduced phase space $((LG_0\times LG_0)/\bt^{diag})\times \A^1_+$.
 This is nothing but the
phase space of the standard full left-right WZW model. We shall not
 make more
explicit the full left-right WZW symplectic structure. The corresponding
formulas are complicated
and not illuminating. Anyway, the canonical quantization of the
 quasitriangular WZW model
will proceed by quantization of its chiral components. This
 is fortunate, because
all important Poisson brackets are written in terms of the
 collection of $r$-matrices.
These $r$-matrices appeared already in the literature in different
context (e.g. KZB equation
\cite{B,FW}) and their quantum counterparts are often known,
too \cite{F}. Of course,
the latter circumstance makes the quantization task more accessible.

\chapter{Conclusions and outlook}

The most important   result of this paper is the explicit description
of the symplectic structure $\om_L^{qWZ}$
and of the Hamiltonian $H_L^{qWZ}$ of the quasitriangular
chiral WZW model.
The $q$-deformation of the standard current algebra then emerged
in the natural way.
We did not compare our $q$-current algebra with that of Reshetikhin
 and
Semenov-Tian-Shansky \cite{RST}. We believe, however, that the
 former is the sort
of the real form of the latter. This belief stems from the
fact that
the formalism of \cite{RST} (and of \cite{ST}) use rather the
 theory of the
complex Heisenberg doubles of the form $D=G\times G$ for an
 appropriate group
$G$. Our doubles $\ttD,\hhD,D$ are not of this form but they
are probably
real forms of such doubles in the sense of the remark in Section 5.2.6.
In any case,
our $q$-current algebra is dictated by the choice of
the affine  Lu-Weinstein-Soibelman
double of the Kac-Moody group.
It should be stressed, in particular,
that our method applies not only for this particular choice
if the double $D$.
 There may be different doubles $D$ of the loop group $LG_0$
and if we succeed to construct the WZW double $\ttD$ (in the
 sense of the Definition 4.10)
then our general  construction of  Section 4.3 again works.

We have achieved more than the results described above
in the sense that
we have built up
the whole approach how to $q$-deform the WZW model and its derived
 products.
In fact, we have no doubts that our construction should be
easily generalizable
to the gauged WZW models by performing an appropriate symplectic
reduction induced
by equating the non-Abelian moment maps to some fixed values.
 Also the quasitriangular
boundary WZW model should be at reach by first constructing the
 central extension
of the group of segments and then picking up some of its
 Drinfeld doubles. Finally,
the supersymmetrization of the quasitriangular geodesical
model should lead also
to the SUSY version of the q-WZW theory.

Among other obtained results, we should mention the universal
description of the
WZW model in terms of the central biextensions. We can thus
argue, that the WZW-like models
can be constructed not only for the loop group case.
The construction of the doubles
$D$, $\hhD$ and $\ttD$ of chapter 4 is another result of this
article which permitted
to deforme the most important case: the loop group WZW model.
The crucial thing
to do was to prove the global decomposability $\ttD=\ti G\ti B=
\ti B\ti G$. We have succeeded
to do that in the series of lemmas of section 4.4.

The most important open problem is the quantization of the
 quasitriangular WZW model.
As in the standard case, the chiral part of the model should
 have the Hilbert
space consisting of the set of quantized dressing orbits of
the Kac-Moody group.
Only the orbits with the integrable highest weight should be present
 (with the multiplicity
1).  Those integrable highest weight correspond to those points in
the alcove
for which the induced symplectic form on the dressing orbit is
 integral.
The quantized dressing orbits should carry the unitary representation
 of the
q-Kac-Moody group. The $q$-vertex operators should be the
q-Kac-Moody tensor
operators fulfilling the quantum version of the relation (5.174).
They should obey
the quantized braiding relation of the type (5.153); the theory of
the quantum
grupoids  \cite{Xu} should be relevant for this story.
It is important
that the  Poisson brackets of the basic observables are of the
$r$-matrix
type. The explicite knowledge of the corresponding quantum
$R$-matrices should
considerably facilitate the quantization program.  Another
 related important task
would be to find the $q$-free field representation of the model.
Perhaps the results
of \cite{LS} might be of use in the $sl_2$ case  although a sort
of analytic continuation in $\k$
is needed
to arrive from their version of  the $q$-current algebra to ours.
We expect that further developments of the ideas presented in this
article
can reveal interesting connections with results of other references,
e.g.
\cite{Co, FR1,FR2,Fro,JKOS}
The important question is the status of the Virasoro generators.
It is remarkable
that the Virasoro group does act  on the classical chiral phase
 space $M_L^{WZ}$ for
every value of $q$. For $q=1$ (or $\e=0$) this action is Hamiltonian.
 For generic $q$,
if the group action is   Poisson-Lie (probably in the
combined Virasoro-Kac-Moody sense),
or even quasi-Poisson, then
the $q$-Virasoro algebra could be interpreted
as the Hopf or quasi-Hopf algebra, respectively. It
would be certainly interesting
to analyze this issue in more detail.

\chapter{Appendices}

\section{The loop group primer}
In what follows, $G_0$ will be always a simple compact  connected
and simply connected
Lie group. A group  $G$ of smooth maps from the circle $S^1$ into
$G_0$ is called
the loop group and it is often denoted as $G=LG_0$. The group
 structure of $G$
is naturally given by the pointwise multiplication, the unit element
 $e$ is the constant
map with value $e_0$, where $e_0$ is the unit element of $G_0$.

The Lie algebra $\G$ of $G$ consists of smooth maps from $S^1$
into $\G_0$ again with
the pointwise commutator. The invariant symmetric nondegenerate
bilinear
form on $\G$ is given by
\be (\xi,\eta\zg={1\over 2\pi}\int d\si ( \xi(\si),
\eta(\si))_{\G_0^\bc},\ee
where $(.,.)_{\G_0^\bc}$ is the standard Killing-Cartan form
on $\G_0^\bc$ renormalized in such
a way that the square of the length of the longest root is
equal to two. For example,
for the Lie algebra $su(N)$, $(.,.)_{\G_0^\bc}$ is given by
the trace taken in the fundamental
representation.
We shall identify the dual $\G^*$ of $\G$ with $\G$ itself via
the bilinear form $(.,.\zg$.
We shall call the corresponding map $\U$. Thus
\be \la \al,\xi\ra=(\U (\al),\xi\zg, \qquad \al\in\G^*,\quad
\xi\in \G.\ee

The construction of the standard central extension of  loop
groups goes as follows \cite{Mick}:
First one considers a group $DG_0$ of smooth maps from the unit
{\it Disc}  (in the complex plane)
 into $G_0$ with
the usual pointwise multiplication. We can now define an extended
 group $\wh {DG_0}$ whose
elements are pairs $(f,\lm)$ where $f\in DG_0$ and $\lm\in U(1)$
and whose multiplication
law reads
\be (f_1,\lm_1)(f_2,\lm_2)=(f_1f_2,\lm_1\lm_2
\exp{[2\pi i\ga (f_1,f_2)]}).\ee
Here  $\ga$ is a real valued 2-cocycle on $DG_0$
given by
\be \ga(f_1,f_2)={1\over 8\pi^2}\int_{Disc}
 (f_1^{-1}df_1\stackrel{\w}{,}df_2 f_2^{-1})_{\G_0^\bc}.\ee
Consider now a subgroup $\d G$ of $DG_0$ consisting of all
smooth maps
from {\it Disc} into $G_0$ such that their value  on the
boundary $\d${\it  Disc}$=S^1$
is the unit element $e_0$ of $G_0$. Any $g\in\d G$ can be
thought of as a map
$g:S^2\to G_0$ by identifying the boundary $S^1$ of {\it Disc}
 with the north pole of $S^2$.
It turns out that there is a homomorphism
$\Th:\d G\to\wh{DG_0}$  defined by
\be \Th(g)=(g,\exp{[-2\pi iC(g)]}),\ee
where
\be C(g)={1\over 24\pi^2}\int_{Ball}(dgg^{-1}\stackrel{\w}{,}
dgg^{-1}\w dgg^{-1})_{\G_0^\bc}.\ee
Here $Ball$ is the unit ball whose boundary is $S^2$ and we
have extended
the map $g:S^2\to G_0$ to a map $g: Ball\to G_0$. It is not
immediately obvious
that the homomorphism $\Th$ is correctly defined, or, in other words,
 that
$\exp{[-2\pi iC(g)]}$ does not depend on the extension of $g$ to $B$.
The standard
argument of the independence of this term on the extension
can be found e.g. in \cite{Mick}. The demonstration
that $\Th$ is indeed a homomorphism is based on the Polyakov-Wiegmann
formula \cite{PW}
which asserts that
\be C(g_1g_2)=C(g_1)+C(g_2)-\gamma(g_1,g_2).\ee
The fact  that the image $\Th(\d G)$ is  a normal subgroup
in $\wh {DG_0}$ follows from the identity
\be C(fgf^{-1})+\gamma(f,g)+\gamma(fg,f^{-1})=C(g).\ee
The latter identity is also the direct consequence of the
Polyakov-Wiegmann formula.

The standard central extension $\hat G$  of the group $G=LG_0$
is defined as the factor group
$\wh {DG_0}/\Th(\d G)$. This group is a (nontrivial) circle bundle
 over the
base space $LG_0=\{g:S^1\to G_0\vert g$ smooth $\}$. The
projection $\pi$ is
 $(g,\lm)\to g\vert_{S^1}$ and the center of $\hat G$ is
represented by the pairs $(1,\lm)\in
\wh {DG_0}$. The projection homomorphism from $\wh {DG_0}$
onto $\hat G$ will be referred to as
$\wp$.

Now we are going to calculate the  following expression
\be \dto \hat g e^{s\hat x}\hat g^{-1}
e^{-s\hat x}\in\hG,\ee where $\hat g\in \hat G$ and
 $\hat x\in Lie(\hat G)\equiv \hG$. From
this computation we shall extract two informations: 1) what is
the commutator in $\hG$;
2) what is the explicit form of the adjoint action of $\hat G$ on
$\hat\G$.

Let us choose two representatives $\wp^{-1}\hat g$ and
$e^{s\wp^{-1}_*\hat x}$
in $\widehat {DG_0}$ of the classes $\hat g$ and
 $e^{s\hat x}$ in $\hat G$. It is clear
that
\be \hat g e^{s\hat x}\hat g^{-1}
e^{-s\hat x}=\wp\biggl((\wp^{-1}\hat g)
e^{s\wp^{-1}_*\hat x}(\wp^{-1}\hat g)^{-1}
e^{-s\wp^{-1}_*\hat x}\biggr).\ee
We shall therefore first calculate an expression
\be \dto (\Gamma,\lm)(e^{s\xi},e^{s\al})
(\Gamma,\lm)^{-1}(e^{-s\xi},e^{-s\al}),\ee
where
\be  (\Gamma,\lm)=\wp^{-1}\hat g;\quad (\xi,\al)=\wp_*^{-1}\hat x.\ee
Using (7.3), we calculate
$$\dto (\Gamma,\lm)(e^{s\xi},e^{s\al})
(\Gamma,\lm)^{-1}(e^{-s\xi},e^{-s\al})=$$ $$=
\dto (\Gamma,1)(e^{s\xi},1) (\Gamma,1)^{-1}(e^{-s\xi},1)$$
$$= \dto(\Gamma e^{s\xi},\exp{is\over 4\pi}\int_{Disc}
 (\Gamma^{-1}d\Gamma\stackrel{\w}{,}d\xi)_{\G_0^\bc})
(\Gamma^{-1} e^{-s\xi},\exp{is\over 4\pi}\int_{Disc}
(d\Gamma\Gamma^{-1}\stackrel{\w}{,}d\xi)_{\G_0^\bc})$$
$$=\dto (\Gamma e^{s\xi}\Gamma^{-1} e^{-s\xi},
\exp{is\over 4\pi}\int_{Disc} \{2(\Gamma^{-1}d\Gamma\stackrel{\w}{,}
d\xi)_{\G_0^\bc}+([\xi,\Gamma^{-1}d\Gamma]\stackrel{\w}{,}
\Gamma^{-1}d\Gamma)_{\G_0^\bc}\})$$
$$ =\dto (\Gamma e^{s\xi}\Gamma^{-1} e^{-s\xi},\exp{{-is\over 2\pi}
\int_{Disc} d(\Gamma^{-1}d\Gamma,\xi)_{\G_0^\bc}})=$$
\be = (\Gamma\xi\Gamma^{-1}-\xi,-{i\over 2\pi}\int_{S^1=\d Disc}
(\Gamma^{-1}d\Gamma,\xi)_{\G_0^\bc}).\ee

From (7.13), one may derive the Lie algebra commutator in
 $Lie(\wh DG_0)=D\G_0+i\br$.
 Indeed, by setting
$\Gamma=e^{t\eta}$ and deriving with respect to $t$ at $t=0$,
 one obtains
\be [(\eta,i\al),(\xi,i\beta)]=([\eta,\xi], {i\over 2\pi}
\int_{S^1\equiv \d Disc}(\eta,d\xi)_{\G_0^\bc}),
\quad \al,\beta\in\br.\ee
Finally, the commutator in the Lie algebra $\hG=Lie(\hat G)$
is given by the same formula
as (7.14) but $\eta$ and $\xi$ are to be considered as  elements
 of $L\G_0$ rather than
those of $D\G_0$. We see also that the map $\iota:\G\to\hG$
(cf. Section 2.1) is simply given by
\be \iota(\xi)=(\xi,0).\ee
Although it should be already  clear  from (7.14) what is the
formula for the
cocycle $\rho$,  we nevertheless write it down  explicitely:
\be \rho(\eta,\xi)={1\over 2\pi}\int_{S^1}(\eta,d\xi)_{\G_0^\bc}.\ee
From (7.13), we obtain also the formula for the adjoint action
\be \hat g(\xi,i\beta)\hat g^{-1}=(g\xi g^{-1},i\beta-{i\over
2\pi}\int_{S^1}
(g^{-1}dg,\xi)_{\G_0^\bc}),\ee
where $g=\pi(\hat g)$.

\section{Cotangent bundle of a group manifold}

Consider a (possibly infinite dimensional) connected Lie group $G$
and its
cotangent bundle $T^*G$.

The points $K$ of $T^*G$ are couples $(P_K,F_K)$ where $P_K$ is a point
in $G$ and $F_K$ is a differential 1-form at the point $P_K$;
in other words
$F_K\in T^*_{P_K}G$. We shall equip $T^*G$ with the standard
 group structure
by introducing the following product $QK$ of two elements $Q,K\in T^*G$:
\be P_{QK}=P_QP_K;\qquad F_{QK}=R^*_{P_K^{-1}}F_Q + L^*_{P_Q^{-1}}F_K.\ee
Here $R^*$ and $L^*$ denote pull-backs with respect to the right and left
translation by elements of  $G$, respectively.
The inverse element $K^{-1}$ is given by
\be P_{K^{-1}}=P_K^{-1},\qquad F_{K^{-1}}=-R^*_{P_K}L^*_{P_K}F_K.\ee
The unit element $E$ fulfills
\be P_E=e,\qquad F_E=0,\ee
where $e$ is the unit element of $G$.
\vskip1pc
\noindent {\bf Remarks}:

{\small \noindent 1) The projection on the base $P:T^*G\to G$
defined by $P(K)=P_K$ is
a morphism of groups according to (7.18).

\noindent 2) Upon the trivialization of the cotangent
 bundle $T^*G$ by right
(or left) translations, the group law (7.18) turns out to
correspond to
the semidirect product of $G$ and its coalgebra $\G^*$. The latter
is viewed as the Abelian group  underlying the
vector space $\G^*$ and $G$ acts on $\G^*$ by means of the
 coadjoint action. However,
in what follows we shall rather use the formula (7.18) because
 the trivialization
 breaks the natural left-right symmetry of the product.}

\vskip1pc

 We  shall denote the Lie algebra of the
group $T^*G$ as $\D$. Clearly, $\D$ can be written as a semidirect sum
of Lie algebras
\be \D=\G+\G^*,\ee
 where $\G$ acts on $\G^*$ in the coadjoint way and $\G^*$ is the
{\it commutative}
Lie subalgebra of $\D$. It turns out, that there exists a
non-degenerate
symmetric bilinear form on $\D$ defined by
\be (x+x^*,y+y^*)_\D=\la x^*,y\ra +\la y^*, x\ra,
\qquad x,y\in\G,\quad x^*,y^*\in\G^*.\ee
Here $\la .,.\ra$ denotes the canonical pairing between the
 Lie coalgebra $\G^*$
and  the Lie algebra $\G$.

It is the matter of a straightforward calculation to see that
the form $(.,.)_\D$
is moreover invariant and the subalgebras $\G$ and $\G^*$ are
both isotropic with respect
to this form. Said in formulas:

\be ([X,Y],Z)_\D+(Y,[X,Z])_\D=0, \qquad (\G,\G)_\D=(\G^*,\G^*)_\D=0,\ee
where $X,Y,Z\in\D$.

The cotangent bundle of any manifold $M$ is equipped with the
canonical symplectic
structure.
The corresponding symplectic $2$-form $\omega$ can be written as
\be \omega =d\theta,\ee
where $\theta$ is a $1$-form on $T^*M$ called the symplectic potential.
 It is
defined in a point $K=(P_K,F_K)\in T^*M$ as
\be \theta =P^*F_K.\ee
In words, $\theta$ is the pullback of the form $F_K$ living in
$P_K\in M$ by
the projection map $P:K\to P_K$.

  It is now convenient  to introduce a set of differential
operators (vector
fields) on $T^*G$. Define (as in \cite{ST})
\be \nl:C^\infty(T^*G)\to C^\infty(T^*G)\otimes \D^*;
\qquad \nr:C^\infty(T^*G)\to C^\infty(T^*G)\otimes \D^*\ee
as follows
\be \la \nl \phi,\al\ra (K)=\dto\phi(e^{s\al}K),\qquad
\la \nr \phi,\al\ra (K)=\dto\phi(K e^{s\al}).\ee
Here $\al\in\D$, $K\in T^*G$ and $\phi\in C^{\infty}(T^*G)$.

 Define also a linear  operator $\R:\D\to\D$ as follows
\be \R(x+x^*)=x-x^*,\qquad x\in\G,\quad x^*\in\G^*.\ee
Paranthetically, this operator is known as the  classical $R$-matrix and
the Lie algebra
$\D$ is the factorizable Baxter-Lie algebra in the sense of \cite{ST}.
 We have now
the following lemma.
\vskip1pc

\noindent {\bf Lemma 7.1} (Semenov-Tian-Shansky \cite{ST2}):
The Poisson bracket
corresponding to the symplectic form (7.24)
on $T^*G$ can be written as follows
\be \{\phi,\psi\}_{T^*G}=
\jp(\nl \phi,\R^*\nl \psi)_{\D^*}+\jp(\nr \phi,\R^*\nr \psi)_{\D^*}.\ee
Here $(.,.)_{\D^*}$ is  the  bilinear form on the dual of $\D$ induced
by the (nondegenerate) bilinear form $(.,.)_\D$ and $\R^*:\D^*\to\D^*$
is the map dual to $\R$. It might be illuminating to write
the bracket (7.29)
in some basis $T^i,t_i;~ i=1,\dots, dimG$ of $\D$ where $T^i$'s
form the basis
of $\G$ and $t_i$'s the corresponding dual basis of $\G^*$. We obtain
$$\{\phi,\psi\}_{T^*G}=
\jp\la \nl \phi,T^i\ra\la\nl \psi, t_i\ra -
\jp\la \nl \phi,t_i\ra\la\nl \psi, T^i\ra +$$
\be + \jp\la \nr \phi,T^i\ra\la\nr \psi, t_i\ra -
\jp\la \nr \phi,t_i\ra\la\nr \psi, T^i\ra,\ee
where the standard Einstein summation convention is used.
\vskip1pc
\noindent {\bf Remark}: {\small
It is important to note, that the canonical Poisson bracket on $T^*G$
can be written entirely in terms of the Lie group structure of $T^*G$.
 This way
of writing this bracket  stands at the basis of our Poisson-Lie
 generalization
of the standard WZW story.}

\vskip1pc
\noindent {\bf Proof}:
First we realize that the left (right) trivialization of the cotangent
 bundle
$T^*G$ gives a diffeomorphism between $T^*G$ and the direct product of
two manifolds $G$ and $\G^*$. In other words, there exist two global
 decompositions
 $T^*G=G\G^*=\G^*G$, where $\G^*$ is the fiber of the cotangent
bundle at the unit
element $e$ of the group $G$. Thus we may write for each $K\in T^*G$:
\be K=(g_L(K),0)(e,\beta_R(K))=(e,\beta_L(K))(g_R(K),0),\ee
where
\be g_L(K)=g_R(K)=P_K;\ee
\be \beta_L(K)=R_{P_K}^*F_K,\qquad \beta_R(K)=L_{P_K}^*F_K.\ee
Here $L^*_{P_K}$ is the pullback map of the differential forms on $G$
 with respect
to the left translation by the element $P_K$ and similarly $R^*_{P_K}$
 is the right
pullback.
Instead of  somewhat cumbersome expressions (7.31), we shall rather
write
\be K=g_L(K)\beta_R(K)=\beta_L(K) g_R(K).\ee

It is  then clear that the functions on $T^*G$ of the form $\Phi(P_K)$
and
$\Psi(\beta_L(K))$ generate the whole algebra of smooth functions on
 $T^*G$ hence
it is enough to compute the Poisson brackets between them. Even more
specially,
instead of arbitrary functions $\Psi$ on $\G^*$ it is sufficient to
 consider
linear ones, i.e. the functions $\la \beta_L,x\ra$ where $x\in\G$.

For the case of the group manifold there exists a
convenient expression for the symplectic potential $\theta$
in terms of  the invariant Maurer-Cartan forms.  Recall their
definitions:
In what follows, the expression $\lm_G$ ($\rho_G$) will denote
the left(right)invariant $\G$-valued Maurer-Cartan form on the
group $G$ defined by
\be \lm_G(X_g)=L_{g^{-1}*}X_g,\quad \rho_G(X_g)=
R_{g^{-1}*}X_g,\quad X_g\in T_gG.\ee
Note that the forms $\lm_G$ and $\rho_G$ are often
written also as
\be \lm_G=g^{-1}dg,\quad \rho_G=dgg^{-1}.\ee

The symplectic potential $\theta$ can be  then simply expressed
in the "coordinates"
$(\beta,g)$
as
\be \theta =\la \beta_L, dg g^{-1}\ra.\ee
Here we have abandonned the subscript $R$ on $g$, since anyway
 $g_R=g_L\equiv g$.
By using the formula
$$d(dgg^{-1})=dgg^{-1}\w dgg^{-1}$$ for the exterior derivative of
the Maurer-Cartan form
$\rho_{ G}$, we have
\be \om =\la d\beta_L\w dgg^{-1}\ra +\la \beta_L,
dgg^{-1}\w dgg^{-1}\ra.\ee
Now pick up a basis $T^i\in\G$ and its dual basis $t_i\in\G^*$.
It is also convenient to use a short-hand notation
$\la \beta_L,T^i\ra\equiv \beta^i$. Then the  form $\omega$ can
be written as
\be \omega=d\beta^i\w R^*_{g^{-1}}t_i+
\jp\beta^i f_i^{~mn}R^*_{g^{-1}}t_m\w R^*_{g^{-1}}t_n,\ee
 where $f_i^{~mn}$ are the structure constants of the
 Lie algebra $\G$.
This expression can be readily  inverted to give the
corresponding Poisson tensor $\Pi$:
\be \Pi=\jp\beta^mf_m^{~ij}{\d\over \d\beta^i}\w {\d\over \d\beta^j}-
{\d\over \d\beta^i}\w R_{g *}T^i.\ee
Since we have that $\la\nabla^L_G,x\ra=R_{g *}x$, for $x\in\G$,
 we obtain from $\Pi$
the following  Poisson brackets
\be  \{\Phi_1(g),\Phi_2(g)\}=0;\ee
\be \{\Phi(g),\la \beta_L,x\ra\}=\dto\Phi(e^{sx}g)\equiv
\la\nabla^L_G\Phi,x\ra;\ee
\be \{\la\beta_L,x\ra,\la \beta_L,y\ra\}=\la\beta_L,[x,y]\ra.\ee
We are going  to prove now that the same set of the Poisson bracket
 can be obtained directly
from the Semenov-Tian-Shansky formula (7.29).

For two functions $\Phi_{1,2}:G\to \br$ we calculate
\be \{\Phi_1(P_K),\Phi_2(P_K)\}_{T^*G}=0.\ee
This follows from the following fact
\be \la \nl\Phi_{j}(P_K),x^*\ra=\dto \Phi_{j}(P_{e^{sx^*}K})=
\dto\Phi_{j}(P_K)=0,\quad j=1,2\ee
and from its right analogue.
Here $x^*$ is an element of $\G^*$ viewed as the element of $\D$.
Typically,
$x^*$ is $t_i$ in (7.45).

The bracket of the type $\{\la\beta_L,x\ra,\la \beta_L,y\ra\}_{T^*G}$
for $x,y\in\G$
is more involved. In order to compute it, we need to prove
the following formulas:
\be\la \nl\la \beta_R(K),x\ra,y\ra=\la \nr\la \beta_L(K),x\ra,y\ra=0;\ee
\be \la \nl\la \beta_L(K),x\ra,y^*\ra=
\la \nr\la \beta_R(K),x\ra,y^*\ra=\la y^*,x\ra;\ee
\be \la \nl\la \beta_R(K),x\ra,y^*\ra= \la y^*,Ad_{P_K}x\ra;\ee
\be \la \nr\la \beta_L(K),x\ra,y^*\ra=\la y^*,Ad_{P_K^{-1}}x\ra;\ee
\be \la \nl\la \beta_L(K),x\ra,y\ra=\la \beta_L(K),[x,y]\ra;\ee
\be \la \nr\la \beta_R(K),x\ra,y\ra=-\la \beta_R(K),[x,y]\ra;\ee
where $x,y\in\G$  and $x^*,y^*\in\G^*$.
We  prove e.g. only the last formula,
 one can prove the others in full analogy. We have
$$ \la \nr\la \beta_R(K),x\ra,y\ra=\dto\la\beta_R(Ke^{sy}),x\ra=$$
$$\dto \la L^*_{P_{(Ke^{sy})}}F_{Ke^{sy}},x\ra=
\dto\la L^*_{e^{sy}}L^*_{P_K}R^*_{e^{-sy}}F_K,x\ra=$$
\be \dto \la \beta_R(K), L_{e^{sy}*}R_{e^{-sy}*}x\ra=
-\la \beta_R(K),[x,y]\ra.\ee
Now with the help of the formulas (7.46) - (7.51),
we calculate directly
\be \{\la\beta_L(K),x\ra,\la \beta_L(K),y\ra\}_{T^*G}=
\la\beta_L(K),[x,y]\ra.\ee
The remaining bracket between $\Phi:G\to\br$ and
$\la\beta_L,x\ra$ can be directly
evaluated again with the help of the formulas (7.46) - (7.51):
$$ \{\Phi(P_K),\la \beta_L(K),x\ra\}_{T^*G}=
\jp\la\nabla_G^L\phi,T^i\ra\la x,t_i\ra
+\jp \la \nabla_G^R\phi,T^i\ra\la t_i,Ad_{P_K^{-1}}x\ra=
$$
\be =\dto\Phi(e^{sx}P_K)
\equiv\la\nabla^L_G \Phi,x\ra.\ee
Here $\nabla^R_G$
is the map from $C^\infty(G)$ into $C^\infty(G)\otimes\G^*$
(cf. (7.26,7.27)). The reader has certainly noticed that
if the operators $\nl$ and $\nr$ appear without an index
 specifying the group it means
that they act on the double $D$. Otherwise we indicate as
above $\nabla^L_G$ or $\nabla^R_G$.

We recognize in the Poisson brackets (7.44),(7.53) and (7.54) the  brackets
(7.41),(7.43) and (7.42) obtained by inverting the symplectic form.

For completeness, we list the brackets involving the right "currents"
$\la\beta_R(K),x\ra$:
\be \{\la\beta_R(K),x\ra,\la \beta_R(K),y\ra\}_{T^*G}=
-\la\beta_R(K),[x,y]\ra;\ee
\be \{\la\beta_R(K),x\ra,\la \beta_L(K),y\ra\}_{T^*G}=0;\ee
\be\{\Phi(P_K),\la \beta_R(K),x\ra\}
=\la\nabla^R_G \Phi,x\ra.\ee
The lemma is proved.

\rightline{\#}
\vskip1pc
\section{The symplectic reduction in the dual language}
 The  function  $\la\hat\beta_R(\hat K),\tF\ra=
\la\hat\beta_L(\hat K),\tF\ra$ defined on $T^*\hat G$
is the moment map that generates the central
circle action on $T^*\hat G$. We can see it  from (7.30)
\be \{\phi,\la\hat\beta_R,\hat T^\infty\ra \}_{T^*\hat G}=
\la \nr\phi,\hat T^\infty\ra
\equiv\dto\phi(\hat Ke^{t\hat T^\infty})=\la \nl\phi,
\hat T^\infty\ra\ee
for every function on $T^*\hat G$.
The symplectic reduction   of Section 2.2.3 can be performed
also  as follows:
 first  fix a submanifold
$\mk$
of $T^*\hat G$  consisting of those
points where the moment map  $\la\hbr,\hat T^\infty\ra$ acquires
 the fixed constant value $\k$.
Note that the same constant $\k$ appears in the definition of
 the Hamiltonian $\hat H$ given by (2.41).
Now the central circle action preserves the submanifold $\mk$,
 hence  we can consider
 the space
of its orbits $\mk/U(1)$. In topologically good cases the latter
is a manifold and, in fact,
it is the reduced
phase space. We can easily calculate the reduced Poisson bracket
 of functions
$\eta$ and $\rho$ living on the reduced symplectic manifold
$\mk/U(1)$. First we take
their pullbacks $\Pi^*\eta$ and $\Pi^*\rho$ with respect to the map
 $\Pi:\mk\to \mk/U(1)$
that associates to a point in $\mk$ the corresponding central
circle orbit. Those
pullbacks are functions living on $\mk$. We extend them  on the
whole
unreduced manifold $T^*\hat G$ in such a way that they be invariant
with respect
to the central circle action on $T^*\hat G$. We compute the
unreduced Poisson bracket
of the extended functions and we restrict the result on $\mk$.
The function on $\mk$
thus obtained is clearly invariant with respect to the central
circle action (this follows
 from the Jacobi identity for the Poisson bracket  ); or,
in other
words, it is constant on the orbits. It is then a pullback of
 some function living
on $\mk/U(1)$. The latter is nothing but the reduced Poisson bracket
$\{\eta,\rho\}_{red}$.

The mechanism of the symplectic reduction is perhaps even more
transparent in the  dual
language liked by noncommutative geometers. The role of the
manifold $T^*\hat G$
is played by the algebra $\A$ of smooth functions on $\thg$.
The algebra of functions
on the reduced phase space $\mk/U(1)$ can be obtained in two steps.
First one considers
the subalgebra $Inv(\A)\subset\A$ consisting of functions in $\A$
whose unreduced
Poisson bracket with the moment map $\la\hbl,\hat T^\infty\ra$ vanishes.
There is a distinguished
ideal $I_\k(\A)$ in $Inv(\A)$ consisting of the functions of the form
$Inv(\A)(\la\hbl,\hat T^\infty\ra -\k)$.
Factorizing $Inv(\A)$ by its ideal $I_k(\A)$ gives
the algebra of functions on the reduced symplectic manifold $\mk/U(1)$.
 The reduced
Poisson bracket of $\eta$ and $\rho$ as above can be computed by
 choosing any
representatives of the classes $\eta$ and $\rho$ in $Inv(\A)$ and
by computing the
unreduced
Poisson bracket of those representatives. The last step consists
 in taking the class
of the result.

Of course, the symplectic reduction of some dynamical system is
a consistent procedure  if the Hamiltonian of the unreduced system
(Poisson) commutes with the moment map. In this case the Hamiltonian
is an element of $Inv(\A)$ and as such it gives rise to some function
on the reduced phase space. This function is called the Hamiltonian
of the reduced system. In our case we have to show that the Hamiltonian
$$\hat H(\hat K)=-{1\over 2\k}(\iota^*(\hat\beta_L),
\iota^*(\hat\beta_L))_{\G^*}
-{1\over 2\k}(\iota^*(\hat\beta_R),\iota^*(\hat\beta_R))_{\G^*}
\eqno(2.41)$$
is invariant function with respect to the central
 circle
action on $T^*\hat G$. It is easy to see this since we have from (7.18)
and (7.33)
$$ \hbr(e^{s\tF}\hat K)=L^*_{\hat P_{(e^{s\tF}\hat K)}}
\hat F_{(e^{s\tF}\hat K)}=
L^*_{\hat P_{\hat K}}L^*_{e^{s\tF}}\hat F_{(e^{s\tF}\hat K)}$$\be =
L^*_{\hat P_{\hat K}}L^*_{e^{s\tF}}L^*_{e^{-s\tF}}\hat F_{\hat K}=
L^*_{\hat P_{\hat K}}\hat F_{\hat K}=\hbr(\hat K).
\ee
In the same way we may check the invariance of $\hbl$ and hence of
the whole Hamiltonian
$H$.

We are going to show that the reduced symplectic manifold $\mk/U(1)$ is
  indeed
diffeomorphic to the cotangent bundle $T^*G$ of the non-extended
 group $G$.
On the other hand, the reduced symplectic structure  does {\it not}
 coincide with the canonical
symplectic structure on $T^*G$ (unless $\k=0$).

\subsection{The map between
$\mk/U(1)$  and $T^*G$.}

There
is a distinguished subgroup of $T^*\hat G$ that we shall
denote $M_0(\hat G)$.
It is formed by those  elements $\hat K$ of $T^*\hat G$ that satisfy
\be \hat F_{\hat K}(L_{\hat P_{\hat K}*}\hat T^\infty) =0\ee
or, equivalently,
\be \la \hat\beta_R(\hat K),\hat T^\infty\ra=0.\ee

Here the hats indicate that we are dealing with the extended group
 $\hat G$.
The map $\hat\beta_R:\hat G\to\hat\G^*$ is defined as in (7.33).
  $L_*$  is the  push-forward map
acting on the vector $\hat T^\infty$ which lives in the Lie algebra
$\hat\G$ of $\hat G$ viewed as
 the tangent space  $T_{\hat e}\hat G$ at the unit element $\hat e$ of
$\hat G$.  The reader may note that
the condition (7.60) is equivalent to
$\hat F_{\hat K}(R_{\hat P_{\hat K}*}\hat T^\infty) =0$.This follows
from the
fact that the  vector $L_{\hat P_{\hat K}*}\hat T^\infty$ corresponds
 to the
$right$ infinitesimal action of the  central circle
(injected in $\hat G$) at the point $\hat P_{\hat K}$ and from the fact
that the
extension is central hence the left action of the $U(1)$
coincides with the right action.

It is the  matter of a direct check that the elements $\hat K$
fulfilling (7.60), (7.61)
form a subgroup of $T^*\hat G$. We shall now show that $M_0 (\hat G)$
is naturally the central extension of the group $T^*G$ by the circle
group $U(1)$.
In order to write down
the corresponding  exact sequence of group homomorphisms
\be 1\to U(1)\to M_0(\hat  G)\stackrel{\Pi_0}{\to} T^*G\to 1,\ee
we have to specify the injection of U(1) into $M_0(\hat G)$ and the
homomorphism $\Pi:M_0(\hat G)\to T^*G$. The injection is clear since
$\hat G$ is  the subgroup of $M_0(\hat G)$ (it is formed by
the elements $\hat K$ of $T^*(\hat G)$ with $\hat F_{\hat K}=0$) hence
we inject $U(1)$ in  $\hat G$ as in (2.1) and this trivially induces the
injection in (7.62).

Now the homomorphism $\Pi_0$ is constructed as follows: one first notes
that
the kernel of the push-forward map $\pi_*$ at some
point $\hat P_{\hat K}\in \hat G$ is linearly generated precisely
 by the vector
$L_{\hat P_{\hat K}*}\hat T^\infty(=R_{\hat P_{\hat K}*}\hat T^\infty)$.
 In other words,
the condition (7.60) implies that there exists an unique 1-form $F_K$
living
in the point $\pi(\hat P_{\hat K})$ such that
\be \pi^*F_K=\hat F_{\hat K}.\ee
Rephrasing differently,  every  form $\hat F_{\hat K}$, living in
the point
$\hat P_{\hat K}$
and satisfying the condition (7.60), is a pull-back by $\pi$
of the  uniquely given form  $F_K$ in the point $\pi(\hat P_{\hat K})$.
But a form  in a point defines  an element of the group $T^*G$;
 we have denoted it  as $K$ in our context.
Now the map $\Pi_0$ is defined by associating to every $\hat K\in
M_0(\hat G)$ the corresponding element $K\in T^*G$. Another way of
writing this is as follows
\be P_{\Pi_0(\hat K)}=\pi(\hat P_{\hat K});\ee
\be \hat F_{\hat K}=\pi^*F_{\Pi_0(\hat K)}.\ee

\noindent {\bf Lemma 7.2:}
$\Pi_0$ is the group homomorphism.

\noindent {\bf Proof:}
One has to verify the validity of two relations:
\be P(\Pi_0(\hat Q\hat K))=P(\Pi_0(\hat Q))P(\Pi_0(\hat K));\ee
\be F_{\Pi_0(\hat Q\hat K)}=R^*_{P_{\Pi_0(\hat K)}^{-1}}F_{\Pi_0(\hat Q)}
 + L^*_{P_{\Pi_0(\hat Q)}^{-1}}F_{\Pi_0(\hat K)}.\ee
The first one (7.66) is simple, one has from (7.64)
\be P(\Pi_0(\hat Q\hat K))=\pi(\hat P(\hat Q\hat K))=
\pi(\hat P(\hat Q))
\pi(\hat P(\hat K))=P(\Pi_0(\hat Q))P(\Pi_0(\hat K)).\ee
To prove the second one, we have from (7.65) and from the
"hat"-version of (7.18)
\be \pi^* F_{\Pi_0(\hat Q\hat K)}=\hat F_{\hat Q\hat K}=
R^*_{\hat P_{\hat K}^{-1}}\pi^* F_{\Pi_0(\hat Q)} +
 L^*_{\hat P_{\hat Q}^{-1}}
\pi^*F_{\Pi_0(\hat K)}.\ee
From the homomorphism property of $\pi:\hat G\to G$,
 it follows for every $\hat k\in \hat G$:
\be \pi^* R^*_{\pi(\hat k)}=R^*_{\hat k}\pi^*\ee
and similarly for the left translation by $\hat k$.
Combining this fact together
with (7.64) and (7.65)
we arrive at
\be \pi^* F_{\Pi_0(\hat Q\hat K)}=\pi^*R^*_{P_{\Pi_0(\hat K)}^{-1}}
F_{\Pi_0(\hat Q)}
 + \pi^* L^*_{P_{\Pi_0(\hat Q)}^{-1}}F_{\Pi_0(\hat K)}.\ee
But this implies the relation (7.67) because the pull-back $\pi^*$
 of a non-zero form on $G$
would be a non-zero form on $\hat G$.

\rightline{\#}

\noindent We shall now again consider submanifold $M_\k(\hat G)$ of
the group
manifold $T^*\hat G$  formed by all points $\hat K\in T^*\hat G$
fulfilling
the condition
\be  \hat F_{\hat K}(L_{\hat P_{\hat K}*}\hat T^\infty) =
\la \hat\beta_R(\hat K),\tF\ra=\k.\ee
For $\k=0$  we recover from $\mk$ the manifold $M_0(\hat G)$
defined above,
hence our
notation is consistent. Note, however, that
if $\k\neq 0$, the manifold $\mk$ is not a subgroup of $T^*\hat G$.

We can construct a
 natural diffeomorphism relating $\mk$ and $\mo$. We proceed as
 follows: first consider a one-parameter subgroup of $T^*\hat G$
consisting of those points $\hat N(s)\in T^*\hat G$ that fulfil
\be \hat P_{\hat N(s)}=\hat e,\qquad  \hat F_{\hat N(s)}=
s\hat t_{\infty},\ee
where $s\in\br$ and $\tf$ is the $1$-form in $\hat e$ satisfying
\be \tf(\hat T^{\infty})=1,\qquad \tf(\iota(\G))=0.\ee
The conditions (7.74) determine $\tf$ unambiguously.

\vskip1pc
\noindent {\bf Remark:} It is easy to check that the group
$\hat N(s)$
normalizes the group $\mo$ (i.e. $\hat N\mo \hat N^{-1}\subset \mo$).
This means
in our context
that $T^*\hat G$ is naturally a semidirect product of $\hat N$ and $\mo$.
\vskip1pc
\noindent The diffeomorphism relating  $\mk$ and $\mo$ is now simply
given by
\be \hat K_0\to \hat N(\k)\hat K_0, \qquad \hat K_0\in\mo.\ee
It is evident that this diffeomorphism commutes with the central
 circle action hence
the spaces of the $U(1)$ orbits on $\mk$ and on $\mo$
are also diffeomorphic. Finally, from the exact sequence (7.62) it
 follows that the space of orbits of $\mo$ is nothing but $T^*G$.

\subsection{The reduced Poisson bracket on $\mk/U(1)$.}

Let us know compute the reduced Poisson bracket on $\mk/U(1)$.
 Since the latter is
diffeomorphic to $T^*G$, it is sufficient to determine the
Poisson brackets of
 a distinguished set of functions  on $T^*G$  of the form $\Phi(P_K)$
 and
$\la \beta_R(K),x\ra$.  Here $\Phi:G\to\br$, $K\in T^*G$ and
$x\in\G$. The functions of this special form  generate
 (via the  diffeomorphism above) the
whole algebra of functions on $\mk/U(1)$.
Recall that the $canonical$ Poisson brackets of those functions
are given by the equations  (7.41-43)). However, the reduced
Poisson bracket
of the same quantities are different as the following theorem says:
\vskip1pc
\noindent {\bf Theorem 7.3:} The reduced symplectic structure on
$\mk/U(1)\cong T^*G$
is fully determined by the following Poisson brackets:

\be \{\Phi_1(P_K),\Phi_2(P_K)\}_{red}=0;\ee
\be \{\Phi(P_K),\la \beta_L(K),x\ra\}_{red}=\dto\Phi(e^{sx}P_K)\equiv
\la\nabla^L_G\Phi,x\ra;\ee
\be \{\la\beta_L(K),x\ra,\la \beta_L(K),y\ra\}_{red}=\la\beta_L(K),
[x,y]\ra+\k\rho(x,y).\ee
Recall that $\rho(x,y)$ is the cocycle characterizing the central
 extension (2.2).
\vskip1pc
\noindent {\bf Remarks}: 1) the reduced brackets (7.76) and (7.77) are
the same
as the canonical brackets (7.41) and (7.42), respectively. However,
the reduced bracket
(7.78) differs from the corresponding canonical bracket by the
cocycle term.
We thus see that (unless $\k=0$)
the reduced symplectic structure does  {\it not} coincides with
the canonical one
 on  $T^*G$.

~~~~~2) It may seem that the adding of the cocycle in (7.78)
does not
represent a "big" change of the Poisson bracket. However,
the reader may check as
an exercise  that
the reduced bracket of the type $\{\la\beta_R,x\ra,\la
\beta_R,y\ra\}_{red}$ is
already much more complicated than its canonical counterpart (7.55)
(We do not list this
calculation here since we shall not need it in this paper).
The reason why the
bracket of the left currents $\beta_L$ is simpler than that
of the right currents $\beta_R$ is due
to the left-right asymmetry in the map relating $\mk/U(1)$
and $T^*G$. The choice of this  map is not
canonical; we have chosen it in this way in order to make
contact with
the standard description of the WZW symplectic structure in \cite{Mad}.
If we change this map (it is also possible to make a left-right
 symmetric  choice)
the reduced Poisson brackets of the left and right currents
$\beta_{L,R}$ change but
the theory does not change. Indeed,  $\beta_{L,R}$ would then
correspond  to {\it different}
functions
on $\mk/U(1)$ if the diffeomorphism has changed. We stress that
the natural dynamical
variables of the problem are $\hat\beta_{L,R}$ and they get
expressed differently
in terms of the observables on $T^*G$ under the change of the
diffeomorphism.

~~~~3) It does not follow from anywhere that the level
$\k$ must be an integer. This constraint
will appear at the quantum level. It can be simply understood
intuitively, since
$k$ is the "momentum" conjugated to the $angle$ variable
 parametrizing the central circle.

In order to prove the theorem, we shall need the following
lemma
\vskip1pc
\noindent {\bf Lemma 7.4:} The pullback $\Pi^*\la\beta_L,x\ra$
of the function $\la\beta_L,x\ra$
on $T^*G$ via the map  $\Pi:\mk\to\mk/U(1)$ is given by the
following formula
\be (\Pi^*\la\beta_L,x\ra) (\hat K)=\la\hbl(\hat K),\iota(x)\ra,
\qquad \hat K\in\mk,\quad x\in\G.\ee
\vskip1pc
\noindent {\bf Remark}: The map $\Pi:\mk\to\mk/U(1)$ has been
 defined at the beginning
of the section 7.3; the statement of the lemma makes sense
since we have established
in section 7.3.1 that $\mk/U(1)$ and $T^*G$ are diffeomorphic.
\vskip1pc
\noindent {\bf Proof}: First we have to   prove  two important
relations:
\be \la\beta_L(\Pi_0(\hat K)),x\ra= \la\hat\beta_L(\hat K),
\iota(x)\ra,\quad \hat K\in M_0(\hat G)\ee
and
\be \la\hat\beta_L(\hat N(s)\hat K),\iota(x)\ra=\la\hat\beta_L(\hat K),
\iota(x)\ra,\qquad s\in\br,
\quad \hat K\in T^*\hat G,\quad x\in\G . \ee
Indeed, the first relation  is implied by (7.33),(2.3),(7.64),(7.70)
 and (7.65)
 as follows
$$\la\beta_L(\Pi_0(\hat K)),x\ra =
\la R^*_{P_{\Pi_0(\hat K)}}F_{\Pi_0(\hat K)},\pi_* (\iota(x))\ra
=\la F_{\Pi_0(\hat K)}, R_{\pi(\hat P_{\hat K})*}\pi_* (\iota(x))\ra=$$
$$=\la F_{\Pi_0(\hat K)}, \pi_*R_{\hat P_{\hat K}*} \iota(x)\ra=
\la \pi^*F_{\Pi_0(\hat K)}, R_{\hat P_{\hat K}*} \iota(x)\ra=$$
\be =\la \hat F_{\hat K},
R_{\hat P_{\hat K}*} \iota(x)\ra=\la R^*_{\hat P_{\hat K}}
\hat F_{\hat K},\iota(x)\ra=
 \la\hat\beta_L(\hat K),\iota(x)\ra.\ee
The second relation in turn follows from (7.18), (7.33) and (7.74):
$$ \la\hbl(\hat N(s)\hat K),\iota(x)\ra
=\la R^*_{\hat P_{\hat N(s)\hat K}}\hat F_{\hat N(s)\hat K},
\iota(x)\ra=$$
\be =\la R^*_{\hat P_{\hat K}}(sR^*_{\hat P_{\hat K}^{-1}}\tf +
\hat F_{\hat K}),\iota(x)\ra=
s\la \tf,\iota(x)\ra +\la \hbl(\hat K),\iota(x)\ra=\la \hbl(\hat K),
\iota(x)\ra.
\ee
Now we have by definition
\be (\Pi^*\la\beta_L,x\ra)(\hat K)=
\la\beta_L(\Pi_0(\hat N(-\k)\hat K))\ra.\ee
Combining (7.80) and (7.81), it follows
\be (\Pi^*\la\beta_L,x\ra) (\hat K)=\la\hbl(\hat K),\iota(x)\ra.\ee
The lemma is proved.

\rightline{\#}

\vskip1pc
\noindent {\bf Proof of the theorem 7.3:}
Consider now an arbitrary pair of functions $\phi,\psi$ on $T^*G$.
 In order to calculate
their reduced Poisson brackets, we first have to pull back them
 to functions on $\mk$ via
our map $\Pi:\mk\to\mk/U(1)$. We have, for instance,
\be (\Pi^*\phi)(\hat K)=\phi(\Pi_0(\hat N(-\k)\hat K)),\qquad
\hat K\in\mk.\ee
Now we have to extend the functions $\Pi^*\phi$ and $\Pi^*\psi$
 to the whole manifold $T^*\hat G$
in such a way that they be invariant with respect to the central
 circle action. The resulting
functions can be referred to as $\phi_{ext}$, $\psi_{ext}$ and
 can be conveniently chosen
as
\be \phi_{ext}(\hat K)=\phi(\Pi_0(\hat N(-\la\beta_R(\hat K),
\tF\ra)\hat K)),
\qquad \hat K\in T^*\hat G\ee
and in the same way for $\psi_{ext}$. Now we should compute
the unreduced canonical
bracket of $\phi_{ext}$ and $\psi_{ext}$.

First we calculate the extensions of two particular functions on
 $T^*G$. First one is
$(\Phi\circ P)(K)\equiv\Phi(P_K)$ where $\Phi:G\to\br$ and the
 second is $\la\beta_L(K),x\ra$
where $x\in\G$.
We obtain from (7.64) and (7.87)
$$ (\Phi\circ P)_{ext}(\hat K)=
\Phi(P_{\Pi_0(\hat N(-\la\beta_R(\hat K),\tF\ra)\hat K)})=$$
\be =\Phi(\pi(\hat P_{\hat N(-\la\beta_R(\hat K),\tF\ra)\hat K}))=
\Phi(\pi(\hat P_{\hat K}))\ee
and from (7.80),(7.81)  and (7.87)
$$ \la\beta_L,x\ra_{ext}(\hat K) =\la \beta_L(\Pi_0(\hat
 N(-\la\beta_R(\hat K),\tF\ra)\hat K),x\ra=$$
\be  \la \hat\beta_L(\hat N(-\la\beta_R(\hat K),\tF\ra)\hat K),
\iota(x)\ra=
\la \hat\beta_L(\hat K),\iota(x)\ra.\ee
Now it is  easy to calculate the unreduced  bracket
$\{\la\beta_L,x\ra_{ext},\la\beta_L,y\ra_{ext}\}_{T^*\hat G}$.
 We have from (7.89), (7.43) and (2.7)
$$ \{\la\beta_L,x\ra_{ext},\la\beta_L,y\ra_{ext}\}_{T^*\hat G}(\hat K)=
\{\la \hat\beta_L(\hat K),\iota(x)\ra,\la \hat\beta_L(\hat K),\iota(y)
\ra\}_{T^*\hat G}=$$
$$ =\la \hbl(\hat K),[\iota(x),\iota(y)]\ra=
\la \hbl(\hat K),\iota([x,y])+\rho(x,y)\tF\ra=$$
\be =\la \hbl(\hat K),\iota([x,y])\ra +\rho(x,y)\la \hbl(\hat K),
\tF\ra.\ee
The resulting function is clearly invariant with respect
to the central circle action.
From Lemma 7.4 and Eq.(7.72) it follows that its restriction to $\mk$ is
  given by the following expression
\be \la \hbl(\hat K),\iota([x,y]\ra +\k\rho(x,y)=
(\Pi^*\la \beta_l,[x,y]\ra)(\hat K)+\k\rho(x,y), \quad\hat K\in\mk.\ee
Hence we obtain for the reduced Poisson bracket
\be \{\la \beta_L(K),x\ra,\la \beta_L(K),y\ra\}_{red}=
\la\beta_L(K),[x,y]\ra +\k\rho(x,y).\ee
This is precisely the relation (7.78).

To prove (7.76) is very easy because from (7.88) and (7.44) we have
immediately
\be \{(\Phi_1\circ P)_{ext}(\hat K),(\Phi_2\circ P)_{ext}
(\hat K)\}_{T^*\hat G}=
\{\Phi_1(\pi(\hat P_{\hat K})),\Phi_2(\pi(\hat
P_{\hat K}))\}_{T^*\hat G}=0.\ee

It remains to verify the relation (7.77). It follows directly from
the following
computation (cf. (7.42))
$$ \{(\Phi\circ P)_{ext}(\hat K),\la\beta_L,x\ra_{ext}(\hat
K)\}_{T^*\hat G}=
\{\Phi(\pi(\hat P_{\hat K})),\la \hat\beta_L(\hat K),\iota(x)\ra
\}_{T^*\hat G}=$$
\be = \dto \Phi(\pi(e^{s\iota(x)}\hat P_{\hat K}))=\dto
 \Phi(e^{sx}\pi(\hat P_{\hat K})).\ee
The theorem is proved.

\rightline{\#}
\section{Proof of the Lemma 5.8}
With the notation of the section 5.2.2, the proposition to be
 proved reads
\vskip1pc
\lem : It  holds
\be \hat m_L(\wti{Ad}_{\hat k}\hat a)=\hat b_L(\wh{Ad}_{\hat k}\hat a),
\quad
\ti g_R(\wti{Ad}_{\hat k}\hat a)=\hat g_R(\wh{Ad}_{\hat k}\hat a).\ee
\vskip1pc
\pro:  Consider any element $(\bar{\bl},\lm)\in\hdgc$.
The elements of $\dgc$ can be viewed as smooth maps from an
 interval $[0,1]$ parametrized
by $r$ (the radius of the disc) into the loop group $\lgc$.
 Since the loop group $\lgc$
is diffeomorphic  to the direct product of the  manifolds
$LG_0$ and $\l+$, it follows
that $ \bar{\bl}$ can be uniquely decomposed as
\be \bar{\bl}=XA,\quad A\in DG_0,\quad X\in\+.\ee
Here $\+$ is the subgroup  $DG_0^\bc$ consisting of the elements
of the form $X(\si,r)$,
where for every fixed  $r=r_0$, $X(\si,r_0)\in\l+$.

 Consider a $\Th^\bc$-representative
 $\bbK\in \br^2\times_{S,Q}\hdgc$ of
an element $\ttK\in\ttD$;
\be \bbK=(\bar{\bl},e^{ip+P},w+is),\ee
where $(\bar{\bl},e^{i\lm+L})\in\hdgc$, $\bar{\bl}\in\dgc$
and $p,P,w,s\in\br$.
  As the consequence of the decomposition (7.96),
also  $\bbK$ can be  decomposed uniquely  as
\be \bbK=(\bar{\bar b},e^t,w) \tilde *(\bar {\bar g},e^{i\phi},is),\ee
where  $\bar{\bar g}\in DG_0$,  $\bar{\bar b}\in\+$,
$\phi,t\in\br$ and the product $\ti *$ is that of $\br^2\times_{S,Q}\hdgc$.

By applying the homomorphism $\hat\wp_\bc$ (cf. Section 4.4.3) on the both
 sides of (7.98),
 we obtain
\be \ttK=(\hat\wp_\bc(\bar{\bar b},e^t),w)(\hat\wp_\bc(\bar
{\bar g},e^{i\phi}),is).\ee
Now we prove that
\be    (\hat\wp_\bc(\bar{\bar b},e^t),w)=\ti b_L(\ttK),  \quad
(\hat\wp_\bc(\bar {\bar g},e^{i\phi}),is)=\ti g_R(\ttK).\ee
This will follow immediately from the uniqueness of the
decomposition $\ttD=\ti B\ti G$,
if we demonstrate that
\be (i)\quad  (\hat\wp_\bc(\bar{\bar b},e^t),w)\in\ti B,\quad (ii)
 \quad
(\hat\wp_\bc(\bar {\bar g},e^{i\phi}),is) \in\ti G.\ee
The statement (ii)  is  the direct consequence of the way how
$\ti G$ is
embedded in $\ttD$ (Lemma 4.18 and its proof); the statement (i)
 requires
a bit more of work, however. Actually we must
 to show that it exists $b\in\l+$   such  that
\be  \hat\wp_\bc({\bar b},e^{t})=
\quad  \hat\wp_\bc(\bar{\bar b},e^t).\ee
Recall that $\bar b$ was defined in Lemma 4.19, as the
map from the $Disc$ into $G_0^\bc$
whose boundary is $b\in\l+$.
Let us show that $b$  given by
\be b=(\Pi_0^\bc\circ\hat\wp_\bc)((\bar{\bar b},e^t))\ee
solves (7.102). This amounts to show that
\be C^\bc(\bar b\bar{\bar b}^{-1})=0.\ee
The last equation follows from the general fact that  $C^\bc$
vanishes on $\+\cap\d G^\bc$.
Indeed, recall that if $\bl\in\delta G^\bc
\cap\+$, then $\bl$ can be interpreted as a map from the
$D$-Riemann sphere
$S^2$ into $G_0^\bc$ (cf. the discussion
of the WZW term in Section 4.4 and in Appendix 7.1) and
extended to a map $\bl_{ext}$
from the unit $Ball$ wrapped by $S^2$
into $G_0^\bc$. Now the map $\bl_{ext}$ can also be
interpreted as a map from a certain
{\it  half-disc}
into $LG_0^\bc$. This half-disc is diffeomorphic to the space
 of  orbits of the azimutal rotation around the axis
connecting the northern with the southern pole of the $Ball$ and it
 can be clearly identified
with the half-disc bounded by the  north-south axis of rotation
and by the half of the
Greenwich Meridian. Points of each such  orbit
are parametrized by the loop parameter $\si$ and the map
$\bl_{ext}$ restricted to any such orbit
is naturally the element of $\lgc$.
It is crucial to remark, that the loop group $\lgc$ can be
diffeomorphically  decomposed
as a product of $LG_0$ and $\l+$ hence 	also the group of
smooth maps from the half-disc
(with apropriate boundary conditions) into $\lgc$ can be
decomposed accordingly. If we
decompose the map $\bl_{ext}$ in this way, we observe that
the $\l+$ part of this decomposition is
 still
the map that extends the original element  $\bl\in\d G^\bc
\cap\+$ into the $Ball$. We refer to  such an extension as
to $\bl^+_{ext}$ and we recall
that it can be understood as the map from the half-disc into $\l+$.
Thus we can calculate the $C^\bc(\bl)$ by using the extension
$\bl^+_{ext}$ in the defining
formula (4.132).
 The result
is that the integral over the loop variable $\si$ can be
converted on the equatorial contour
integral
on the $\si$-Riemann sphere, where the integrated function
in the
$z$-variable is everywhere holomorphic
on the southern hemisphere.   The contour can be therefore
shrunk
to the point without encountering any singularity,
hence $C^\bc(\bl)=0$.

We recapitulate, what we have learned so far: having
an element $\ttK\in\ttD$, we can
find $\ti b_L(\ttK)$ and $\ti g_R (\ttK)$ by picking up any
$\Th^\bc$-representative  $\bbK\in \br^2\times_{S,Q}\hdgc$,
 decomposing $\bbK$
as in (7.98) and evaluating
$$ \ti b_L(\ttK)=   (\hat\wp_\bc(\bar{\bar b},e^t),w),  \quad
\ti g_R(\ttK) =(\hat\wp_\bc(\bar {\bar g},e^{i\phi}),is).$$
 Similarly, we can prove also that
\be \hat b_L(\hhK)=   (\wp_\bc(\bar{\bar b},1),w),  \quad
\hat g_R(\hhK) =(\wp_\bc(\bar {\bar g},e^{i\phi}),0),\ee
Here  $\bar K=({\bl},e^{ip},w)$ is any $\Th^\bc_\br$-representative
of $\hhK\in\hhD$ with a
 decomposition
\be \bar K=(\bar{\bar b},1,w)\hat *(\bar {\bar g},e^{i\phi},0).\ee
Of course,
  $({\bl},e^{ip})\in\rdgc$, $\bar{\bl}\in\dgc$, $\bar{\bar g}\in DG_0$,
  $\bar{\bar b}\in\+$,
$p,w,\phi\in\br$ and the product $\hat *$ is taken in $\br\times_Q\rdgc$.

Now we need the formula for $\hat m_L\equiv \hat m\circ\ti b_L$.
 Clearly,
\be \hat  m_L(\ttK)=((\Pi_0^\bc\circ\hat\wp_\bc)(\bar{\bar b},e^t),w).\ee
In particular, it is immediate to realize that $\hat  m_L(\ttK)$ does
 not depend
on $t$.

We have shown so far, how to calculate the maps $\hat  m_L $,$\ti b_L$,
 $\hat b_L$,
$\ti g_R$ and $\hat g_R$ by using  the "Iwasawa" decomposition (7.96)  of
the $\Th^\bc$-representatives
of $\ttK$ (or $\Th^\bc_\br$-representatives of $\hhK$).
We can now view  the elements $\hat k\in\hat G$,$\hat a\in\hat B$ as
 elements of
$\hhD$ but also as elements of $\ttD$. Their representatives can be
 chosen as
\be\bar{\bar k}=(\bar k,e^{i\psi},0),\quad \bar{\bar a}=
(\bar a,1,a^\infty).\ee
Now it is crucial to notice , that $\bar{\bar k},\bar{\bar a}$
make sense as the
elements
of $\br^2\times_{S,Q}\hdgc$ and also as the elements of
$\br\times_Q\rdgc$. If we want
to calculate the quantities $\ti b_L(\wti{Ad}_{\hat k}\hat a)$
and
$\hat b_L(\wh{Ad}_{\hat k}\hat a)$, we should understand
 $\bar{\bar k},\bar{\bar a}$ respectively in the former and the
latter sense.

Consider now any n elements $\bbK_j$, $j=1,\dots n$ of
$\br^2\times_{S,Q}\hdgc$ of the form
\be \bbK_j=(\bar{\bl_j},e^{ip_j},w_j).\ee
As we already  know, we can view $\bbK_j$ of this form also as
the elements of
$\br\times_Q\rdgc$. Now we denote by $\ti *$ the product in
$\br^2\times_{S,Q}\hdgc$
and by $\hat *$ the one in $\br\times_Q\rdgc$. Using   (4.108)
 and (4.129), we arrive immediately at
\be \bbK_1\hat *\bbK_2\hat *\dots \hat *\bbK_n=(1,e^{t(\bbK_1,\bbK_2,
\dots \bbK_n)},0)
\ti *(\bbK_1\ti *\bbK_2\ti *\dots\ti *\bbK_n), \ee
where l.h.s is viewed as the element of $\br^2\times_{S,Q}\hdgc$ and
 $t(.,\dots,.)$
is some real function whose explicit form is not needed for the proof
of the lemma. From the
equation (7.110) for $\bbK_1=\bar{\bar k}$, $\bbK_2=\bar {\bar a}$ and
$\bbK_3={\bar{\bar k}}^{-1}$,
we conclude that
\be \wh{Ad}_{\bar{\bar k}}{\bar{\bar a}}=(1,e^{t(\bar {\bar k},
\bar{\bar a},{\bar{\bar k}}^{-1})},0)
\ti *(\wti {Ad}_{\bar{\bar k}}{\bar{\bar a}} ).\ee
Now we have the "Iwasawa" decomposition (7.106)
\be \wh{Ad}_{\bar{\bar k}}{\bar{\bar a}} =(\bar{\bar b},1,w)
\hat *(\bar {\bar g},e^{i\phi},0),\ee
for some $\bar{\bar b}\in\+$, $\bar {\bar g}\in DG_0$,
 $\phi,w\in\br$.
From (7.110) and (7.111), we obtain
\be \wti {Ad}_{\bar{\bar k}}{\bar{\bar a}}=(\bar{\bar b},
e^{t'(\bar{\bar k},\bar{\bar a})},w)
\ti *(\bar {\bar g},e^{i\phi},0), \ee
where $t'(\bar{\bar k},\bar{\bar a})$ is some real function
 whose form is irrelevant for us.
Now from (7.100), (7.105), (7.107), (7.111), (7.112)  and (7.113), we
 conclude immediately
$$\hat m_L(\wti{Ad}_{\hat k}\hat a)=
\hat b_L(\wh{Ad}_{\hat k}\hat a),\quad
\ti g_R(\wti{Ad}_{\hat k}\hat a)=\hat g_R(\wh{Ad}_{\hat k}\hat a).$$
The lemma is proved.

\rightline{\#}


\begin{thebibliography}{19}
\bi{AKM} {A. Alekseev, Y. Kosmann-Schwarzbach and E. Meinrenken,
{\it Quasi-Poisson manifolds}, math.DG/0006168}
\bibitem{AM} {A. Alekseev and A. Malkin, {\it Commun. Math. Phys.}
 {\bf 162} (1994) 147}
\bi{AS} {A. Alekseev and S. Shatashvili, {\it Commun. Math. Phys.}
{\bf 128} (1990) 197}
\bi{AT}{A. Alekseev and  I. Todorov,{\it Nucl.Phys.} {\bf B421}
(1994) 413}
\bi{Ba}{O. Babelon, {\it Phys. Lett} {\bf B215} (1988) 523}
\bi{Mad}{J. Balog, L. Feher  and  L. Palla, {\it On the chiral
 WZNW phase space,
exchange r-matrices and Poisson-Lie groupoids}, hep-th/9912173}
\bi{B} {D. Bernard, {\it Nucl. Phys.} {\bf B303} (1988) 77}
\bi{BT}{M. Blau and G. Thompson, {\it Lectures on 2d gauge
 theories:
Topological aspects
and path integral techniques}, in {\it Proceedings of the 1993
Trieste Summer School
on High Energy Physics and Cosmology} (eds. E. Gava et al.),
World Scientific, Singapore (1994)
175, hep-th/9310144}
\bi{BT2} { M. Blau and  G. Thompson,
   { Commun. Math. Phys.}{\bf 171} (1995) 639}
\bibitem{Bl}{B. Blok, {\it Phys.Lett.} {\bf B233} (1989) 359}
\bi{Byts} {A. G. Bytsko, {\it    Tensor operators in R-matrix approach},
 q-alg/9512030}
 \bi{BS} { A. G. Bytsko and  V. Schomerus, {\it  Commun. Math. Phys.}
{\bf 191} (1998) 87}
 \bi{CL} {L. Caneschi and M. Lysiansky, {\it Nucl. Phys.} {\bf B505}
 (1997) 701}
\bibitem{CGO}{M. Chu, P. Goddard, I. Halliday, D. Olive and A.
 Schwimmer, {\it Phys. Lett.}
{\bf B266} (1991) 71}
\bibitem{CG}{M. Chu and  P. Goddard, {\it Nucl. Phys.} {\bf B445}
 (1995) 145}
\bi{Co} {A. Connes, {\it Noncommutative geometry}, London,
Academic Press, 1994}
\bi{QWZW} {L. Dabrowski, T. Krajewski and  G.  Landi,
 {\it Int. J. Mod. Phys.}
{\bf  B14} (2000) 2367}
\bi{EV} {P. Etingof and A. Varchenko, {\it
Geometry and classification of solutions of the Classical
 Dynamical Yang-Baxter Equation},
        q-alg/9703040 }
\bi{Fad}{L.D. Faddeev, {\it Commun. Math. Phys.} {\bf 132}
(1990) 131}

\bi{FG}{F. Falceto and K. Gaw\c edzki, {\it J.Geom.Phys.}
 {\bf 11} (1993) 251}

\bi{F} {G. Felder, {\it Conformal field theory and integrable
 systems  associated
to elliptic curves}, in {\it Proceedings of the International
Congress
of Mathematicians}, Zurich, 1994}
\bi{FW} {G. Felder and C. Wieczerkowski, {\it  Commun. Math. Phys.}
 {\bf 176} (1996) 133}
\bi{FFS} {J. Figueroa - O' Farrill and S. Stanciu, {\it Phys. Lett.}
{\bf  B327}  (1994) 40-46}
\bi{FHT} {P. Furlan, L. Hadjiivanov and  I.Todorov,
      {\it  Nucl.Phys.} {\bf B474} (1996) 497}
\bi{FR1} {E. Frenkel and N. Reshetikhin, {\it Quantum affine
algebras and deformations of the Virasoro and $W$-algebras},
q-alg/9505025}
\bi{FR2} {I. Frenkel and N. Reshetikhin, {\it Commun. Math. Phys.}
{\bf 146} (1992) 1}
\bi{Fro} {C. Fronsdal, {\it Exact deformations of quantum groups;
applications to the affine case}, q-alg/9602034}
\bi{GawP} {K. Gaw\c edzki, {\it Lectures on CFT},  preprint
 IHES-P-97-2 (1997)}
\bi{Gaw}{K. Gaw\c edzki, {\it Conformal field theory: a case study},
 hep-th/9904145}
\bi{Ga0}{K. Gaw\c edzki, {\it Commun. Math. Phys.} {\bf 139} (1991) 201}
\bi{GKO} {P. Goddard, A. Kent and  D. Olive, {\it
    Commun. Math. Phys.} {\bf 103} (1986) 105}
\bi{JKOS} {M. Jimbo, H. Konno, S. Odake and J. Shiraishi,
{\it Quasi-Hopf twistors for elliptic quantum groups}, q-alg/9712029}
\bi{MR}{A. Medina and P. Revoy, {\it Ann. Scient. ENS} {\bf 18}
(1985) 553, in French}
\bi{Mick}{J. Mickelsson, {\it Current algebras and groups}, New York,
Plenum Press, 1989}
\bibitem{KS1}{C. Klim\v c\'\i k and P. \v Severa, {\it Phys. Lett.}
{\bf B351}
 (1995) 455; C. Klim\v c\'\i k, {\it Nucl. Phys. (Proc. Suppl.)} {\bf
B46} (1996) 116; P. \v Severa,
{\it Minim\'alne plochy a dualita}, Diploma thesis, 1995, in Slovak}
\bi{KS8} {C. Klim\v c\'\i k and  P. \v Severa, {\it T-duality and
the moment map},
preprint IHES-P-96-70,  hep-th/9610198,
In Carg\`ese 1996, {\it  Quantum fields and quantum space time}, 323-329}
\bi{KZ} {V. G. Knizhnik and A. B. Zamolodchikov, {\it Nucl.Phys.}
{\bf B247} (1984) 83}
\bi{LW}{J.-H. Lu and A. Weinstein, {\it J. Diff. Geom.} {\bf 31}
 (1990) 510}
\bi{LS} { S. Lukyanov and  S. Shatashvili,
      {\it  Phys. Le} {\bf   B298} (1993) 111}
\bibitem{PW}{A.Polyakov and P.B.Wiegmann, {\it Phys. Lett.}{\bf B311}
 (1983) 549}
 \bi{PS} {A. Pressley and G. Segal, {\it Loop groups}, Oxford,
 Clarendon Press, 1986}
\bi{RST} {N. Yu. Reshetikhin and M. A. Semenov-Tian-Shansky,
 {\it
Lett. Math. Phys. } {\bf 19} (1990) 133}
\bi{ST}{M. Semenov-Tian-Shansky, {\it Theor. Math. Physics}
{\bf 93} (1992) 302}
\bi{ST2} {M. Semenov-Tian-Shansky, {\it Publ.RIMS} {\bf 21},
Kyoto Univ. (1985) 1237}
\bi{S} {Ya.S. Soibelman, {\it Algebra Analiz} {\bf 2}(1990) 190}
\bi{WHW} {E. Whittaker and G. Watson, {\it A course of modern analysis},
Cambridge, Cambridge University Press,  1969, page 489}
\bibitem{Wi} {E. Witten, {\it Commun. Math. Phys.} {\bf 92} (1984) 455}
\bi{Xu}{P. Xu, {\it Quantum grupoids}, math.QA/9905192}
\bibitem{Zel}{D. Zhelobenko and A. Stern, {\it Representations of
 Lie groups},
Moscow, Nauka, 1983, in Russian, page 117}
\end{thebibliography}
\end{document}